\documentclass[11pt]{article}

\usepackage{geometry}
\usepackage[dvipsnames]{xcolor}
\usepackage{amsfonts}
\usepackage{amsmath}
\usepackage{amssymb}
\usepackage{bbold}
\usepackage{BOONDOX-cal}
\usepackage{longtable}
\usepackage{lscape}
\usepackage{graphicx}
\usepackage{multirow}
\usepackage{makecell}
\usepackage{wrapfig}
\usepackage[normalem]{ulem}

\usepackage{caption}
\usepackage{subcaption}

\usepackage[colorlinks=true,pdfstartview=FitV, linkcolor=blue, citecolor=blue,urlcolor=blue]{hyperref} 
\usepackage[sort&compress,numbers,colon,merge]{natbib}

\newlength{\bibitemsep}
\newlength{\bibparskip}
\let\oldthebibliography\thebibliography
\renewcommand\thebibliography[1]{%
  \oldthebibliography{#1}%
  \setlength{\parskip}{\bibitemsep}%
  \setlength{\itemsep}{\bibparskip}%
}

\allowdisplaybreaks
\newcommand{\braket}[1]{\left\langle#1\right\rangle}

\begin{document}
\pagenumbering{Alph}

\begin{titlepage}
\vspace*{-1cm}
\phantom{hep-ph/***}
\flushright
\hfil{IFIC/22-19}
\hfil{FTUV-22-0620.7524}
\hfil{CPPC-2022-07}

\vskip 1.5cm
\begin{center}
\mathversion{bold}
{\LARGE\bf
Flavour anomalies meet flavour symmetry
}\\[3mm]
\mathversion{normal}
\vskip .3cm
\end{center}
\vskip 0.5  cm
\begin{center}
{\large Innes Bigaran}$^{1}$,
{\large Tobias Felkl}$^{2}$,
{\large Claudia Hagedorn}$^{3,4}$,
{\large Michael A.~Schmidt}$^{2}$
\\
\vskip .7cm
{\footnotesize
$^{1}$ ARC Centre of Excellence for Dark Matter Particle Physics, School of Physics, The University of Melbourne, Victoria 3010, Australia\\[0.3cm]
$^{2}$ Sydney Consortium for Particle Physics and Cosmology, School of Physics, The University of New South Wales, Sydney, NSW 2052, Australia\\[0.3cm]
$^{3}$ Instituto de F\'isica Corpuscular, Universidad de Valencia and CSIC,
Edificio Institutos Investigaci\'on, Catedr\'atico Jos\'e Beltr\'an 2, 46980 Paterna, Spain\\[0.3cm]
$^{4}$ Istituto Nazionale di Fisica Nucleare, Sezione di Padova, Via F. Marzolo 8, 35131 Padua, Italy\\[0.3cm]

\vskip .5cm
\begin{minipage}[l]{.9\textwidth}
\begin{center}
\textit{E-mail:}
\tt{ibigaran@fnal.gov}, \tt{t.felkl@unsw.edu.au}, \tt{claudia.hagedorn@ific.uv.es}, \tt{m.schmidt@unsw.edu.au}
\end{center}
\end{minipage}
}
\end{center}
\vskip 1cm
\begin{abstract}
We construct an extension of the Standard Model with a scalar leptoquark $\phi\sim (3,1,-\tfrac13)$ and the discrete flavour symmetry $G_f=D_{17}\times Z_{17}$
to explain anomalies observed in charged-current semi-leptonic $B$ meson decays and in the muon anomalous magnetic moment, together with the charged fermion masses and quark mixing. 
 The symmetry $Z_{17}^{\rm diag}$, contained in $G_f$, remains preserved by the leptoquark couplings, at leading order, and efficiently suppresses couplings of the leptoquark to the first generation of quarks and/or electrons, thus avoiding many stringent experimental bounds. The strongest constraints on the parameter space are imposed by the radiative charged lepton flavour violating decays $\tau\to\mu\gamma$ and $\mu\to e\gamma$. A detailed analytical and numerical study demonstrates the feasibility to simultaneously explain the data on the lepton flavour universality ratios $R(D)$ and $R(D^\star)$ and the muon anomalous magnetic moment, while passing the experimental bounds from all other considered flavour observables.  
\end{abstract}
\end{titlepage}
\pagenumbering{arabic}
\tableofcontents

\newpage

\setcounter{footnote}{0}

%%%%%%%%%%%%%%%%%%%%%

%%%%%%%%%%%%%%%%%%%%%%%%%%%%%%%%%%%%%%%%%%%%%%%%%%%%%%
\section{Introduction}
\label{sec:intro}
%%%%%%%%%%%%%%%%%%%%%%%%%%%%%%%%%%%%%%%%%%%%%%%%%%%%%
The Standard Model (SM) has been very successful in 
describing the gauge interactions involving SM fermions, the Higgs and gauge bosons. 
  However, the observed values of fermion masses
and mixing can only be accommodated with a judicious choice of free parameters, appearing in the Yukawa matrices,  
and cannot be predicted. In particular, the strong hierarchy among charged fermion masses, the potentially different type of mass spectrum in the neutrino
sector, as well as the fact that only the Cabibbo angle is sizeable among quarks, while two
of the mixing angles in the lepton sector are large, necessitate a profound explanation. 

Given the success of symmetries in describing the gauge 
interactions of the SM particles, it is tempting to also employ a symmetry $G_f$, acting on the flavour (or generation) space, in order to 
explain the features of fermion masses and mixing. Abelian symmetries, such as a $U(1)$ group~\cite{Froggatt:1978nt}, have turned out to be sufficient
in order to correctly accommodate the hierarchy among charged fermion masses by an appropriate choice of the $U(1)$ charges of the different
generations of the species of SM fermions. However, fermion mixing, especially the striking difference between the mixing among quarks and leptons
as well as the possibility to predict a certain mixing pattern (e.g.~tri-bimaximal mixing among leptons~\cite{Harrison:2002er,Harrison:2002kp,Xing:2002sw,Harrison:2003aw}), points towards a non-abelian, discrete group
 as flavour symmetry which can be broken non-trivially. For reviews about the application of these groups in high energy particle physics, see~references~\cite{Ishimori:2010au,King:2013eh,Feruglio:2019ybq,Grimus:2011fk}. 

In recent years there have been several anomalous measurements in flavour physics which deviate from the SM predictions and hint at a non-trivial flavour structure. BaBar~\cite{BaBar:2012obs,BaBar:2013mob}, Belle~\cite{Belle:2015qfa,Belle:2019rba,Belle:2016dyj,Belle:2017ilt} and LHCb~\cite{LHCb:2015gmp,LHCb:2017smo,LHCb:2017rln} have measured the ratios\footnote{For brevity, we do not indicate antiparticles by overbars unless required for clarity.}
\begin{align}
	R(D^{(\star)}) = \frac{\Gamma(B\to D^{(\star)} \tau\nu)}{\Gamma(B \to D^{(\star)}\ell\nu)}
\end{align}
with $\ell=e,\mu$ which are sensitive probes of 
lepton flavour universality~(LFU).
The result of the combined fit leads to larger values for $R(D)$ and $R(D^{\star})$ and exhibits a tension with the SM prediction at the $3.4\,\sigma$ level~\cite{Amhis:2019ckw}.
There is also a long-standing discrepancy between the measured value~\cite{Muong-2:2006rrc,Muong-2:2021ojo} and the theoretical prediction~\cite{Davier:2017zfy,Keshavarzi:2018mgv,Colangelo:2018mtw,Hoferichter:2019mqg,Davier:2019can,Keshavarzi:2019abf,Kurz:2014wya,FermilabLattice:2017wgj,
Budapest-Marseille-Wuppertal:2017okr,RBC:2018dos,Giusti:2019xct,Shintani:2019wai,FermilabLattice:2019ugu,Gerardin:2019rua,Aubin:2019usy,Giusti:2019hkz,Melnikov:2003xd,Masjuan:2017tvw,
Colangelo:2017fiz,Hoferichter:2018kwz,Gerardin:2019vio,Bijnens:2019ghy,Colangelo:2019uex,Pauk:2014rta,Danilkin:2016hnh,Jegerlehner:2017gek,Knecht:2018sci,
Eichmann:2019bqf,Roig:2019reh,Colangelo:2014qya,Blum:2019ugy,
Aoyama:2012wk,Aoyama:2019ryr,Czarnecki:2002nt,Gnendiger:2013pva,
Aoyama:2020ynm} 
of the anomalous magnetic moment~(AMM) of the muon, $a_\mu =(g-2)_\mu/2$. 
The combined fit to the experimental data shows a $4.2\,\sigma$ tension~\cite{Muong-2:2021ojo} in $\Delta a_\mu = a_\mu^{\rm exp} - a_\mu^{\rm SM}$.\footnote{There is an ongoing debate about the theoretical prediction of the hadronic vacuum polarisation. While the current determination of the leading-order hadronic vacuum polarisation is obtained using dispersion relations, c.f. reference~\cite{Aoyama:2020ynm}, recent lattice calculations~\cite{Borsanyi:2020mff,Ce:2022kxy,Alexandrou:2022amy,Davies:2022epg} predict a value consistent with the experimental result of the AMM of the muon.} These three anomalies are summarised in table \ref{table:anomalies}.
 \begin{table}[tbh!]\centering
  \def\arraystretch{1.3} 
	\begin{tabular}{|l|cl|cl|c|}
 \hline
 \multicolumn{6}{|c|}{\textsc{ Anomalies}}\\
 \hline
{Observable} & \multicolumn{2}{c|}{SM prediction} &  \multicolumn{2}{c|}{Experiment} & Significance \\
\hline
$R(D)$ & $0.297\pm0.008$ & \cite{Straub:2018kue,david_straub_2021_5543714,Bordone:2019vic}  & $0.340 \pm 0.027 \pm 0.013$ & \cite{Amhis:2019ckw} & $1.4\;\sigma$ \\
$R(D^\star)$ & $0.245\pm0.008$ & \cite{Straub:2018kue,david_straub_2021_5543714,Bordone:2019vic}  & $0.295 \pm 0.010 \pm 0.010$ & \cite{Amhis:2019ckw} &  $2.9\;\sigma$\\
$\Delta a_\mu$ &0&& $(2.51 \pm 0.59) \times 10^{-9}$ & \cite{Muong-2:2021ojo,Aoyama:2020ynm} &  $4.2\;\sigma$\\ 
\hline
 \end{tabular}
\caption{\textbf{Overview of the three anomalies to be addressed in this work and their present significance.} The experimental values for $R(D)$ and $R(D^\star)$ are quoted from the Heavy Flavour Averaging Group (HFLAV) fit \emph{circa} 2021, and the 
combined significance of these two anomalies is $3.4\;\sigma$, with a correlation of $\rho=-0.38$~\cite{Amhis:2019ckw}.
} \label{table:anomalies}
\end{table}

In reference~\cite{Bauer:2015knc} Bauer and Neubert have proposed a simultaneous explanation of the flavour anomalies
in terms of the scalar leptoquark (LQ) $\phi$ transforming as $(3,1,-\tfrac13)$ under the SM gauge group. The importance of LQ couplings to right-handed (RH) fermions has been emphasised in reference~\cite{Cai:2017wry}  and it has been demonstrated that the LQ $\phi$ cannot explain the discrepancies in $b\to s\mu\mu$ which requires the introduction of additional particles, see e.g.~references~\cite{Crivellin:2017zlb,Buttazzo:2017ixm,Marzocca:2018wcf,Bigaran:2019bqv,Balaji:2019kwe,Crivellin:2019dwb,Saad:2020ucl,Saad:2020ihm,Gherardi:2020qhc,Bordone:2020lnb,Julio:2022bue,Chen:2022hle}. 
In the vast majority of these studies only the couplings which are needed to explain the flavour anomalies are introduced, while all other couplings are set to zero without providing any explanation for the vanishing 
couplings nor for the size of the non-zero ones. 

In this work, we construct a model with a discrete flavour symmetry to explain the observed flavour anomalies in $R(D)$, $R(D^{\star})$ and in the AMM of the muon. This model is also capable of correctly describing  
the strong hierarchy among charged fermion masses as well as the quark mixing, leaving aside neutrino masses and lepton mixing. 
Given this focus, the three generations of SM fermions are (mostly) assigned to a doublet and a singlet of $G_f$. 
For this reason, we choose a dihedral group as flavour symmetry.
Both single-valued dihedral groups, $D_n$, as well as double-valued dihedral groups, $D_n^\prime$, form series of groups that feature one- and two-dimensional irreducible representations in case the index $n$ of the group $D_n$ ($D_n^\prime$) is at least $n=3$ ($n=2$), see e.g.~references~\cite{Grimus:2003kq,Grimus:2005mu,Blum:2007jz,Lam:2007qc} for their application to fermion mixing. 
A thorough analysis shows that a model with the flavour group $G_f = D_{17} \times Z_{17}$  can pass all requirements, e.g.~coming from the non-observation of charged lepton flavour violating (cLFV) decays 
such as $\tau\to \mu\gamma$. The residual symmetry $Z_{17}^{\rm diag}$, the diagonal subgroup of $G_f$, which is preserved by the LQ couplings to the SM fermions, at leading order, is crucial in order to 
appropriately suppress those to the first generation of quarks and/or electrons. The breaking of the flavour symmetry is achieved with the help of four spurions that acquire 
a certain vacuum expectation value (VEV), given in terms of the expansion parameter $\lambda$, $\lambda \approx 0.22$, of the model.
 For related studies on the use of flavour symmetries to explain the anomalies observed in semi-leptonic $B$ meson decays,
 see references~\cite{deMedeirosVarzielas:2015yxm,deMedeirosVarzielas:2018bcy,deMedeirosVarzielas:2019lgb}.

The paper is organised as follows. In section~\ref{sec:setup} the model is introduced, the choice of $G_f$, its residual symmetry, and the particle assignment are explained as well as the spurions necessary in order to achieve viable textures for the LQ couplings and the charged fermion mass matrices are specified. The explicit form of the mass matrices and the LQ couplings in both the interaction basis and the charged fermion mass basis is derived in section~\ref{sec:Yukawas_LQcouplings}. Analytical expressions for charged fermion masses and quark mixing are also given. Section~\ref{sec:phenomenology} serves as introduction to the phenomenological study which includes the analytical estimates, the numerical scan of the primary observables in section~\ref{sec:primary} and the comprehensive numerical analysis of all observables in section~\ref{sec:secondarytertiary}. We summarise and give an outlook in section~\ref{sec:concl}. Technical details and supplementary material are collected in appendices~\ref{app:D17} to~\ref{app:supp6}.

%%%%%%%%%%%%%%%%%%%%%%%%%%%%%%%%%%%%%%%%%%%%%%%%%%%%%%
\section{Setup of model}
\label{sec:setup}
%%%%%%%%%%%%%%%%%%%%%%%%%%%%%%%%%%%%%%%%%%%%%%%%%%%%%%

 In section~\ref{subsec:choiceGffermions}, we first argue for the choice of the flavour symmetry to be a dihedral group, and establish assignments under this group for the three generations of  different SM fermion species. We continue in section~\ref{subsec:SetupLQcouplings} with the introduction of the LQ and its relevant couplings. In section~\ref{subsec:LQreferenceCoup}, we focus on particular textures of the LQ couplings and further specify the transformation properties of the fields of the model, as well as the employed flavour symmetry and its breaking. In section~\ref{subsec:structureYukcouplings}, we turn to the Yukawa sector and ensure that the observed charged fermion mass hierarchies and the Cabibbo-Kobayashi-Maskawa~(CKM) mixing matrix are correctly generated. For convenience, in section~\ref{subsec:summarychoices}, we summarise the choice of the flavour symmetry $G_f$, all fields and their transformation properties under $G_f$, as well as the employed spurions and their assumed VEVs.

%%%%%%%%%%%%%%%%%%%%%%%%%%%%%%%%%%%%%%%%%%%%%%%%%%%%%%
\subsection{Choice of flavour symmetry and fermion assignment}
\label{subsec:choiceGffermions}
%%%%%%%%%%%%%%%%%%%%%%%%%%%%%%%%%%%%%%%%%%%%%%%%%%%%%%

We choose a member of the series of dihedral groups $D_n$ with $n \geq 3$ as candidate flavour symmetry, as these groups contain several inequivalent one- and two-dimensional irreducible representations. This permits two distinct assignments of the three generations of SM fermions: either ${\bf 1}+{\bf 1}+{\bf 1}$ or ${\bf 2}+{\bf 1}$. Both assignments prove to be useful for our purposes.

For the charged fermions, we are motivated to use the assignment ${\bf 2}+{\bf 1}$ as much as possible, since the heaviest masses are associated with the third generation, and the mixing between the first (second) and third generations of quarks is small. The doublets and singlets used for the different fermion species, left-handed~(LH) quark doublets $Q_i$, RH down-type quarks $d_{Ri}$, LH lepton doublets $L_i$ and RH charged leptons $e_{Ri}$, are in general inequivalent.\footnote{The fields $d_{Li}$ ($u_{Li}$) denote the LH down-type (up-type) quarks that are the lower (upper) component of the LH quark doublets $Q_i$. Similarly, the fields $e_{Li}$ ($\nu_{Li}$) are the LH charged leptons
(neutrinos), being the lower (upper) component of the LH lepton doublets $L_i$.} We, consequently, expect that the index $n$ of the dihedral group should be at least $n = 9$ in order to offer a minimum of four inequivalent two-dimensional representations.

For the RH up-type quarks $u_{Ri}$, we choose to assign each generation to a singlet, ${\bf 1}+{\bf 1}+{\bf 1}$, which may or may not be inequivalent.\footnote{Whether or not all RH up-type quarks can be assigned to inequivalent one-dimensional representations of the
dihedral group depends on whether the index $n$ of the chosen group is even or odd, since in the case of even $n$ the group has four inequivalent singlets, while $D_n$ with $n$ odd only comprises
two inequivalent singlets~\cite{Blum:2007jz}. One could also consider double-valued dihedral groups $D_n^\prime$, $n$ integer, which all provide four inequivalent one-dimensional representations~\cite{Blum:2007jz}.} This can facilitate the accommodation of the very pronounced mass hierarchy among the up-type quarks. Such an assignment also simplifies the achievement of the desired texture of the LQ coupling to RH charged leptons $e_{Ri}$ and up-type quarks $u_{Ri}$, ${\bf y}$, see eq.~(\ref{eq:texturesxylambda_ex}). Thus, we are able to partially unify the three generations of four of the five different SM fermion species.

We do not discuss neutrino masses nor lepton mixing in this work, which may otherwise hint at a different assignment of the three generations of LH leptons under the flavour symmetry.

Furthermore, we consider a Two Higgs Doublet Model~(2HDM) of type-II~\cite{Hall:1981bc,Donoghue:1978cj}, in which one of the Higgs fields, $H_u$, is responsible for the masses of up-type quarks, while the other one, $H_d$, provides the masses of down-type quarks and charged leptons.\footnote{We assume the decoupling limit, in which the lightest Higgs is SM-like and the further scalars are decoupled. This can be achieved, e.g.~in case one of the VEVs is induced~\cite{Haber:1989xc}. The potential of the two Higgs doublets $H_u$ and $H_d$ is not discussed and might require adding further scalar fields and/or terms, softly breaking the imposed symmetries, in order to correctly achieve the VEVs of $H_u$ and $H_d$, shown in eq.~(\ref{eq:HdHuVEVsapprox}).} This simplifies the search for and, at the same time, amplifies the choice of a suitable flavour symmetry, as we see below. 

We can, thus, write the Lagrangian containing the Yukawa couplings of the charged fermions as follows
\begin{equation}
\label{eq:LagYukcouplings}
{\cal L}_{\mathrm{Yuk}} =  - Y^u_{ij} \, \overline{Q_i} \, H_u \, u_{Rj} - Y^d_{ij} \, \overline{Q_i} \, H_d \, d_{Rj} - Y^e_{ij} \, \overline{L_i} \, H_d \, e_{Rj} + \mathrm{h.c.}
\end{equation}
with the Yukawa coupling matrices $Y^u$, $Y^d$ and $Y^e$ being, in general, complex three-by-three matrices.

%%%%%%%%%%%%%%%%%%%%%%%%%%%%%%%%%%%%%%%%%%%%%%%%%%%%%%
\subsection{Leptoquark couplings}
\label{subsec:SetupLQcouplings}
%%%%%%%%%%%%%%%%%%%%%%%%%%%%%%%%%%%%%%%%%%%%%%%%%%%%%%

The main topic of this study is, however, \emph{not} the correct description of charged fermion masses and quark mixing with the help of a flavour symmetry. Rather, it is exploring the possibilities of capturing the main features of a particular flavour structure of the couplings of the LQ, $\phi$, to the SM fermions, which satisfactorily explains (some of) the present flavour anomalies, while also passing existing phenomenological constraints. 

For this reason, we begin with the Lagrangian containing the two relevant LQ couplings, before electroweak symmetry breaking
\begin{equation}
\label{eq:LagintLQcouplings}
{\cal L}^{\mathrm{int}}_{\mathrm{LQ}} = \hat{x}_{ij} \, \overline{L^c_i} \, \phi^\dagger \, Q_j + \hat{y}_{ij} \, \overline{e^c_{Ri}} \, \phi^\dagger \, u_{Rj} + \mathrm{h.c.} \; ,
\end{equation}
where $\hat{x}_{ij}$ and $\hat{y}_{ij}$ are, in general, complex numbers.
We define ${\bf \hat{x}}$ and ${\bf \hat{y}}$ as complex three-by-three matrices whose elements are denoted as $\hat{x}_{ij}$ and $\hat{y}_{ij}$, respectively. The hatted notation $\;\hat{}\;$ is used to indicate that these LQ couplings are given in the interaction basis of the SM fermions.

The quantum numbers of $\phi^\dagger$ coincide with those of the scalar LQ conventionally denoted $S_1$~\cite{Dorsner:2016wpm}, i.e.~under the SM gauge group, $\phi\sim(3,1,-\tfrac13)$. In contrast to reference~\cite{Dorsner:2016wpm}, we omit the possible coupling to RH neutrinos which are absent in this model, as well as diquark couplings, since the latter can induce proton decay if not appropriately constrained. 
 Imposing baryon number conservation forbids such diquark couplings.
Additionally, we neglect possible couplings between the LQ $\phi$ and the Higgs doublets $H_u$ and $H_d$ and assume that $\phi$ does not acquire a non-vanishing VEV. In this way, the LQ does not impact the potential of $H_u$ and $H_d$.

In the following, we use the results of reference~\cite{Cai:2017wry} to derive suitable textures of the LQ couplings as a starting point
for this model. In order to match the convention of reference~\cite{Cai:2017wry}, we change to the charged fermion mass basis
\begin{equation}
\label{eq:LagmassLQcouplings}
{\cal L}^{\mathrm{mass}}_{\mathrm{LQ}} = x_{ij} \, \overline{(\nu^\mathrm{m}_{Li})^c} \, \phi^\dagger \, d_{Lj}^\mathrm{m} + y_{ij} \, \overline{(e^\mathrm{m}_{Ri})^c} \, \phi^\dagger \, u_{Rj}^\mathrm{m} 
- z_{ij} \, \overline{(e^\mathrm{m}_{Li})^c} \, \phi^\dagger \, u_{Lj}^\mathrm{m} + \mathrm{h.c.} \; .
\end{equation}
The fields with the superscript ${}^\mathrm{m}$, $u_{Li}^\mathrm{m}$, $u_{Ri}^\mathrm{m}$, $d_{Li}^\mathrm{m}$, $d_{Ri}^\mathrm{m}$, $e_{Li}^\mathrm{m}$, $e_{Ri}^\mathrm{m}$ and $\nu_{Li}^\mathrm{m}$, represent
the SM fermion fields in the mass basis. These are related to the fields, 
$u_{Li}$, $u_{Ri}$, $d_{Li}$, $d_{Ri}$, $e_{Li}$, $e_{Ri}$ and $\nu_{Li}$, in the interaction basis as follows, formulated in matrix-vector notation,
\begin{eqnarray}
\label{eq:LRrotations1}
&&u_{L} = L_u \, u_{L}^\mathrm{m} \; , \;\; u_{R} = R_u \, u_{R}^\mathrm{m} \; , \;\; d_{L} = L_d \, d_{L}^\mathrm{m} \; , \;\; d_{R} = R_d \, d_{R}^\mathrm{m} \; , \;\; 
\\ \label{eq:LRrotations2}
&&e_{L} = L_e \, e_{L}^\mathrm{m} \; , \;\; e_{R} = R_e \, e_{R}^\mathrm{m} \;\; \mbox{and} \;\; \nu_{L} = L_e \, \nu_{L}^\mathrm{m} \; .
\end{eqnarray}
We reiterate that the basis change of LH neutrinos coincides with the one of LH charged leptons, 
since neutrinos are massless in this model and thus lepton mixing is unphysical. The couplings $x_{ij}$, $y_{ij}$ and $z_{ij}$ in eq.~(\ref{eq:LagmassLQcouplings}) are, in general, complex numbers, like $\hat{x}_{ij}$ and $\hat{y}_{ij}$,
and we define the LQ couplings ${\bf x}$, ${\bf y}$ and ${\bf z}$ as complex three-by-three matrices with elements $x_{ij}$, $y_{ij}$ and $z_{ij}$, respectively.
The two LQ couplings ${\bf x}$ and ${\bf z}$ in eq.~(\ref{eq:LagmassLQcouplings})
both stem from the LQ coupling ${\bf \hat{x}}$ in eq.~(\ref{eq:LagintLQcouplings}) and are, consequently, related by the quark mixing matrix $V_\mathrm{CKM}$.

%%%%%%%%%%%%%%%%%%%%%%%%%%%%%%%%%%%%%%%%%%%%%%%%%%%%%%%%%%%%%%%%
\subsection{Viable textures of leptoquark couplings}
\label{subsec:LQreferenceCoup}
%%%%%%%%%%%%%%%%%%%%%%%%%%%%%%%%%%%%%%%%%%%%%%%%%%%%%%%%%%%%%%%%

Possible textures of the LQ couplings ${\bf x}$ and ${\bf y}$ that permit an explanation of the flavour anomalies in $R(D)$, $R(D^\star)$
and in the AMM of the muon have been proposed and studied in numerous publications. The study in reference~\cite{Cai:2017wry} has performed two separate scans of these couplings, each assuming a slightly different texture for ${\bf y}$. We use the results of the scan in which the form of the LQ couplings ${\bf x}$ and ${\bf y}$
has been fixed to
\begin{equation}
\label{eq:scan2texturesxy}
{\bf x}= \left(
\begin{array}{ccc}
 0 & 0 & 0\\
 0 & x_{22} & x_{23}\\
0 & x_{32} & x_{33}
\end{array}
\right) \;\, \mbox{and} \;\,
{\bf y}= \left(
\begin{array}{ccc}
 0 & 0 & 0\\
 0 & 0 & y_{23}\\
 0 & y_{32} & 0
\end{array}
\right)
 \; ,
\end{equation}
where, a priori, all non-vanishing entries of ${\bf x}$ and $y_{32}$ can be of order one (or even larger), while $y_{23}$ is bounded as $|y_{23}| \leq 0.05$.
A detailed analysis of the results of this scan shows that for $\hat{m}_\phi \lesssim 5$, where $\hat{m}_\phi$ measures the mass $m_\phi$ of the LQ $\phi$ in TeV, the textures of ${\bf x}$ and ${\bf y}$ can be expressed in terms of the expansion
parameter $\lambda$
\begin{equation}
\label{eq:deflambda}
\lambda \approx 0.2 \; .
\end{equation}
One viable set of textures is 
\begin{equation}
\label{eq:texturesxylambda_ex}
{\bf x} \sim \left(
\begin{array}{ccc}
0 & 0 & 0\\
0 & \lambda^3 & \lambda\\
0 & \lambda^2 & 1
\end{array}
\right) \;\, \mbox{and} \;\,
{\bf y} \sim \left(
\begin{array}{ccc}
 0 & 0 & 0\\
 0 & 0 & \lambda^3\\
 0 & 1 & 0
\end{array}
\right) \; ,
\end{equation}
where each non-zero element is accompanied by a complex order-one number.\footnote{We note that many phenomenological analyses take the elements of the LQ couplings ${\bf x}$ and ${\bf y}$
to be real for simplicity. We refrain from doing so, since we do not include a CP symmetry in this model.} 
 In particular, $x_{33}\sim 1$ and $y_{32}\sim 1$ facilitate the explanation of the anomalies in $R(D)$ and $R(D^\star)$, while $z_{23} \ y_{23} \sim x_{23} \, y_{23}\sim\lambda^4$ helps to achieve 
 $\Delta a_\mu\sim 10^{-9}$.   
 We concentrate on achieving these textures of ${\bf x}$ and ${\bf y}$, with the zeros denoting elements (much) smaller than $\lambda^4 \sim 10^{-3}$.

These couplings are specified in the basis where the down-type quark mass matrix, $M_d$,
and the charged lepton mass matrix, $M_e$, are (nearly) diagonal, whereas the up-type quark mass matrix, $M_u$, is the origin of the CKM mixing matrix. The unitary transformation
associated with the RH up-type quarks is assumed to be (close to) the identity matrix in flavour space. Furthermore, all fermion masses are canonically ordered, so that no additional permutations of columns and/or rows of the mass matrices are necessary.

In order to proceed with the assignment of the particles to representations of the flavour symmetry, we first fix the transformation properties of the LQ $\phi$. We choose it to be in the trivial singlet of the entire flavour symmetry $G_f$. The elements $x_{33}$ and $y_{32}$ are both of order one, and thus should be non-zero in the limit of an unbroken flavour symmetry. More generally, this should be true for all couplings, including the Yukawa couplings of the charged fermions
of order one, e.g.~the Yukawa coupling that gives rise to the top quark mass. Otherwise large flavour symmetry breaking effects would be needed, which are difficult to control. We discuss this issue, when addressing the charged fermion mass matrices in section~\ref{subsec:structureYukcouplings}.

To achieve $x_{33} \sim 1$ constrains us to assign the third generation of LH lepton
doublets, $L_3$, and of LH quark doublets, $Q_3$, to complex conjugated representations of the flavour symmetry. Furthermore, $y_{32} \sim 1$ requires that the third generation of RH charged leptons, $e_{R3}$, and the second generation of
RH up-type quarks, $u_{R2}$, also transform as complex conjugated representations. Note that they should be in complex conjugated representations given the form of the LQ couplings in eq.~(\ref{eq:LagintLQcouplings}).
However, as all representations of (single-valued) dihedral groups are real, complex conjugation refers to an external $Z_N$ symmetry with $N>2$, whose purpose becomes clear in the following.
Indeed, we can fix, without loss of generality, $L_3 \sim {\bf 1_1}$,
$Q_3 \sim {\bf 1_1}$, $e_{R3} \sim {\bf 1_1}$, and $u_{R2} \sim {\bf 1_1}$ under the dihedral group.

%%%%%%%%%%%%%%%%%%%%%%%%%%%%%%%%%%%%%%%%%%%%%%%%%%%%%%%%%%%%%%%%
\subsubsection{Flavour symmetry breaking in leptoquark couplings}
\label{subsubsec:SymmetryBreakingLQCouplings}
%%%%%%%%%%%%%%%%%%%%%%%%%%%%%%%%%%%%%%%%%%%%%%%%%%%%%%%%%%%%%%%%

The other non-zero elements of the LQ couplings ${\bf x}$ and ${\bf y}$ are achieved by breaking the flavour symmetry with some spurion, acquiring a suitably aligned VEV.
The largest non-zero element of ${\bf x}$ and ${\bf y}$ that is not of order one is of order $\lambda$. This determines the size of the symmetry breaking parameter, at least for the LQ couplings. We follow a minimalistic approach by generating all elements, $x_{22}$, $x_{23}$, $x_{32}$ and $y_{23}$, with the help of a single spurion, called $S$.
Since these non-zero elements have different orders of magnitude in $\lambda$,
we expect that $x_{ij}, y_{ij} \sim \lambda^k$ arise from the insertion of $k$ powers of the spurion.\footnote{Since we work in a non-supersymmetric model, it can also be the conjugated spurion and/or some suitable combination of both.}

The insertion of a single spurion $S$ for $x_{23} \sim \lambda$ forces us to assign the spurion $S$ to the same (real) two-dimensional representation of the dihedral group as the first two generations of LH lepton doublets, $L$. This two-dimensional representation can be chosen without loss of generality as ${\bf 2_1}$ of the dihedral group. Clearly, the VEV of the spurion also needs to be aligned in a specific way in order to only generate the element $x_{23} \sim \lambda$, and not $x_{13}$ at the same or similar level. We come back to this point in section~\ref{subsubsec:LQrefResidual}.

Knowing that $x_{32} \sim \lambda^2$ and thus is due to two spurion insertions, we have to have the first two generations of LH quark doublets, $Q$, in a representation different from ${\bf 2_1}$ and, indeed, a suitable choice is ${\bf 2_2}$,
since the product of ${\bf 2_1}$ with itself contains as irreducible two-dimensional representation ${\bf 2_2}$, see appendix~\ref{app:D17}. Choosing $L \sim {\bf 2_1}$, but $Q \sim {\bf 2_2}$, also ensures that no large elements are generated
among $x_{11}$, $x_{12}$, $x_{21}$ and $x_{22}$. 

At the level of three spurion insertions, $S^3$, however, $x_{22} \sim \lambda^3$ can be generated, as desired, compare eq.~(\ref{eq:texturesxylambda_ex}). This is possible, since the product of ${\bf 2_1} \times {\bf 2_1} \times {\bf 2_1}$ can contain the doublet ${\bf 2_3}$.

Finally, we note that also $y_{23}$ should be generated at order $\lambda^3$. So, the combination of the first two generations of RH charged leptons,
$e_R$, and of the RH up-type quark $u_{R3}$ should transform as the same two-dimensional representation ${\bf 2_3}$ (and possibly with an appropriate charge under an external $Z_N$ symmetry). We arrive at the conclusion that
$e_R$ has to be in ${\bf 2_3}$. We remind that it remains to be checked explicitly that only the element $y_{23} \sim \lambda^3$ is generated, and not $y_{13}$ as well. Whether or not this happens, depends on the alignment of the VEV of the
three spurion insertion and the relevant Clebsch-Gordan coefficients, see appendix~\ref{app:D17}.

At the same time, we have to ensure that  the elements $y_{21}$ and $y_{22}$ (and also $y_{11}$ and $y_{12}$) are not generated at order $\lambda^3$ or larger. The best option in order to achieve this goal is to assign different charges under an external $Z_N$ symmetry to the RH up-type quarks. This is in general also required in order to keep $y_{32} \sim 1$ and $y_{31}$ and $y_{33}$ (much) suppressed, since the dihedral group might not
offer enough inequivalent one-dimensional representations to achieve this.

In summary, we are able to generate all non-zero elements of the LQ couplings ${\bf x}$ and ${\bf y}$ of their correct order in $\lambda$. Since we are in a non-supersymmetric context, also the conjugated spurion $S^\dagger$ can couple. Indeed, this cannot be avoided by the dihedral group as part of the flavour symmetry, since it only provides real representations. This is one of the arguments for considering as flavour symmetry the direct product
of a dihedral group $D_n$ and an external $Z_N$ symmetry with $N>2$.

Due to the size of the symmetry breaking parameter, $\lambda \approx 0.2$, the spurion $S$ might not be suitable for generating charged fermion masses. Since these follow a stronger hierarchy, this would not be possible unless
we could achieve this by multiple insertions of the spurion. As we see in section~\ref{subsec:structureYukcouplings}, it is necessary to introduce three further spurions, $T$, $U$ and $W$, with different transformation properties under the flavour symmetry and with different VEVs (in size and/or alignment), for correctly describing the charged fermion masses and quark mixing.

%%%%%%%%%%%%%%%%%%%%%%%%%%%%%%%%%%%%%%%%%%%%%%%%%%%%%%%%%%%%%%%%
\subsubsection{Protecting textures of leptoquark couplings with a residual symmetry}
\label{subsubsec:LQrefResidual}
%%%%%%%%%%%%%%%%%%%%%%%%%%%%%%%%%%%%%%%%%%%%%%%%%%%%%%%%%%%%%%%%

It is well-known that the zero elements in the first column and row of the LQ couplings ${\bf x}$ and ${\bf y}$ should be preserved to a high degree. As these couplings induce interactions involving the first generation of leptons and/or quarks, experimental bounds on them are particularly strong. In this model, we ensure such a suppression by a residual symmetry, i.e. the vanishing elements in ${\bf x}$ and ${\bf y}$ in eq.~(\ref{eq:scan2texturesxy}) are protected from
becoming non-zero, so long as the residual symmetry is intact.

This residual symmetry is a subgroup of the flavour symmetry $D_n \times Z_N$ of the model. As residual symmetry, we use an abelian symmetry because a non-abelian one can easily become too constraining. A type of residual symmetry which has been successfully employed in approaches with flavour symmetries of the form $X \times Z_N$ is $Z_N^\mathrm{diag}$. The symmetry $Z_N^\mathrm{diag}$ corresponds to the diagonal subgroup of a $Z_N$ symmetry, contained in the non-abelian group $X$, and the external $Z_N$ symmetry~\cite{Hagedorn:2011un, Hagedorn:2011pw}. We, hence, choose $N=n$ in the following, i.e.~
\begin{equation}
\label{eq:Gfstructure}
G_f=D_n \times Z_n \;\; \mbox{with a single index} \;\; n \; .
\end{equation}
Furthermore, we assume that the $Z_n$ symmetry contained in $D_n$ is generated by the generator $a$ of $D_n$~--~see appendix~\ref{app:D17} for the generators of the dihedral group $D_{17}$.
Since the residual symmetry $Z_n^\mathrm{diag}$ should be preserved by the textures of the LQ couplings ${\bf x}$ and ${\bf y}$, all non-vanishing elements of ${\bf x}$ and ${\bf y}$ in eq.~(\ref{eq:scan2texturesxy})
should correspond to combinations of SM fermions (and the LQ $\phi$) with zero charge under $Z_n^\mathrm{diag}$. At the same time, the spurion $S$ should acquire a VEV that is compatible with the preservation of this residual symmetry. 

We make the eventual choice\footnote{We do not comment further about this choice. However, we mention that we have studied different values of the index $n$ of the dihedral group $D_n$ with regard
to the possibility to generate all operators needed for the LQ couplings ${\bf x}$ and ${\bf y}$ and for a valid description of charged fermion masses and quark mixing and, at the same time, not to give rise to other contributions to the LQ couplings and the charged fermion mass matrices which strongly perturb the leading-order results.}
\begin{equation}
\label{eq:choicen17}
n=17\;.
\end{equation}
It is straightforward to derive a set of charges under the external $Z_{17}$ symmetry in order to ensure the preservation of the residual symmetry in the LQ coupling ${\bf{x}}$, given that we have already fixed that $L_3 \sim {\bf 1_1}$, $Q_3 \sim {\bf 1_1}$,
$L \sim {\bf 2_1}$, $Q \sim {\bf 2_2}$, as well as $S \sim {\bf 2_1}$. 

Let us set the charge of $L_3$ under the external $Z_{17}$ symmetry to $1$.\footnote{Other choices are possible at this point. A valid choice is determined by the requirement that no contribution to the LQ couplings ${\bf x}$ and ${\bf y}$ nor to the charged fermion mass matrices, when including the spurions $T$, $U$ and $W$ beyond $S$ as well as their conjugated fields, is generated which perturbs the leading-order structure of these and thus leads to unacceptably large flavour violation and/or wrong results for charged fermion masses and quark mixing. The presented set of charges under the external $Z_{17}$ symmetry is such a valid choice.} Then, we have to have that $Q_3$ carries the $Z_{17}$ charge $16$. Since $L_3$ and $Q_3$ are both singlets under $D_{17}$, their charge under the external $Z_{17}$ symmetry coincides with their charge under the residual symmetry $Z_{17}^\mathrm{diag}$.
Furthermore, the fact that both $x_{23}$ and $x_{32}$ in the LQ coupling ${\bf x}$ should be allowed as well, compare eq.~(\ref{eq:scan2texturesxy}), requires $L_2$ and $Q_2$ to transform in the same way under the residual
symmetry $Z_{17}^\mathrm{diag}$ as $L_3$ and $Q_3$, respectively. Knowing this, we can compute the charge of $L \sim {\bf 2_1}$, whose second component is $L_2$, under the external $Z_{17}$ symmetry, and arrive at
$2$ as $Z_{17}$ charge for $L$. Similarly, we have for $Q \sim {\bf 2_2}$, whose second component is $Q_2$, that its charge under the external $Z_{17}$ symmetry is $1$. Then, automatically also the element $x_{22}$ in the LQ coupling ${\bf x}$
is invariant under the residual symmetry $Z_{17}^\mathrm{diag}$. Additionally, we can check that $L_1$ and $Q_1$ both have the charge $3$ under the residual symmetry $Z_{17}^\mathrm{diag}$ and hence none of the elements of
the first column and row of the LQ coupling ${\bf x}$ is allowed in the limit of the residual symmetry $Z_{17}^\mathrm{diag}$ being preserved. 

In order to couple the spurion $S$ in an $Z_{17}$-invariant way to the combination
$\overline{L^c} \, \phi^\dagger \, Q_3$ we have to assign the $Z_{17}$ charge $16$ to $S$. For $S \sim {\bf 2_1}$ thus its first component $S_1$ carries no charge under the residual symmetry $Z_{17}^\mathrm{diag}$. Consequently, this
has to be the component which acquires a non-zero VEV, while the VEV of the other one, $S_2$, has to vanish
\begin{equation}
\label{eq:SVEV}
 \braket{S}  = \begin{pmatrix} \lambda \\ 0\end{pmatrix} \; .
\end{equation}
By explicit computation one can check that the other two operators, $\overline{L^c_3} \, \phi^\dagger \, Q \, S^2$ as well as $\overline{L^c} \, \phi^\dagger \, Q \, S^3$, generating $x_{32}$ and $x_{22}$ of the appropriate size, are invariant as
well.\footnote{The spurions are treated as dimensionless flavour symmetry breaking fields. Thus, we do not need to introduce a cutoff scale in order to restore the correct mass dimension of the operators.}

In order to protect the vanishing elements in the LQ coupling ${\bf y}$, we appeal to the residual symmetry $Z_{17}^\mathrm{diag}$ as well. To do so,  we have to assign appropriate charges to the RH fermions $u_{Ri}$, $e_{R}$ and $e_{R3}$. Given that only two elements are non-zero in the limit of unbroken $Z_{17}^\mathrm{diag}$, we can only fix a certain combination of $Z_{17}$ charges. Therefore, the $Z_{17}$ charges of the $D_{17}$ singlets $e_{R3}$ and $u_{R2}$ should be opposite, e.g.~for $9$ being the $Z_{17}$ charge of $e_{R3}$, that of $u_{R2}$ should be $8$. Also, the charges of $e_{R2}$, the second component of $e_R \sim {\bf 2_3}$, and of the $D_{17}$ singlet $u_{R3}$ should be opposite under the residual symmetry. A possible choice is that $e_{R2}$ has the $Z_{17}^\mathrm{diag}$ charge $16$, since $e_R$ has the charge $2$ under the external $Z_{17}$ symmetry and, thus, $u_{R3}$ carries the $Z_{17}^\mathrm{diag}$
charge $1$~--~which also corresponds to its charge under the external $Z_{17}$ symmetry. 

%%%%%%%%%%%%%%%%%%%%%%%%%%%%%%%%%%%%%%%%%%%%%%%%%%%%%%%%%%%%%%%%
 \begin{table}[t!]\centering
  \def\arraystretch{1.3} 
  \begin{tabular}{|l|c||l|c||l|c||l|c||l|c|}
 \hline
Field & $Z_{17}^\mathrm{diag}$ & Field & $Z_{17}^\mathrm{diag}$ & Field & $Z_{17}^\mathrm{diag}$& Field & $Z_{17}^\mathrm{diag}$& Field & $Z_{17}^\mathrm{diag}$\\
\hline
$Q_1$ & $3$ & $d_{R1}$ & $5$ & $e_{R1}$ & $5$ & $S_1$ & $0$ & $W_1$ & $14$\\
$Q_2$ & $16$ & $d_{R2}$ & $14$ & $e_{R2}$ & $16$ & $S_2$ & $15$ & $W_2$ & $10$\\
$Q_3$ & $16$ & $d_{R3}$ & $7$ & $e_{R3}$ & $9$ & $T_1$ & $10$ & &\\
$u_{R1}$ & $13$ & $L_1$ & $3$ & $H_u$ & $15$ & $T_2$ & $6$ & &\\
$u_{R2}$ & $8$ & $L_2$ & $1$ & $H_d$ & $9$ & $U_1$ & $10$ & &\\
$u_{R3}$ & $1$ & $L_3$ & $1$ & $\phi$ & $0$ & $U_2$ & $6$& &\\
\hline
 \end{tabular}
\caption{\mathversion{bold}{\small {\bf Charge under residual symmetry $Z_{17}^\mathrm{diag}$}.\mathversion{normal} We list the charge of the different fermions, scalar fields and spurions under the residual
 symmetry $Z_{17}^\mathrm{diag}$, preserved by the leading-order structure of the LQ couplings ${\bf x}$ and ${\bf y}$, see eq.~(\ref{eq:texturesxylambda_ex}).
This residual symmetry $Z_{17}^\mathrm{diag}$ is the diagonal subgroup of the $Z_{17}$ symmetry, contained in $D_{17}$ and generated by the generator $a$, compare appendix~\ref{app:D17}, and the external
$Z_{17}$ symmetry.}} \label{table:residualsymmetry}
\end{table}
%%%%%%%%%%%%%%%%%%%%%%%%%%%%%%%%%%%%%%%%%%%%%%%%%%%%%%%%%%%%%%%%

With this $Z_{17}$ charge assignment, we can check that apart from the elements $y_{32}$ and $y_{23}$ no other element of the second and third columns
of the LQ coupling ${\bf y}$ is invariant under the residual symmetry $Z_{17}^\mathrm{diag}$. We can, furthermore, explicitly check that the operator $\overline{e^c_{R}} \, \phi^\dagger \, u_{R3} \, S^3$ is invariant under the external $Z_{17}$
symmetry. In order to avoid letting any element of the first column of ${\bf y}$ be invariant under $Z_{17}^\mathrm{diag}$, we choose the charge of the $D_{17}$ singlet $u_{R1}$ under the external $Z_{17}$ symmetry to be $13$.

The presented choice of $Z_{17}$ charge assignments also takes constraints into account from the requirement of a correct description of charged fermion masses and quark mixing.  It, furthermore, suppresses flavour violation, in particular involving the first lepton and/or quark generation, by limiting the contributions to the LQ coupling ${\bf y}$ from operators involving the other spurions $T$, $U$, $W$, and their conjugated fields.

In table~\ref{table:residualsymmetry}, the chosen charges of the different fermions, scalar fields and spurions under the residual symmetry $Z_{17}^\mathrm{diag}$ have been collected.

%%%%%%%%%%%%%%%%%%%%%%%%%%%%%%%%%%%%%%%%%%%%%%%%%%%%%%
\subsection{Flavour structure of Yukawa couplings}
\label{subsec:structureYukcouplings}
%%%%%%%%%%%%%%%%%%%%%%%%%%%%%%%%%%%%%%%%%%%%%%%%%%%%%%

In the next step, we turn to the construction of the charged fermion mass matrices and complete the assignment of the SM fermions under the flavour symmetry $G_f=D_{17} \times Z_{17}$ by also fixing the transformation properties of the
RH down-type quarks. We introduce three further spurions, $T$, $U$ and $W$, all transforming as doublets under $D_{17}$. These are responsible for the generation of the correct charged fermion mass hierarchy and the Cabibbo angle $\theta_C$
of the order of $\lambda$. By contrasting eqs.~(\ref{eq:TVEV}, \ref{eq:UVEV}, \ref{eq:WVEV}) with eq.~(\ref{eq:SVEV}), we note that the size of their VEVs is (significantly) smaller than that of the spurion $S$.

We recall that the LQ couplings ${\bf x}$ and ${\bf y}$ are given in the mass basis of charged leptons and down-type quarks. This means $M_e$ and $M_d$ should be (almost) diagonal, while the CKM mixing matrix should arise from the up-type quark mass matrix $M_u$, and the unitary transformation relating the interaction and mass bases of the RH up-type quarks should be (close to) the identity matrix. The approximate form of the mass matrices $M_e$, $M_d$ and $M_u$ in terms of $\lambda$ is therefore
 \begin{eqnarray}
 \label{eq:MeMdapprox}
&&M_e \sim \left(
 \begin{array}{ccc}
 \lambda^4 & 0 & 0\\
 0 & \lambda^2 & 0\\
 0 & 0 & 1
 \end{array}
 \right) \, \braket{ H_d^0 } \; , \;\;
 M_d \sim \left(
 \begin{array}{ccc}
 \lambda^4 & 0 & 0\\
 0 & \lambda^2 & 0\\
 0 & 0 & 1
 \end{array}
 \right) \, \braket{H_d^0} \; , \;\;
 \\[0.1in]
 \label{eq:Muapprox}
 \mbox{and}&&M_u \sim \left(
 \begin{array}{ccc}
 \lambda^8 & \lambda^5 & \lesssim\lambda^3 \\
 0 & \lambda^4 & \lambda^2\\
 0 & 0 & 1
 \end{array}
 \right) \, \braket{ H_u^0}.
 \end{eqnarray}
In eqs.~(\ref{eq:MeMdapprox}, \ref{eq:Muapprox}) all non-vanishing elements are accompanied by complex order-one coefficients, and the vanishing elements imply that these entries are (strongly) suppressed.

 As discussed in section~\ref{subsec:choiceGffermions}, we work in a 2HDM: $H_u$ gives masses to up-type quarks, and $H_d$ to down-type quarks as well as charged leptons. Therefore, the hierarchy between the bottom quark (tau lepton) mass and the mass of the top quark can be generated via an appropriate hierarchy among the VEVs of the two Higgs doublets $H_d$ and $H_u$.
 Typical values of these VEVs are
 \begin{equation}
 \label{eq:HdHuVEVsapprox}
 \braket{H_d^0}= \frac{v_d}{\sqrt{2}}\sim 2.4 \; \mathrm{GeV} \;\; \mbox{and} \;\; \braket{H_u^0} = \frac{v_u}{\sqrt{2}}\sim 174 \; \mathrm{GeV}
 \end{equation}
 so that $v_d^2 +v_u^2 = v^2 \sim (246 \; \mathrm{GeV})^2$. Both Higgs doublets transform as trivial singlet ${\bf 1_1}$ under $D_{17}$, but need to carry a non-trivial charge under the external $Z_{17}$ symmetry to allow for the generation of the top quark and tau lepton mass at tree level.   
  
  %%%%%%%%%%%%%%%%%%%%%%%%%%%%%%%%%%%%%%%%%%%%%%%%%%%%%%%%%%%%%%%%
 \subsubsection{Generation of down-type quark and charged lepton masses}
 \label{subsubsec:YukawaHd} 
 %%%%%%%%%%%%%%%%%%%%%%%%%%%%%%%%%%%%%%%%%%%%%%%%%%%%%%%%%%%%%%%%
 
The invariance of the operator $\overline{L_3} \, H_d \, e_{R3}$ under the external $Z_{17}$ symmetry requires that the charge of $H_d$ is $9$.\footnote{We do not consider it an issue that the VEV of $H_d$~(and also of $H_u$) spontaneously breaks the external $Z_{17}$ symmetry because it is broken anyway (at a higher scale) by the VEVs of the spurions $S$, $T$, $U$ and $W$.} Since the size of the bottom quark mass is similar to that of the tau lepton, we also require that the operator $\overline{Q_3} \, H_d \, d_{R3}$ is invariant. This fixes the transformation properties of $d_{R3}$ to $d_{R3}\sim{\bf 1_1}$ under $D_{17}$, and the charge of $d_{R3}$ under the external $Z_{17}$ symmetry to be $7$.

In order to generate the mass of the muon and of the strange quark, we invoke a further spurion, $T$. One relevant operator is thus $\overline{L} \, H_d \, e_R \, T$, requiring that the spurion $T$ carries the charge $8$ under the external $Z_{17}$
symmetry. We can determine the transformation properties of $T$ under $D_{17}$ to be $T \sim {\bf 2_2}$ by noting that the second component of the covariant in ${\bf 2_2}$ is $\overline{L_2} \, H_d \, e_{R2}$. Furthermore, we know then also that the first component of $T$ should acquire a non-zero VEV of order $\lambda^2$
  \begin{equation}
 \label{eq:TVEV}
 \braket{T}  = \begin{pmatrix} \lambda^2 \\ 0\end{pmatrix} \; .
 \end{equation}
For details about the necessary Clebsch-Gordan coefficients, see appendix~\ref{app:D17}. We can check that this VEV breaks the residual symmetry $Z_{17}^\mathrm{diag}$, invoked to protect the form of the LQ couplings ${\bf x}$ and ${\bf y}$.
This is not unexpected, but indicates that couplings of this spurion to the LQ $\phi$ should be appropriately suppressed by the flavour symmetry $G_f=D_{17} \times Z_{17}$. 

At the same time, the spurion $T$ should generate the strange quark mass, i.e.~the operator $\overline{Q} \, H_d \, d_R \, T$ should be invariant under $G_f$. For this to work, we have to fix the transformation properties of $d_R$ accordingly. Its charge under the external $Z_{17}$ symmetry should be $1$, and $d_R \sim {\bf 2_4}$ under $D_{17}$ such that the second component of the covariant in ${\bf 2_2}$ reads $\overline{Q_2} \, H_d \, d_{R2}$. This completes the fixing of the transformation properties of the three generations of all SM fermion species.

The simplest way to generate the mass of the electron and of the down quark would be to modify the VEV of the spurion $T$ so that its second component acquires a VEV of order $\lambda^4$. We do not pursue this possibility and instead introduce a further spurion, $U$, which transforms in the same way under $G_f$ as $T$, but acquires a VEV of the form
  \begin{equation}
 \label{eq:UVEV}
 \braket{U}  = \begin{pmatrix} 0 \\ \lambda^4\end{pmatrix} \; .
 \end{equation}
Similar to the VEV of $T$, this VEV does not preserve the residual symmetry $Z_{17}^\mathrm{diag}$, maintained in the LQ couplings ${\bf x}$ and ${\bf y}$ (at leading order). The two relevant operators for the mass of the electron and of the down quark are $\overline{L} \, H_d \, e_R \, U$ and $\overline{Q} \, H_d \, d_R \, U$, respectively.

The reason for employing the further spurion $U$ is twofold. Firstly, it allows the undetermined order-one coefficients accompanying the aforementioned operators to be used to correctly achieve the masses of both the electron and the down quark. This would not be possible with only the spurion $T$. Secondly, in this way the computation of higher-order operators with several insertions of the spurions $T$ and $U$ is simplified, and their number can be controlled better.

This concludes the discussion of the generation of the charged lepton and the down-type quark mass matrices. In the end, both mass matrices are not exactly diagonal, since further operators are always induced.\footnote{One example is the operator $\overline{L} \, H_d \, e_{R3} \, S^\dagger$. It is invariant because $\overline{L_3} \, H_d \, e_{R3}$ is generated at tree level, $L$ transforms as the same doublet of $D_{17}$ as the spurion $S^{(\dagger)}$ and the piece
$\overline{L_2} \, H_d \, e_{R3}$ is like $\overline{L_3} \, H_d \, e_{R3}$ invariant under the residual symmetry $Z_{17}^\mathrm{diag}$ which is also left unbroken by the VEV of the spurion $S^{(\dagger)}$. We, hence, already know that
the element $M_{e,23}$ of the charged lepton mass matrix must arise at the order $\lambda \, \braket{H_d^0}$ from the operator $\overline{L} \, H_d \, e_{R3} \, S^\dagger$.\label{foot1}}
We show in the next section that this neither poses a problem for the charged fermion mass matrices nor for achieving the textures of the LQ couplings ${\bf x}$ and ${\bf y}$. 

%%%%%%%%%%%%%%%%%%%%%%%%%%%%%%%%%%%%%%%%%%%%%%%%%%%%%%%%%%%%%%%%
 \subsubsection{Generation of up-type quark masses and quark mixing matrix}
 \label{subsubsec:YukawaHu} 
 %%%%%%%%%%%%%%%%%%%%%%%%%%%%%%%%%%%%%%%%%%%%%%%%%%%%%%%%%%%%%%%%

 The invariance of the operator $\overline{Q_3} \, H_u \, u_{R3}$ under the external $Z_{17}$ symmetry requires that the charge of $H_u$ is $15$. In the limit of unbroken flavour symmetry, the only up-type quark mass generated is the one of the top quark. In order to arrive at a non-zero mass for the charm quark, and to generate the Cabibbo angle of the correct order of magnitude, we introduce a last spurion, $W$. It should couple to $\overline{Q} \, H_u \, u_{R2}$ and, hence, $W$ has to carry the charge $12$ under the external $Z_{17}$ symmetry and transform like $Q$ under $D_{17}$,  $W\sim{\bf 2_2}$. Since the purpose of introducing $W$ is to generate the charm quark mass as well as the Cabibbo angle,
both its components should acquire a non-vanishing VEV
\begin{equation}
 \label{eq:WVEV}
 \braket{W}  = \begin{pmatrix} \lambda^5 \\ \lambda^4\end{pmatrix} \; .
 \end{equation}
Indeed, given the structure of the covariant $\overline{Q} \, H_u \, u_{R2}$, the lower component of the VEV of $W$ generates the charm quark mass, while the upper one is responsible for the size of the Cabibbo angle, which is
$\theta_C \approx \lambda$. As can be shown, the VEV of the spurion $W$ also breaks the residual symmetry $Z_{17}^\text{diag}$. This completes the set of spurions we use in this model.

Two points still need to be addressed, namely the generation of the two smaller quark mixing angles and of the mass of the up quark. The former issue can be solved by noting that in this non-supersymmetric model also the conjugated spurions contribute to the charged fermion mass matrices and LQ couplings ${\bf x}$ and ${\bf y}$. Indeed, one can check that with the assigned transformation properties the operator $\overline{Q} \, H_u \, u_{R3} \, (S^\dagger)^2$ leads to an
element $M_{u,23}$ in the up-type quark mass matrix which is of the order $\lambda^2 \, \braket{H_u^0}$ and thus can correctly generate $\theta_{23} \sim \lambda^2$.\footnote{We note that also
$\overline{Q} \, H_d \, d_{R3} \, (S^\dagger)^2$ is invariant, since the combinations $H_u \, u_{R3}$ and $H_d \, d_{R3}$ transform in the same way. This is due to the fact that both operators $\overline{Q_3} \, H_u \, u_{R3}$ and
$\overline{Q_3} \, H_d \, d_{R3}$ are generated at tree level. As we show in the analysis of quark masses and mixing, this does not pose a problem. It is also not an obstacle for achieving the texture of the LQ coupling ${\bf x}$, as shown in eq.~(\ref{eq:texturesxylambda_ex}). From the viewpoint of the residual symmetry $Z_{17}^\mathrm{diag}$, one can argue that not only the mass of the top and of the bottom quark are invariant under $Z_{17}^\mathrm{diag}$,
but also the elements $M_{u,23}$ and $M_{d,23}$ of the up-type and down-type quark mass matrix.\label{foot2}}
 At the same time, this induces $\theta_{13} \sim\lambda^3$, where the additional suppression factor $\lambda$ arises from the Cabibbo angle.\footnote{This leads to issues with generating a large enough value for $V_{td}$ and for the Jarlskog invariant $J_{\mathrm{CP}}$~\cite{Jarlskog:1985ht}
as well as to a too tight relation between the CKM mixing matrix elements $V_{us}$, $V_{ub}$ and $V_{cb}$, as we comment
 in section~\ref{subsubsec:quarkmixing}. However, generating $\theta_{13}$ directly through an element $M_{u,13}$ of the up-type quark mass matrix of the order $\lambda^3 \, \braket{H_u^0}$ would have as immediate consequence that also
 an element $M_{d,13}$ of the order $\lambda^3 \, \braket{H_d^0}$ is produced in the down-type quark mass matrix, since the combination of fields $\overline{Q} \, H_u \, u_{R3}$
 and $\overline{Q} \, H_d \, d_{R3}$ transforms in the same way. Such a large element in the down-type quark mass matrix leads, upon re-diagonalisation of the latter, to
 rather large elements in the first column of the LQ coupling ${\bf x}$. For this
 reason, we prefer to neither generate the element $M_{u,13}$ nor $M_{d,13}$ through operators, respecting all symmetries of the model.\label{foot3}} 
 
 The correct order of the up quark mass arises from the operator $\overline{Q} \, H_u \, u_{R1} \, T^2 \, U$ which is automatically invariant under
the flavour symmetry $G_f$ and leads to the term $\overline{u_{L1}} \, \braket{H_u^0} \, u_{R1} \, \lambda^8$, after flavour and electroweak symmetry breaking.

%%%%%%%%%%%%%%%%%%%%%%%%%%%%%%%%%%%%%%%%%%%%%%%%%%%%%%
\subsection{Summary of flavour symmetry, particle content and spurions}
\label{subsec:summarychoices}
%%%%%%%%%%%%%%%%%%%%%%%%%%%%%%%%%%%%%%%%%%%%%%%%%%%%

\noindent In this section, we briefly summarise the essential information about the model. The flavour symmetry is
\begin{equation}
G_f = D_{17} \times Z_{17} \; 
\end{equation}
and the particle content is given in table~\ref{tab:particles}. The VEVs of the spurions are 
\begin{align}
\label{eq:spurionVEVs}
  \braket{S} & = \begin{pmatrix} \lambda \\ 0\end{pmatrix}\; , &
  \braket{T} & = \begin{pmatrix} \lambda^2 \\ 0\end{pmatrix}\; , &
  \braket{U} & = \begin{pmatrix} 0 \\ \lambda^4\end{pmatrix}\; , &
  \braket{W} & = \begin{pmatrix} \lambda^5 \\ \lambda^4\end{pmatrix}\; .
\end{align}
Since we do not address the potential of these spurions and thus also not how their VEVs can be correctly aligned, we also do not discuss possible perturbations of these VEVs and their potential
impact on the results for charged fermion masses, quark mixing and the form of the LQ couplings ${\bf x}$ and ${\bf y}$ in terms of the symmetry breaking parameter $\lambda$. 

%%%%%%%%%%%%%%%%%%%%%%%%%%%%%%%%%%%%%%%%%%%%%%%%%%%%%%%%%%%%%%%%
\begin{table}[th!]\centering
  \def\arraystretch{1.4} 
   \begin{tabular}{|l|ccc|cc|}
  \hline
    Field & SU(3) & SU(2) & U(1) & $D_{17}$ & $Z_{17}$ \\\hline
    $Q={\left( \begin{array}{c} Q_1 \\ Q_2 \end{array}
    \right)}$ & 3 & 2 & $\tfrac16$ & ${\bf 2_2}$ & 1 \\[0.02in]
    $Q_{3}$ & 3 & 2 & $\tfrac16$ & ${\bf 1_1}$ &  16 \\[0.02in]
    $u_{R1}$ & 3 & 1 & $\tfrac23$ & ${\bf 1_2}$ &13 \\[0.02in]
    $u_{R2}$ & 3 & 1 & $\tfrac23$ & ${\bf 1_1}$ & 8\\[0.02in]
    $u_{R3}$ & 3 & 1 & $\tfrac23$ & ${\bf 1_1}$ & 1\\[0.02in]
    $d_{R}={\left( \begin{array}{c} d_{R1} \\ d_{R2} \end{array}
    \right)}$ & 3 & 1 & $-\tfrac13$ & ${\bf 2_4}$ & 1\\[0.02in]
    $d_{R3}$ & 3 & 1 & $-\tfrac13$ & ${\bf 1_1}$ &  7 \\[0.02in]
    $L={\left( \begin{array}{c} L_1 \\ L_2 \end{array}
    \right)}$ & 1 & 2 & $-\tfrac12$ & ${\bf 2_1}$ & 2 \\[0.02in]
    $L_{3}$ & 1 & 2 & $-\tfrac12$ & ${\bf 1_1}$ & 1 \\[0.02in]
    $e_{R}={\left( \begin{array}{c} e_{R1} \\ e_{R2} \end{array}
    \right)}$ & 1 & 1 & $-1$ & ${\bf 2_3}$ & 2 \\[0.02in]
    $e_{R3}$ & 1 & 1 & $-1$ & ${\bf 1_1}$ & 9\\[0.02in]
    \hline
    $H_u$ &  1 & 2& $-\tfrac12$ & ${\bf 1_1}$ & 15 \\[0.02in]
    $H_d$ &  1 & 2& $\tfrac12$ & ${\bf 1_1}$ & 9 \\[0.02in]
    $\phi$ &  $3$ & 1 & $-\tfrac13$ & ${\bf 1_1}$ & $0$ \\[0.02in]
    \hline
    $S={\left( \begin{array}{c} S_1 \\ S_2 \end{array}
    \right)}$ & 1 & 1 & $0$ & ${\bf 2_1}$ & 16 \\[0.02in]
    $T={\left( \begin{array}{c} T_1 \\ T_2 \end{array}
    \right)}$ & 1 & 1 & $0$ & ${\bf 2_2}$ & 8 \\[0.02in]
    $U={\left( \begin{array}{c} U_1 \\ U_2 \end{array}
    \right)}$ & 1 & 1 & $0$ & ${\bf 2_2}$ & 8 \\[0.02in]
    $W={\left( \begin{array}{c} W_1 \\ W_2 \end{array}
    \right)}$ & 1 & 1 & $0$ & ${\bf 2_2}$ & 12\\
    \hline
  \end{tabular}
  \caption{{\small {\bf Particle content of the model}. The fermions, scalar fields and spurions (flavour symmetry breaking fields) and their transformation properties under the SM gauge group $\text{SU}(3)\times\text{SU}(2)\times\text{U}(1)$~as well as the
		  flavour symmetry $G_f =D_{17} \times Z_{17}$ are given. Particles in a   
		  two-dimensional irreducible representation of $D_{17}$ are evidenced as two-component vector.
\label{tab:particles}}}
\end{table}
%%%%%%%%%%%%%%%%%%%%%%%%%%%%%%%%%%%%%%%%%%%%%%%%%%%%%%%%%%%%%%%%

%%%%%%%%%%%%%%%%%%%%%%%%%%%%%%%%%%%%%%%%%%%%%%%%%%%%%%
\section{Mass matrices and leptoquark couplings}
\label{sec:Yukawas_LQcouplings}
%%%%%%%%%%%%%%%%%%%%%%%%%%%%%%%%%%%%%%%%%%%%%%%%%%%%%%

In this section, we list the operators contributing to the charged fermion mass matrices $M_u$, $M_d$ and $M_e$, and to the LQ couplings ${\bf \hat{x}}$ and ${\bf \hat{y}}$ from eq.~(\ref{eq:LagintLQcouplings}). We do so for operators that contribute up to and including order $\lambda^{12}$ in the symmetry breaking parameter, and assume that the VEVs of the spurions $S$, $T$, $U$ and $W$ are of the form given in eq.~(\ref{eq:spurionVEVs}). Each operator is accompanied by a complex order-one coefficient. The lists of operators are usually ordered according to the number of spurion insertions. We further emphasise that the spurions are treated as dimensionless flavour symmetry breaking fields. Thus, no cutoff scale needs to be introduced to achieve the correct mass dimension of the operators.

In the lists, the operators stand for all possible combinations of the involved fields which lead to an invariant of the flavour symmetry $G_f$. Thus, they can correspond to more than one independent contribution to the charged fermion mass matrices $M_u$, $M_d$ and $M_e$ or the LQ couplings ${\bf \hat{x}}$ and ${\bf \hat{y}}$. We take this into account in the computation, and signal them by using primed coefficients in those instances, e.g.~$\alpha^d_8$, $(\alpha^d_8)^\prime$~--~see eq.~(\ref{eq:Mdpararel}) in appendix~\ref{app:relatepara}. 

We generally omit all operators with insertions of powers of $S \, S^\dagger$, $T \, T^\dagger$, $U \, U^\dagger$ and $W \, W^\dagger$, and products thereof. Typically, these duplicate the contribution from the operator without this insertion, but have at least an additional suppression of order $\lambda^2$, $\lambda^4$ and $\lesssim \lambda^8$, respectively. There are two exceptions to this rule: $a)$ the subleading-order (in $\lambda$,  SLO) contribution to the elements $M_{u,12}$ and $M_{u,22}$ of the up-type quark mass matrix involving $S \, S^\dagger$, and $b)$ operators involving the insertion $W \, W^\dagger$. For exception $a)$, the elements $M_{u,12}$ and $M_{u,22}$ carry the same parameter dependence at leading order (in $\lambda$, LO), generated by the second operator in eq.~(\ref{eq:upopsLO}), but they receive partially different SLO corrections at relative order $\lambda^2$ from the first and second operators in eq.~(\ref{eq:upopsSLO}), seen explicitly in eq.~(\ref{eq:Mupararel}) in appendix~\ref{app:relatepara}. For exception $b)$, $W \, W^\dagger$ also contains a covariant in ${\bf 2_4}$ with a non-vanishing VEV, so some operators with this insertion can lead to non-redundant contributions~--~although they are always suppressed by at least~$\lambda^8$ with respect to contributions from operators without this insertion. 

After listing the operators, we present the form of the charged fermion mass matrices $M_u$, $M_d$ and $M_e$, analytic formulae for charged fermion masses, and the unitary matrices for LH and RH fermions, needed in order to arrive at the charged fermion mass basis. The latter are necessary to compute the LQ couplings ${\bf x}$ and ${\bf y}$ in eq.~(\ref{eq:LagmassLQcouplings}) from ${\bf \hat{x}}$
and ${\bf \hat{y}}$, respectively, in eq.~(\ref{eq:LagintLQcouplings}). We also explicitly detail the form of the LQ coupling ${\bf z}$, appearing in eq.~(\ref{eq:LagmassLQcouplings}). 

All matrices, $M_u$, $M_d$, $M_e$, ${\bf \hat{x}}$, ${\bf \hat{y}}$,
${\bf x}$, ${\bf y}$ and ${\bf z}$, are given in an effective parametrisation, where the parameters are related to the coefficients of the contributing operators. For completeness, these relations can be found in appendix~\ref{app:relatepara}. For the analytic computations in this section we assume all parameters to be real, but note that they are taken to be complex-valued in subsequent phenomenological studies.

When computing the CKM mixing matrix $V_\text{CKM}$, we find analytically and numerically that this model (as outlined so far) cannot be in
full agreement with experimental data. Specifically, we find that a large enough value for $V_{td}$ and for the Jarlskog invariant $J_{\mathrm{CP}}$, as well as a correct value of $V_{ub}$ (with $V_{us}$ and $V_{cb}$ already fixed) cannot be produced. We comment on this in section~\ref{subsec:quarksector}, and point out how this issue can be solved with a slight change in the form of the up-type quark mass matrix $M_u$, i.e.~by enhancing the
element $M_{u,13}$ to be of the order $\lambda^3 \, \braket{H_u^0}$. We call the results obtained in this model \emph{without} modification of $M_u$ `scenario A', and those \emph{with} the modification of $M_u$ `scenario B'. The form of the LQ couplings ${\bf x}$, ${\bf y}$ and ${\bf z}$ is computed in both scenarios.

The quark sector is discussed in section~\ref{subsec:quarksector}, while the charged lepton sector is addressed in section~\ref{subsec:chargedleptonsector}. Section~\ref{subsec:LQcouplings} is dedicated to the LQ couplings. 

%%%%%%%%%%%%%%%%%%%%%%%%%%%%%%%%%%%%%%%%%%%%%%%%%%%%%%
\subsection{Quark sector}
\label{subsec:quarksector}
%%%%%%%%%%%%%%%%%%%%%%%%%%%%%%%%%%%%%%%%%%%%%%%%%%%%%%

Here we discuss the results for the up-type quark mass matrix $M_u$ and the down-type quark mass matrix $M_d$. We then move on to address the CKM mixing matrix in the aforementioned scenarios A and B.

%%%%%%%%%%%%%%%%%%%%%%%%%%%%%%%%%%%%%%%%%%%%%%%%%%%%%%
\subsubsection{Up quark sector}
\label{subsubsec:upquarksector}
%%%%%%%%%%%%%%%%%%%%%%%%%%%%%%%%%%%%%%%%%%%%%%%%%%%%%%

In the up quark sector, we consider four operators at LO that generate the up-type quark masses and the three quark mixing angles.
These four operators read
\begin{equation}
\label{eq:upopsLO}
{\cal L}_{\mathrm{Yuk, LO}}^u=\alpha^u_1 \, \overline{Q_3} \, H_u  \, u_{R3}+ \alpha^u_2 \, \overline{Q}  \, H_u \, u_{R2} \, W + \alpha^u_3 \, \overline{Q} \, H_u \, u_{R3}  \, (S^\dagger)^2 + \alpha^u_4 \, \overline{Q}  \, H_u \, u_{R1} \, T^2 \, U  .
\end{equation}
At SLO, the following operators give contributions up to and including $\lambda^{12}$ to the up-type quark mass matrix
\begin{eqnarray}
\label{eq:upopsSLO}
{\cal L}_{\mathrm{Yuk, SLO}}^u&=&\alpha^u_5 \,  \overline{Q}  \, H_u \, u_{R2} \, S \, S^\dagger \, W + \alpha^u_6 \, \overline{Q}\, H_u \, u_{R2}  \, (S^\dagger)^4 \, T +\alpha^u_7 \, \overline{Q_3} \, H_u \, u_{R2} \, (S^\dagger)^2 \, T  \;\;\;\;
\\ \nonumber
&+&\alpha^u_8 \, \overline{Q_3} \, H_u \, u_{R2} \, (W^\dagger)^2+ \alpha^u_9 \, \overline{Q}  \, H_u \, u_{R2} \, W^2 \, W^\dagger + \alpha^u_{10} \, \overline{Q} \, H_u \, u_{R1}  \, T \, U^2
\\ \nonumber
&+&\alpha^u_{11} \, \overline{Q_3}  \, H_u \, u_{R2} \, S^2 \, W +\alpha^u_{12} \, \overline{Q} \, H_u \, u_{R2} \, T^\dagger \, U \, W + \alpha^u_{13} \, \overline{Q}  \, H_u\, u_{R2} \, T \, U^\dagger \, W
\\ \nonumber
&+& \alpha^u_{14} \, \overline{Q} \, H_u \, u_{R3}  \, T^\dagger \, (W^\dagger)^2 +\alpha^u_{15} \, \overline{Q} \, H_u \, u_{R3} \, (S^\dagger)^2 \, T\, U^\dagger +\alpha^u_{16} \, \overline{Q} \, H_u\, u_{R3}  \, S^2 \, T^\dagger \, W
\\ \nonumber
&+& \alpha^u_{17} \, \overline{Q} \, H_u\, u_{R3} \, S^2 \, U^\dagger \, W  +\alpha^u_{18} \, \overline{Q} \, H_u\, u_{R2}  \, S^2 \, T^\dagger \, W^\dagger +\alpha^u_{19} \, \overline{Q} \, H_u \, u_{R2} \, S^2 \, U^\dagger \, W^\dagger
\\ \nonumber
&+&\alpha^u_{20} \, \overline{Q} \, H_u\, u_{R2}  \, (S^\dagger)^2 \, (W^\dagger)^2  + \alpha^u_{21} \, \overline{Q} \, H_u \, u_{R3} \, (S^\dagger)^2 \, W \, W^\dagger
\\ \nonumber
&+&\alpha^u_{22} \, \overline{Q_3} \, H_u \, u_{R1}  \, S^2 \, T \, U^2+ \alpha^u_{23} \, \overline{Q_3} \, H_u \, u_{R1}  \, (S^\dagger)^4 \, (U^\dagger)^2
\\ \nonumber
&+&\alpha^u_{24} \, \overline{Q_3} \, H_u \, u_{R3}  \, S^4 \, T^\dagger \, W +\alpha^u_{25} \, \overline{Q_3}\, H_u \, u_{R3} \, (S^\dagger)^4 \, T \, W^\dagger
\\ \nonumber
&+&\alpha^u_{26} \, \overline{Q_3} \, H_u\, u_{R2}  \, S^4 \, T^\dagger \, W^\dagger + \alpha^u_{27} \, \overline{Q} \, H_u\, u_{R2}  \, (S^\dagger)^4 \, T^2 \, U^\dagger
\\ \nonumber
&+& \alpha^u_{28} \, \overline{Q} \, H_u\, u_{R3}  \, S^4 \, (T^\dagger)^2 \, W^\dagger + \alpha^u_{29} \, \overline{Q} \, H_u \, u_{R2}  \, S^6 \, (T^\dagger)^2
\\ \nonumber
&+&\alpha^u_{30} \, \overline{Q}\, H_u  \, u_{R3} \, (S^\dagger)^6 \, T \, W^\dagger \; .
\end{eqnarray}
\noindent Of the operators in eq.~(\ref{eq:upopsSLO}), the first two are the most important, since they contribute at relative order $\lambda^2$ to the elements $M_{u, 12}$ and $M_{u, 22}$. The operators with the coefficients $\alpha^u_6$ and $\alpha^u_7$ are examples of operators that appear automatically once the field content of the LO operators is determined. We note that several of these operators lead to two independent contributions to the up-type quark mass matrix $M_u$. The operator with the coefficient $\alpha^u_5$ induces contributions of order $\lambda^6$ to the element $M_{u,22}$ and of $\lambda^7$ to the element $M_{u,12}$, but with a different relative sign; the one with $\alpha^u_{14}$ gives contributions of order $\lambda^{10}$ and $\lambda^{11}$; the one with $\alpha^u_{16}$ yields contributions of order $\lambda^8$ and $\lambda^9$; the one with $\alpha^u_{18}$ leads to contributions of order $\lambda^8$ and $\lambda^9$; finally, the operator with the coefficient $\alpha^u_{20}$ gives rise to two independent contributions of order $\lambda^{10}$ and $\lambda^{11}$, respectively.

The up-type quark mass matrix can thus be effectively parametrised, up to and including order $\lambda^{12}$, as
\begin{equation}
\label{eq:Mupara}
M_u = \left(
\begin{array}{ccc}
f_{11} \, \lambda^8 & f_{12} \, \lambda^5 & f_{13} \, \lambda^8\\
f_{21} \, \lambda^{10} & f_{22} \, \lambda^4 & f_{23} \, \lambda^2\\
f_{31} \, \lambda^{12} & f_{32} \, \lambda^4 & f_{33}
\end{array}
\right) \; \braket{H_u^0} \; ,
\end{equation}
where $f_{ij}$ are generally independent, complex order-one numbers, apart from $f_{12}$ and $f_{22}$. The latter fulfil the relation
\begin{equation}
\label{eq:f12f22rel}
f_{12}-f_{22} \sim c \, \lambda^2
\end{equation}
with $c$ being complex.\footnote{In order to reflect this relation better in the effective parametrisation of $M_u$, one can express the element $M_{u,12}$ as $(f_{22} + \tilde{f}_{12} \lambda^2)\; \lambda^5 \braket{H_u^0}$ instead, where $\tilde{f}_{12}$ is a complex order-one number.} As mentioned above, the first two operators in eq.~(\ref{eq:upopsSLO}), with the coefficients $\alpha^u_5$ and $\alpha^u_6$, are the source of this difference~--~see also eq.~(\ref{eq:Mupararel}) in appendix~\ref{app:relatepara}. The expressions for the other parameters $f_{ij}$ in terms of the coefficients $\alpha^u_i$ are given in eq.~(\ref{eq:Mupararel}) in appendix~\ref{app:relatepara} as well.

From the effective parametrisation of $M_u$, we can derive expressions for the up-type quark masses. Note that in order to clearly show these results here (and in the following), we only explicitly mention the most relevant terms. Thus, the quark masses can be expressed as
\begin{eqnarray}
\label{eq:mumasses}
m_u&=&  \, \left| f_{11} \, \lambda^8 + \mathcal{O}(\lambda^{10}) \right| \, \braket{H_u^0} \; ,
\\ \nonumber
m_c&=&  \left| f_{22} \, \lambda^4 + \left( \frac{f_{12}^2}{2 \, f_{22}} - \frac{f_{23} f_{32}}{f_{33}} \right) \, \lambda^6 + \mathcal{O}(\lambda^8) \right| \, \braket{H_u^0} \; ,
\\ \nonumber
m_t&=& \left| f_{33} + \frac{f_{23}^2}{2 \, f_{33}} \, \lambda^4 + \mathcal{O}(\lambda^8) \right| \, \braket{H_u^0} \; .
\end{eqnarray}
We confirm that the dominant contributions to the three different masses come from the first, second and fourth operator in eq.~(\ref{eq:upopsLO}), as expected from the construction of the model.
The matrices $L_u$ and $R_u$ transforming LH and RH up-type quarks from the interaction to the mass basis read, up to and including order $\lambda^{12}$,
\begin{equation}
\label{eq:Luform}
L_u = \left(
\begin{array}{ccc}
1 - \frac{f_{12}^2}{2 \, f_{22}^2} \, \lambda^2 + \mathcal{O} (\lambda^4)& \frac{f_{12}}{f_{22}} \, \lambda + \mathcal{O}(\lambda^3) & \frac{f_{13}}{f_{33}} \, \lambda^8 + \mathcal{O}(\lambda^9)\\
-\frac{f_{12}}{f_{22}} \, \lambda + \mathcal{O}(\lambda^3)& 1- \frac{f_{12}^2}{2 \, f_{22}^2} \, \lambda^2 + \mathcal{O}(\lambda^4)& \frac{f_{23}}{f_{33}} \, \lambda^2 + \mathcal{O}(\lambda^6)\\
\frac{f_{12} f_{23}}{f_{22} f_{33}} \, \lambda^3 + \mathcal{O}(\lambda^5) & -\frac{f_{23}}{f_{33}} \, \lambda^2 +\mathcal{O}(\lambda^4) & 1- \frac{f_{23}^2}{2 \, f_{33}^2} \, \lambda^4 + \mathcal{O}(\lambda^8)
\end{array}
\right) 
\end{equation}
and
\begin{equation}
\label{eq:Ruform}
R_u= \left(
\begin{array}{ccc}
1+ \mathcal{O}(\lambda^{10})& \frac{f_{11} f_{12}}{f_{22}^2} \, \lambda^5 + \mathcal{O}(\lambda^6)& \frac{f_{21} f_{23} + f_{31} f_{33}}{f_{33}^2} \, \lambda^{12} + \mathcal{o}(\lambda^{12})\\[0.1in]
- \frac{f_{11} f_{12}}{f_{22}^2} \, \lambda^5 + \mathcal{O}(\lambda^6)& 1 + \mathcal{O}(\lambda^8)& \frac{f_{32}}{f_{33}} \, \lambda^4 + \mathcal{O}(\lambda^6)\\[0.1in]
\frac{f_{11} f_{12} f_{32}}{f_{22}^2 f_{33}} \, \lambda^9 + \mathcal{O} (\lambda^{10})& -\frac{f_{32}}{f_{33}} \, \lambda^4 + \mathcal{O}(\lambda^6)& 1+ \mathcal{O}(\lambda^8)
\end{array}
\right) \, .
\end{equation}

\noindent We note that $L_u$ is the primary source of the CKM mixing matrix, whereas the matrix $R_u$ should be close to the identity matrix~--~in accordance with the basis in which the textures of the LQ couplings ${\bf x}$ and ${\bf y}$ are given in  eq.~(\ref{eq:texturesxylambda_ex}). The obtained forms of $L_u$ and $R_u$ fulfil these requirements to a good degree. 

The largest deviation of $R_u$ from the identity matrix is of order $\lambda^4$,  due to the operator with the coefficient $\alpha^u_7$ that appears automatically. As can be seen below, this deviation in $R_u$ is partly responsible for the generation of the element $y_{33}$ of the LQ coupling ${\bf \hat{y}}$ in the charged fermion mass basis of order $\lambda^4$~--~compare eq.~(\ref{eq:yparaA}) and eq.~(\ref{eq:ypararelA}) in appendix~\ref{app:relatepara}. Furthermore, $R_{u,21} \sim \lambda^5$ together with $\hat{y}_{32} \sim 1$, see eq.~(\ref{eq:hatypara}), has particular relevance for the texture of the same LQ coupling ${\bf y}$, leading to $y_{31} \sim \lambda^5$, see eq.~(\ref{eq:yparaA}) and eq.~(\ref{eq:ypararelA}) in appendix~\ref{app:relatepara}.

%%%%%%%%%%%%%%%%%%%%%%%%%%%%%%%%%%%%%%%%%%%%%%%%%%%%%%
\subsubsection*{Introducing scenario B}
\label{subsubsubsec:introscenarioB}
%%%%%%%%%%%%%%%%%%%%%%%%%%%%%%%%%%%%%%%%%%%%%%%%%%%%%%

One can already infer from the form of the matrix $L_u$ in eq.~(\ref{eq:Luform}) that the CKM mixing matrix element $V_{td}$ as well as the Jarlskog invariant, $J_{\mathrm{CP}}$, are likely to be very suppressed. This 
suppression originates from the entry $L_{u,13}$, which is only of order $\lambda^8$.
Furthermore, the tight relation between the elements $L_{u,21}$, $L_{u,31}$ and $L_{u,32}$ leads to a too-strong correlation between $V_{us}$, $V_{ub}$ and $V_{cb}$. These points are discussed further in section~\ref{subsubsec:quarkmixing}.

A simple way to resolve these issues is to enhance the element $M_{u,13}$ in the up-type quark mass matrix $M_u$, namely
\begin{equation}
\label{eq:MuparaB}
M_u = \left(
\begin{array}{ccc}
f_{11} \, \lambda^8 & f_{12} \, \lambda^5 & \tilde{f}_{13} \, \lambda^3\\
f_{21} \, \lambda^{10} & f_{22} \, \lambda^4 & f_{23} \, \lambda^2\\
f_{31} \, \lambda^{12} & f_{32} \, \lambda^4 & f_{33}
\end{array}
\right) \; \braket{H_u^0}
\end{equation}
with $\tilde{f}_{13}$ being a complex order-one number. Adding \emph{ad hoc} a further contribution to the element $M_{u,13}$ is not explained by an appropriate operator in the context of this model. From the arguments given in footnote~\ref{foot3},
it is, however, likely that such a contribution can only be generated by either an operator which explicitly breaks the flavour symmetry $G_f$ or by changing at least part of the fermion assignment and/or $G_f$.

The up-type quark masses are mostly unaffected by this change, except that the SLO term in the top quark mass, see eq.~(\ref{eq:mumasses}), is slightly enhanced and of order $\lambda^6$. The matrices $L_u$
and $R_u$ read
\begin{equation}
\label{eq:LuformB}
L_u= \left(
\begin{array}{ccc}
1 - \frac{f_{12}^2}{2 \, f_{22}^2} \, \lambda^2 + \mathcal{O} (\lambda^4)& \frac{f_{12}}{f_{22}} \, \lambda + \mathcal{O}(\lambda^3) & \frac{\tilde{f}_{13}}{f_{33}} \, \lambda^3 + \mathcal{O}(\lambda^7)\\[0.05in]
-\frac{f_{12}}{f_{22}} \, \lambda + \mathcal{O}(\lambda^3)& 1- \frac{f_{12}^2}{2 \, f_{22}^2} \, \lambda^2 + \mathcal{O}(\lambda^4)& \frac{f_{23}}{f_{33}} \, \lambda^2 + \mathcal{O}(\lambda^6)\\[0.05in]
\Big(\frac{f_{12} f_{23}}{f_{22} f_{33}} - \frac{\tilde{f}_{13}}{f_{33}}\Big) \, \lambda^3 + \mathcal{O}(\lambda^5) & -\frac{f_{23}}{f_{33}} \, \lambda^2 +\mathcal{O}(\lambda^4) & 1- \frac{f_{23}^2}{2 \, f_{33}^2} \, \lambda^4 + \mathcal{O}(\lambda^6)
\end{array}
\right)
\end{equation}
and
\begin{equation}
\label{eq:RuformB}
R_u= \left(
\begin{array}{ccc}
1+ \mathcal{O}(\lambda^{10})& \frac{f_{11} f_{12}}{f_{22}^2} \, \lambda^5 + \mathcal{O}(\lambda^6)& \frac{f_{11} \tilde{f}_{13}}{f_{33}^2} \, \lambda^{11} + \mathcal{O}(\lambda^{12})\\[0.1in]
- \frac{f_{11} f_{12}}{f_{22}^2} \, \lambda^5 + \mathcal{O}(\lambda^6)& 1 + \mathcal{O}(\lambda^8)& \frac{f_{32}}{f_{33}} \, \lambda^4 + \mathcal{O}(\lambda^6)\\[0.1in]
\frac{f_{11} f_{12} f_{32}}{f_{22}^2 f_{33}} \, \lambda^9 + \mathcal{O} (\lambda^{10})& -\frac{f_{32}}{f_{33}} \, \lambda^4 + \mathcal{O}(\lambda^6)& 1+ \mathcal{O}(\lambda^8)
\end{array}
\right) \; .
\end{equation}

\noindent As expected, the element $L_{u,13}$ in the matrix $L_u$ is now of order $\lambda^3$ and the tight relation between the elements $L_{u,21}$, $L_{u,31}$ and $L_{u,32}$ is relaxed. In this way, all mentioned short-comings of the resulting CKM mixing matrix are remedied~--~see further discussion in section~\ref{subsubsec:quarkmixing}.
The matrix $R_u$ is very mildly affected by this change in $M_u$, since only the element $R_{u,13}$ is enhanced to order $\lambda^{11}$.

%%%%%%%%%%%%%%%%%%%%%%%%%%%%%%%%%%%%%%%%%%%%%%%%%%%%%%
\subsubsection{Down quark sector}
\label{subsubsec:downquarksector}
%%%%%%%%%%%%%%%%%%%%%%%%%%%%%%%%%%%%%%%%%%%%%%%%%%%%%%

At LO there are three operators responsible for the generation of the down-type quark masses: one arising at tree level, and the other two requiring the insertion of one spurion, $T$ or $U$. Thus, we have
\begin{equation}
\label{eq:downopsLO}
{\cal L}_{\mathrm{Yuk, LO}}^d=\alpha^d_1 \, \overline{Q_3} \, H_d \, d_{R3} + \alpha^d_2 \, \overline{Q} \, H_d \, d_R \, T + \alpha^d_3 \, \overline{Q} \, H_d \, d_R \, U \; .
\end{equation}
At SLO, we find several more operators
\begin{eqnarray}
\label{eq:downopsSLO}
{\cal L}_{\mathrm{Yuk, SLO}}^d&=& \alpha^d_4 \, \overline{Q} \, H_d \, d_{R3} \, (S^\dagger)^2 + \alpha^d_5 \, \overline{Q_3} \, H_d \, d_R  \, S^2 \, T + \alpha^d_6 \, \overline{Q} \, H_d \, d_R \, T^\dagger \, U^2
\\ \nonumber
&+& \alpha^d_7 \, \overline{Q}  \, H_d \, d_R \, T^2 \, U^\dagger + \alpha^d_8 \, \overline{Q} \, H_d \, d_{R3} \, T^\dagger \, (W^\dagger)^2 + \alpha^d_9 \, \overline{Q} \, H_d  \, d_R \, T \,  W \, W^\dagger
\\ \nonumber
&+&\alpha^d_{10} \, \overline{Q} \, H_d \, d_{R3} \, (S^\dagger)^2 \, T \, U^\dagger +\alpha^d_{11} \, \overline{Q} \, H_d \, d_{R3} \, S^2 \, T^\dagger \, W+\alpha^d_{12} \, \overline{Q} \, H_d \, d_{R3} \, S^2 \, U^\dagger \, W
\\ \nonumber
&+&\alpha^d_{13} \, \overline{Q} \, H_d \, d_R \, S^2 \, (W^\dagger)^2 +\alpha^d_{14} \, \overline{Q} \, H_d \, d_{R3} \, (S^\dagger)^2 \, W \, W^\dagger +\alpha^d_{15} \, \overline{Q_3} \, H_d \, d_R  \, S^2 \, T^\dagger \, U^2
\\ \nonumber
&+&\alpha^d_{16} \, \overline{Q} \, H_d \, d_R \, S^4 \, W +\alpha^d_{17} \, \overline{Q_3} \, H_d \, d_R \, (S^\dagger)^2 \, T^2 \, W^\dagger +\alpha^d_{18} \, \overline{Q_3} \, H_d \, d_{R3} \, S^4 \, T^\dagger \, W
\\ \nonumber
&+&\alpha^d_{19} \, \overline{Q_3} \, H_d \, d_{R3} \, (S^\dagger)^4 \, T \, W^\dagger +\alpha^d_{20} \, \overline{Q_3} \, H_d \, d_R \, S^6 \, W +\alpha^d_{21} \, \overline{Q} \, H_d \, d_R \, (S^\dagger)^4 \, T^2 \, W^\dagger
\\ \nonumber
&+&\alpha^d_{22} \, \overline{Q} \, H_d \, d_{R3} \, S^4 \, (T^\dagger)^2 \, W^\dagger +\alpha^d_{23} \, \overline{Q} \, H_d \, d_{R3}  \, (S^\dagger)^6 \, T \, W^\dagger
\\ \nonumber
&+&\alpha^d_{24} \, \overline{Q} \, H_d \, d_R \, S^6 \, T^\dagger \, W^\dagger +\alpha^d_{25} \, \overline{Q} \, H_d \, d_R \, (S^\dagger)^7 \, (T^\dagger)^2.
\end{eqnarray}
\noindent The first operator in eq.~(\ref{eq:downopsSLO}) has been discussed already, since it is automatically present once the corresponding operator in the up quark sector is considered. Similarly, the existence of the second operator in this list (with the coefficient $\alpha^d_5$) is automatic, once we have accounted for the LO operators generating the dominant structures in the charged fermion mass matrices and in the LQ couplings ${\bf \hat{x}}$ and ${\bf \hat{y}}$. 
 
 We note that the operators with the following coefficients lead to more than one independent contraction, and hence contribution, to the down-type quark mass matrix. The operator with $\alpha^d_8$ leads to two independent contributions of order $\lambda^{10}$ and $\lambda^{11}$; the one with $\alpha^d_9$ yields two contributions, both of order $\lambda^{11}$; the operator with $\alpha^d_{11}$ gives two contributions of order $\lambda^8$ and $\lambda^9$; the one with $\alpha^d_{13}$ leads to three contributions of order $\lambda^{10}$, $\lambda^{11}$
and $\lambda^{12}$; finally, the operator with the coefficient $\alpha^d_{16}$ leads to two contributions of order $\lambda^8$ and $\lambda^9$.

The effective parametrisation of the down-type quark mass matrix, including all contributions up to and including order $\lambda^{12}$, therefore reads
\begin{equation}
\label{eq:Mdpara}
M_d = \left(
\begin{array}{ccc}
d_{11} \, \lambda^4 & d_{12} \, \lambda^8 & d_{13} \, \lambda^8\\
d_{21} \, \lambda^{10} & d_{22} \, \lambda^2 & d_{23} \, \lambda^2\\
d_{31} \, \lambda^{12} & d_{32} \, \lambda^4 & d_{33}
\end{array}
\right) \, \braket{H_d^0}
\end{equation}
with $d_{ij}$ being, in general, independent complex order-one numbers, related to the coefficients $\alpha^d_i$ as shown in eq.~(\ref{eq:Mdpararel}) in appendix~\ref{app:relatepara}. Furthermore, we arrive at the down-type quark masses 
\begin{eqnarray}
\label{eq:mdmasses}
m_d&=& \left|d_{11} \, \lambda^4 + \mathcal{O} (\lambda^{12}) \right| \, \braket{H_d^0} \; ,
\\ \nonumber
m_s&=& \left| d_{22} \, \lambda^2 - \frac{d_{23} (d_{22} d_{23}+2 \, d_{32} d_{33})}{2 \, d_{33}^2} \, \lambda^6 + \mathcal{O} (\lambda^{10}) \right| \, \braket{H_d^0} \; ,
\\ \nonumber
m_b&=& \left| d_{33} + \frac{d_{23}^2}{2 \, d_{33}} \, \lambda^4 + \mathcal{O}  (\lambda^8) \right| \, \braket{H_d^0} \; ,
\end{eqnarray}
with the dominant contributions, arising from the three operators in eq.~(\ref{eq:downopsLO}), as expected from the construction of the model.

For the matrices $L_d$ for LH, and $R_d$ for RH down-type quarks we have up to and including order $\lambda^{12}$
\begin{equation}
\label{eq:Ldform}
L_d= \left(
\begin{array}{ccc}
1 - \frac{d_{12}^2}{2 \, d_{22}^2} \, \lambda^{12} + \mathcal{o} (\lambda^{12})& \frac{d_{12}}{d_{22}} \, \lambda^6 + \mathcal{O} (\lambda^{10})& \frac{d_{13}}{d_{33}} \, \lambda^8 + \mathcal{O}(\lambda^{12})\\
-\frac{d_{12}}{d_{22}} \, \lambda^6 + \mathcal{O} (\lambda^{10})& 1 - \frac{d_{23}^2}{2 \, d_{33}^2} \, \lambda^4 +\mathcal{O} (\lambda^8) & \frac{d_{23}}{d_{33}} \, \lambda^2 + \mathcal{O} (\lambda^6)\\
- \frac{(d_{13} d_{22}-d_{12} d_{23})}{d_{22} d_{33}} \, \lambda^8+ \mathcal{O} (\lambda^{12})& -\frac{d_{23}}{d_{33}} \, \lambda^2 + \mathcal{O} (\lambda^6)& 1 - \frac{d_{23}^2}{2 \, d_{33}^2}\, \lambda^4 + \mathcal{O}(\lambda^8)
\end{array}
\right)
\end{equation}
and\\[0.05in]
$
R_d=
$
\begin{equation}
 \label{eq:Rdform}
\!\!\!\left(
\begin{array}{ccc}
1 + \mathcal{o}(\lambda^{12})& \frac{(d_{11} d_{12} + d_{21} d_{22})}{d_{22}^2} \, \lambda^8 +\mathcal{O}(\lambda^{12})& \frac{d_{11} d_{13}+d_{21} d_{23}+d_{31} d_{33}}{d_{33}^2} \, \lambda^{12} + \mathcal{o}(\lambda^{12})\\
- \frac{(d_{11} d_{12} + d_{21} d_{22})}{d_{22}^2} \, \lambda^8 +\mathcal{O}(\lambda^{12})& 1 + \mathcal{O} (\lambda^8)& \frac{(d_{22} d_{23} + d_{32} d_{33})}{d_{33}^2} \, \lambda^4 + \mathcal{O}(\lambda^8)\\
\mathcal{O}(\lambda^{12})& - \frac{(d_{22} d_{23} + d_{32} d_{33})}{d_{33}^2} \, \lambda^4 + \mathcal{O}(\lambda^8)& 1 + \mathcal{O}(\lambda^8)
\end{array}
\right) \; .
\\[0.1in]
\end{equation}

We can see that both matrices, $L_d$ and $R_d$, are close to the identity matrix, except for the (23)-block in $L_d$ where a rotation of order $\lambda^2$ is present. This result has been anticipated in the preceding
section, see footnote~\ref{foot2}. The effect of this rotation is twofold. On the one hand, it leads to an additional contribution to the quark mixing angle $\theta_{23}$, which is of the same order as the contribution arising from the up quark sector, see $L_u$ in eq.~(\ref{eq:Luform}) and eq.~(\ref{eq:LuformB}) and compare the form of the CKM mixing matrix in eq.~(\ref{eq:VCKMA}) (scenario A) and eq.~(\ref{eq:VCKMB}) (scenario B). On the other hand, it induces contributions to the elements $x_{22}$ and $x_{32}$ of the LQ coupling ${\bf \hat{x}}$ in the charged fermion mass basis, which are of the same order as the elements $\hat{x}_{22}$ and $\hat{x}_{32}$ of the LQ coupling ${\bf \hat{x}}$ itself,
see eq.~(\ref{eq:xpara}) and eq.~(\ref{eq:xpararel}) in appendix~\ref{app:relatepara}. 

%%%%%%%%%%%%%%%%%%%%%%%%%%%%%%%%%%%%%%%%%%%%%%%%%%%%%%
\subsubsection{Quark mixing}
\label{subsubsec:quarkmixing}
%%%%%%%%%%%%%%%%%%%%%%%%%%%%%%%%%%%%%%%%%%%%%%%%%%%%%%
We first present the CKM mixing matrix, $V_{\rm CKM}$, as obtained from the matrices $L_u$ and $L_d$, shown in eq.~(\ref{eq:Luform}) and eq.~(\ref{eq:Ldform}). This reflects the result 
 of the model without modification of the 
up-type quark mass matrix $M_u$, i.e. in scenario A. Here, we find \\

$
V_{\rm CKM}=L_u^\dagger \, L_d =
$\\[0.02in]
\begin{equation}
\label{eq:VCKMA}
\!\!\!\!\!\!\!\!\!\!\!\!\!\!\!\!\!\!\!\!\!\!\!\!\!\!\left(
\begin{array}{ccc}
1-\frac{f_{12}^2}{2 \, f_{22}^2} \, \lambda^2 + \mathcal{O}(\lambda^4)& -\frac{f_{12}}{f_{22}} \, \lambda + \mathcal{O} (\lambda^3)& \frac{f_{12} (d_{33} f_{23}-d_{23} f_{33})}{d_{33} f_{22} f_{33}} \, \lambda^3 +\mathcal{O}(\lambda^5)\\[0.05in]
\frac{f_{12}}{f_{22}} \, \lambda + \mathcal{O} (\lambda^3)& 1-\frac{f_{12}^2}{2 \, f_{22}^2} \, \lambda^2 + \mathcal{O}(\lambda^4)& \left( \frac{d_{23}}{d_{33}} -\frac{f_{23}}{f_{33}}\right) \, \lambda^2 +\mathcal{O}(\lambda^4)\\[0.05in]
\frac{d_{22} d_{33} f_{13}-d_{12} d_{33} f_{23}-d_{13} d_{22} f_{33}+d_{12} d_{23} f_{33}}{d_{22} d_{33} f_{33}} \, \lambda^8 + \mathcal{O}(\lambda^9)& -\left( \frac{d_{23}}{d_{33}} -\frac{f_{23}}{f_{33}}\right) \, \lambda^2 +\mathcal{O}(\lambda^6)& 1-\frac{(d_{33} f_{23}-d_{23} f_{33})^2}{2 \, d_{33}^2 f_{33}^2} \, \lambda^4 + \mathcal{O}(\lambda^8)
\end{array}
\right) \; .
\\[0.1in]
\end{equation}
\noindent There is an obvious suppression of the CKM mixing matrix element $V_{td} \sim \lambda^8$ compared to its experimentally measured value, $|V_{td}|=0.00854^{+0.00023}_{-0.00016}\sim\lambda^3$~\cite{ParticleDataGroup:2020ssz}.
Furthermore, assuming that the effective parameters $d_{ij}$ and $f_{ij}$ are complex, we can estimate the size of the Jarlskog invariant $J_{\mathrm{CP}}$ and see that it is of order $J_{\mathrm{CP}}= \mathrm{Im} \left( V_{ud} \, V_{ub}^* \, V_{td}^* \, V_{tb} \right) \sim \lambda^{11}$. This is in conflict with the measured value, $J_{\mathrm{CP}}=\Big( 3.00^{+0.15}_{-0.09} \Big) \times 10^{-5} \sim \lambda^6$~\cite{ParticleDataGroup:2020ssz}.

In addition, we note that the relation between $V_{us}$, $V_{cb}$ and $V_{ub}$ is too tight to accommodate all three CKM mixing matrix elements in accordance with the experimental data~\cite{ParticleDataGroup:2020ssz}. In this model we have
\begin{equation}
\label{eq:VusVcb}
|V_{us}| \approx \left| \frac{f_{12}}{f_{22}} \right| \, \lambda \sim \lambda \;\; \mbox{and} \;\; |V_{cb}| \approx \left| \frac{d_{23}}{d_{33}} -\frac{f_{23}}{f_{33}} \right| \, \lambda^2 \sim \lambda^2
\end{equation}
and as well
\begin{equation}
\label{eq:Vubrel}
|V_{ub}| \approx \left| \frac{f_{12}}{f_{22}} \, \left( \frac{d_{23}}{d_{33}} - \frac{f_{23}}{f_{33}} \right) \right| \, \lambda^3 \approx |V_{us}| \, |V_{cb}|
\end{equation}
which leads with $|V_{us}| =0.22650$ and $|V_{cb}|=0.04053$~\cite{ParticleDataGroup:2020ssz} to
$|V_{ub}| \approx 0.0092$. This is about a factor of $2.5$ wrong with respect to the experimental best-fit value of $V_{ub}$, $|V_{ub}|=0.00361^{+0.00011}_{-0.00009}$~\cite{ParticleDataGroup:2020ssz}, and clearly outside the range preferred at the $3 \,\sigma$ level.

We have confirmed these findings with a chi-squared fit~--~while the charged fermion masses are fitted well at the scale $\mu=1 \, \mathrm{TeV}$~\cite{Xing:2007fb},
quark mixing cannot be brought into full agreement with experimental data~\cite{ParticleDataGroup:2020ssz}.

%%%%%%%%%%%%%%%%%%%%%%%%%%%%%%%%%%%%%%%%%%%%%%%%%%%%%%
\subsubsection*{In scenario B}
\label{subsubsubsec:scenarioB}
%%%%%%%%%%%%%%%%%%%%%%%%%%%%%%%%%%%%%%%%%%%%%%%%%%%%%%

Using instead the matrix $L_u$ as given in eq.~(\ref{eq:LuformB}), we have for the CKM mixing matrix 
 \begin{equation}
 \begin{aligned}
\label{eq:VCKMB}
&V_{\rm CKM} =\\
&\left( \begin{array}{ccc}
1-\frac{f_{12}^2}{2 \, f_{22}^2} \, \lambda^2 + \mathcal{O}(\lambda^4)& -\frac{f_{12}}{f_{22}} \, \lambda + \mathcal{O} (\lambda^3)& \Big(\frac{f_{12} (d_{33} f_{23}-d_{23} f_{33})}{d_{33} f_{22} f_{33}} - \frac{\tilde{f}_{13}}{f_{33}}\Big) \, \lambda^3 +\mathcal{O}(\lambda^5)\\[0.05in]
\frac{f_{12}}{f_{22}} \, \lambda + \mathcal{O} (\lambda^3)& 1-\frac{f_{12}^2}{2 \, f_{22}^2} \, \lambda^2 + \mathcal{O}(\lambda^4)& \left( \frac{d_{23}}{d_{33}} -\frac{f_{23}}{f_{33}}\right) \, \lambda^2 +\mathcal{O}(\lambda^4)\\[0.05in]
\frac{\tilde{f}_{13}}{f_{33}} \, \lambda^3 + \mathcal{O}(\lambda^7)& -\left( \frac{d_{23}}{d_{33}} -\frac{f_{23}}{f_{33}}\right) \, \lambda^2 +\mathcal{O}(\lambda^6)& 1-\frac{(d_{33} f_{23}-d_{23} f_{33})^2}{2 \, d_{33}^2 f_{33}^2} \, \lambda^4 + \mathcal{O}(\lambda^6)
\end{array}
\right).
\end{aligned}
\end{equation}
As we can clearly see, the anticipated changes in the CKM mixing matrix are achieved: the enhancement of $V_{td}$, which is now of order $\lambda^3$, and in turn the enhancement of the Jarlskog invariant to
$J_{\mathrm{CP}} \sim \lambda^6$, as well as the loosening of the tight relation between $V_{us}$, $V_{cb}$ and $V_{ub}$,
\begin{equation}
\label{eq:VubB}
|V_{ub}| \approx \left| \frac{f_{12}}{f_{22}} \, \left(\frac{d_{23}}{d_{33}} -\frac{f_{23}}{f_{33}} \right)+ \frac{\tilde{f}_{13}}{f_{33}} \right| \, \lambda^3 \; .
\end{equation}
Indeed, a chi-squared fit shows that scenario B leads to an excellent agreement with the experimental data~--~not only of the quark mixing parameters~\cite{ParticleDataGroup:2020ssz},
but also all charged fermion masses are fitted very well at the scale $\mu=1 \, \mathrm{TeV}$~\cite{Xing:2007fb}.

%%%%%%%%%%%%%%%%%%%%%%%%%%%%%%%%%%%%%%%%%%%%%%%%%%%%%%
\subsection{Charged lepton sector}
\label{subsec:chargedleptonsector}
%%%%%%%%%%%%%%%%%%%%%%%%%%%%%%%%%%%%%%%%%%%%%%%%%%%%%%

Like we did for the quark sector, here we first present the list of operators. We then give the form of the charged lepton mass matrix $M_e$ in the effective parametrisation, and extract analytical formulae for the charged lepton masses and the 
 matrices $L_{e}$ and $R_{e}$ of LH and RH charged leptons, respectively, needed in order to arrive at the mass basis. 

Three operators are mainly responsible for the generation of the charged lepton masses, like in the case of the down-type quark masses, namely
\begin{equation}
\label{eq:chlepopsLO}
{\cal L}_{\mathrm{Yuk, LO}}^e= \alpha^e_1 \, \overline{L_3}  \, H_d \, e_{R3} + \alpha^e_2 \, \overline{L} \, H_d \, e_{R}  \, T + \alpha^e_3 \, \overline{L} \, H_d \, e_{R}  \, U \; .
\end{equation}
As envisaged in the construction of this model, these have the analogous form as those found in the down quark sector, compare eq.~(\ref{eq:downopsLO}).
The operators, arising at SLO, differ in general
\begin{eqnarray}
\label{eq:chlepopsSLO}
{\cal L}_{\mathrm{Yuk, SLO}}^e&=&\alpha^e_4 \, \overline{L}  \, H_d \, e_{R3} \, S^\dagger + \alpha^e_5 \, \overline{L_3} \, H_d \, e_{R} \, S \, T + \alpha^e_6 \, \overline{L} \, H_d \, e_{R} \, T \, W \, W^\dagger
\\ \nonumber
&+& \alpha^e_7 \, \overline{L_3} \, H_d \, e_{R} \, S^\dagger \, (T^\dagger)^2 \, W^\dagger + \alpha^e_8 \, \overline{L_3} \, H_d \, e_{R} \, S^\dagger \, T^\dagger \, U^\dagger \, W^\dagger
+ \alpha^e_9 \, \overline{L} \, H_d \, e_{R} \, S^2 \, (W^\dagger)^2
\\ \nonumber
&+&\alpha^e_{10} \, \overline{L} \, H_d \, e_{R3} \, S \, T^\dagger \, (W^\dagger)^2 +\alpha^e_{11} \, \overline{L} \, H_d \, e_{R} \, S^2 \, (T^\dagger)^3 +\alpha^e_{12} \, \overline{L} \, H_d \, e_{R} \, S^2 \, T^\dagger \, (U^\dagger)^2
\\ \nonumber
&+&\alpha^e_{13} \, \overline{L} \, H_d \, e_{R} \, S^4 \, W +\alpha^e_{14} \, \overline{L} \, H_d \, e_{R3} \, S^3 \, T^\dagger \, W
+\alpha^e_{15} \, \overline{L_3} \, H_d \, e_{R}  \, (S^\dagger)^3 \, T^\dagger \, W
\\ \nonumber
&+& \alpha^e_{16} \, \overline{L_3} \, H_d \, e_{R} \, (S^\dagger)^3 \, U^\dagger \, W +\alpha^e_{17} \, \overline{L} \, H_d \, e_{R} \, (S^\dagger)^2 \, (T^\dagger)^2 \, W^\dagger
\\ \nonumber
&+&\alpha^e_{18} \, \overline{L_3} \, H_d  \, e_{R}\, S^3 \, (W^\dagger)^2+ \alpha^e_{19} \, \overline{L_3}\, H_d \, e_{R} \, S^3 \, (T^\dagger)^3 +\alpha^e_{20} \, \overline{L_3} \, H_d \, e_{R} \, S^5 \, W
\\ \nonumber
&+&\alpha^e_{21} \, \overline{L_3} \, H_d  \, e_{R3} \, S^4 \, T^\dagger \, W + \alpha^e_{22} \, \overline{L} \, H_d \, e_{R} \, (S^\dagger)^4 \, T^\dagger \, W +\alpha^e_{23} \, \overline{L} \, H_d \, e_{R} \, (S^\dagger)^4 \, U^\dagger \, W
\\ \nonumber
&+&\alpha^e_{24} \, \overline{L_3} \, H_d \, e_{R3} \, (S^\dagger)^4 \, T \, W^\dagger + \alpha^e_{25} \, \overline{L}_3 \, H_d \, e_{R} \, (S^\dagger)^3 \, T^2 \, W^\dagger
\\ \nonumber
&+&\alpha^e_{26} \, \overline{L} \, H_d \, e_{R3} \, (S^\dagger)^5 \, T \, W^\dagger + \alpha^e_{27} \, \overline{L} \, H_d \, e_{R} \, (S^\dagger)^4 \, T^2 \, W^\dagger \; .
\end{eqnarray}
We briefly comment on the first two of these operators. The first one with the coefficient $\alpha^e_4$ has already been identified in the preceding section, compare footnote~\ref{foot1}. The second one with $\alpha^e_5$
 also turns out to be an operator that is automatically induced, once the field content of the LO operators, responsible for the dominant contributions to the charged fermion mass matrices and the LQ
couplings ${\bf \hat{x}}$ and ${\bf \hat{y}}$, has been fixed. We note that only the operator with the coefficient $\alpha^e_9$ leads to two independent contributions to the charged lepton mass matrix $M_e$:
one of order $\lambda^{11}$ and another one of $\lambda^{12}$, compare also eq.~(\ref{eq:Mepararel}) in appendix~\ref{app:relatepara}.

For the charged lepton mass matrix $M_e$ the following effective parametrisation is found
\begin{equation}
\label{eq:Mepara}
M_e = \left(
\begin{array}{ccc}
e_{11} \, \lambda^4 & e_{12} \, \lambda^{12} & \mathcal{o} (\lambda^{12})\\
e_{21} \, \lambda^8 & e_{22} \, \lambda^2 & e_{23} \, \lambda\\
e_{31} \, \lambda^9 & e_{32} \, \lambda^3 & e_{33}
\end{array}
\right) \, \braket{H_d^0}
\end{equation}
with $e_{ij}$ being complex order-one numbers that are related to the coefficients $\alpha^e_i$ as shown in eq.~(\ref{eq:Mepararel}) in appendix~\ref{app:relatepara}.
We emphasise that the element $M_{e,13}$ is only generated at an order higher than $\lambda^{12}$.  

From $M_e$ in eq.~(\ref{eq:Mepara}), we can derive for the charged lepton masses
\begin{eqnarray}
\label{eq:memasses}
m_e&=& \left| e_{11} \, \lambda^4 + \mathcal{o}(\lambda^{12})\right| \, \braket{H_d^0} \; ,
\\ \nonumber
m_\mu&=& \left| e_{22} \, \lambda^2 - \frac{e_{23} (e_{22} e_{23}+2 \, e_{32} e_{33})}{2 \, e_{33}^2} \, \lambda^4 + \mathcal{O} (\lambda^6)\right| \, \braket{H_d^0}  \; ,
\\ \nonumber
m_\tau&=& \left| e_{33} + \frac{e_{23}^2}{2 \, e_{33}} \, \lambda^2 + \mathcal{O}(\lambda^4) \right| \, \braket{H_d^0}  \; .
\end{eqnarray}
These results match the expectations from the construction of the model, since the three operators in eq.~(\ref{eq:chlepopsLO}) dominantly generate the three different charged lepton masses. 
We note that, in particular, the muon mass can receive sizeable contributions from the LQ at one-loop level, if the observed value of the AMM of the muon is explained in this model. 
These contributions can be compensated by adjusting the effective parameter $e_{22}$ appropriately, see eq.~(\ref{eq:memasses}). 
For formulae and estimates of these contributions, see section~\ref{subsubsec:p_ana_AMMamu} and appendix~\ref{app:charged-lepton-mass-correction}.

The matrices $L_e$ and $R_e$ read
\begin{equation}
\label{eq:Leform}
L_e=  \left(
\begin{array}{ccc}
1 +\mathcal{o}(\lambda^{12})& \frac{e_{11} e_{21}}{e_{22}^2} \, \lambda^8 + \mathcal{O}(\lambda^{10})& \mathcal{o}(\lambda^{12})\\
- \frac{e_{11} e_{21}}{e_{22}^2} \, \lambda^8 + \mathcal{O}(\lambda^{10})& 1-\frac{e_{23}^2}{2 \, e_{33}^2} \, \lambda^2 + \mathcal{O}(\lambda^4)& \frac{e_{23}}{e_{33}} \, \lambda +\mathcal{O} (\lambda^3)\\
\frac{e_{11} e_{21} e_{23}}{e_{22}^2 e_{33}} \, \lambda^9 +\mathcal{O}(\lambda^{11})&-\frac{e_{23}}{e_{33}} \, \lambda +\mathcal{O} (\lambda^3)& 1-\frac{e_{23}^2}{2 \, e_{33}^2} \, \lambda^2 +\mathcal{O}(\lambda^4)
\end{array}
\right)
\end{equation}
and\\[0.05in]
$
R_e=
$
\begin{equation}
\label{eq:Reform}
\!\!\!\left(
\begin{array}{ccc}
1-\frac{e_{21}^2}{2 \, e_{22}^2} \, \lambda^{12} +\mathcal{o}(\lambda^{12})& \frac{e_{21}}{e_{22}} \, \lambda^6 + \mathcal{O}(\lambda^8)& \frac{(e_{21} e_{23}+e_{31} e_{33})}{e_{33}^2} \, \lambda^9 + \mathcal{O}(\lambda^{11})\\
-\frac{e_{21}}{e_{22}} \, \lambda^6 + \mathcal{O}(\lambda^8)& 1 - \frac{(e_{22} e_{23}+e_{32} e_{33})^2}{2 \, e_{33}^4} \, \lambda^6 + \mathcal{O}(\lambda^8)& \frac{(e_{22} e_{23} +e_{32} e_{33})}{e_{33}^2} \, \lambda^3 + \mathcal{O}(\lambda^5)\\
-\frac{(e_{22} e_{31}-e_{21} e_{32})}{e_{22} e_{33}} \, \lambda^9 +\mathcal{O}(\lambda^{11})&-\frac{(e_{22} e_{23} +e_{32} e_{33})}{e_{33}^2} \, \lambda^3 + \mathcal{O}(\lambda^5)& 1 -\frac{(e_{22} e_{23}+e_{32} e_{33})^2}{2 \, e_{33}^4} \, \lambda^6 + \mathcal{O}(\lambda^8)
\end{array}
\right) \; .\\[0.1in]
\end{equation}
We reiterate that the matrix $L_e$ is also applied to the LH neutrinos in order to transform from the interaction to the mass basis, since neutrinos are massless in this model and thus lepton mixing is unphysical.

In both matrices, $L_e$ and $R_e$, the $(23)$-block deviates from being close to the identity matrix. These deviations are induced by the operators with the coefficients $\alpha^e_4$ and $\alpha^e_5$
which have been identified as automatically allowed, if the LO operators for the charged fermion mass matrices and the LQ couplings ${\bf \hat{x}}$ and ${\bf \hat{y}}$ are accounted for.

 The effect of the rotation of order $\lambda$ in $L_e$ is to also contribute to the elements $x_{22}$ and $x_{23}$ of the LQ coupling ${\bf \hat{x}}$ in the charged fermion mass basis and to do so at the same order
  as the elements $\hat{x}_{22}$ and $\hat{x}_{23}$ of the LQ coupling ${\bf \hat{x}}$ itself, compare eq.~(\ref{eq:xpara}) and eq.~(\ref{eq:xpararel}) in appendix~\ref{app:relatepara}.
 The impact of the rotation of order $\lambda^3$ in $R_e$ is to also generate the element $y_{22}$ of the LQ coupling ${\bf \hat{y}}$ in the charged fermion mass basis of order $\lambda^3$ in addition to the element $\hat{y}_{22} \sim \lambda^3$ of the LQ coupling ${\bf\hat{y}}$ itself ~--~
  see eq.~(\ref{eq:yparaA}) and eq.~(\ref{eq:ypararelA}) in appendix~\ref{app:relatepara}.

%%%%%%%%%%%%%%%%%%%%%%%%%%%%%%%%%%%%%%%%%%%%%%%%%%%%%%
\subsection{Leptoquark couplings}
\label{subsec:LQcouplings}
%%%%%%%%%%%%%%%%%%%%%%%%%%%%%%%%%%%%%%%%%%%%%%%%%%%%%%

We first list the operators, contributing to the LQ couplings ${\bf \hat{x}}$ and ${\bf \hat{y}}$, up to and including order $\lambda^{12}$, and then
discuss the form of ${\bf x}$, ${\bf y}$ and ${\bf z}$, the LQ couplings ${\bf \hat{x}}$ and ${\bf \hat{y}}$ in the charged fermion mass basis, in the
two different scenarios, scenario A and scenario B.

%%%%%%%%%%%%%%%%%%%%%%%%%%%%%%%%%%%%%%%%%%%%%%%%%%%%%%
\subsubsection{Couplings in interaction basis}
\label{subsubsec:LQcoupinter}
%%%%%%%%%%%%%%%%%%%%%%%%%%%%%%%%%%%%%%%%%%%%%%%%%%%%%%

We begin with the LO operators, responsible for the main structure of the LQ coupling ${\bf \hat{x}}$. There are four of them
\begin{equation}
\label{eq:hatxopsLO}
{\cal L}^{\mathrm{int}}_{\mathrm{{\bf \hat{x}},LO}} = \beta^L_1 \, \overline{L_3^c} \, \phi^\dagger \, Q_3 + \beta^L_2 \, \overline{L^c} \, \phi^\dagger \, Q_3 \, S + \beta^L_3 \, \overline{L_3^c}  \, \phi^\dagger \, Q \, S^2
+ \beta^L_4 \, \overline{L^c} \, \phi^\dagger \, Q \, S^3 \; .
\end{equation}
They coincide with those, anticipated in the construction of the model in the preceding section. At SLO, there are several more operators
\begin{eqnarray}
\label{eq:hatxopsSLO}
{\cal L}^{\mathrm{int}}_{\mathrm{{\bf \hat{x}},SLO}}&=&\beta^L_5 \, \overline{L_3^c} \, \phi^\dagger \, Q  \, T \, W^2 + \beta^L_6 \, \overline{L^c} \, \phi^\dagger \, Q  \, S^\dagger \, T \, W^\dagger
+ \beta^L_7 \, \overline{L^c} \, \phi^\dagger \, Q \, S^\dagger \, U \, W^\dagger
\\ \nonumber
&+& \beta^L_8 \, \overline{L_3^c} \, \phi^\dagger\, Q  \, S^2 \, T^\dagger \, U + \beta^L_9 \, \overline{L^c} \, \phi^\dagger  \, Q \, S^\dagger \, (T^\dagger)^2 \, W
+ \beta^L_{10} \, \overline{L^c} \, \phi^\dagger \, Q  \, S^\dagger \, T^\dagger \, U^\dagger \, W
\\ \nonumber
&+& \beta^L_{11} \, \overline{L^c} \, \phi^\dagger  \, Q \, S \, T \, W^2 + \beta^L_{12} \, \overline{L^c} \, \phi^\dagger \, Q_3  \, S^\dagger \, T \, W^2
+ \beta^L_{13} \, \overline{L_3^c} \, \phi^\dagger \, Q  \, (S^\dagger)^2 \, T \, W^\dagger
\\ \nonumber
&+& \beta^L_{14} \, \overline{L_3^c} \, \phi^\dagger \, Q \, (S^\dagger)^2 \, U \, W^\dagger + \beta^L_{15} \, \overline{L_3^c} \, \phi^\dagger\, Q  \, S^2 \, W \, W^\dagger
+ \beta^L_{16} \, \overline{L^c} \, \phi^\dagger\, Q  \, S^3 \, T^\dagger \, U
\\ \nonumber
&+& \beta^L_{17} \, \overline{L^c} \, \phi^\dagger \, Q_3 \, (S^\dagger)^3 \, T \, W^\dagger+ \beta^L_{18} \, \overline{L^c} \, \phi^\dagger \, Q  \, S \, (T^\dagger)^3 \, W^\dagger
+ \beta^L_{19} \, \overline{L^c} \, \phi^\dagger  \, Q \, S^3 \, W \, W^\dagger
\\ \nonumber
&+& \beta^L_{20} \, \overline{L^c} \, \phi^\dagger  \, Q \, (S^\dagger)^5 \, U^\dagger+ \beta^L_{21} \, \overline{L^c} \, \phi^\dagger  \, Q \, (S^\dagger)^3 \, T^2 \, W
+ \beta^L_{22} \, \overline{L_3^c} \, \phi^\dagger \, Q_3  \, S^4 \, T^\dagger \, W
\\ \nonumber
&+& \beta^L_{23} \, \overline{L_3^c}  \, \phi^\dagger \, Q_3 \, (S^\dagger)^4 \, T \, W^\dagger+ \beta^L_{24} \, \overline{L_3^c} \, \phi^\dagger \, Q  \, (S^\dagger)^4 \, T^2 \, W
+ \beta^L_{25} \, \overline{L^c} \, \phi^\dagger \, Q_3  \, S^5 \, T^\dagger \, W
\\ \nonumber
&+& \beta^L_{26} \, \overline{L_3^c} \, \phi^\dagger \, Q \, S^6 \, T^\dagger \, W \; .
\end{eqnarray}
All couplings $\beta^L_i$ are complex order-one coefficients. Like before, we note that several of these operators lead to two independent contributions to the LQ coupling ${\bf \hat{x}}$. The operator with the coefficient $\beta^L_5$ leads to contributions of order $\lambda^{10}$ and $\lambda^{11}$; the one with $\beta^L_6$ gives contributions of order $\lambda^7$ and $\lambda^8$;
the operator with $\beta^L_{10}$ induces two of order $\lambda^{11}$ and $\lambda^{12}$; the one with $\beta^L_{11}$  yields contributions of order $\lambda^{11}$ and $\lambda^{12}$; the one with
$\beta^L_{13}$ gives rise to two of order $\lambda^8$ and $\lambda^9$; finally, the operator with the coefficient $\beta^L_{21}$ leads to two independent contributions of order $\lambda^{11}$ and $\lambda^{12}$.

From the contributions of these operators, we can deduce the form of the LQ coupling ${\bf \hat{x}}$, up to and including order $\lambda^{12}$,
\begin{equation}
\label{eq:hatxpara}
{\bf\hat{x}} = \left(
\begin{array}{ccc}
\hat{a}_{11} \, \lambda^9 & \hat{a}_{12} \, \lambda^{12} & \mathcal{o}(\lambda^{12})\\
\hat{a}_{21} \, \lambda^8 & \hat{a}_{22} \, \lambda^3 & \hat{a}_{23} \, \lambda\\
\hat{a}_{31} \, \lambda^8 & \hat{a}_{32} \, \lambda^2 & \hat{a}_{33}
\end{array}
\right)
\end{equation}
with the effective parameters $\hat{a}_{ij}$ being, in general, complex order-one numbers. How these are related to the coefficients $\beta^L_i$ can be found in eq.~(\ref{eq:xhatpararel}) in appendix~\ref{app:relatepara}.
We note that the element $\hat{x}_{13}$ is only generated at an order higher than $\lambda^{12}$.

Although not yet in the charged fermion mass basis, we can already compare this form of the LQ coupling with the
texture, envisaged in eq.~(\ref{eq:texturesxylambda_ex}). We clearly see that the elements of the first column and row are protected well by the residual symmetry $Z_{17}^\mathrm{diag}$, while the elements
in the $(23)$-block of the LQ coupling ${\bf \hat{x}}$ all have the desired order of magnitude in $\lambda$, see the texture in eq.~(\ref{eq:texturesxylambda_ex}).

In the end, we also discuss the operators, contributing to the LQ coupling ${\bf \hat{y}}$, up to and including order $\lambda^{12}$. We identify only two operators as LO ones
\begin{equation}
\label{eq:hatyopsLO}
{\cal L}^{\mathrm{int}}_{\mathrm{{\bf \hat{y}},LO}} = \beta^R_1 \, \overline{e_{R3}^c} \, \phi^\dagger \, u_{R2} + \beta^R_2 \, \overline{e_{R}^c} \, \phi^\dagger \, u_{R3} \, S^3 \; ,
\end{equation}
which are expected from the construction of the model. At SLO, several more operators are found
\begin{eqnarray}
\label{eq:hatyopsSLO}
{\cal L}^{\mathrm{int}}_{\mathrm{{\bf \hat{y}},SLO}}&=&\!\!\!\!\!\beta^R_3 \, \overline{e_{R}^c}  \, \phi^\dagger \, u_{R2}\, S \, T +\beta^R_4 \, \overline{e_{R3}^c} \, \phi^\dagger \, u_{R3} \, S^2 \, T^\dagger
+ \beta^R_5 \, \overline{e_{R3}^c} \, \phi^\dagger \, u_{R3} \, W^2
\\ \nonumber
&+&\!\!\!\!\!\!\beta^R_6 \, \overline{e_{R}^c} \, \phi^\dagger \, u_{R1}  \, S^\dagger \, (U^\dagger)^2 +\beta^R_7 \, \overline{e_{R}^c} \, \phi^\dagger \, u_{R1} \, S \, U \, W +\beta^R_8 \, \overline{e_{R3}^c} \, \phi^\dagger \, u_{R3} \, (S^\dagger)^2 \, W^\dagger
\\ \nonumber
&+&\!\!\!\!\!\!\beta^R_9 \, \overline{e_{R}^c} \, \phi^\dagger \, u_{R3}  \, S^\dagger \, T \, W^\dagger +\beta^R_{10} \, \overline{e_{R}^c} \, \phi^\dagger \, u_{R3} \, S^\dagger\, (T^\dagger)^2 \, W
+\beta^R_{11} \, \overline{e_{R}^c} \, \phi^\dagger \, u_{R3} \, S^\dagger\, T^\dagger \, U^\dagger \, W
\\ \nonumber
&+&\!\!\!\!\!\!\beta^R_{12} \, \overline{e_{R}^c} \, \phi^\dagger \, u_{R3}  \, S \, T \, W^2 +\beta^R_{13} \, \overline{e_{R}^c} \, \phi^\dagger \, u_{R2}  \, S^\dagger\, (T^\dagger)^2 \, W^\dagger
+\beta^R_{14} \, \overline{e_{R}^c}  \, \phi^\dagger \, u_{R2} \, S^\dagger\, T^\dagger \, U^\dagger \, W^\dagger
\\ \nonumber
&+&\!\!\!\!\!\!\beta^R_{15} \, \overline{e_{R}^c} \, \phi^\dagger \, u_{R1}\, (S^\dagger)^3 \, T \, U +\beta^R_{16} \, \overline{e_{R3}^c} \, \phi^\dagger \, u_{R1}  \, (S^\dagger)^2 \, T^\dagger \, (U^\dagger)^2
+\beta^R_{17} \, \overline{e_{R}^c} \, \phi^\dagger \, u_{R2}  \, (S^\dagger)^3\, T^\dagger \, W
\\ \nonumber
&+&\!\!\!\!\!\!\beta^R_{18} \, \overline{e_{R}^c} \, \phi^\dagger \, u_{R2} \, (S^\dagger)^3\, U^\dagger \, W +\beta^R_{19} \, \overline{e_{R}^c} \, \phi^\dagger \, u_{R3}  \, S \, (T^\dagger)^3 \, W^\dagger
+\beta^R_{20} \, \overline{e_{R}^c} \, \phi^\dagger \, u_{R2} \, S^3 \, (W^\dagger)^2
\\ \nonumber
&+&\!\!\!\!\!\!\beta^R_{21} \, \overline{e_{R}^c} \, \phi^\dagger \, u_{R2} \, S^3 \, (T^\dagger)^3 +\beta^R_{22} \, \overline{e_{R3}^c}\, \phi^\dagger \, u_{R1}  \, S^4 \,  U^2 +\beta^R_{23} \, \overline{e_{R}^c}\, \phi^\dagger \, u_{R3}  \, (S^\dagger)^5 \,  U^\dagger
\\ \nonumber
&+&\!\!\!\!\!\!\beta^R_{24} \, \overline{e_{R}^c}\, \phi^\dagger \, u_{R2} \, S^5 \,  W +\beta^R_{25} \, \overline{e_{R3}^c}  \, \phi^\dagger \, u_{R3} \, (S^\dagger)^4 \, T \,  W +\beta^R_{26} \, \overline{e_{R}^c} \, \phi^\dagger \, u_{R3} \, (S^\dagger)^3 \, T^2 \,  W
\\ \nonumber
&+&\!\!\!\!\!\!\beta^R_{27} \, \overline{e_{R3}^c} \, \phi^\dagger \, u_{R2}  \, S^4 \, T^\dagger \,  W +\beta^R_{28} \, \overline{e_{R3}^c} \, \phi^\dagger \, u_{R2}  \, (S^\dagger)^4 \, T \,  W^\dagger
+\beta^R_{29} \, \overline{e_{R}^c} \, \phi^\dagger \, u_{R2}  \, (S^\dagger)^3 \, T^2 \,  W^\dagger \; ,
\end{eqnarray}
with all coefficients $\beta^R_i$ being complex order-one numbers. The presence of the first and the second operator is automatic after having fixed the transformation properties of the fields
which are relevant for the LO terms of the charged fermion mass matrices $M_u$, $M_d$, $M_e$ and the LQ couplings ${\bf \hat{x}}$ and ${\bf \hat{y}}$. We note that all listed operators give rise to a single (independent)
contribution to the LQ coupling ${\bf \hat{y}}$.

We arrive at the effective parametrisation for ${\bf \hat{y}}$ to be of the form
\begin{equation}
\label{eq:hatypara}
{\bf\hat{y}} = \left(
\begin{array}{ccc}
\hat{b}_{11} \, \lambda^9 & \hat{b}_{12} \, \lambda^9 & \hat{b}_{13} \, \lambda^9\\
\hat{b}_{21} \, \lambda^9 & \hat{b}_{22} \, \lambda^3 & \hat{b}_{23} \, \lambda^3\\
\hat{b}_{31} \, \lambda^{12} & \hat{b}_{32} & \hat{b}_{33} \, \lambda^4
\end{array}
\right) \; .
\end{equation}
The parameters $\hat{b}_{ij}$ are, in general, complex order-one numbers and are related to the coefficients $\beta^R_i$ as shown in eq.~(\ref{eq:yhatpararel}) in appendix~\ref{app:relatepara}. 

We may already compare this result to the texture of the LQ coupling ${\bf y}$ in eq.~(\ref{eq:texturesxylambda_ex}) and see that it contains the same two dominant terms, $\hat{y}_{32}$ and $\hat{y}_{23}$, like the texture
with $y_{32} \sim 1$ and $y_{23} \sim \lambda^3$. At the same time, however, the form of the LQ coupling ${\bf \hat{y}}$ in eq.~(\ref{eq:hatypara}) also has rather large elements $\hat{y}_{22} \sim \lambda^3$ and $\hat{y}_{33} \sim \lambda^4$.
 These arise from the operators with the coefficients $\beta^R_3$ and $\beta^R_4$ which have been identified as automatically allowed, once the LO operators, contributing to the charged fermion mass matrices
  and LQ couplings ${\bf \hat{x}}$ and ${\bf \hat{y}}$ and their particle content are fixed.
We note that none of the elements of the first column and row of the LQ coupling ${\bf \hat{y}}$ is larger than $\lambda^9$, showing the effectiveness of the residual symmetry $Z_{17}^\mathrm{diag}$. Couplings to electrons and/or up quarks are thus suppressed.

%%%%%%%%%%%%%%%%%%%%%%%%%%%%%%%%%%%%%%%%%%%%%%%%%%%%%%
\subsubsection{Couplings in charged fermion mass basis}
\label{subsubsec:LQcoupmass}
%%%%%%%%%%%%%%%%%%%%%%%%%%%%%%%%%%%%%%%%%%%%%%%%%%%%%%

In this section, we display the results for the LQ couplings ${\bf x}$, ${\bf y}$ and ${\bf z}$, namely the LQ couplings ${\bf \hat{x}}$ and ${\bf \hat{y}}$ in the charged fermion mass basis, compare eq.~(\ref{eq:LagmassLQcouplings}).
The LQ coupling ${\bf x}$ is obtained by applying the matrices $L_e$ and $L_d$ to ${\bf \hat{x}}$, while ${\bf z}$ by applying $L_e$ and $L_u$ to ${\bf \hat{x}}$. The LQ coupling ${\bf y}$ is generated from ${\bf \hat{y}}$
by applying the matrices $R_e$ and $R_u$.
In doing so, we distinguish between the two different scenarios, scenario A and scenario B, for the LQ couplings ${\bf z}$ and ${\bf y}$.

We use the matrices $L_e$ and $L_d$ in eqs.~(\ref{eq:Ldform},\ref{eq:Leform}) and arrive at the LQ coupling ${\bf x}$. This matrix can be
parametrised as
\begin{equation}
\label{eq:xpara}
{\bf x} = L_e^T \, {\bf \hat{x}} \, L_d = \left(
\begin{array}{ccc}
a_{11} \, \lambda^9 & a_{12} \, \lambda^{11} & a_{13} \, \lambda^9\\
a_{21} \, \lambda^8 &a_{22} \, \lambda^3 & a_{23} \, \lambda\\
a_{31} \, \lambda^8 &a_{32} \, \lambda^2 & a_{33}
\end{array}
\right) \; ,
\end{equation}
where the effective parameters $a_{ij}$ are related to the parameters $\hat{a}_{ij}$, $d_{ij}$ and $e_{ij}$, found in the matrix ${\bf \hat{x}}$ in eq.~(\ref{eq:hatxpara}), $M_d$ in eq.~(\ref{eq:Mdpara})
and $M_e$ in eq.~(\ref{eq:Mepara}), respectively. The explicit form of these relations is given in eq.~(\ref{eq:xpararel}) in appendix~\ref{app:relatepara}. In general, they can also be expected to be complex order-one numbers.

Comparing the form of the LQ coupling ${\bf x}$ in eq.~(\ref{eq:xpara}) to the texture of ${\bf x}$ in eq.~(\ref{eq:texturesxylambda_ex}), we clearly see that all elements of the first column and row
are suppressed, i.e.~none of them is larger than $\lambda^8$. At the same time, the elements $x_{33}$, $x_{23}$, $x_{32}$ and $x_{22}$ have the expected order of magnitude in $\lambda$. 

%%%%%%%%%%%%%%%%%%%%%%%%%%%%%%%%%%%%%%%%%%%%%%%%%%%%%%
\subsubsection*{In scenario A}
\label{subsubsubsec:LQscenarioA}
%%%%%%%%%%%%%%%%%%%%%%%%%%%%%%%%%%%%%%%%%%%%%%%%%%%%%%

In scenario A, i.e.~the model without any modification of the up-type quark mass matrix $M_u$, we find the form of the LQ coupling ${\bf z}$, when
 applying the matrices $L_e$ and $L_u$, see eqs.~(\ref{eq:Leform},\ref{eq:Luform}), to the LQ coupling ${\bf \hat{x}}$ in eq.~(\ref{eq:hatxpara}). It is
 \begin{equation}
\label{eq:zparaA}
{\bf z} = L_e^T \, {\bf \hat{x}} \, L_u= \left(
\begin{array}{ccc}
c_{11} \, \lambda^9 & c_{12} \, \lambda^{10} & c_{13} \, \lambda^9\\
c_{21} \, \lambda^4 &c_{22} \, \lambda^3 & c_{23} \, \lambda\\
c_{31} \, \lambda^3 &c_{32} \, \lambda^2 & c_{33}
\end{array}
\right)\; .
\end{equation}
 The effective parameters $c_{ij}$ are related to $\hat{a}_{ij}$, $e_{ij}$ and $f_{ij}$ from eqs.~(\ref{eq:hatxpara},\ref{eq:Mepara},\ref{eq:Mupara}).
Again, the explicit form of these relations can be found in appendix~\ref{app:relatepara}, see eq.~(\ref{eq:zpararelA}).

We note that it might be useful to evidence the strong correlation between the LQ couplings ${\bf x}$ and ${\bf z}$ by using a different parametrisation for ${\bf z}$, namely
\begin{equation}
\label{eq:zxparaA}
{\bf z} = \left(
\begin{array}{ccc}
(a_{11} - \frac{(c^A_{12})^2}{2 \, a_{11}} \, \lambda^2 + c^A_{11} \, \lambda^3) \, \lambda^9 & c^A_{12} \, \lambda^{10} & a_{13} \, \lambda^9\\
(-\frac{c^A_{12}}{a_{11}} \, (a_{22}+a_{23} \, \tilde{c}) + c^A_{21} \, \lambda) \, \lambda^4 &(a_{22} + a_{23} \, \tilde{c} + c^A_{22} \, \lambda^2) \, \lambda^3 & (a_{23}+ c^A_{23} \, \lambda^4) \, \lambda\\
(-\frac{c^A_{12}}{a_{11}} \, (a_{32}+a_{33} \, \tilde{c}) + c^A_{31} \, \lambda) \, \lambda^3 &(a_{32}+ a_{33} \, \tilde{c} + c^A_{32} \, \lambda^2) \, \lambda^2 & a_{33} + c^A_{33} \, \lambda^4
\end{array}
\right) \; ,
\end{equation}
\noindent where $a_{ij}$ are the same parameters as in ${\bf x}$ in eq.~(\ref{eq:xpara}). The new effective parameters $c^A_{ij}$ and $\tilde{c}$ are rather involved expressions in the other parameters so that we just take them
to be complex order-one numbers, apart from
\begin{equation}
\label{eq:zxparaAadd}
c^A_{12}= \frac{\hat{a}_{11} f_{12}}{f_{22}} + \mathcal{O} (\lambda)\;\; \mbox{and} \;\; \tilde{c} = \frac{d_{23}}{d_{33}} - \frac{f_{23}}{f_{33}} \; .
\end{equation}
We use the matrices $R_e$ and $R_u$ in eqs.(\ref{eq:Reform},\ref{eq:Ruform}) and ${\bf \hat{y}}$ in eq.~(\ref{eq:hatypara}) in order to arrive at the form of the LQ coupling ${\bf \hat{y}}$ in the charged fermion mass basis
 \begin{equation}
\label{eq:yparaA}
{\bf y} = R_e^T \, {\bf \hat{y}} \, R_u =\left(
\begin{array}{ccc}
b_{11} \, \lambda^9 & b_{12} \, \lambda^9 & b_{13} \, \lambda^9\\
b_{21} \, \lambda^8 & b_{22} \, \lambda^3 & b_{23} \, \lambda^3\\
b_{31} \, \lambda^5 & b_{32}  & b_{33} \, \lambda^4
\end{array}
\right) \, .
\end{equation}
The effective parameters $b_{ij}$ are related to $\hat{b}_{ij}$ from the LQ coupling ${\bf\hat{y}}$ in eq.~(\ref{eq:hatypara}), to $e_{ij}$ of the charged lepton mass matrix
$M_e$ in eq.~(\ref{eq:Mepara}) and to $f_{ij}$ of the up-type quark mass matrix $M_u$ in eq.~(\ref{eq:Mupara}).
These relations are given in eq.~(\ref{eq:ypararelA}) in appendix~\ref{app:relatepara}.

Comparing this form of the LQ coupling ${\bf y}$ with the texture in eq.~(\ref{eq:texturesxylambda_ex}), we see that in the charged fermion mass basis not only the elements $y_{22} \sim \lambda^3$ and $y_{33} \sim \lambda^4$
turn out to be larger, but also the element $y_{31} \sim \lambda^5$. As we see in section~\ref{subsec:primary_analytic}, these couplings do not enter the analytic estimates for the strongest (primary) constraints on this model. We find that they generally lead to subleading contributions to these estimates, or appear in the estimates for secondary/tertiary observables, discussed further in section~\ref{sec:secondarytertiary}. For example, $y_{22}$ contributes to the process $b\to c\mu \nu$, relevant for subdominant contributions to the LFU ratios $R(D)$  and $R(D^\star)$, and for the secondary observables $R_D^{\mu/e}$  and $R_{D^\star}^{e/\mu}$. The  LQ coupling $y_{33}$ relates the top quark to the tau lepton, and shows up in subdominant loop-level contributions to tau lepton decays including $\tau\to\mu\gamma$.  The LQ coupling $y_{31}$ is relevant for subleading contributions to the decay $B \to \tau\nu$, representing a secondary observable, and to  tau lepton decays to light mesons, e.g.~$\tau \to\pi \mu$, which correspond to tertiary observables.
 The LQ couplings $y_{1j}$ involving the electron are still suppressed.

%%%%%%%%%%%%%%%%%%%%%%%%%%%%%%%%%%%%%%%%%%%%%%%%%%%%%%
\subsubsection*{In scenario B}
\label{subsubsubsec:LQscenarioB}
%%%%%%%%%%%%%%%%%%%%%%%%%%%%%%%%%%%%%%%%%%%%%%%%%%%%%%

In scenario B, where the element $M_{u,13}$ of the up-type quark mass matrix $M_u$ is enhanced, see eq.~(\ref{eq:MuparaB}), the matrices $L_u$ and $R_u$ are found in eqs.~(\ref{eq:LuformB},\ref{eq:RuformB}). When using these in order to compute the form of the LQ coupling ${\bf z}$, we find the following: while the order of magnitude in $\lambda$ of the different elements
of ${\bf z}$ is not changed with respect to the matrix shown in eq.~(\ref{eq:zparaA}), the relations of the effective parameters $c_{ij}$ to the parameters $\hat{a}_{ij}$, $e_{ij}$, $f_{ij}$ and $\tilde{f}_{13}$ are to some extent altered. If we compare to the ones given in eq.~(\ref{eq:zpararelA}) in appendix~\ref{app:relatepara}, we now have for $c_{21}$ and $c_{31}$
\begin{eqnarray}
\label{eq:zpararelB}
c_{21}&=& -\frac{f_{12}}{e_{33} f_{22} f_{33}} \, \left( \hat{a}_{33} e_{23} f_{23} -\hat{a}_{23} e_{33} f_{23} - \hat{a}_{32} e_{23} f_{33} + \hat{a}_{22} e_{33} f_{33}\right)
\\ \nonumber
&& - \frac{\tilde{f}_{13}}{f_{33}} \, \left( \hat{a}_{23} - \frac{\hat{a}_{33} e_{23}}{e_{33}} \right) + \mathcal{O}(\lambda^2)\; ,
\\ \nonumber
c_{31}&=& \frac{f_{12} (\hat{a}_{33} f_{23}-\hat{a}_{32} f_{33})}{f_{22} f_{33}}  - \frac{\tilde{f}_{13}}{f_{33}} \, \hat{a}_{33} + \mathcal{O} (\lambda^2)\; .
\end{eqnarray}
As a consequence, the correlation between the LQ couplings ${\bf x}$ and ${\bf z}$ leads to a slightly different parametrisation than the one, displayed in eq.~(\ref{eq:zxparaA}), i.e. 
 \begin{equation}
\label{eq:zxparaB}
{\bf z} = \left(
\begin{array}{ccc}
(a_{11} - \frac{(c^B_{12})^2}{2 \, a_{11}} \, \lambda^2 + c^B_{11} \, \lambda^3) \, \lambda^9 & c^B_{12} \, \lambda^{10} & (a_{13} + c^B_{13} \, \lambda^3) \, \lambda^9\\
(-\frac{c^B_{12}}{a_{11}} \, (a_{22}+a_{23} \, \tilde{c}) - a_{23} \, \bar{c} + c^B_{21} \, \lambda) \, \lambda^4 &(a_{22} + a_{23} \, \tilde{c} + c^B_{22} \, \lambda^2) \, \lambda^3 & (a_{23}+ c^B_{23} \, \lambda^4) \, \lambda\\
(-\frac{c^B_{12}}{a_{11}} \, (a_{32}+a_{33} \, \tilde{c}) - a_{33}  \, \bar{c} + c^B_{31} \, \lambda) \, \lambda^3 &(a_{32}+ a_{33} \, \tilde{c} + c^B_{32} \, \lambda^2) \, \lambda^2 & a_{33} + c^B_{33} \, \lambda^4
\end{array}
\right)\; . \\[0.05in]
\end{equation}
\noindent Most of the parameters $c^B_{ij}$ are complex order-one numbers. Their expressions in terms of the other parameters are rather lengthy,\footnote{The parameter $c^B_{13}$ is new with respect to the parametrisation of ${\bf z}$ in eq.~(\ref{eq:zxparaA}).}
apart from $c^B_{12}=c^A_{12}$, $\tilde{c}$ and $\bar{c}$. The former two are still of the form as given in eq.~(\ref{eq:zxparaAadd}), while the further parameter $\bar{c}$ is defined as
\begin{equation}
\label{eq:zxparaBadd}
\bar{c} = \frac{\tilde{f}_{13}}{f_{33}} \; .
\end{equation}
Coming to the form of the LQ coupling ${\bf y}$, when using $R_u$ in eq.~(\ref{eq:RuformB}), we see that neither its form, found in eq.~(\ref{eq:yparaA}), nor the definition of the effective parameters $b_{ij}$, given in
eq.~(\ref{eq:ypararelA}) in appendix~\ref{app:relatepara}, are altered. Nevertheless, the change in the up-type quark mass matrix $M_u$ in scenario B also leaves a slight imprint at higher order in $\lambda$ on the LQ coupling ${\bf y}$
with the maximum change in $y_{31}$ at order $\lambda^7$.

%%%%%%%%%%%%%%%%%%%%%%%%%%%%%%%%%%%%%%%%%%%%%%%%%%%%%%
\section{Outline of phenomenological study}
\label{sec:phenomenology}
%%%%%%%%%%%%%%%%%%%%%%%%%%%%%%%%%%%%%%%%%%%%%%%%%%%%%%

In the following, we outline the strategy for the phenomenological study of the aforementioned model. In particular, we highlight the important features common to the studies detailed in sections~\ref{sec:primary} and~\ref{sec:secondarytertiary}.

\paragraph{Classification of observables.} We classify all analysed observables to one of the following three categories: primary, secondary or tertiary observables. The primary observables comprise the anomalies in $R(D)$, $R(D^\star)$ and in the AMM of the muon, as well as the observables for which contributions generated in this model can (substantially) violate the current experimental bounds and/or are accessible in upcoming experiments. Examples of the former are the radiative cLFV decays  $\mu\to e\gamma$ and $\tau\to\mu\gamma$, while processes such as $\mu \to 3 \, e$ and $\mu-e$ conversion in aluminium belong to the latter. These observables are studied analytically in section~\ref{sec:primary}, and numerically  in sections~\ref{sec:primary} and~\ref{sec:secondarytertiary}. Secondary observables, for instance $B\to\tau\nu$, do not presently provide any competitive constraint, but are expected to offer an opportunity to further test this model in the mid-term future. 
These are discussed both analytically and numerically in section~\ref{sec:secondarytertiary}. 
 Tertiary observables, such as the AMM of the electron, do not lead to any restriction on the parameter space of the model given the present experimental status. We find analytically that they do not deviate significantly from the SM predictions. The projected sensitivity for these observables is thus not sufficient to probe a signal consistent with this model. However, if a deviation from the SM prediction is observed, this could challenge the model. We mention them in section~\ref{sec:secondarytertiary} and appendix~\ref{app:tertiary}, and incorporate them in the second numerical scan.

\paragraph{Implemented model setup.} As mentioned in section~\ref{subsec:choiceGffermions}, the presented model contains two Higgs doublets, $H_u$ and $H_d$, that give
masses to up-type quarks as well as to down-type quarks and charged leptons, respectively,  upon electroweak symmetry breaking. Nevertheless, we simplify the model in the phenomenological study and consider it as model with one SM-like Higgs doublet, i.e.~we only take into account one 
SM-like Higgs, ignoring effects due to scalars other than the LQ $\phi$, and appropriately rescale the effective parameters $f_{ij}$, $d_{ij}$ and $e_{ij}$, contained in the up-type quark, down-type quark
and charged lepton mass matrices $M_u$, $M_d$ and $M_e$, respectively.

Since only in scenario B the results for quark mixing are in full agreement with experimental data, c.f.~section~\ref{subsubsec:quarkmixing}, 
we focus on this scenario in the phenomenological study. For scenario A, we note that only the form of the effective parameters $c_{21}$
and $c_{31}$ is slightly different, see section~\ref{subsubsec:LQcoupmass}. According to the analytic results, the parameter $c_{21}$ only contributes at SLO to $\mu - e$ conversion in nuclei,
see section~\ref{subsubsec:p_ana_mueconv} and also table~\ref{table:primaryconstraints_coupling_coeffs}, while $c_{31}$ is relevant for the computation of the secondary observable $B \to\tau\nu$, compare 
section~\ref{subsec:secondary_analytic}. 
We, thus, do not expect any significant differences in the phenomenological results for these two scenarios. This expectation is, indeed, confirmed with a smaller 
data sample of the first numerical scan.
 
\paragraph{Bases of LQ couplings.} The form of the LQ couplings is presented in two different bases, the interaction basis as well as the charged fermion mass basis, see section~\ref{subsec:LQcouplings}.
The former basis refers to the \emph{hatted} LQ couplings $\bf{\hat{x}}$ and $\bf{\hat{y}}$ with effective parameters $\hat{a}_{ij}$ and $\hat{b}_{ij}$, see definition in eq.~(\ref{eq:LagintLQcouplings}) and explicit forms in eq.~\eqref{eq:hatxpara} and eq.~\eqref{eq:hatypara}, while the latter basis corresponds to the \emph{unhatted} LQ couplings $\bf{x}$, $\bf{y}$ and $\bf{z}$ with effective parameters $a_{ij}$, $b_{ij}$ and 
$c_{ij}$ or $c_{ij}^B$, $\tilde{c}$ and $\bar{c}$, see definition in eq.~(\ref{eq:LagmassLQcouplings}) and explicit forms in eqs.~\eqref{eq:xpara}, \eqref{eq:yparaA} and \eqref{eq:zparaA} or \eqref{eq:zxparaB} (for scenario B).  Each of the parameters $a_{ij}$, $b_{ij}$ and $c_{ij}$ is (at LO) given by a linear combination of some of the effective parameters $\hat{a}_{ij}$ and $\hat{b}_{ij}$ with coefficients constituted by $f_{ij}$, $d_{ij}$ and $e_{ij}$, which parametrise the mass matrices $M_u$, $M_d$ and $M_e$, respectively. The explicit relations between the parameters in the two bases can be found in appendix~\ref{app:relatepara}.

While the interaction basis directly reflects the impact of the imposed flavour symmetry, the charged fermion mass basis is usually employed in phenomenological
studies that focus on the effects of the LQ. For this reason, unhatted LQ couplings are used in analytic computations with ${\bf z}$ being parametrised in terms of the effective parameters $c_{ij}$, see sections~\ref{subsec:primary_analytic} and~\ref{subsec:secondary_analytic}, as well as in the first numerical scan with the LQ coupling ${\bf z}$ given in terms of $c_{ij}^B$, $\tilde{c}$ and $\bar{c}$, 
 c.f.~section~\ref{subsec:primary_numeric}. On the other hand, the second numerical scan 
 is performed in the interaction basis, see section~\ref{sec:secondarytertiary} and appendix~\ref{app:supp6}. 

\paragraph{Strategy of numerical scans.} In order to study the phenomenology of the model in depth, we perform two numerical scans. In the following, we give details about the employed strategy.

For the first scan, discussed in section~\ref{subsec:primary_numeric}, we only consider primary observables, and thus refer to it as the \emph{primary} scan. 
Since the LQ couplings in the model span a parameter space of high dimensionality, it is reasonable to first establish which of the effective parameters prove most relevant for the induced phenomenology.  
The main purpose of constructing the model is generating textures of the LQ couplings which are suitable to explain the currently observed flavour anomalies in $R (D)$, $R (D^\star)$ and in the AMM of the muon. 
So, as a first step we deem it sufficient to only consider the effective parameters, contained in the LQ couplings $\textbf{x}$, $\textbf{y}$ and $\textbf{z}$, without making explicit reference to the interaction basis. 
 We investigate the capability of the model to explain the mentioned anomalies and how the imposed current experimental bounds shape the viable parameter space. 
We also establish biases on the relevant effective parameters $a_{ij}$ and $b_{ij}$ that are applied in the second numerical scan, see section~\ref{ssec:prelimBias}. The contributions to the relevant observables are computed with the help of the analytic expressions given in appendix~\ref{app:pheno}. In addition, we use \texttt{Wilson}~\cite{Aebischer:2018bkb} to account for renormalisation group~(RG) running under QCD.

For the second scan, detailed in section~\ref{sec:secondarytertiary}, we take into account all observables, primary, secondary, and tertiary, and thus refer to it as \emph{comprehensive}. 
 In particular, we include secondary observables and outline how they can provide tangible signals for this model in the future, see section~\ref{subsec:resultsSecondary}.  
Tertiary observables are also cross-checked and the generated ranges for these observables are summarised in appendix~\ref{app:tertiary}.
In order to ensure that this model accommodates charged fermion masses and quark mixing, we fix the effective parameters $f_{ij}$, $d_{ij}$ and $e_{ij}$ by performing a chi-squared fit, see 
details in section~\ref{ssec:cferm}. 
 Furthermore, we vary most of the effective parameters, contained in the LQ couplings ${\bf \hat{x}}$ and ${\bf \hat{y}}$, in the ranges laid out in eq.~(\ref{eq:unbiased_magnitude}) and eq.~(\ref{eq:unbiased_phase}), apart from the ones which are 
 identified as playing a dominant role for the phenomenology of the model. 
  For these effective parameters, we apply a suitable biasing in order to more efficiently target the parameter space preferred by the primary 
 observables, as detailed in section~\ref{ssec:prelimBias}.  
  As computational tools, we use \texttt{SARAH}, \texttt{SPheno}~\cite{Porod:2014xia,Porod:2011nf} and \texttt{flavio}~\cite{Straub:2018kue,david_straub_2021_5543714} in the comprehensive scan.

\paragraph{Range of LQ couplings.} In agreement with the expansion in $\lambda$, we assume the magnitude of an unbiased parameter $s_{ij}$ to be in the range 
\begin{equation}
\label{eq:unbiased_magnitude}
\lambda = 0.2265 \leq |s_{ij}|\leq \frac 1\lambda \approx 4.42 \; ,
\end{equation}
and its phase to lie in the interval
\begin{equation}
\label{eq:unbiased_phase}
0  \leq \mathrm{Arg} (s_{ij}) < 2 \, \pi \; .
\end{equation}
Here, $s_{ij}$ corresponds to any of the parameters $a_{ij}$, $b_{ij}$ and $c_{ij}$ or $c_{ij}^B$ (except for $c_{12}^B$, see below), while working in the charged fermion mass basis, and  
 to any of the parameters $\hat{a}_{ij}$ and $\hat{b}_{ij}$, in the case of the interaction basis. In two instances, different choices for certain parameters are made. In the primary scan, employing the
 charged fermion mass basis, smaller ranges for the magnitudes of the effective parameters $c_{12}^B$, $\bar c$ and $\tilde c$ are used in order to better approximate the relation
 of the LQ couplings ${\bf x}$ and ${\bf z}$ that is determined by the CKM mixing matrix, see section~\ref{subsubsec:primary_preliminaries}. In the comprehensive scan, using the 
 interaction basis, biases on certain effective parameters $a_{ij}$ and $b_{ij}$ are imposed that are derived from the results of the primary scan, see section~\ref{ssec:prelimBias}.

In the analytic study, it is assumed that all effective parameters vary as indicated in eq.~(\ref{eq:unbiased_magnitude}) and eq.~(\ref{eq:unbiased_phase}). 
Depending on the studied observable, we either give an approximate relation based on the LO in $\lambda$ or an estimate accounting only for the correct order of magnitude. 
  
Inspecting the relations between the parameters in the interaction and charged fermion mass basis that are given in appendix~\ref{app:relatepara}, we conclude that 
 the value of an unhatted parameter can significantly differ from the value of the corresponding hatted parameter.
 Consequently, the results of the primary scan over the LQ couplings $\bf{x}$, $\bf{y}$ and $\bf{z}$ do not entirely agree with those obtained from the comprehensive scan over the LQ couplings 
 $\bf{\hat{x}}$ and $\bf{\hat{y}}$, see discussion in section~\ref{ssec:primaryresultsComprehensive}.

\paragraph{Range of LQ masses.} We consider the following three values of the LQ mass $m_\phi$ as benchmarks
\begin{equation}
\label{eq:hatmphi_num}
\hat{m}_\phi = \frac{m_\phi}{\text{TeV}}=2 \, , 4 \; \mbox{and} \; 6 \; .
\end{equation}
{These choices are compatible with current constraints from direct searches for LQs. The flavour structure of the LQ couplings predicts the dominant decays to be to $\tau t$, $\tau c$ and $\nu b$, while branching ratios (BRs) of decays with muons and electrons as final states are suppressed by at least a further $\lambda^2$. ATLAS~\cite{ATLAS:2021jyv} has constrained LQ masses to fulfil $\hat m_\phi \gtrsim 1.2$ at 95\% C.L. for BR($\phi\to t \tau)\sim\mathrm{BR}(\phi\to b\nu)$. The choice of benchmark values for the LQ masses is even consistent with the strongest present limits on LQ masses from searches for LQs exclusively coupling to muons~(electrons) which are constrained to 
$\hat{m}_\phi>1.7 \, (1.8)$ at 95\% C.L., with minimal dependence on the coupled quark flavour~\cite{ATLAS:2020dsk}; see also \cite{CMS:2021far}.}

%%%%%%%%%%%%%%%%%%%%%%%%%%%%%%%%%%%%%%%%%%%%%%%%%%%%%%
\section{Primary observables: anomalies and constraints}
\label{sec:primary}
%%%%%%%%%%%%%%%%%%%%%%%%%%%%%%%%%%%%%%%%%%%%%%%%%%%%%%

In this section, we first present analytic estimates that help to identify the most relevant LQ couplings for each of the primary observables in section~\ref{subsec:primary_analytic}.
Then, we turn to a numerical study for scenario B in section~\ref{subsec:primary_numeric}.

%%%%%%%%%%%%%%%%%%%%%%%%%%%%%%%%%%%%%%%%%%%%%%%%%%%%%%
\subsection{Analytic estimates}
\label{subsec:primary_analytic}
%%%%%%%%%%%%%%%%%%%%%%%%%%%%%%%%%%%%%%%%%%%%%%%%%%%%%%

The analytic estimates, derived in the following, are expressed in terms of the effective parameters in the charged fermion mass basis. The underlying complete formulae can be found in appendix~\ref{app:pheno}. Note, in particular, that the low-energy effective theory (LEFT) Wilson coefficients are given in the Jenkins-Manohar-Stoffer (JMS) basis~\cite{Jenkins:2017jig}.

%%%%%%%%%%%%%%%%%%%%%%%%%%%%%%%%%%%%%%%%%%%%%%%%%%%%%%

 \begin{table}[t!]\centering
	\def\arraystretch{1.3} 
	\resizebox{\linewidth}{!}{\begin{tabular}{|l|cll|cl|}
	\hline
	 \multicolumn{6}{|c|}{\textsc{ List of Primary Observables}}\\
			\hline
			\multirow{2}{*}{Observable} & \multicolumn{5}{c|}{Experiment}  \\
			\cline{2-6}
			& \multicolumn{3}{c|}{Current constraint/measurement} & \multicolumn{2}{c|}{Future reach} \\
			\hline
			$R (D)$ & 
			$0.339 \pm 0.026 \pm 0.014$
			& at $1\, \sigma$ level &
			\cite{Amhis:2019ckw} &
			$\pm0.016 \, (0.008)$ for $5 \, (50)\,\text{ab}^{-1}$
			& \cite{Forti:2022mti}
			\\
			$R(D^\star)$ & 
			$0.295 \pm 0.010 \pm 0.010$
			& at $1\, \sigma$ level & \cite{Amhis:2019ckw} &
			$\pm0.009 \, (0.0045)$ for $5 \, (50)\,\text{ab}^{-1}$
			& \cite{Forti:2022mti}
			\\
			$\Delta a_\mu$ & $(2.51 \pm 0.59) \times 10^{-9}$ & at $1 \, \sigma$ level & \cite{Muong-2:2021ojo,Aoyama:2020ynm} & $\pm 0.4\times 10^{-9}$ & \cite{Muong-2:2015xgu} \\
			\hline
			$\mathrm{BR}(\tau\to\mu\gamma)$ & $4.2 \times 10^{-8}$ & at 90\% C.L. & \cite{Belle:2021ysv} & 
			$6.9 \times 10^{-9}$ & \cite{Banerjee:2022xuw} \\
			$\mathrm{BR}(\mu\to e\gamma)$ & $4.2 \times 10^{-13}$ & at 90\% C.L. & \cite{MEG:2016leq} & $6 \times 10^{-14}$ & \cite{MEGII:2021fah} \\
			$\mathrm{BR}(\tau \to 3\, \mu)$ & $2.1 \times 10^{-8}$ & at 90\% C.L. & \cite{Hayasaka:2010np} & 
			$3.6\times 10^{-10}$ & \cite{Banerjee:2022xuw} \\
			$\mathrm{BR}(\tau \to \mu e\bar e)$ & $1.8 \times 10^{-8}$ & at 90\% C.L. & \cite{Hayasaka:2010np} & 
			$2.9\times 10^{-10}$ & \cite{Banerjee:2022xuw} \\
			$\mathrm{BR}(\mu\to 3 \, e)$ & $1.0 \times 10^{-12}$ & at 90\% C.L. & \cite{SINDRUM:1987nra} & $20 \, (1) \times 10^{-16}$ & \cite{Blondel:2013ia} \\
			CR($\mu - e;\;$Al) & & & & $2.6 \, (2.9) \times 10^{-17}$ & \cite{COMET:2018auw,Mu2e:2014fns}\\
			$R^\nu_{K^{\star}}{}
			$ & 2.7 & at 90\% C.L. & \cite{Belle:2017oht} &
			$1.0\pm 0.25 \, (0.1)$ for $5 \, (50)\,\text{ab}^{-1}$
			& \cite{Belle-II:2018jsg} \\
			$g_{\tau_{A}}/g_{A}^{\rm SM}$ & $ 1.00154 \pm0.00128$   & at $1\,\sigma$ level & \cite{ALEPH:2005ab,Crivellin:2020mjs} & $\pm 7.5 \, (0.75) \times 10^{-5}$& \cite{Crivellin:2020mjs,Baer:2013cma, FCC:2018evy} \\
			\hline
			$\tau_{B_c}^{\text{SM}}$
			& $0.52^{+0.18}_{-0.12}$ ps & at $1\,\sigma$ level & \cite{
				Beneke:1996xe}
			& 
			& 
			\\
			\hline
			
			$c\overline{c}\to\tau\bar \tau$ & $|b_{32}| < 2.6$ ($\hat{m}_\phi = 2$) & & \cite{ATLAS:2017eiz,Angelescu:2018tyl} & & 
			\\[0.03in]
			\hline
	\end{tabular}}
	\caption{{\small \textbf{List of primary observables}. We list the observables that dominantly constrain this model together with the current experimental constraint/measurement and the future reach. The values for $R(D)$ and $R(D^\star)$ reflect the 2021 averages from the HFLAV collaboration. 
	The future reach for BR($\mu\to 3e$) (in parentheses) is for Phase 1 (2) of the Mu3e experiment. For CR($\mu - e$; Al), the first (second) value is the future reach of COMET (Mu2e).
	The future reach for $R_{K^\star}^\nu$ assumes a result which is consistent with the SM expectation~\cite{Belle-II:2018jsg}. For the  future projections of $g_{\tau_{A}}/g_{A}^{\rm SM}$, we have assumed that the measurements of $g_{\tau_{A}}$ are improved by the same factor as $\sin^2 \theta_\text{eff}$~\cite{Crivellin:2020mjs}; the unbracketed projection is that for the International Linear Collider~(ILC)~\cite{Baer:2013cma}, and the bracketed value is for the Future Circular Collider~(FCC)~\cite{,FCC:2018evy}. The current experimental constraint on the $B_c$ lifetime is $\tau_{B_c}^{\text{exp}} = (0.510\pm 0.009)\;\text{ps}$ \cite{ParticleDataGroup:2020ssz,Amhis:2019ckw}. Note that the constraint arising from high-$p_T$ lepton searches differs from the other constraints, since it is directly imposed in the primary scan via an adequate restriction of the range for $|b_{32}|$ as indicated. }}
	\label{table:primaryconstraints}
\end{table}

\mathversion{bold}
\subsubsection{\texorpdfstring{$R (D)$ and $R (D^\star)$}{RD and RDstar}}
\mathversion{normal}
\label{subsubsec:p_ana_RDRDstar}
\begin{figure}[t!]
	\centering
	\includegraphics[width=0.35\textwidth]{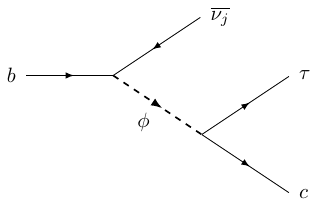}
	\includegraphics[width=0.25\textwidth]{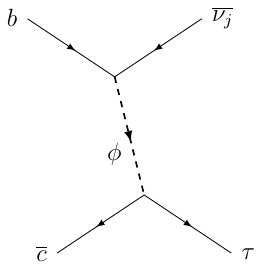}
	\includegraphics[width=0.35\textwidth]{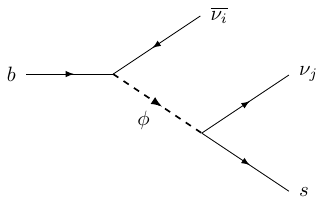}
	\caption{Feynman diagrams for tree-level contributions to charged-current $b\to c$ processes (left, centre) and neutral-current $b\to s$ processes (right) via an intermediate LQ, $\phi$. }
	\label{fig:btoctaunu}
\end{figure}
The LFU ratios $R(D)$ and $R(D^\star)$ are observables of high importance for this study. Taking into account only tree-level corrections induced by the LQ $\phi$, schematically depicted in the left of figure~\ref{fig:btoctaunu}, we find the following terms in the relevant effective semi-leptonic charged-current Lagrangian
\begin{align}
	\mathcal{L} & \supset C_{\nu e du,ij32}^{VLL} (\overline{\nu_i} \gamma^\mu P_Le_j) (\bar b \gamma_\mu P_L c)
	+ C_{\nu e du,ij32}^{SRR} (\overline{\nu_i} P_Re_j) (\bar b P_R c) \\
	& \quad + C_{\nu e du,ij32}^{TRR} (\overline{\nu_i} \sigma^{\mu\nu} P_Re_j) (\bar b \sigma_{\mu\nu} P_R c) 
	+\mathrm{h.c.}\;. \nonumber
\end{align}
The relation between the scalar and tensor Wilson coefficients, $C^{SRR}_{\nu edu,ij32}(m_\phi) = -4 \,C^{TRR}_{\nu edu,ij32}(m_\phi)$, which becomes $C^{SRR}_{\nu edu,ij32}(\mu_B)\approx-8\,C^{TRR}_{\nu edu,ij32}(\mu_B)$ at the hadronic scale $\mu = \mu_B = 4.8$ GeV due to RG running, indicates that contributions from the tensor operator only play a role, if they are enhanced via form factors or the phase-space configuration.

We derive analytic expressions for $R(D)$ and $R(D^\star)$ from the requirement of (approximate) agreement with the results obtained from \texttt{flavio}~\cite{Straub:2018kue,david_straub_2021_5543714}, v2.3, that is, we also use the values $R(D)_{\text{SM}} = 0.297\pm0.008$ and $R(D^\star)_{\text{SM}} = 0.245\pm0.008$ given by \texttt{flavio}. In particular, the latter exhibits a tension with experimental data at the $\sim 3\,\sigma$ level, see table~\ref{table:anomalies} and section~\ref{subsubsec:prelimscan_anomalies}.\footnote{Since v2.0, \texttt{flavio} uses the form factors of reference~\cite{Bordone:2019vic} which are determined via Heavy-Quark Effective Theory. Furthermore, the implementation is based on the helicity formalism~\cite{Gratrex:2015hna} which has been extensively tested as a general framework.}

The LQ $\phi$ modifies $R(D)$ and $R(D^\star)$ dominantly via contributions to the tau lepton channel, that is $j = 3$ in the above formulae. As expected, the largest correction occurs for a tau neutrino $\nu_\tau$ in the final state which allows for interference with the SM contribution. Nonetheless, we generically also account for the lepton flavour violating~(LFV) contribution with a muon neutrino $\nu_\mu$ in the estimates of the relevant observables in this section (see e.g.~in eq.~\eqref{eq:RD_estimate} and eq.~\eqref{eq:RDs_estimate} the rightmost terms), as this may have an appreciable impact. On the contrary, the channel with an electron neutrino $\nu_e$ can always be neglected, since the involved couplings are very small as a result of the residual symmetry $Z^{\text{diag}}_{17}$, that is, $x_{11}, x_{13}\sim\lambda^9$ and $x_{12}\sim\lambda^{11}$, see eq.~(\ref{eq:xpara}).

Corrections to $R(D)$ are mainly due to the interference between the scalar-operator contribution and the SM one
\begin{eqnarray}
\label{eq:RD_estimate}
\frac{R(D)}{R(D)_{\text{SM}}} & \approx 1
- 1.17\;\text{Re}\left(\hat C^{SRR}_{\nu edu,3332}(\mu_B)\right) + 0.63\left(\left|\hat C^{SRR}_{\nu edu,3332}(\mu_B)\right|^2 + \left|\hat C^{SRR}_{\nu edu,2332}(\mu_B)\right|^2\right) \\
& \quad + 0.72\;\text{Re}\left(\hat C^{TRR}_{\nu edu,3332}(\mu_B)\right) + 0.37\left(\left|\hat C^{TRR}_{\nu edu,3332}(\mu_B)\right|^2 + \left|\hat C^{TRR}_{\nu edu,2332}(\mu_B)\right|^2\right)\nonumber
\\
& \approx 1
+ 1.07\;\frac{|a_{33}b_{32}|}{\hat{m}^2_\phi}\cos\big(\text{Arg}(a_{33}) - \text{Arg}(b_{32})\big) + 0.46\;\frac{|a_{33}b_{32}|^2}{\hat{m}^4_\phi} + 0.02\;\frac{|a_{23}b_{32}|^2}{\hat{m}^4_\phi}
\;.\nonumber
\end{eqnarray}
Here, we have introduced dimensionless Wilson coefficients $\hat C = C\,\times\,\text{TeV}^2$ for convenience. The dominant corrections to $R(D^\star)$ are sourced by the interference between the tensor operator and the SM in this model
\begin{eqnarray}
\label{eq:RDs_estimate}
\frac{R(D^\star)}{R(D^\star)_{\text{SM}}} & \approx 1
+ 0.10\;\text{Re}\left(\hat C^{SRR}_{\nu edu,3332}(\mu_B)\right) + 0.03\;\left(\left|\hat C^{SRR}_{\nu edu,3332}(\mu_B)\right|^2 + \left|\hat C^{SRR}_{\nu edu,2332}(\mu_B)\right|^2\right) \\
& \quad + 4.21\;\text{Re}\left(\hat C^{TRR}_{\nu edu,3332}(\mu_B)\right) + 8.60\left(\left|\hat C^{TRR}_{\nu edu,3332}(\mu_B)\right|^2 + \left|\hat C^{TRR}_{\nu edu,2332}(\mu_B)\right|^2\right)\nonumber
\\
& \approx 1
+ 0.36\;
\frac{|a_{33}b_{32}|}{\hat{m}^2_\phi}
\cos\big(\text{Arg}(a_{33}) - \text{Arg}(b_{32})\big)
+ 0.12\;\frac{|a_{33}b_{32}|^2}{\hat{m}^4_\phi} + 0.01\;\frac{|a_{23}b_{32}|^2}{\hat{m}^4_\phi}
\;.\nonumber
\end{eqnarray}
Note that contributions from the vector operator to $R(D^{(\star)})$ are suppressed because of the hierarchy $z_{32}/y_{32} \sim \lambda^2$, see eq.~(\ref{eq:yparaA}) and eq.~(\ref{eq:zxparaB}).

%%%%%%%%%%%%%%%%%%%%%%%%%%%%%%%%%%%%%%%%%%%%%%%%%%%%%%

%%%%%%%%%%%%%%%%%%%%%%%%%%%%%%%%%%%%%%%%%%%%%%%%%%%%%%

\begin{figure}[t!]
\centering
\includegraphics[scale=1.2]{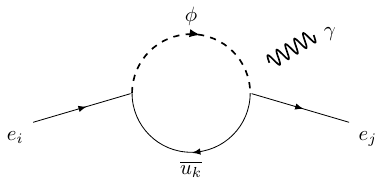}
\caption{Feynman diagram for the  one-loop contribution to the process $e_i\to e_j\gamma$ via an intermediate LQ ($\phi$) and an up-type quark ($u_k$).}
\label{fig:eiejgamma}
\end{figure}

%%%%%%%%%%%%%%%%%%%%%%%%%%%%%%%%%%%%%%%%%%%%%%%%%%%%%%
\subsubsection{Anomalous magnetic moment of muon and muon mass correction}
\label{subsubsec:p_ana_AMMamu}
%%%%%%%%%%%%%%%%%%%%%%%%%%%%%%%%%%%%%%%%%%%%%%%%%%%%%%

Given that the deviation from the SM prediction for the AMM of the muon, $\Delta a_\mu$, is of $4.2~\sigma$ significance~\cite{Aoyama:2020ynm,Muong-2:2021ojo}, we discuss the viability of this model in explaining this anomaly. The leading-order LQ contribution is generated by the one-loop diagram shown in figure~\ref{fig:eiejgamma}. In particular, it is dominated by the contribution in which the chirality flip occurs via a mass insertion on the internal quark leg -- and can therefore be enhanced by the mass of this quark (here denoted $m_{u_k}$, consistent with figure~\ref{fig:eiejgamma}). For the full calculation of the leptonic AMM, we refer to appendix~\ref{app:dipole}. 

Taking the dominant contribution to be the one with the top quark in the loop, and  $m_\phi \gg m_{u_k}$, eq.~\eqref{eq:Cegamma} and eq.~\eqref{eq:EDMAMM} in appendix~\ref{app:dipole} reduce to the following
\begin{align}\label{eq:gm2_estimate}
    \Delta a_\mu \approx -\frac{ 2\,{\rm{Re}}(b_{23}  c_{23}^*)}{\hat{m}_{\phi}^2}\times 10^{-9} = -\frac{ 2 \, |b_{23}c_{23}|}{\hat{m}_{\phi}^2} \, \cos\left( \text{Arg}(b_{23})- \text{Arg}(c_{23})\right)\times 10^{-9} 
	\;.
\end{align}
Contrasting this with table~\ref{table:primaryconstraints},  the order of magnitude of the AMM of the muon generated by this model can be
consistent with the present experimental average.

Requiring that this anomaly is addressed tightly constrains the parameter space of the order-one coefficients  $c_{23} $ and $b_{23}$. To satisfy the current experimental value at the $n$-sigma level requires them to obey the following relations
\begin{align}\left| \left\{
\begin{array}{*{2}{c}}
0.890\;{{\rm{Re}}(b_{23} c_{23}^*)}+2.51, & \hat{m}_\phi = 2 \\
0.307\;{{\rm{Re}}(b_{23} c_{23}^*)}+2.51,& \hat{m}_\phi = 4 \\
0.159\;{{\rm{Re}}(b_{23} c_{23}^*)}+2.51,& \hat{m}_\phi = 6 \\
\end{array} \right\}\right| \lesssim n\times 0.59\;[0.4].
\end{align}
Here, the current experimental bound is shown with the prospective sensitivity given in square brackets. The latter assumes that the
best-fit value
remains fixed but with the target precision listed in table~\ref{table:primaryconstraints}. Results for the AMM of the electron and of the tau lepton are discussed in section~\ref{subsec:tertiary}.

Given that we generally consider complex LQ couplings, there is scope to generate both an AMM and an electric dipole moment~(EDM) for charged leptons, as discussed in appendix~\ref{app:dipole}. The leptonic EDMs do not presently provide competitive constraints on the parameter space of this model and, therefore, we defer the discussion of these to section~\ref{sec:secondarytertiary}. 

Through a diagram similar to that shown in figure~\ref{fig:eiejgamma}, the LQ also introduces a correction to the muon mass. Adapting the result from reference~\cite{Crivellin:2020mjs}, the full expression for this correction can be found in appendix~\ref{app:charged-lepton-mass-correction}.  At LO this contribution reduces to
\begin{align}
	m_{\mu}& \approx \left| m_\mu^{\rm tree} 
	-\frac{3}{16\pi^2}  m_{t} b_{23} c_{23}^*\lambda^4 (1+t_t \ln t_t)\right|\lesssim \left| m_\mu^{\rm tree} \right|
	+\left|\frac{3}{16\pi^2}  m_{t} b_{23} c_{23}^*\lambda^4 (1+t_t \ln t_t)\right| \, ,
\end{align}
where $m_\mu^{\rm tree}$ denotes the tree-level muon mass, the upper bound follows from the triangle inequality, and $t_X$ denotes the mass squared of particle $X$ normalised to the LQ mass squared, i.e.
\begin{equation}\label{eq:tX}
	t_X = \frac{m_X^2}{m_\phi^2}\;.
\end{equation}
In the region of parameter space consistent with explaining the AMM of the muon, we observe numerically that the correction can be significant at the order of 80 percent, with this value being extracted from the data output  of the comprehensive scan discussed in section~\ref{sec:secondarytertiary}. However, as stated in section~\ref{subsec:chargedleptonsector}, it is always possible to absorb this correction by redefining the effective parameter $e_{22}$.\footnote{ An alternative approach to addressing the correction of the muon mass would be to implement a constraint based on its size, as is done for example in references~\cite{Athron:2021iuf,Bigaran:2021kmn}.}

%%%%%%%%%%%%%%%%%%%%%%%%%%%%%%%%%%%%%%%%%%%%%%%%%%%%%%
\mathversion{bold}
\subsubsection{\texorpdfstring{Radiative charged lepton flavour violating decays $e_i \to e_j \gamma$}{Radiative charged lepton flavour violating decays ei->ej gamma}}
\mathversion{normal}
\label{subsubsec:p_ana_eiejgamma}
%%%%%%%%%%%%%%%%%%%%%%%%%%%%%%%%%%%%%%%%%%%%%%%%%%%%%%

Similarly to the AMM of the muon, cLFV decays of the form $e_i \to e_j \gamma$ proceed at LO via the one-loop diagram given in figure~\ref{fig:eiejgamma}. Notably, the diagram for the contribution to the AMM of the muon shares a common vertex with both the ones for the cLFV decays $\tau \to \mu \gamma$ and $\mu \to e\gamma$. Therefore, we expect these two decays to provide competitive constraints on the explanation of the former anomaly. From table~\ref{table:primaryconstraints}, we see that the present experimental bound on BR$(\mu\to e\gamma)$ is five orders of magnitude more stringent than BR$(\tau\to\mu\gamma)$. However, the former provides a weaker constraint due to the efficient suppression of the LQ coupling $y_{13}$, $y_{13}=b_{13} \, \lambda^9$, thanks to the residual symmetry $Z^{\text{diag}}_{17}$, see eq.~(\ref{eq:yparaA}).

Following from eq.~\eqref{eq:Cegamma} and eq.~\eqref{eq:eiejgamma} in appendix~\ref{app:dipole}, we arrive at the following expressions for the LO contributions to these BRs, parametrising the contributions from loops containing the top quark
\begin{align}\label{eq:mueg_estimate}
\text{BR}(\mu\to e
    \gamma) \sim  \frac{|b_{13}c_{23}|^2}{\hat{m}_\phi^4} \times 10^{-11},
\end{align}
and
\begin{align}\label{eq:taumug_estimate}
\text{BR}(\tau\to \mu
    \gamma) \sim  
    \frac{|b_{23}c_{33} |^2}{\hat{m}_\phi^4} \times 10^{-5}.
\end{align} 
Comparing these with the constraints quoted in table~\ref{table:primaryconstraints}, these estimates show that significant rates for both decay modes can be generated. We, thus, use these to constrain the relevant couplings as follows, where the current experimental bound is shown with the prospective sensitivity mentioned in square brackets. For $\mu\to e \gamma$, we have 
\begin{align}\label{eq:approx_bound_mueg}
|b_{13} c_{23}|
\lesssim  \left\{
\begin{array}{*{2}{c}}
0.444\; [0.168], & \hat{m}_\phi = 2 \\
1.264\;[0.477],& \hat{m}_\phi = 4 \\
5.915\;[0.845],& \hat{m}_\phi = 6 \\
\end{array} \right\} .
\end{align}
\noindent This shows that this constraint is especially strong for smaller LQ masses. As indicated above, the parameter $c_{23}$ appears in the expressions for both $\text{BR}(\mu\to e \gamma)$ and the AMM of the muon, which makes the constraint from BR($\mu\to e \gamma$) important for refining the parameter space which could explain the measured value of the AMM of the muon. Similarly, for $\tau\to \mu
 \gamma$ we find
\begin{align}\label{eq:approx_bound_taumug}
|b_{23}c_{33}|
 \lesssim \left\{
\begin{array}{*{2}{c}}
0.259\;[0.105], & \hat{m}_\phi = 2 \\
1.037\;[0.420],& \hat{m}_\phi = 4 \\
2.333\;[0.946],& \hat{m}_\phi = 6 \\
\end{array} \right\}.
\end{align}
Here, the effective parameter $b_{23}$ appears, but also the parameter $c_{33} \approx a_{33}$.  The latter plays an important role in the generation of the corrections to $R({D})$ and $R({D^\star})$ in this model, as discussed in section~\ref{subsubsec:p_ana_RDRDstar}.

 %%%%%%%%%%%%%%%%%%%%%%%%%%%%%%%%%%%%%%%%%%%%%%%%%%%%%%
\mathversion{bold}
\subsubsection{\texorpdfstring{Trilepton decays $e_i \to e_j e_l \overline{e_l}$}{Trilepton ei to ej el el*}}
\mathversion{normal}
\label{subsubsec:p_ana_trilepton}
%%%%%%%%%%%%%%%%%%%%%%%%%%%%%%%%%%%%%%%%%%%%%%%%%%%%%%

\begin{figure}\centering
\includegraphics{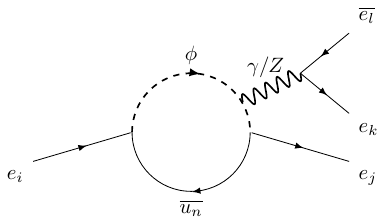}
\hspace{1cm}
\includegraphics{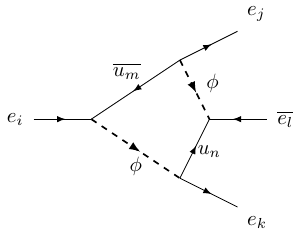}
\caption{Representative Feynman diagrams for the process $e_i\to e_j e_k \overline{e_l}$ mediated by the LQ $\phi$ at one-loop order, where the up-type quarks $u_n$ and $u_m$ run in the loop.}
\label{fig:trilepton}
\end{figure}

Trilepton cLFV decays provide another sensitive probe for new physics, particularly in light of several relevant future sensitivities. Representative Feynman diagrams are shown in figure~\ref{fig:trilepton}. 
The dominant contributions to the three most sensitive processes are
\begin{align}
	\label{eq:muto3e_estimate}
	\mathrm{BR}(\mu\to 3\,e) &\sim \frac{|b_{13} c_{23}|^2}{\hat m_\phi^4} \times 10^{-14}\;, 
	\\
	\label{eq:tauto3mu_estimate}
	\mathrm{BR}(\tau\to 3\,\mu)&\sim \frac{|b_{23}c_{33}|^2+
		0.07\, |c_{23}c_{33}|^2}{\hat m_\phi^4} \times 10^{-7}\;,
	\\
	\label{eq:tautomuee_estimate}
	\mathrm{BR}(\tau\to \mu e \bar e)&\sim \frac{|b_{23}c_{33}|^2+
		0.05\, |c_{23}c_{33}|^2}{\hat m_\phi^4} \times 10^{-7}\;.
\end{align}
Besides the respective long-distance $\gamma$-penguin contribution with the chirality flip due to an internal top quark, for tau lepton decays we also take into account the $Z$-penguin contribution with two internal top quark mass insertions which, although suppressed, becomes relevant for some regions of the parameter space. The full expressions for the BRs can be retrieved from eq.~(\ref{eq:trilepton_jjj}) and eq.~(\ref{eq:trilepton_jkk}) in appendix \ref{sec:trilepton} for the relevant flavour combinations.
For $\gamma$-penguin dominance, one finds \cite{Kuno:1999jp,Crivellin:2013hpa}
\begin{equation} \label{eq:tau_photonpen}
	\frac{\text{BR}(\tau\to 3\,\mu)}{\text{BR}(\tau\to\mu\gamma) } \approx \frac{\alpha_{\rm em}}{3\pi}\left(\ln\left(\frac{m_\tau^2}{m_\mu^2}\right) - \frac{11}{4}\right) \approx \frac{1}{400}\;,
\end{equation}
and thus the existing experimental bound on $\text{BR}(\tau\to\mu\gamma)$ implies that no signal of $\tau\to 3\,\mu$ can be expected at Belle II. Still, sufficiently large $Z$-penguin contributions can render the decays $\tau\to\mu e_i \overline{e_i}$ potentially observable at Belle II.

The upper bounds on the BRs can be translated into constraints on the effective parameters. While the experimental limit on the BR of $\mu\to 3\,e$ is currently less sensitive compared to the one on $\mu\to e\gamma$, the Mu3e experiment~\cite{Blondel:2013ia} is expected to provide a competitive sensitivity, i.e.~
\begin{align}
	|b_{13} c_{23}|
& \lesssim \left\{
\begin{array}{*{2}{c}}
	0.298, & \hat{m}_\phi = 2 \\ 
0.840,& \hat{m}_\phi = 4 \\ 
1.61,& \hat{m}_\phi = 6 \\ 
\end{array} \right\}
\end{align}
using the value given for Phase 2, see table \ref{table:primaryconstraints}.
The decays $\tau\to 3\,\mu$ and $\tau\to \mu e \bar e$ are both mainly sensitive to $|b_{23} c_{33}|$ and lead to similar constraints on the combination. In the regime of $\gamma$-penguin dominance the BRs are closely related which results in
\begin{align} \label{eq:tau_photonpen2}
\frac{\text{BR}(\tau\to3\,\mu)}{\text{BR}(\tau\to\mu e \bar e)} \approx \frac{2\ln(m_\tau/m_\mu) - 11/4}{2\ln(m_\tau/m_\mu) - 3} \approx 1.09 \;.
\end{align}
As the decays are mainly sensitive to small LQ masses, we only present the constraints for $\hat m_\phi=2$. Currently, $\tau\to \mu e\bar e$ imposes~\cite{Hayasaka:2010np} 
\begin{align}
	|b_{23}c_{33}|^2+ 0.0561 \, |c_{23} c_{33}|^2\lesssim	4.30 \qquad\qquad \text{for}\; \hat{m}_\phi = 2\;, 
\end{align}
while in the future the sensitivity of Belle II~\cite{Belle-II:2018jsg} allows to probe 
\begin{align}
	|b_{23}c_{33}|^2+ 0.0806\, |c_{23} c_{33}|^2\lesssim	0.0655 \qquad\,\,\, \text{for}\; \hat{m}_\phi = 2 \, ,
 \end{align}
assuming the absence of a signal.
Due to suppressed LQ couplings, other cLFV trilepton decays do not provide any strong constraints and neither achieve a competitive sensitivity at Belle II.

%%%%%%%%%%%%%%%%%%%%%%%%%%%%%%%%%%%%%%%%%%%%%%%%%%%%%%
\mathversion{bold}
\subsubsection{\texorpdfstring{$\mu-e$ conversion in nuclei}{mue conversion in nuclei}}
\mathversion{normal}
\label{subsubsec:p_ana_mueconv}
%%%%%%%%%%%%%%%%%%%%%%%%%%%%%%%%%%%%%%%%%%%%%%%%%%%%%%

\begin{figure}\centering
\includegraphics{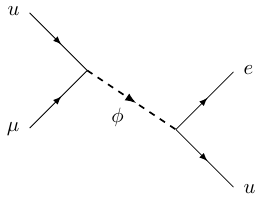}
\hspace{0.5cm}
\includegraphics{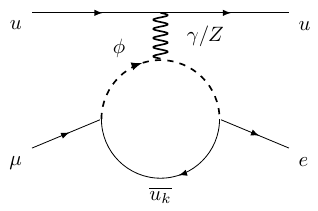}
\hspace{0.5cm}
\includegraphics{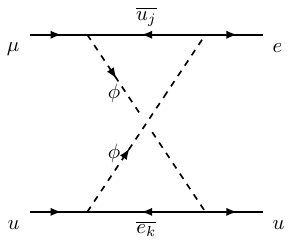}
\caption{Representative Feynman diagrams for $\mu-e$ conversion in nuclei mediated by the LQ $\phi$ at tree level and one-loop order with up-type quarks $u_k$ and $u_j$ and the charged lepton
$e_k$ running in the loop.
}
\label{fig:mueconv_diagram}
\end{figure}

There are relevant contributions to $\mu-e$ conversion in nuclei mediated by the LQ $\phi$ both at tree level and one-loop order. Representative Feynman diagrams are displayed in figure~\ref{fig:mueconv_diagram}.
The dominant contribution to $\mu-e$ conversion in nuclei originates from the long-range $\gamma$-penguin. We find that also tree-level scalar contributions become relevant in some part of the parameter space. Using this approximation the conversion rate~(CR) can be written as, see reference~\cite{Kitano:2002mt} and also appendix~\ref{app:mue_conversion},
\begin{align}\label{eq:mueconvrate_approx}
	\omega_{\rm conv} &=
	\left|-\frac{C_{e\gamma}^{12}}{2m_\mu}D
	\right|^2
	+ \left|- \frac{C_{e\gamma}^{21*}}{2m_\mu} D + \tilde g_{RS}^{(p)} S^{(p)} + \tilde g_{RS}^{(n)} S^{(n)} \right|^2
\end{align}
which is expressed in terms of the relevant dipole-operator Wilson coefficients in the JMS basis~\cite{Jenkins:2017jig} and the effective scalar contribution 
\begin{align}
	C_{e\gamma}^{21} & \approx \frac{e\,m_t}{64\pi^2 m_\phi^2}(7+4\ln t_t)  c_{23}^* b_{13} \lambda^{10}
	\;,
&
\tilde g_{RS}^{(N)} & \approx   G_S^{u,N}  \frac{c_{21} b_{11}^*}{2m_\phi^2} \lambda^{13}  
\;,
\\\nonumber
	C_{e\gamma}^{12} & \approx \frac{e\,m_t}{64\pi^2 m_\phi^2}(7+4\ln t_t)  c_{13}^* b_{23} \lambda^{12}
\end{align}
with $G_{S}^{u,p}=5.1$, $G_S^{u,n}=4.3$~\cite{Kosmas:2001mv} and $N=p,n$. Nuclear physics effects are parametrised by $D$ and $S^{(N)}$ and the numerical values for aluminium are~\cite{Kitano:2002mt} $D=0.0362 \,m_\mu^{5/2}$, $S^{(p)}=0.0155 \,m_\mu^{5/2}$ and $S^{(n)}=0.0167 \,m_\mu^{5/2}$.
Experiments generally report the CR normalised to the muon capture rate, $\mathrm{CR}=\omega_{\rm conv}/\omega_{\rm capt}$, with the latter being $\omega_{\rm capt}=0.7054\times 10^6\, \mathrm{s}^{-1}$ for aluminium~\cite{Kitano:2002mt}.

Although BR($\mu\to e\gamma$) currently leads to stronger constraints, $\mu-e$ conversion in aluminium can provide an excellent probe for the $\mu-e$ transition.
From the expected future reach of COMET to CR($\mu - e$; Al) in table~\ref{table:primaryconstraints}, we derive 
\begin{align}
\left| c_{23} b_{13}^* 
+ 
\left\{
\begin{array}{*{2}{c}}
0.00486\\ 
0.00344\\
0.00293\\
\end{array} \right\} 
c_{21}b_{11}^*\right|^2
+ 
0.00263\,|c_{13}^* b_{23}|^2
	\lesssim
\left\{
\begin{array}{*{2}{c}}
0.00373, & \hat m_\phi =2\\ 
0.0300, & \hat m_\phi =4\\
0.110, & \hat m_\phi =6\\
\end{array} \right\} 
\end{align}
under the assumption of no signal.
The dominant contribution is constituted by the combination $c_{23}b_{13}^* $ which is also constrained by the non-observation of $\mu\to e\gamma$. 
	In fact, if all other contributions are neglected, the CR exhibits the strict correlation $\mathrm{CR}(\mu - e ; \mathrm{Al})  \approx 0.0027 \,\mathrm{BR}(\mu\to e\gamma)$.

%%%%%%%%%%%%%%%%%%%%%%%%%%%%%%%%%%%%%%%%%%%%%%%%%%%%%%
\mathversion{bold}
\subsubsection{\texorpdfstring{$B_c \to \tau \, \nu$}{Bc taunu}}
\mathversion{normal}
\label{subsubsec:p_ana_bctaunu}
%%%%%%%%%%%%%%%%%%%%%%%%%%%%%%%%%%%%%%%%%%%%%%%%%%%%%% 

In this model, the LQ $\phi$ contributes to the leptonic decay $B_c\to\tau\nu$ and therefore modifies the lifetime of the $B_c$ meson via the process illustrated in the centre of figure~\ref{fig:btoctaunu}, see reference~\cite{Alonso:2016oyd}. In line with this approach, we employ a constraint on the $B_c$ lifetime in the SM in this work.
We equate the measured decay width with the sum of the contributions of the SM and from the LQ $\phi$, i.e.~
\begin{equation}\label{eq:fullBcLifetime}
\Gamma^\text{exp}_{B_c} = \Gamma^\text{SM}_{B_c} + \Gamma^\phi_{B_c}.
\end{equation}
Here, we fix $\tau^\text{exp}_{B_c} = 1/\Gamma^\text{exp}_{B_c} = 0.510\pm0.009$ ps~\cite{ParticleDataGroup:2020ssz} to the best-fit value while $\Gamma^\phi_{B_c}$ accounts for the tree-level process $bc\to\tau\nu$ induced by $\phi$. The decay width $\Gamma^\phi_{B_c}$ can be calculated by subtracting the SM contribution to $\Gamma(B_c\to \tau \nu)$, see eq.~(\ref{eq:LeptonicMesonDecay}) and eq.~(\ref{eq:tauBcNPcontribution}). Hence, it also captures interference effects. $\Gamma^\text{SM}_{B_c}$ takes into account all SM contributions to the $B_c$ decay width.

We do not attempt a calculation of $\Gamma^\text{SM}_{B_c} = 1/\tau^\text{SM}_{B_c} $, but instead indirectly infer it from eq.~(\ref{eq:fullBcLifetime}) and confront this inferred value with the result $\tau^\text{SM}_{B_c} \in [0.4,0.7]$ ps~\cite{Beneke:1996xe}, see also table~\ref{table:primaryconstraints} and eq.~(\ref{eq:tauBcSM}) for the complete expression.\footnote{For more recent calculations of the $B_c$ lifetime in the SM, see 
references~\cite{Aebischer:2021ilm,Aebischer:2021eio}.}
The LQ $\phi$ mainly sources the channel with a tau neutrino $\nu_\tau$ in the final state. Upon rearranging eq.~(\ref{eq:fullBcLifetime}), one approximately finds
\begin{align}
	\frac{\tau^\text{SM}_{B_c}}{\tau^\text{exp}_{B_c}} & = \left[1 - \frac{\Gamma^\phi_{B_c}}{\Gamma^\text{exp}_{B_c}}\right]^{-1} \approx 1 + \frac{\Gamma^\phi_{B_c}}{\Gamma^\text{exp}_{B_c}} \\
	& \approx 1 - 0.13\,\frac{\text{Re}(a_{33}b_{32})}{\;\hat{m}_\phi^2} + 0.19\,\frac{|a_{33}b_{32}|^2}{\hat{m}_\phi^4} + 0.01\,\frac{|a_{23}b_{32}|^2}{\hat{m}_\phi^4}\label{eq:tauBcSM_estimateLine1}
	\\
	& = 1 - 0.13\,\frac{|a_{33}b_{32}|}{\hat{m}_\phi^2}\,\cos\big(\text{Arg}(a_{33}) - \text{Arg}(b_{32})\big) + 0.19\,\frac{|a_{33}b_{32}|^2}{\hat{m}_\phi^4} + 0.01\,\frac{|a_{23}b_{32}|^2}{\hat{m}_\phi^4}
	\label{eq:tauBcSM_estimate}
\end{align}
where the rightmost term in eq.~(\ref{eq:tauBcSM_estimateLine1}) and eq.~(\ref{eq:tauBcSM_estimate}) represents the LFV contribution with a muon neutrino $\nu_\mu$ in the final state. Eq.~(\ref{eq:fullBcLifetime}) is also equivalent to the following relation
\begin{equation}\label{eq:relationBc_branchingratios}
\text{BR}(B_c\to \tau\nu) = \text{BR}(B_c\to \tau\nu)_{\text{SM}} - \left(\frac{\tau^{\text{exp}}_{B_c}}{\tau^{\text{SM}}_{B_c}} - 1\right).
\end{equation}
Thereby, imposing an upper bound on the BR, say BR$(B_c\to\tau\nu)\lesssim 0.3$~\cite{Alonso:2016oyd} or BR$(B_c\to\tau\nu)\lesssim 0.1$~\cite{Akeroyd:2017mhr}, which takes into account the (semi)tauonic contributions in the SM and from new physics, is equivalent to $\tau^{\text{SM}}_{B_c}\lesssim 0.70$ ps or $\tau^{\text{SM}}_{B_c}\lesssim 0.55$ ps, respectively.

%%%%%%%%%%%%%%%%%%%%%%%%%%%%%%%%%%%%%%%%%%%%%%%%%%%%%%
\mathversion{bold}
\subsubsection{\texorpdfstring{$R^\nu_{K^{(\star)}}$}{RKnu}}
\mathversion{normal}
\label{subsubsec:p_anaRKnu}
%%%%%%%%%%%%%%%%%%%%%%%%%%%%%%%%%%%%%%%%%%%%%%%%%%%%%% 

We consider the decay $B\to K^{(\star)}\nu\overline{\nu}$ and normalise it to the respective SM prediction in the ratio $R^\nu_{K^{(\star)}}$. Eq.~(\ref{eq:RKnu}) contains the full expression for this contribution, which is derived following reference~\cite{Buras:2014fpa}. The dominant contributions arise via the process illustrated in figure~\ref{fig:btoctaunu}, and give
\begin{eqnarray}
\label{eq:RKstar}
R^\nu_{K^{(\star)}} &\approx& 1 + 1.69\,
\frac{|a_{33}a_{32}|}{\hat{m}_\phi^2}\cos\big(\text{Arg}(a_{33}) - \text{Arg}(a_{32})\big) + 2.15\,\frac{|a_{33}a_{32}|^2}{\hat{m}_\phi^4}  \\
&&+ 0.09\,
\frac{|a_{23}a_{22}|}{\hat{m}_\phi^2}\cos\big(\text{Arg}(a_{23}) - \text{Arg}(a_{22})\big) + 0.01\,\frac{|a_{23}a_{22}|^2}{\hat{m}_\phi^4}\nonumber \\
&&+ 0.11\,\left(\frac{|a_{23}a_{32}|^2}{\hat{m}_\phi^4} + \frac{|a_{33}a_{22}|^2}{\hat{m}_\phi^4}\right)
\;.\nonumber
\end{eqnarray}
The first line of eq.~\eqref{eq:RKstar} represents the contribution from the tau neutrino-antineutrino pair $\nu_\tau \overline{\nu_\tau}$ in the final state, while the second line encodes the contribution from the combination $\nu_\mu \overline{\nu_\mu}$, and the last line contains the LFV contribution from the combinations $\nu_\tau \overline{\nu_\mu}$ and $\nu_\mu \overline{\nu_\tau}$, respectively.
As the contributions to RH vector currents are negligible in this model, we have $R^\nu_K = R^\nu_{K^{\star}}$ and thus the more stringent experimental bound, $R^\nu_{K^{\star}} < 2.7$ at 90\% C.L. \cite{Belle:2017oht},  acts as a primary constraint.

%%%%%%%%%%
%%%%%%%%%%%%%%%%%%%%%%%%%%%%%%%%%%%%%%%%%%%%%%%%%%%%%%
\mathversion{bold}
\subsubsection{\texorpdfstring{$Z\to \tau{\tau}$}{Ztautau}}
\mathversion{normal}
\label{subsubsec:p_ana_leptoZdecays}
%%%%%%%%%%%%%%%%%%%%%%%%%%%%%%%%%%%%%%%%%%%%%%%%%%%%%%

\begin{figure}[t!]
\centering
\includegraphics[scale=1.2]{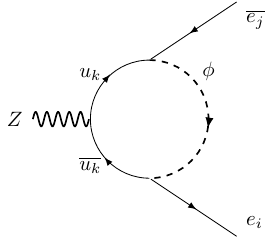}\hspace{1cm}
\includegraphics[scale=1.2]{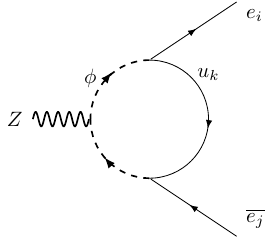}
\caption{Dominant LQ, $\phi$, contribution to the process $Z \to{e_i} \overline{e_j}$, arising at one-loop order with the up-type quark $u_k$ running in the loop.}
\label{fig:ZtautauFD}
\end{figure}

Inducing contributions to $b\to c \tau\nu$ in this model, as discussed in section~\ref{subsubsec:p_ana_RDRDstar}, requires that the LQ coupling to the bottom quark and (in particular) the tau neutrino, encoded in $a_{33}$, is enhanced. This effective parameter is related via the CKM mixing matrix to $c_{33}$ which describes the coupling between the top quark and the tau lepton. Therefore, an explanation of the flavour anomalies in $R({D})$ and $R({D^\star})$ may be associated with large corrections via a top-quark loop to $Z\to \tau{\tau}$. These contributions are illustrated in figure~\ref{fig:ZtautauFD}. 

Following from reference~\cite{Arnan:2019olv}, we parametrise these contributions by considering the effective axial-vector couplings of the $Z$ boson to fermions, where the full expressions for these contributions can be found in appendix~\ref{app:Zdecays}. At LO for the effective $\tau\tau$ coupling, eq.~\eqref{deltag_ZZ} in appendix~\ref{app:Zdecays} reduces to
\begin{align}
  \delta g_{\tau_{A}}\approx \left\{
\begin{array}{*{2}{c}}
2.28\,, & \hat m_\phi =2\\ 
0.75\,, & \hat m_\phi =4\\
0.38\,, & \hat m_\phi =6\\
\end{array} \right\}  |c_{33}|^2 \times  10^{-4}\;.
\end{align}
Following from appendix~\ref{app:Zdecays} for the definition of $g_A^{\rm SM} (<0)$, taking lepton flavour to be conserved for SM couplings (i.e. $g_A^{\rm SM}$ is the same for all lepton flavours) this yields
\begin{align}\label{eq:Ztau_coupling_approx}
 g_{\tau_{A}}/g_A^{\rm SM} \approx 1-  \left[\left\{
\begin{array}{*{2}{c}}
4.54\,, & \hat m_\phi =2\\ 
1.50\,, & \hat m_\phi =4\\
0.75\,, & \hat m_\phi =6\\
\end{array} \right\}\; |c_{33}|^2 \times  10^{-4} \right].
\end{align}
Therefore, allowing for a $3 \, \sigma$ margin about the best-fit value quoted in table~\ref{table:primaryconstraints}, we obtain the following constraints on  $|c_{33}|$
\begin{align}
  |c_{33}| \lesssim \left\{\begin{array}{*{2}{c}}
2.25\,, & \hat m_\phi =2\\ 
3.92\,, & \hat m_\phi =4\\
5.50\,, & \hat m_\phi =6\\
\end{array} \right\}\;. \label{eq:c33taugestimate}
\end{align}
Note that if the present best-fit value for  $g_{\tau_{A}}/g_A^{\rm SM}$ remains the same, but if either of the projected sensitivities to this observable mentioned in table~\ref{table:primaryconstraints} are reached, then this  model would not be capable of addressing this deviation from the SM value. Thus, we do not give a future reach for the bound on $|c_{33}|$ in eq.~\eqref{eq:c33taugestimate}.  This is seen explicitly from the results presented in section~\ref{subsubsec:Hadronic_primary} and in section~\ref{sec:secondarytertiary}.

%%%%%%%%%%
%%%%%%%%%%%%%%%%%%%%%%%%%%%%%%%%%%%%%%%%%%%%%%%%%%%%%%
\mathversion{bold}
\subsubsection{\texorpdfstring{High-$p_T$ dilepton searches}{High pT dilepton searches}}
\mathversion{normal}
\label{subsubsec:s_ana_highpT}
%%%%%%%%%%%%%%%%%%%%%%%%%%%%%%%%%%%%%%%%%%%%%%%%%%%%%%   

Several recent studies~\cite{Fuentes-Martin:2020lea,Angelescu:2020uug,Angelescu:2018tyl,Greljo:2018tzh} have placed constraints on effective operators using LHC data.
In reference~\cite{Angelescu:2018tyl}, the process $q\bar q \to \tau\bar \tau$ has been considered for the LQ $\phi$, among other ones, and the ATLAS analysis in reference~\cite{ATLAS:2017eiz} has been reinterpreted in order to put a constraint on the LQ couplings for masses $1 \leq \hat m_\phi \leq 3$.
Reading off from the top-right of figure 4 in reference~\cite{Angelescu:2018tyl} and using the fact that the LHC does not distinguish between chiralities, we find an upper bound for the LQ coupling 
involving a RH tau lepton and a charm quark
\begin{align}
|y_{32}| = |b_{32}|  <  \hat m_\phi + 0.6
\;.
\end{align}
Similarly, in reference~\cite{Greljo:2018tzh} the process $b+ c \to \tau +\nu$ has been considered and two analyses~\cite{ATLAS:2018ihk,CMS:2018fza} by ATLAS and CMS have been recast to place a constraint on the charged-current effective operators. The resulting constraints, under the assumption of the dominance of a single operator, are found in table II of reference~\cite{Greljo:2018tzh}. In terms of the effective couplings at the LQ mass scale, they read 
\begin{align}
\sqrt{|a_{33}c_{32}|} < 3.5\, \hat m_\phi
\;\;\;\;
\text{and} \;\;\;\;
\sqrt{|a_{33} b_{32}|}  < 0.70\, \hat m_\phi \, ,
\end{align}
where we have included RG corrections due to QCD using \texttt{RunDec}~\cite{Chetyrkin:2000yt,Herren:2017osy}. Still, these constraints are automatically respected in the model, if the experimental bounds on other primary observables are imposed.
  
\begin{table}[t!]\centering
	\resizebox{13cm}{!}{
  \def\arraystretch{1.4} 
	\begin{tabular}{|c|c|}
			\hline
			Observable & Effective parameters \\
			\hline
			$R (D)$ & $a_{33}$, $b_{32}$, ($a_{23}$) \\
			$R(D^\star)$ & $a_{33}$, $b_{32}$, ($a_{23}$) \\
			$\Delta a_\mu$ & $b_{23}$, $c_{23}$  \\
			\hline
			$\mathrm{BR}(\tau\to\mu\gamma)$ & $b_{23}$, $c_{33}$
			\\
			$\mathrm{BR}(\mu\to e\gamma)$ & $b_{13}$, $c_{23}$ \\
			$\mathrm{BR}(\tau \to 3\, \mu)$ & $b_{23}$, $c_{33}$, ($c_{23}$) \\
			$\mathrm{BR}(\tau \to \mu e\bar e)$ & $b_{23}$, $c_{33}$, ($c_{23}$) \\[0.03in] 
			\hline
	\end{tabular}
	\hspace{0.5cm}
	\begin{tabular}{|c|c|}
		\hline
		Observable & Effective parameters \\
		\hline
		$\mathrm{BR}(\mu\to 3 \, e)$ & $b_{13}$, $c_{23}$ \\
		CR($\mu - e;\;$Al) & $b_{13}$, $c_{23}$, ($b_{11}$, $b_{23}$, $c_{13}$, $c_{21}$) \\
		$R^\nu_{K^{\star}}$ & $a_{32}$, $a_{33}$, ($a_{22}$, $a_{23}$) \\
		$g_{\tau_{A}}/g_{A}^{\rm SM}$ & $c_{33}$ \\
		\hline
		$\tau_{B_c}^{\text{SM}}$
		& $a_{33}$, $b_{32}$, ($a_{23}$) \\
		\hline
		$c\overline{c}\to\tau\bar\tau$ & $b_{32}$ \\
		$bc\to\tau\nu$
		& $a_{33}$, $b_{32}$, ($c_{32}$) \\[0.03in] 
		\hline
	\end{tabular}

}
	\caption{{\small\textbf{List of primary observables and
	relevant effective parameters}. We list the observables that dominantly constrain this model together with the effective parameters of LQ couplings in the charged fermion mass basis, see section~\ref{subsubsec:LQcoupmass}, which capture the most relevant contributions, respectively, in line with the analytic estimates performed in section~\ref{subsec:primary_analytic}.
	The parameters listed in round brackets refer to subdominant contributions.
	}} \label{table:primaryconstraints_coupling_coeffs}
\end{table}

%%%%%%%%%%%%%%%%%%%%%%%%%%%%%%%%%%%%%%%%%%%%%%%%%%%%%%
\subsection{Numerical study}
\label{subsec:primary_numeric}
%%%%%%%%%%%%%%%%%%%%%%%%%%%%%%%%%%%%%%%%%%%%%%%%%%%%%%  

In this section, we present and discuss the results of a numerical scan taking into account the primary observables.
The focus rests on studying the way in which the imposed current experimental bounds shape the parameter space compatible with the model, and how this affects the possibility to explain the observed flavour  anomalies in $R(D^{(\star)})$ and in the AMM of the muon, see table~\ref{table:primaryconstraints}. Furthermore, the results help to establish biases for
the 
comprehensive scan 
discussed in section~\ref{sec:secondarytertiary}.

%%%%%%%%%%%%%%%%%%%%%%%%%%%%%%%%%%%%%%%%%%%%%%%%%%%%%%  

\subsubsection{Preliminaries}
\label{subsubsec:primary_preliminaries}

According to the strategy outlined in section~\ref{sec:phenomenology},
the following discussion refers to scenario B only.
The effective parameters\footnote{We remind that $c_{ij}$ and $c_{ij}^B$ are in general two inequivalent sets of effective parameters, see section~\ref{subsubsec:LQcoupmass}.
}
$a_{ij}$, $b_{ij}$ and
$c_{ij}^B$ (except for $c_{12}^B$, see below) of the LQ couplings ${\bf x}$, ${\bf y}$ and ${\bf z}$ are independently varied within the ranges given in eqs.~(\ref{eq:unbiased_magnitude}) and (\ref{eq:unbiased_phase}).
Note that we make a different choice in the case of $|b_{32}|$ for $\hat{m}_\phi = 2$, see table~\ref{table:primaryconstraints}.
Furthermore, combining eqs.~(\ref{eq:zxparaAadd}), (\ref{eq:zxparaBadd}) and the first line in eq.~(\ref{eq:xpararel}) with the structure of the CKM mixing matrix in scenario B, see eqs.~(\ref{eq:VusVcb}) and (\ref{eq:VCKMB}), one finds
\begin{equation}
	|V_{us}| \approx \left|\frac{c_{12}^B}{a_{11}}\right|\lambda, \quad |V_{cb}| \approx |\tilde{c}|\lambda^2 \quad \text{and} \quad |V_{td}| \approx |\bar{c}|\lambda^3
\end{equation}
up to corrections of higher order in $\lambda$. Comparing these predictions to the best-fit values of the experimentally inferred CKM mixing matrix elements, $|V_{us}| =\lambda = 0.22650^{+0.00048}_{-0.00048}$, $|V_{cb}|=0.04053^{+0.00083}_{-0.00061}$, and $|V_{td}|=0.00854^{+0.00023}_{-0.00016}$~\cite{ParticleDataGroup:2020ssz}, we conclude that the relation between the LQ couplings ${\bf x}$ and ${\bf z}$ is, indeed, to a good approximation given by the CKM mixing matrix, if we constrain the parameters $c_{12}^B$, $\tilde{c}$ and $\bar{c}$ as follows 
\begin{equation}
\label{eq:c_num}
c_{12}^B = a_{11}
\, \alpha \, e^{i \, \omega_1} \; , \;\; \tilde{c} = \beta \, e^{i \, \omega_2} \; , \;\; \bar{c} = \gamma \, e^{i \, \omega_3}
\end{equation}
with
\begin{equation}
\label{eq:albegaom_num}
0.5 \leq \alpha , \, \beta , \, \gamma \leq 1.5 \;.
\end{equation}
The phases are varied in the full range
\begin{equation}
	0 \leq \omega_i < 2 \, \pi \;\; \mbox{for} \;\; i=1,2,3\;.
\end{equation}
Here, no information about CP phases, captured by the Jarlskog invariant, has been taken into account.

The subsequent discussion including the figures is based on a sample comprising $4(3)[2]\times10^6$ points for $\hat{m}_\phi = 2(4)[6]$. The hadronic observables $R(D)$, $R(D^\star)$ and $\tau^{\text{SM}}_{B_c}$ exhibit RG running under QCD and so we evaluate them at the scale $\mu = \mu_B = 4.8$ GeV, as detailed in appendix~\ref{sec:TreeLevelMatching}. On the contrary, the remaining leptonic observables are evaluated at $\mu = m_\phi$, that is, we neglect the smaller contributions from QED running in this section.
We impose the current experimental bounds on BR($\tau\to\mu\gamma$), BR($\mu\to e\gamma$), BR($\tau\to3\,\mu$),
BR($\tau\to\mu e \bar e$), BR($\mu\to 3\,e$), $R^\nu_{K^{(\star)}}$, $\tau^{\text{SM}}_{B_c}$ and $g_{\tau_A}/g^{\text{SM}}_A$, see table~\ref{table:primaryconstraints}. For completeness, we also track the contributions to the scalar charged-current Wilson coefficient $C^{SRR}_{\nu edu,3332}$ and provide a brief discussion in appendix~\ref{app:plots_wilsoncoeff}.

In general, for the scatter plots in this section we use round sample points to indicate the violation of at least one of the imposed experimental bounds, and the ones with a specific shape (star, plus, cross) show that all considered current bounds are respected.
The employed colours as well as shapes allow to distinguish well between the results for the different LQ masses, $\hat{m}_\phi  = 2, 4, 6$, as displayed in the plot legends. Furthermore, solid lines generally refer to current experimental data at a given confidence level, whereas a dashed line indicates a prospective bound or a future sensitivity. Gray shadings are used in order to better distinguish the regions in parameter space compatible with current data at different confidence levels. Besides, the green shaded regions, as well as the black cross in the bottom-left plot of figure~\ref{fig:only_anomalies}, indicate the SM prediction for $R(D)$ and $R(D^\star)$ at the $1\,\sigma$ level, respectively.

%%%%%%%%%%%%%%%%%%%%%%%%%%%%%%%%%%%%%%%%%%%%%%%%%%%%%%  
\mathversion{bold}
\subsubsection{\texorpdfstring{$R(D)$, $R(D^\star)$ and anomalous magnetic moment of muon}{RD, RD* and anomalous magnetic moment of muon}}
\mathversion{normal}
\label{subsubsec:prelimscan_anomalies}

\begin{figure}[t!]
	\includegraphics[scale=0.52]{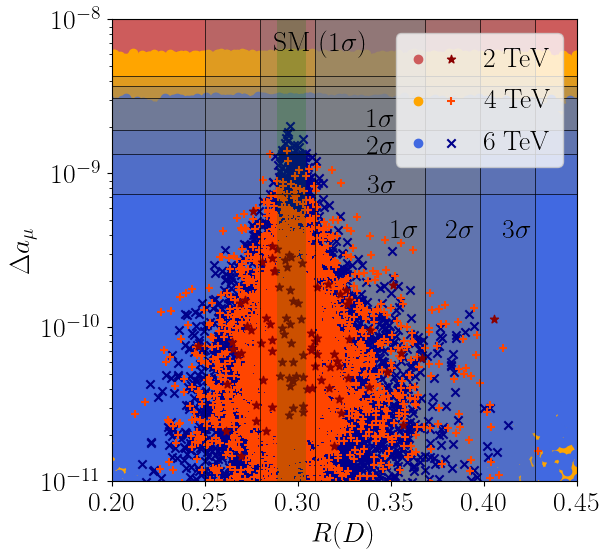}
	\includegraphics[scale=0.52]{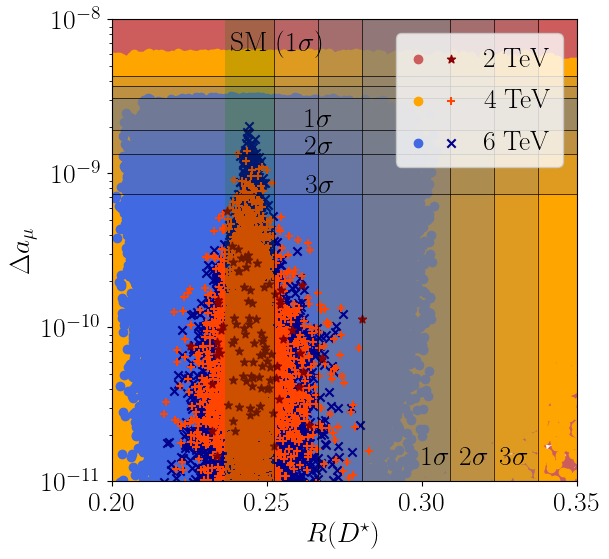}
	\\
	\begin{minipage}[c]{0.5\textwidth}
	\includegraphics[scale=0.52]{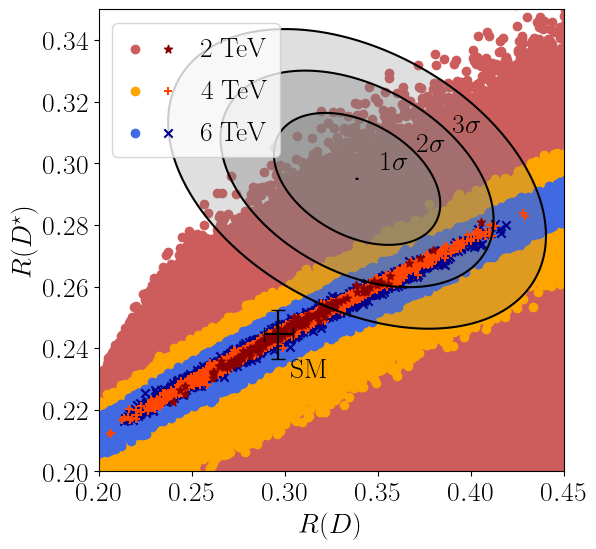}
	\end{minipage}\hfill
	\begin{minipage}[c]{0.5\textwidth}
		\caption{\linespread{1.1}\small{\textbf{Predictions for \mathversion{bold}$R(D)$, $R(D^\star)$ and $\Delta a_\mu$.\mathversion{normal}} The regions marked by solid lines are compatible with the current experimental world averages for $R(D^{(\star)})$ \cite{Amhis:2019ckw} and $\Delta a_\mu$ \cite{Muong-2:2021ojo,Aoyama:2020ynm}, respectively, at the indicated confidence level, see table~\ref{table:primaryconstraints}. We use the values output by \texttt{flavio}, v2.3 for the SM predictions for $R(D^{(\star)})$ at the $1\,\sigma$ level~\cite{Straub:2018kue,david_straub_2021_5543714,Bordone:2019vic}, see appendix~\ref{subsec:RDRDstar}. These are shown by the green shaded bands (top) and the black cross (bottom-left).
		The round points (geometric shapes) indicate that current experimental bounds are violated (respected), see also the main text of section~\ref{subsubsec:primary_preliminaries}.} 
		}
		\label{fig:only_anomalies}
	\end{minipage}
\end{figure}

\paragraph{Addressing the anomalies.} The capability of this model to explain the anomalies in $R(D)$, $R(D^\star)$ and in the AMM of the muon, as found in the primary scan, is illustrated in figure~\ref{fig:only_anomalies}. A priori, a value up to $\Delta a_\mu\approx 3 \times 10^{-9}$ or larger can be achieved, depending on the LQ mass, in accordance with the analytic estimate in eq.~(\ref{eq:gm2_estimate}) in the case of large LQ couplings. Still, after imposing the experimental bounds of all primary observables, a result of the order $\Delta a_\mu \sim 10^{-9}$ is not generic, but instead we find a suppression by one or two orders of magnitude for about 90 percent of the viable sample points with positive $\Delta a_\mu$ generated in the primary scan, irrespective of the LQ mass. 
We remark that imposing these experimental bounds does not lead to a preference for either sign of $\Delta a_\mu$, as is expected, since none of the primary observables exhibits a particular sensitivity to the phase of $b_{23}$ or $c_{23}\approx a_{23}$.\footnote{As can be seen in eq.~(\ref{eq:zxparaB}), the effective parameters $c_{23}$ and $a_{23}$ as well as $c_{33}$ and $a_{33}$ agree up to $\mathcal{O}(\lambda^4)$, respectively. Since $c_{23}$ and $c_{33}$ are not varied directly in the primary scan, the implications for these are mainly discussed in terms of $a_{23}$ and $a_{33}$ in this section.}
Nevertheless, the results hint towards the possibility of explaining $\Delta a_\mu$ at the $2\,\sigma$ level or better in this model,
see the top in figure~\ref{fig:only_anomalies}.

Furthermore, as expected from eq.~(\ref{eq:gm2_estimate}), i.e.~$\Delta a_\mu\propto |b_{23}c_{23}|$, and $R(D^{(\star)})$ mainly controlled by $|a_{33}b_{32}|$, see eqs.~(\ref{eq:RD_estimate}, \ref{eq:RDs_estimate}), these observables are a priori not (strongly) correlated in this model. The distribution of viable sample points in figure~\ref{fig:only_anomalies} is due to the experimental constraint on BR($\tau\to\mu\gamma$), see section~\ref{subsubsec:prelimscan_eiejgamma} for more details. In particular, this entails a tension between explaining the flavour anomaly in $R(D^\star)$ at the $3\,\sigma$ level or better and generating $\Delta a_\mu\sim 10^{-9}$. 

Using \texttt{flavio} \cite{Straub:2018kue,david_straub_2021_5543714,Bordone:2019vic} (since v2.0), one finds that the SM prediction $R(D)_{\text{SM}} = 0.297\pm0.008$ is compatible with the current experimental world average
at the $2\,\sigma$ level, that is, the anomaly is primarily constituted by the discrepancy between $R(D^\star)_{\text{SM}} = 0.245\pm0.008$ and the corresponding experimental value
\cite{Amhis:2019ckw} which overlap only at the $3\,\sigma$ level.\footnote{Since the values for $R(D)_{\text{SM}}$ and $R(D^\star)_{\text{SM}}$ that are generated by \texttt{flavio} differ from those quoted in reference~\cite{Amhis:2019ckw}, the significances are not in exact correspondence with the ones in table~\ref{table:anomalies}.} Thus, a combined explanation of the anomalies in $R(D)$, $R(D^\star)$ and in the AMM of the muon at a confidence level of $3\,\sigma$ or better is challenging in the primary scan, in particular due to the correlation between the latter two observables. We refer to section~\ref{ssec:primaryresultsComprehensive} for a revision of these trends.

The observables $R(D)$ and $R(D^\star)$ are linearly correlated in the model by construction. As is visible in the bottom-left plot in figure~\ref{fig:only_anomalies}, only in the case $\hat{m}_\phi = 6$ a combined explanation of the anomalies in $R(D^{(\star)})$ at the $1\,\sigma$ level is a priori impossible.
 Imposing the experimental bounds results in a quite pronounced correlation, namely $R(D^\star) \approx 0.30\,R(D)+0.15$, and a combined explanation of $R(D)$ and $R(D^\star)$ is possible at the $2\,\sigma$ level for all considered LQ masses.

\begin{figure}[t!]
	\includegraphics[scale=0.31]{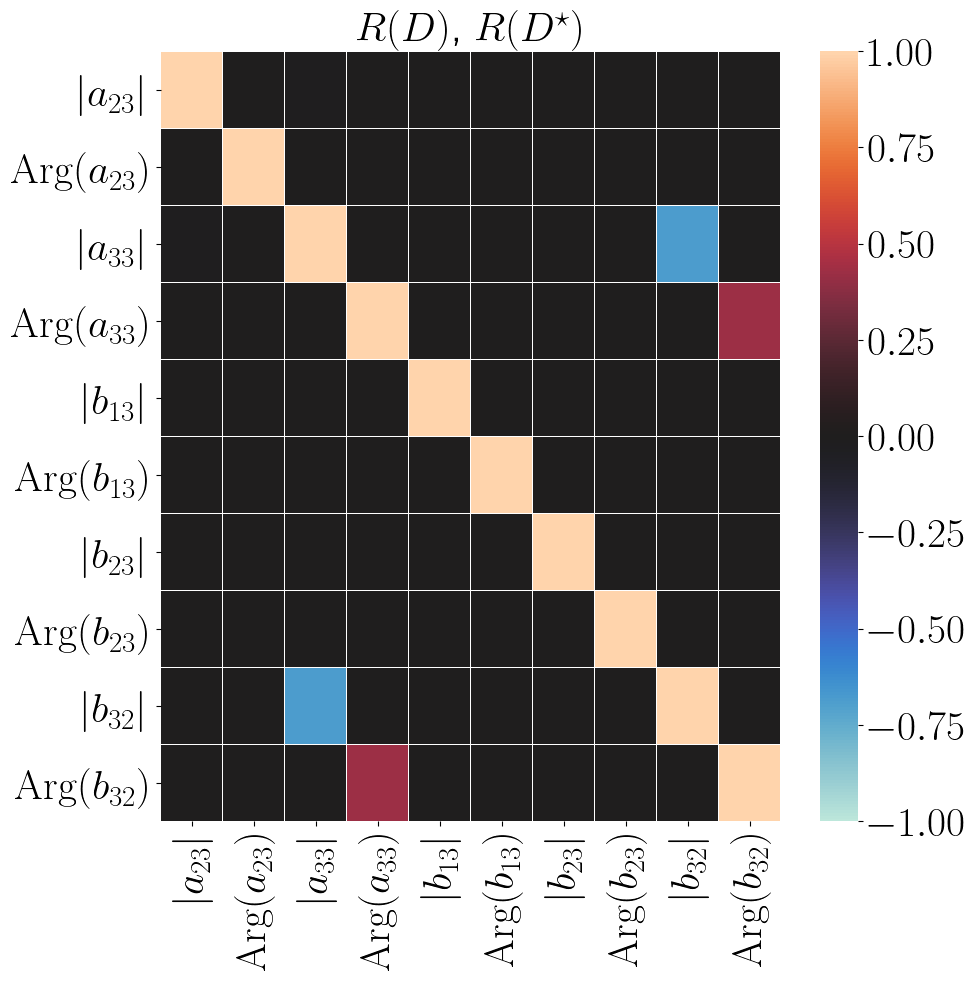}
	\includegraphics[scale=0.31]{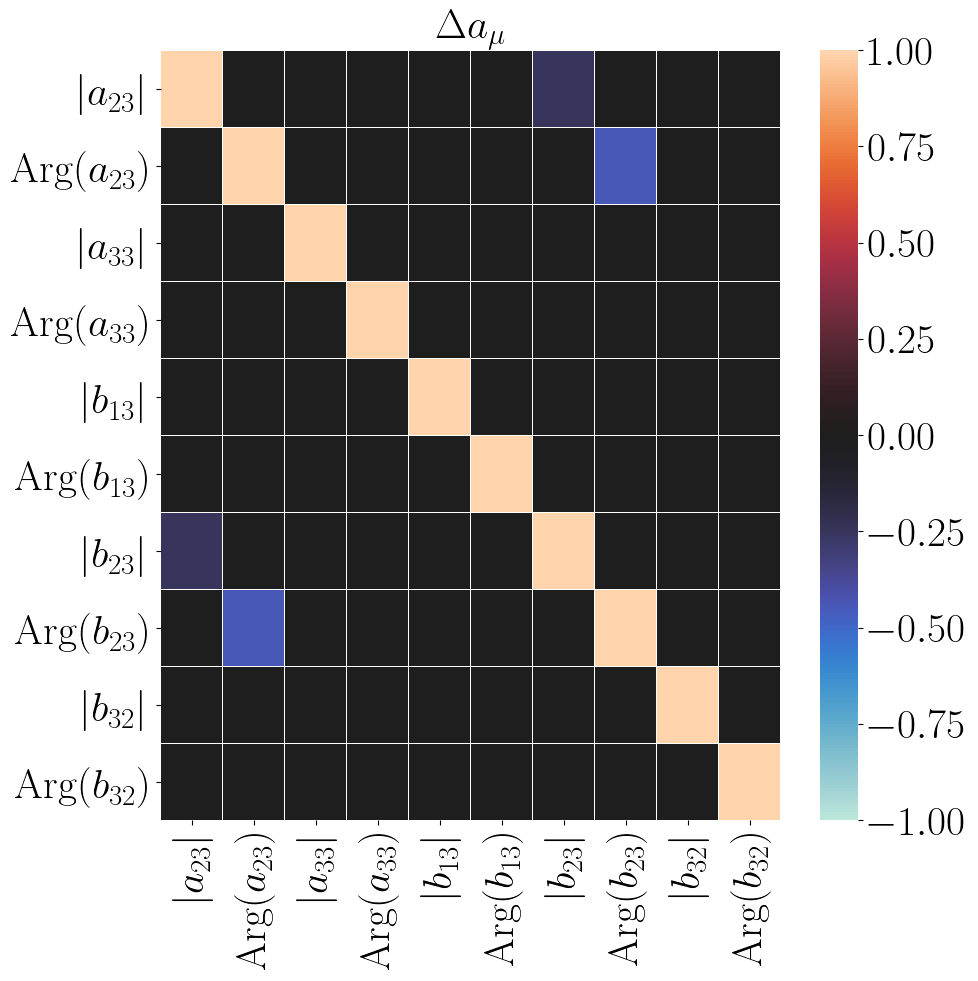}
	\caption{\linespread{1.1}\small{\textbf{Correlation plots for \mathversion{bold}$\hat{m}_\phi = 4$ based on the sample points which explain $R(D^{(\star)})$ (left) and $\Delta a_\mu$ (right) at $2\,\sigma$ and $1\,\sigma$ level, respectively. \mathversion{normal}} The plots visualise Spearman's rank correlation coefficient, calculated via the library \texttt{seaborn} \cite{Waskom2021}. A negative (positive) correlation among, e.g., the magnitudes of two effective parameters indicates that if one of them increases, the other one tends to decrease (also increase).
			Note that sample points not respecting all experimental bounds on the primary observables are taken into account here as well.
			}}
	\label{fig:correlation_plots}
\end{figure}

\paragraph{Correlations between parameters.} In order to substantiate these results, we have checked for all LQ couplings whether they display some non-trivial correlation, if the flavour anomalies in $R(D)$ and $R(D^\star)$ are explained, or in the case the measured value of the AMM of the muon is explained, assuming a certain confidence level in each case. For that purpose, we make use of an algorithm to calculate  Spearman's rank correlation coefficient, as provided by the library \texttt{seaborn}~\cite{Waskom2021}.

The correlation plots in figure~\ref{fig:correlation_plots} show the effective parameters, separated in magnitude and argument, which display a non-zero correlation, if $R(D^{(\star)})$ (left) and $\Delta a_\mu$ (right) are explained, respectively. Here, all sample points for which the respective anomaly is explained are taken into account, regardless of whether all experimental bounds on the primary observables are respected or not. The effective parameter $b_{13}$ is included as well for the sake of comparison, because it is sensitive to the experimental bounds. Its effects are detailed in sections~\ref{subsubsec:prelimscan_eiejgamma} and~\ref{subsubsec:prelimscan_trilepton_mueconv}. We have chosen $\hat{m}_\phi = 4$ and the confidence level of $2\,\sigma$ as illustrative example. Nevertheless, the results are not appreciably different for the other considered LQ masses and confidence levels. 
A negative (positive) correlation is shown in blueish (reddish) colour.
The points entering the correlation plot for $R(D^{(\star)})$ and $\Delta a_\mu$ comprise roughly 10 and 15 percent of the entire sample for $\hat{m}_\phi = 4$, respectively.

As evidenced by the analytic estimates for $R(D)$ and $R(D^\star)$ in eqs.~(\ref{eq:RD_estimate}) and (\ref{eq:RDs_estimate}), the result for either observable is largely controlled by the product $|a_{33}b_{32}|$ which has to fall in an appropriate range to explain the anomalies. Furthermore, the arguments of the (complex) effective parameters $a_{33}$ and $b_{32}$ have to be positively correlated, implying that their difference should be close to zero and thus the cosines appearing in eqs.~(\ref{eq:RD_estimate}) and (\ref{eq:RDs_estimate}) take values close to one. This shows that explaining the flavour anomalies in $R(D)$ and $R(D^\star)$ requires the contribution linear in $|a_{33}b_{32}|$ to be 
 positive, that is, the contribution quadratic in $|a_{33}b_{32}|$ is generically too small to yield a dominant effect.

Similarly, as explaining the anomaly in the AMM of the muon needs positive $\Delta a_\mu$, the difference of $\text{Arg}(c_{23})\approx\text{Arg}(a_{23})$ and Arg($b_{23}$) is necessarily close to $\pi$ so that the sign of the cosine appearing in eq.~(\ref{eq:gm2_estimate}) can cancel the negative overall sign. Thus, the right plot in figure~\ref{fig:correlation_plots} indicates a (moderate) negative correlation, both in the case of $|a_{23}|$ and $|b_{23}|$ as well as for the arguments. Note that the negative correlation of the magnitudes is less pronounced than in the case of $|a_{33}|$ and $|b_{32}|$, see left plot in figure~\ref{fig:correlation_plots}. We interpret this as being due to the fact that the product $|a_{33}b_{32}|$ more directly determines the result for $R(D^{(\star)})$, since there is not only the contribution arising from the interference with the SM, but also the (smaller) contribution proportional to $|a_{33}b_{32}|^2$, which is unaffected by $\text{Arg}(a_{33}) - \text{Arg}(b_{32})$, cf. eqs.~(\ref{eq:RD_estimate}) and (\ref{eq:RDs_estimate}). For the dominant contribution to the AMM of the muon instead, a too large value of $|a_{23}b_{23}|$ can be easily compensated by an appropriate value of $\text{Arg}(a_{23}) - \text{Arg}(b_{23})$. Thus, in the case of the AMM of the muon the sensitivities to the magnitudes and arguments of $a_{23}$ and $b_{23}$ are more similar.

%%%%%%%%%%%%%%%%%%%%%%%%%%%%%%%%%%%%%%%%%%%%%%%%%%%%%%  

\mathversion{bold}
\subsubsection{\texorpdfstring{Radiative charged lepton flavour violating decays $\tau\to\mu\gamma$ and $\mu\to e\gamma$}{Radiative charged lepton flavour violating decays tau->mu gamma and mu->e gamma}}
\mathversion{normal}
\label{subsubsec:prelimscan_eiejgamma}

\begin{figure}[t!]
	\includegraphics[scale=0.52]{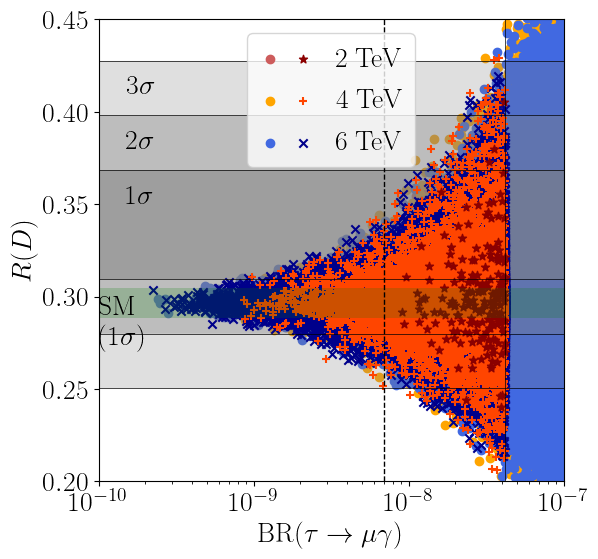}
	\includegraphics[scale=0.52]{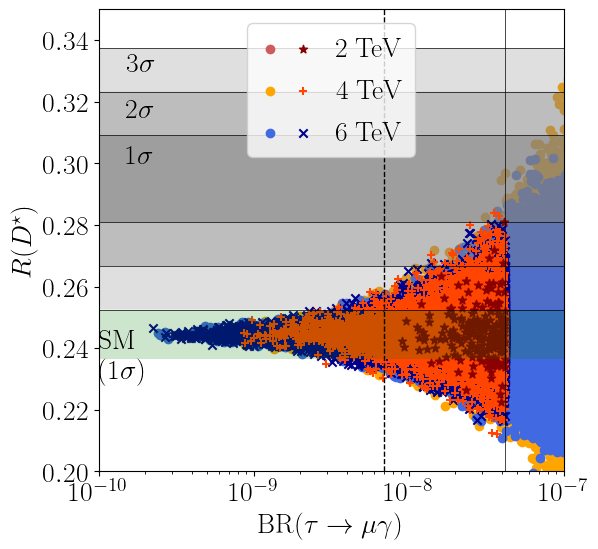}
	\\
	\includegraphics[scale=0.52]{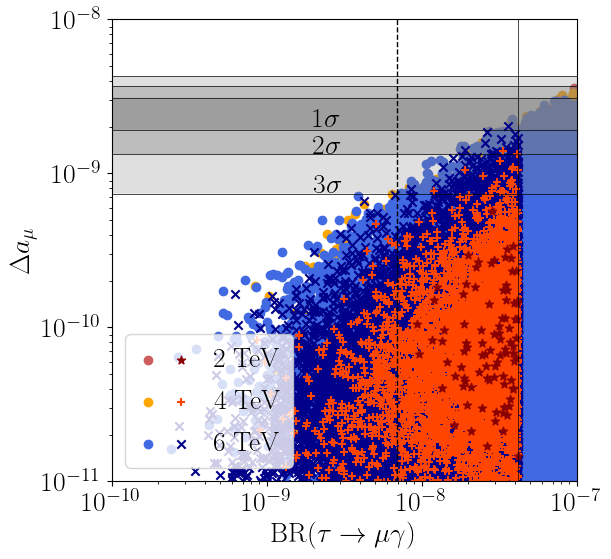}
	\includegraphics[scale=0.52]{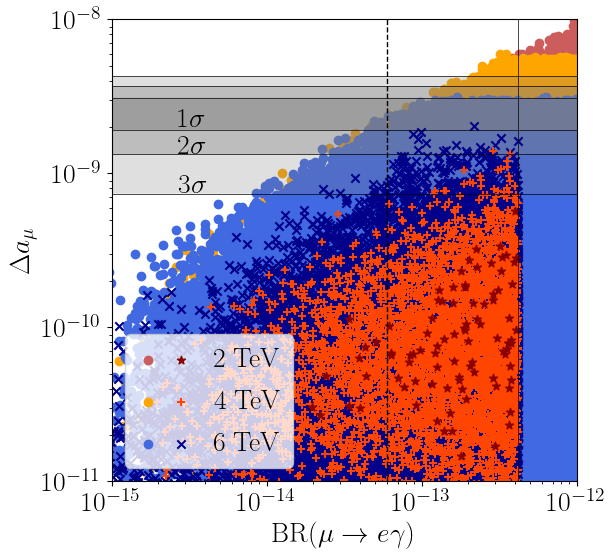}
	\caption{\linespread{1.1}\small{\textbf{Constraining power and future reach of \mathversion{bold}$\tau\to\mu\gamma$ and $\mu\to e\gamma$. \mathversion{normal}}
			The vertical solid (dashed) lines indicate the current bound on (future sensitivity of) BR($\tau\to\mu\gamma$) \cite{Belle:2021ysv,Belle-II:2018jsg} in the upper plots and the bottom-left one, and the current bound on (future sensitivity of) BR($\mu\to e\gamma$) \cite{MEG:2016leq,MEGII:2021fah} in the bottom-right plot, see table~\ref{table:primaryconstraints}.
			The round points (geometric shapes) indicate that current experimental bounds are violated (respected), see also the main text of section~\ref{subsubsec:primary_preliminaries}.}
	}
	\label{fig:constrain_power_taumug}
\end{figure}

\paragraph{Shaping the parameter space.} We move on to the discussion of the primary observables acting as constraints on the model, starting with the radiative cLFV decays $\tau\to\mu\gamma$ and $\mu\to e\gamma$. The interplay between the corresponding BRs, $R(D^{(\star)})$ and $\Delta a_\mu$ is shown in figure~\ref{fig:constrain_power_taumug}. As is expected from the analytic estimates, there are correlations between these observables: BR($\tau\to\mu\gamma$) is intertwined with $R(D)$ and $R(D^\star)$ via the effective parameter $|c_{33}|\approx|a_{33}|$, and with $\Delta a_\mu$ through $|b_{23}|$, while the latter observable also largely depends on $|c_{23}|\approx|a_{23}|$, which is, on the other hand, constrained by the experimental bound on BR($\mu\to e\gamma$), see sections~\ref{subsubsec:p_ana_RDRDstar}, \ref{subsubsec:p_ana_AMMamu} and \ref{subsubsec:p_ana_eiejgamma}.
One typically generates large contributions to BR$(\tau\to\mu\gamma)$, 
also depending on the LQ mass. Thus, this observable represents a strong constraint on the parameter space of this model.
Still, the experimental bound on BR$(\mu\to e\gamma)$ can be easily saturated as well.

This implies that both the flavour anomalies in $R(D)$ and $R(D^\star)$ can individually be explained at least at the $2\,\sigma$ level for all considered LQ masses, while passing the current experimental bound on BR($\tau\to\mu\gamma$), see figure~\ref{fig:constrain_power_taumug}. Still, the result for $R(D^\star)$ turns out to be always smaller than the experimental best-fit value. Note, though, that even in the case of a non-observation of $\tau\to\mu\gamma$ at Belle II~\cite{Belle-II:2018jsg}, an explanation of $R(D)$ within the $1\,\sigma$ level would still be possible, whereas an accommodation of the anomaly in $R(D^\star)$ would be disfavoured in that case. 

Furthermore, the shape of the viable parameter space in figure~\ref{fig:only_anomalies} can be understood by noticing the role of the experimental bound on BR($\tau\to\mu\gamma$). As indicated in section~\ref{subsec:primary_analytic}, the deviation of $R(D)/R(D)_{\text{SM}}$ and $R(D^\star)/R(D^\star)_{\text{SM}}$ from one can be approximated as a quadratic function in $|a_{33}|$, respectively, see eqs.~(\ref{eq:RD_estimate}) and (\ref{eq:RDs_estimate}). Together with $\Delta a_\mu \propto |b_{23}|$, see eq.~(\ref{eq:gm2_estimate}), and the experimental bound on BR($\tau\to\mu\gamma$) constraining the product $|b_{23}c_{33}|\approx|b_{23}a_{33}|$ according to eq.~(\ref{eq:taumug_estimate}), this bounds $R(D)/R(D)_{\text{SM}}$ and $R(D^\star)/R(D^\star)_{\text{SM}}$ from above as a function of $\Delta a_\mu$.

The upcoming searches for $\tau\to\mu\gamma$ and $\mu\to e\gamma$ \cite{MEGII:2021fah} will both probe large parts of the currently viable parameter space. In particular, for $\hat{m}_\phi = 2$ 
 the search for $\tau\to\mu\gamma$ is expected to provide a relevant test for this model.
The bottom-left plot in figure~\ref{fig:constrain_power_taumug} also indicates that current data on $\tau\to\mu\gamma$ implies an upper limit on the AMM of the muon, $\Delta a_\mu\lesssim 3\times 10^{-9}$, in this model. This can readily be recovered from combining the estimates in eq.~(\ref{eq:gm2_estimate}) and (\ref{eq:taumug_estimate}) with the current experimental bound on BR$(\tau\to\mu\gamma)$, BR$(\tau\to\mu\gamma)_{\text{exp}} < 4.2\times 10^{-8}$ \cite{Belle:2021ysv}.
In addition, both the future search for $\tau\to\mu\gamma$ at Belle II and the one for $\mu\to e\gamma$ at MEG II will test the capability of the model to explain the measured value of $\Delta a_\mu$ and potentially render an explanation of this flavour anomaly unlikely, see also the discussion in section~\ref{ssec:primaryresultsComprehensive}.

\begin{figure}[t!]
	\begin{minipage}[c]{0.5\textwidth}
		\includegraphics[scale=0.31]{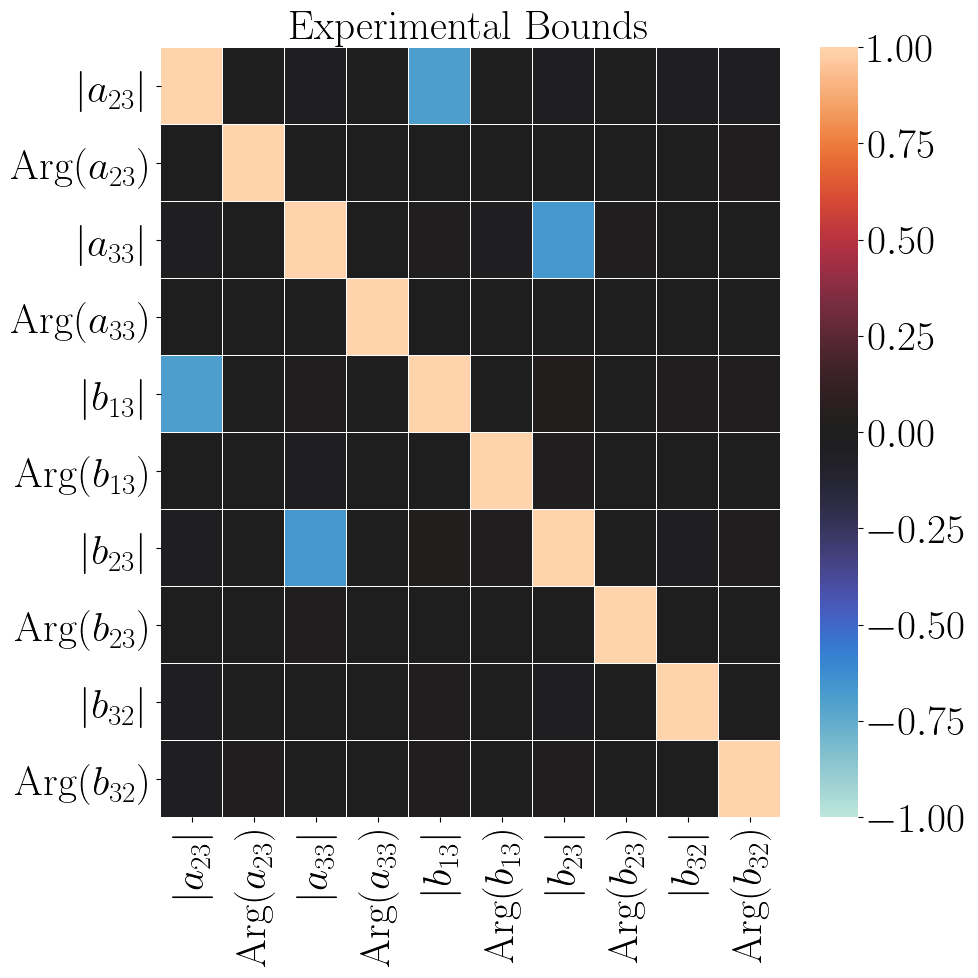}
	\end{minipage}\hfill
	\begin{minipage}[c]{0.45\textwidth}
		\caption{\linespread{1.1}\small{\textbf{\mathversion{bold}Correlation plot for $\hat{m}_\phi = 4$ based on the sample points which respect the experimental bounds of all primary constraints.\mathversion{normal}} The plot visualises Spearman's rank correlation coefficient, calculated via the library \texttt{seaborn} \cite{Waskom2021}, see caption of figure~\ref{fig:correlation_plots} for more details.
		We remind that the LO contributions to BR$(\tau\to\mu\gamma)$ and BR$(\mu\to e\gamma)$ are proportional to $|a_{33}b_{23}|^2$ and $|a_{23}b_{13}|^2$, respectively, see section~\ref{subsubsec:p_ana_eiejgamma}.}}
		\label{fig:correlation_plots_bounds}
	\end{minipage}
\end{figure}

\paragraph{Correlations between parameters.} We, thus, find that the available parameter space of this model is dominantly constrained by the experimental bounds on the radiative cLFV decays $\tau\to\mu\gamma$ and $\mu\to e\gamma$. This is further evidenced by the correlation plot in figure~\ref{fig:correlation_plots_bounds} which shows the effective parameters that display non-zero correlations, if the experimental bounds of all primary constraints are imposed. As for the correlation plots discussed in section~\ref{subsubsec:prelimscan_anomalies}, the mass $\hat{m}_\phi = 4$ is chosen as illustrative example and the results are not appreciably different for the other considered LQ masses. We also include the effective parameter $b_{32}$ in order to contrast the findings to the case of explaining the experimental anomalies in figure~\ref{fig:correlation_plots}. Note that only 0.35 percent of the generated sample points respect all imposed bounds for $\hat{m}_\phi = 4$ and thus constitute the plot in figure~\ref{fig:correlation_plots_bounds}.

Imposing an adequate negative correlation between the magnitudes $|a_{23}|$ and $|b_{13}|$ as well as $|a_{33}|$ and $|b_{23}|$, respectively, is sufficient in the primary scan to render a sample point compatible with every experimental constraint taken into account. This is in very good agreement with the findings of section~\ref{subsubsec:p_ana_eiejgamma}.
Generally, at least one of the two BRs, $\text{BR}(\tau\to\mu\gamma)$ and $\text{BR}(\mu\to e\gamma)$, is larger than its corresponding current experimental bound in the primary scan, if a bound on one of the other primary constraints is violated. Thus, the latter appear to be considerably less competitive. Still, this observation is partly revised in section~\ref{ssec:primaryresultsComprehensive}.

%%%%%%%%%%%%%%%%%%%%%%%%%%%%%%%%%%%%%%%%%%%%%%%%%%%%%%  

\mathversion{bold}
\subsubsection{\texorpdfstring{Trilepton decays $\mu\to 3\,e$, $\tau\to 3\,\mu$, $\tau\to \mu e \bar e$ and $\mu-e$ conversion in aluminium}{Trilepton decays mu->3e, tau->3mu, tau->mu ee and mu-e conversion in aluminium}}
\mathversion{normal}
\label{subsubsec:prelimscan_trilepton_mueconv}

In this section, we discuss the findings of the primary scan for several cLFV trilepton decays and $\mu-e$ conversion in aluminium.\footnote{We note that, relatively independently of the target nucleus, the model can generate contributions of $\mathcal{O}(10^{-13})$ to the respective CRs for $\hat{m}_\phi = 4$ and 6, and contributions of $\mathcal{O}(10^{-12})$ for $\hat{m}_\phi = 2$. These are, however, ruled out due to the stringent bound on and the strong correlation with BR$(\mu\to e\gamma)$ in this regime. Thus, the current experimental bounds, CR$(\mu - e;\,\text{Ti}[\text{Au}]\{\text{Pb}\})_{\text{exp}} < 0.061[0.070]\{4.6\}\times 10^{-11}$,~\cite{Wintz:1998rp,SINDRUMII:1996fti,SINDRUMII:2006dvw} do not impose relevant constraints on the model. In addition, the reach of future searches for $\mu-e$ conversion in aluminium~\cite{COMET:2018auw,Mu2e:2014fns} is projected to be three to four orders of magnitude better than for carbon targets~\cite{Teshima:2019orf}.} 

\begin{figure}[t!]
	\includegraphics[scale=0.52]{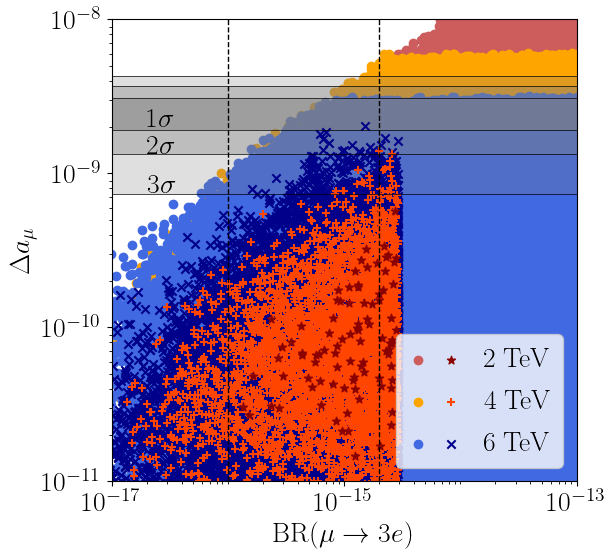}
	\includegraphics[scale=0.52]{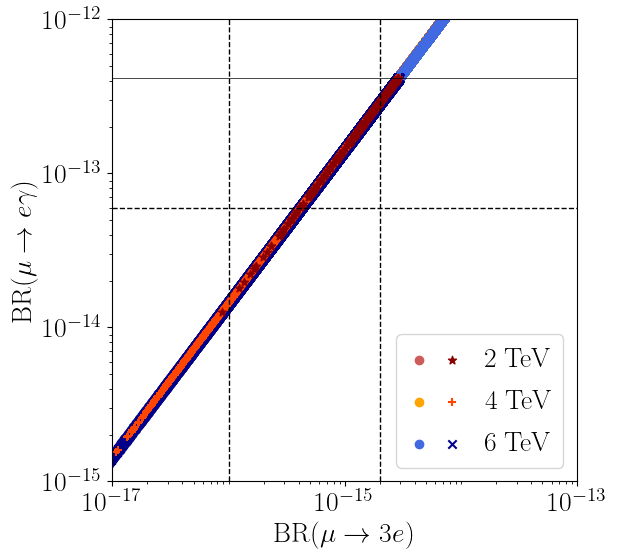}
	\\
	\caption{\linespread{1.1}\small{\textbf{\mathversion{bold}Constraining power and future reach of $\mu\to 3\,e$. \mathversion{normal}} The vertical dashed lines indicate the respective projected reach of Phase 1 and Phase 2 of the Mu3e experiment~\cite{Blondel:2013ia}, see table~\ref{table:primaryconstraints}. The round points (geometric shapes) indicate that current experimental bounds are violated (respected), see also the main text of section~\ref{subsubsec:primary_preliminaries}.} 
	}
	\label{fig:mu3e}
\end{figure}

The results, shown in the left plot in figure~\ref{fig:mu3e}, indicate that the reach of Phase 2 of the Mu3e experiment~\cite{Blondel:2013ia} may render an explanation of the anomaly in the AMM of the muon in this model
unlikely.
The right plot in figure~\ref{fig:mu3e} verifies that $\mu\to 3\,e$ is entirely dominated by long-range contributions from $\gamma$-penguin diagrams. Thus, one can effectively establish a one-to-one correspondence with the BR of $\mu\to e\gamma$ in the model, as stated in section~\ref{subsubsec:p_ana_trilepton}.

\begin{figure}[t!]
	
	\includegraphics[scale=0.52]{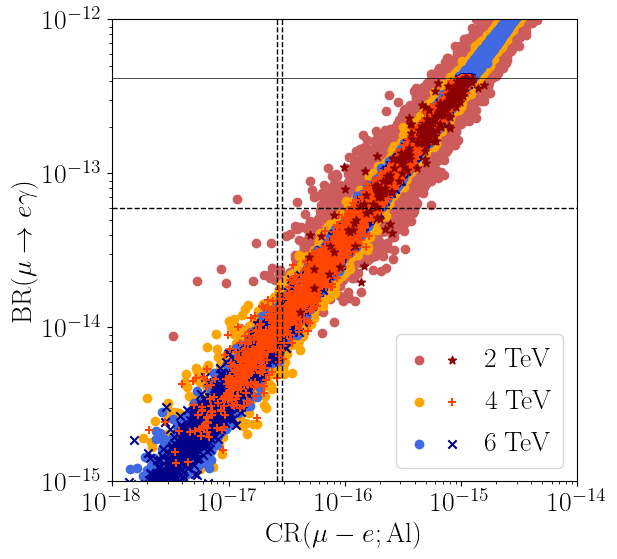}
	\includegraphics[scale=0.52]{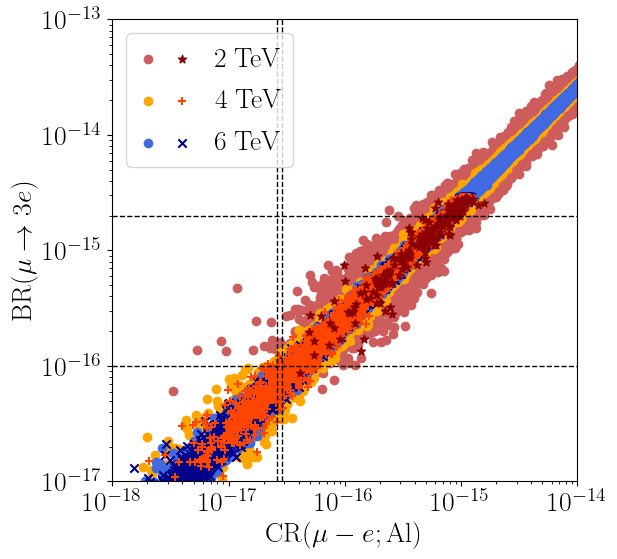}
	\begin{minipage}[c]{0.5\textwidth}
		\includegraphics[scale=0.52]{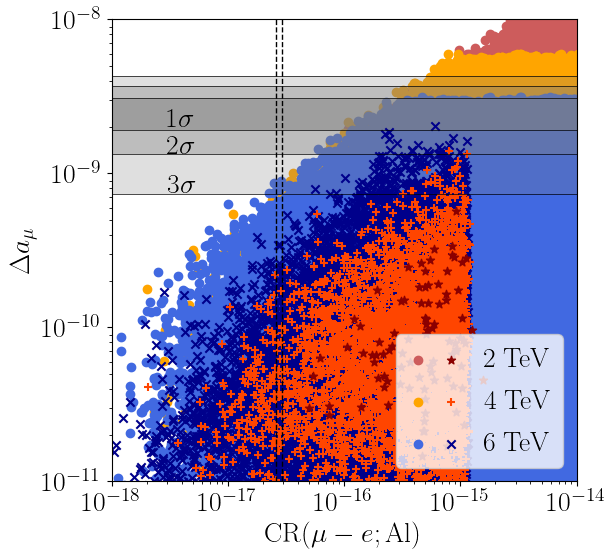}
	\end{minipage}\hfill
	\begin{minipage}[c]{0.5\textwidth}
	\caption{\linespread{1.1}\small{\textbf{\mathversion{bold}Future reach of $\mu-e$ conversion in Al.\mathversion{normal}} The vertical dashed lines indicate the future sensitivity of $\mu-e$ conversion in Al as anticipated by COMET~\cite{COMET:2018auw} and Mu2e~\cite{Mu2e:2014fns}, see table~\ref{table:primaryconstraints}. The round points (geometric shapes) indicate that current experimental bounds are violated (respected), see also the main text of section~\ref{subsubsec:primary_preliminaries}.}
		}
		\label{fig:mueconv}
	\end{minipage}
\end{figure}

If $\mu -e$ conversion in nuclei was similarly dominated by long-range $\gamma$-penguins, the plots in the top in figure~\ref{fig:mueconv} would also just feature a straight line in the centre of the coloured region. Due to subdominant contributions, see section~\ref{subsubsec:p_ana_mueconv}, the result can generically deviate from the $\gamma$-penguin approximation by a factor two or three.
 Still, the future search for $\mu-e$ conversion in aluminium can be expected to complement the one for $\mu\to 3\,e$, as can be seen in the top-right plot in figure~\ref{fig:mueconv}. The experiments COMET~\cite{COMET:2018auw} and Mu2e~\cite{Mu2e:2014fns} are both projected to efficiently probe the possibility of explaining the measured value of $\Delta a_\mu$ in this model, see the bottom-left plot in figure~\ref{fig:mueconv}.

\begin{figure}[t!]
	\includegraphics[scale=0.52]{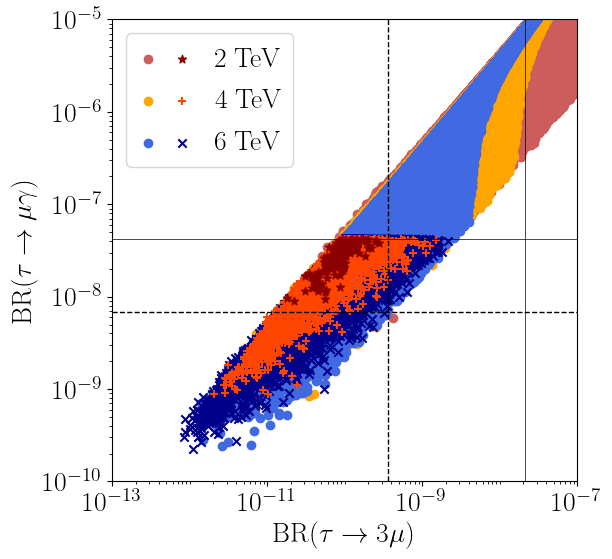}
	\includegraphics[scale=0.52]{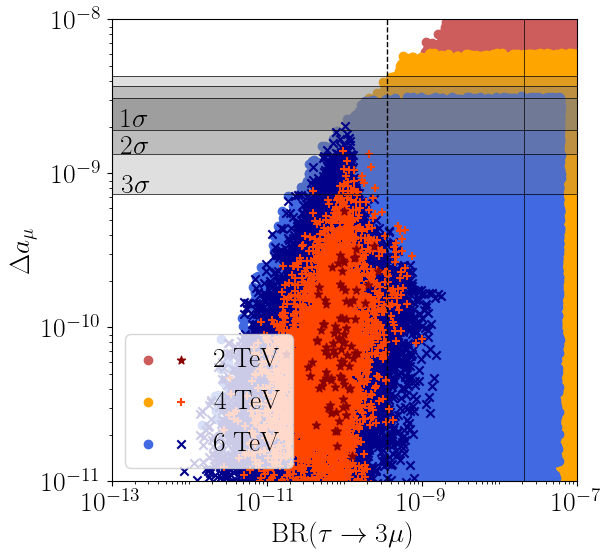}
	\\
	\includegraphics[scale=0.52]{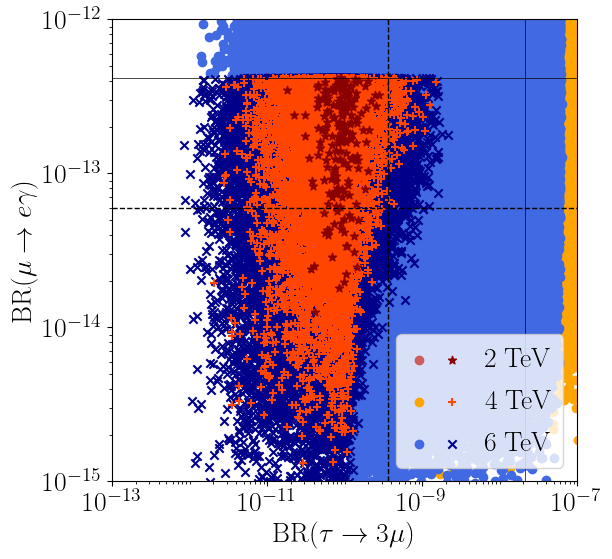}
	\includegraphics[scale=0.52]{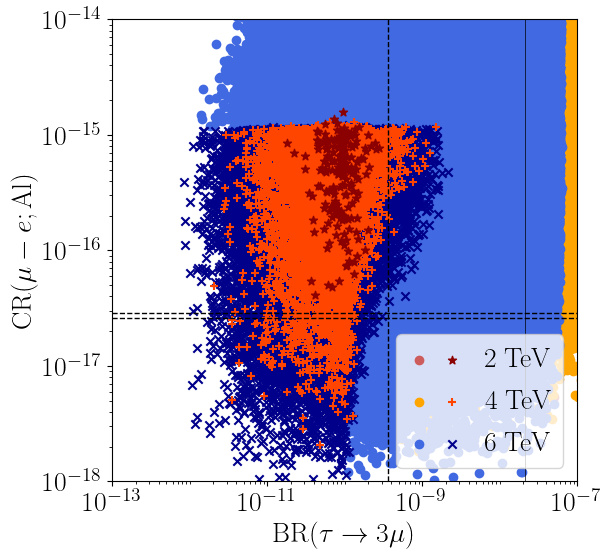}
	\caption{\linespread{1.1}\small{\textbf{\mathversion{bold}Constraining power and future reach of $\tau\to 3\,\mu$.\mathversion{normal}} The vertical solid (dashed) lines indicate the current bound on (future sensitivity of) $\tau\to3\,\mu$ \cite{Hayasaka:2010np,Belle-II:2018jsg}, see table~\ref{table:primaryconstraints}. The round points (geometric shapes) indicate that current experimental bounds are violated (respected), see also the main text of section~\ref{subsubsec:primary_preliminaries}.}
	}
	\label{fig:tau3mu}
\end{figure}

In the following, we only discuss plots involving $\tau\to3\,\mu$. However, the obtained BRs for $\tau\to3\,\mu$ and $\tau\to\mu e \bar e$ are almost identical in the primary scan and thus the inferred statements also apply to BR($\tau\to\mu e \bar e$). There can be, nevertheless, appreciable differences between the two in the comprehensive scan, see section~\ref{sec:Comprehensive_leptonic}. Figure~\ref{fig:tau3mu} confirms that the upcoming search for $\tau\to 3\,\mu$ at Belle II \cite{Belle-II:2018jsg} can be expected to probe a region of the parameter space which is compatible with current constraints. As demonstrated in section~\ref{subsubsec:p_ana_trilepton}, this region corresponds to sufficiently large $Z$-penguin contributions. Indeed, if only long-range $\gamma$-penguins were present, the top-left plot involving $\text{BR}(\tau\to\mu\gamma)$ would display a straight line located at the upper edge of the coloured region.

The hierarchy $|c_{23}|\approx|a_{23}| \gg |b_{23}|$ required for large $Z$-penguin contributions also suppresses the product of the magnitudes of the two effective parameters and thus the contribution to the AMM of the muon, see eqs.~(\ref{eq:gm2_estimate}) and~(\ref{eq:tauto3mu_estimate}). As a consequence, observing $\tau\to 3\,\mu$ at Belle II would indicate that an explanation of the measured value of $\Delta a_\mu$ is very unlikely for $\hat{m}_\phi = 4,6$. For these LQ masses, conversely, the largest contributions to $\Delta a_\mu$ are generated, if BR$(\tau\to 3\,\mu)$ remains below the prospective sensitivity. This upper bound on $\Delta a_\mu$, $\Delta a_\mu\propto|b_{23}|$, as a function of BR$(\tau\to3\,\mu$), BR$(\tau\to3\,\mu)\propto|c_{33}|^2$, is again mainly due to the experimental constraint on BR$(\tau\to\mu\gamma$), BR$(\tau\to\mu\gamma)\propto|b_{23}c_{33}|^2$.

In the model, a signal in $\tau\to 3\,\mu$ effectively enforces a signal in $\tau\to\mu\gamma$, but the reverse is not true in general. Furthermore, the plots in the bottom of figure~\ref{fig:tau3mu} suggest that a result BR$(\tau\to 3\,\mu)\gtrsim\mathcal{O}(10^{-10})$ becomes increasingly disfavoured, if the contributions to cLFV $\mu\to e$ transitions shrink. Since $|b_{13}| \gtrsim \lambda$ in the primary scan, this shrinkage mostly relies on small values for $|c_{23}|\approx|a_{23}|$, see sections~\ref{subsubsec:p_ana_eiejgamma} and~\ref{subsubsec:p_ana_mueconv}, and so the $Z$-penguin contributions to $\tau\to 3\,\mu$ become more suppressed. Hence, BR$(\tau\to 3\,\mu)$ is more tightly correlated with BR$(\tau\to\mu\gamma)$ in this case, and it is more difficult to respect the stringent experimental bound on the latter. In turn, if $\tau\to 3\,\mu$ is observable at Belle II, $|c_{23}|\approx|a_{23}|$ must be rather large and therefore one generates an enhancement of BR$(\mu\to e\gamma)$ and CR$(\mu - e;\text{Al})$. Note that this interplay is far less pronounced in the comprehensive scan, see section~\ref{ssec:primaryresultsComprehensive}.

%%%%%%%%%%%%%%%%%%%%%%%%%%%%%%%%%%%%%%%%%%%%%%%%%%%%%%  

\mathversion{bold}
\subsubsection{\texorpdfstring{$B_c\to \tau\nu$, $R^\nu_{K^{(\star)}}$ and $Z\to\tau\tau$}{Bc -> tau nu, R^nu_K(*) and Z->tau tau}}
\mathversion{normal}
\label{subsubsec:Hadronic_primary}
\begin{figure}[t!]
	\includegraphics[scale=0.52]{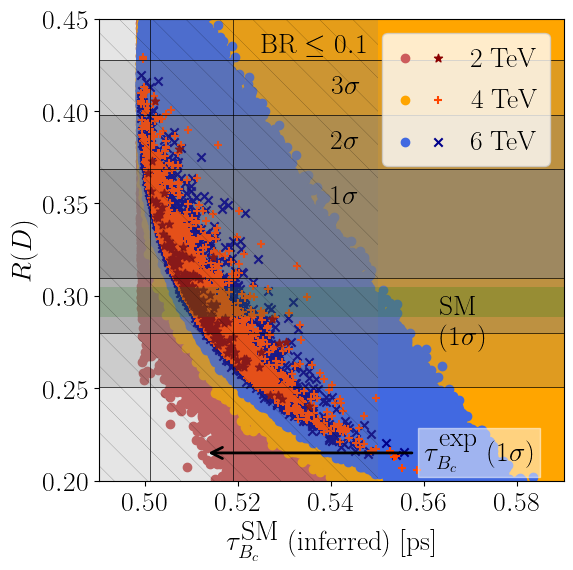}
	\includegraphics[scale=0.52]{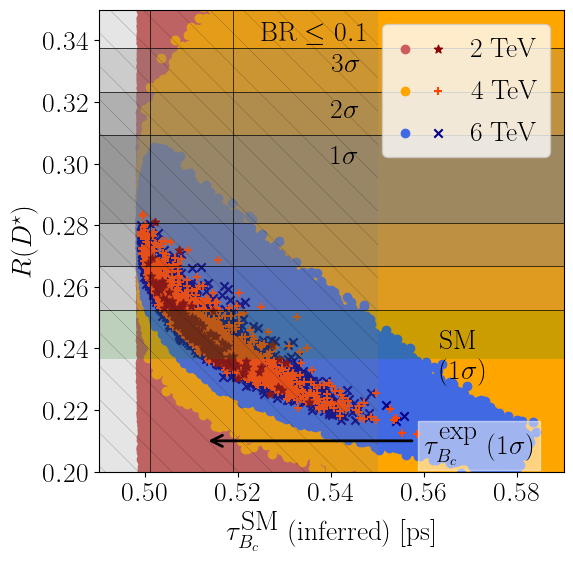}
	\\
	\includegraphics[scale=0.52]{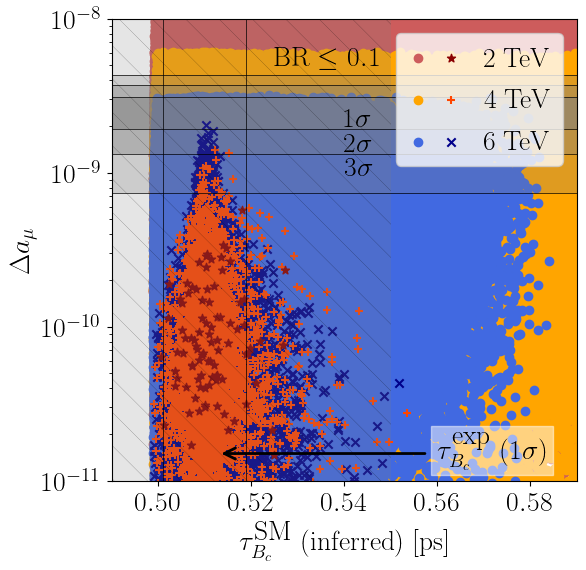}
	\includegraphics[scale=0.52]{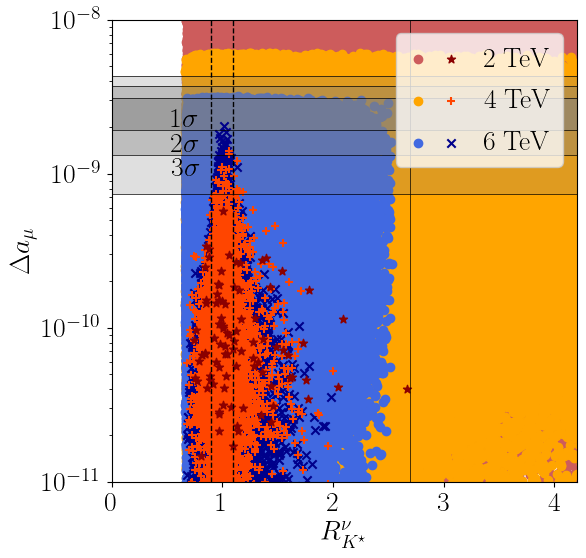}
	\caption{\linespread{1.1}\small{\textbf{\mathversion{bold}Constraining power and future reach of $\tau_{B_c}^\text{SM}$  and $R^\nu_{K^\star}$.\mathversion{normal}} In the top and the bottom-left plot, the vertical solid lines indicate the region in which the inferred SM contribution to the $B_c$ lifetime agrees with the experimental best-fit value at the $1\,\sigma$ level~\cite{ParticleDataGroup:2020ssz}, and the hatched area marks the region in which BR($B_c\to\tau\nu$) remains smaller than 0.1, as given via eq.~(\ref{eq:relationBc_branchingratios}). In the bottom-right plot, the vertical solid line shows (dashed lines show) the region compatible with the current experimental bound on (future reach of) $R^\nu_{K^\star}$\cite{Belle:2017oht,Belle-II:2018jsg}, see also table~\ref{table:primaryconstraints}. For the future reach, an SM-like value and an uncertainty of ten percent are assumed. The round points (geometric shapes) indicate that current experimental bounds are violated (respected), see also the main text of section~\ref{subsubsec:primary_preliminaries}.}
	}
	\label{fig:tauBc_RK}
\end{figure}
We proceed with a discussion of further hadronic observables as well as the axial-vector coupling of $Z$ bosons to tau leptons.
As illustrated in figure~\ref{fig:tauBc_RK}, a large contribution from the LQ to the lifetime of the $B_c$ meson is incompatible with the imposed experimental bounds.\footnote{We stress again that we have not attempted to perform a calculation of the SM contribution to the $B_c$ lifetime, but have indirectly inferred it from the requirement that the combined contribution from the SM and the LQ agrees with the experimentally determined lifetime.} In particular, the model can accommodate the current best-fit value of $R(D)_{\text{exp}} = 0.339 \pm 0.026 \pm 0.014$, even if that contribution vanished. Besides, the results suggest that $\tau^{\text{SM}}_{B_c}$ would still be close to agreeing with the measured lifetime $\tau_{B_c}^{\text{exp}} = (0.510\pm 0.009)\;\text{ps}$ \cite{ParticleDataGroup:2020ssz} at the $1\,\sigma$ level for the largest value of $R(D^\star)$ achievable in this model (which would also be closest to the best-fit value). In case of larger LQ masses, a substantial contribution to the lifetime of the $B_c$ meson from the LQ only arises, if $R(D)$ and $R(D^\star)$ become smaller than in the SM, respectively, which is in agreement with the opposite signs of the respective contributions linear in $|a_{33}b_{32}|$ in the analytic estimates in eqs.~(\ref{eq:RD_estimate}), (\ref{eq:RDs_estimate}) and (\ref{eq:tauBcSM_estimate}). This interdependence can get (partly) lifted, if the channel with a muon neutrino $\nu_\mu$ in the final state becomes more relevant, see section~\ref{ssec:primaryresultsComprehensive} for details.

Nevertheless, a deviation of $\tau^{\text{SM}}_{B_c}$ from the best-fit value of $\tau^{\text{exp}}_{B_c}$ by more than ten percent is incompatible with the considered constraints. This implies that the BR for $B_c\to\tau\nu$ remains below 0.1 in most cases, and can potentially exceed this limit only to a very small degree. In line with eq.~(\ref{eq:relationBc_branchingratios}), imposing the upper bound BR$(B_c\to\tau\nu)\lesssim 0.1$ constrains the SM contribution to the lifetime to fulfil $\tau^{\text{SM}}_{B_c} \lesssim 0.55$ ps, indicated by the hatched region in the top and the bottom-left plot in figure~\ref{fig:tauBc_RK}. Therein, the vertical solid lines indicate the region in which the SM prediction agrees with the measured lifetime of the $B_c$ meson at the $1\,\sigma$ level. Furthermore, we recall that BR$(B_c\to\tau\nu)\lesssim 0.3$ corresponds to $\tau^{\text{SM}}_{B_c} \lesssim 0.7$ ps.

If the measured value of $\Delta a_\mu$ is explained at the $3\,\sigma$ level or better in this model, we find that a substantial deviation of $\tau^{\text{SM}}_{B_c}$ from the measured $B_c$ lifetime is very unlikely for $\hat{m}_\phi = 4, 6$. This reflects the fact that an explanation of $R(D)$ and $R(D^\star)$ competes with an explanation of $\Delta a_\mu$, see sections~\ref{subsubsec:prelimscan_anomalies} and \ref{subsubsec:prelimscan_eiejgamma}. The results of the primary scan also suggest that $R^\nu_{K^{(\star)}}$ is close to one in that case. We refer to section~\ref{ssec:primaryresultsComprehensive} for a discussion of the results of the comprehensive scan. Thus, the prospective measurement of $B\to K^{(\star)} + \text{invisible}$ at Belle II~\cite{Belle-II:2018jsg} provides a promising avenue to test this model, since the primary scan prompts the expectation that the non-appearance of a substantial excess on top of the SM expectation implies the best chances for an explanation of the observed anomaly in the AMM of the muon. As is the case for $R(D)/R(D)_{\text{SM}}$ and $R(D^\star)/R(D^\star)_{\text{SM}}$, the observables $R^\nu_{K^{(\star)}}$ and $\tau^{\text{SM}}_{B_c}/\tau^{\text{exp}}_{B_c}$ can also be approximated as quadratic functions in $|a_{33}|$, respectively, and are therefore correlated with $\Delta a_\mu$, $\Delta a_\mu\propto |b_{23}|$, via the experimental bound on BR$(\tau\to\mu\gamma$), BR$(\tau\to\mu\gamma) \propto |a_{33}b_{23}|^2$, see the relevant estimates in section~\ref{subsec:primary_analytic}.

\begin{figure}[t!]
	\includegraphics[scale=0.52]{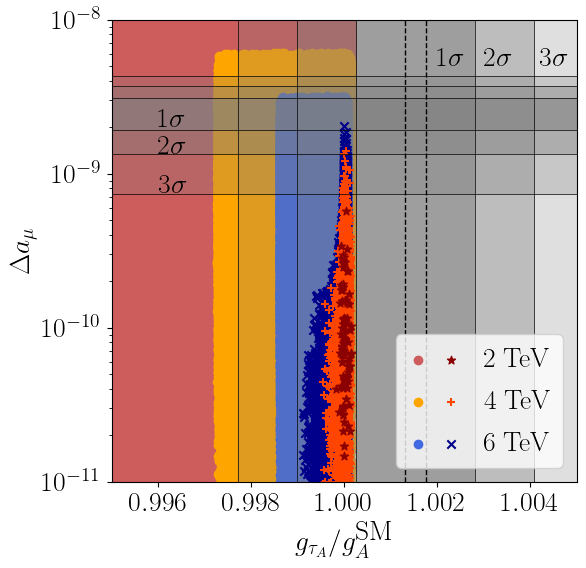}
	\includegraphics[scale=0.52]{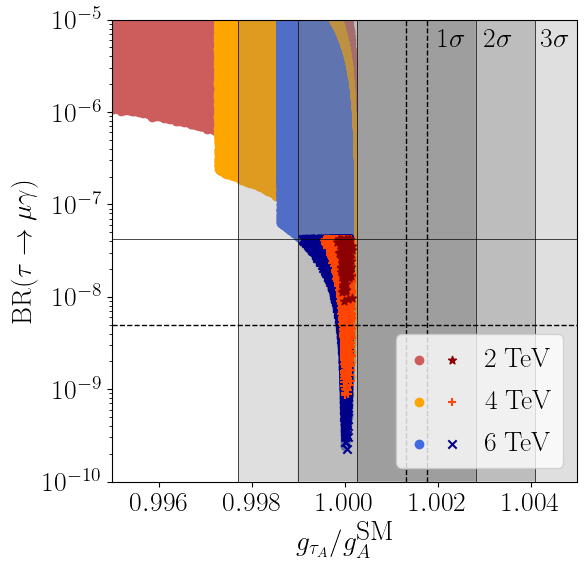}
	\\
	\caption{\linespread{1.1}\small{\textbf{\mathversion{bold}Constraining power and future reach of $g_{\tau_A}/g^{\text{SM}}_A$.\mathversion{normal}}
			The regions indicated by vertical solid (dashed) lines are compatible with the current experimental world averages for (future sensitivity of) $g_{\tau_A}/g^{\text{SM}}_A$~\cite{Muong-2:2021ojo,Aoyama:2020ynm} at the shown confidence level~\cite{ALEPH:2005ab,Crivellin:2020mjs} (at the $3\,\sigma$ level~\cite{Crivellin:2020mjs,Baer:2013cma, FCC:2018evy}), see table~\ref{table:primaryconstraints}. The round points (geometric shapes) indicate that current experimental bounds are violated (respected), see also the main text of section~\ref{subsubsec:primary_preliminaries}.}}
	\label{fig:Z_tautau}
\end{figure}

As evidenced in figure~\ref{fig:Z_tautau}, the contributions to the AMM of the muon, $\Delta a_\mu\propto|a_{23}b_{23}|$, and $g_{\tau_A}/g^{\text{SM}}_A$ are not per se correlated, as is expected from eq.~(\ref{eq:Ztau_coupling_approx}) according to which the difference of $g_{\tau_A}/g^{\text{SM}}_A$ from one is essentially only controlled by $|c_{33}|\approx|a_{33}|$. Still, if the experimental constraints of all primary observables are imposed, the axial-vector coupling of $Z$ bosons to tau leptons is necessarily SM-like, if the measured value of $\Delta a_\mu$ is explained at the $3\,\sigma$ level. In particular, the deviation from LFU would be constrained to be much smaller than 0.1 percent. This correlation is established through the bound on BR$(\tau\to\mu\gamma$), BR$(\tau\to\mu\gamma)\propto|a_{33}b_{23}|^2$, which is illustrated by the plot on the right in figure~\ref{fig:Z_tautau}. It is clearly visible that a deviation of $g_{\tau_A}/g^{\text{SM}}_A$ from the current experimental world average~\cite{ALEPH:2005ab,Crivellin:2020mjs} by more than $2\,\sigma$ is incompatible with the constraint on BR$(\tau\to\mu\gamma)$. Furthermore, the future search for $\tau\to\mu\gamma$ at Belle II~\cite{Belle-II:2018jsg} can conclusively test the capability of the model to induce a significant deviation from LFU in axial-vector couplings.

%%%%%%%%%%%%%%%%%%%%%%%%%%%%%%%%%%%%%%%%%%%%%%%%%%%%%%
\section{Comprehensive study}
\label{sec:secondarytertiary}
%%%%%%%%%%%%%%%%%%%%%%%%%%%%%%%%%%%%%%%%%%%%%%%%%%%%%%

In this section, we conduct a comprehensive scan over the parameter space of this model, only considering the phenomenology of scenario B. 
In order to do so, we first perform a chi-squared fit to the charged fermion masses and quark mixing, as detailed in section~\ref{ssec:cferm}. This fixes (a subset of) the effective parameters $d_{ij}, e_{ij}$ and $f_{ij}$. Furthermore, we obtain the unitary matrices necessary in order to transform the LQ couplings ${\bf \hat{x}}$ and ${\bf \hat{y}}$ to the charged fermion mass basis. Then, we bias this numerical study to account for the parameter space, preferred by the primary observables, as has been revealed by the analysis in section~\ref{subsec:primary_numeric}. We can thereby focus on important regions of parameter space and extract the most useful information from this multi-dimensional parameter scan. 
The details of the biasing can be found in section~\ref{ssec:prelimBias}. Otherwise we vary all parameters with flat priors, and in the ranges specified in eqs.~\eqref{eq:unbiased_magnitude} and \eqref{eq:unbiased_phase}. For more information about the scan procedure, we refer to appendix~\ref{app:supp6_scanmethod}.

To perform this numerical study, we use a combination of different computational software.
 We encode this model in \texttt{SARAH}~\cite{Porod:2014xia,Porod:2011nf}. The program \texttt{SARAH} generates an output module for use with \texttt{SPheno}~\cite{Porod:2011nf}, which can calculate the Wilson coefficients, decay rates and a set of flavour observables, defined by \texttt{FlavorKit}~\cite{Porod:2014xia}.\footnote{ A comprehensive discussion of this can be found in reference~\cite{Vicente:2015zba}.} We also make use of \texttt{flavio}~\cite{Straub:2018kue} to process analytically defined sets of Wilson coefficients, where appropriate, and use these to calculate a broader class of flavour observables. The running of Wilson coefficients in \texttt{flavio} is implemented using the \texttt{Wilson} package~\cite{Aebischer:2018bkb}. In this way, it is possible to construct an efficient multi-dimensional parameter scan, the results of which are discussed in the following. Information regarding the conventions of the shown plots is given in section~\ref{ssec:comprehensivePlots}.

We consider not only the primary observables in section~\ref{ssec:primaryresultsComprehensive} and appendix~\ref{app:addfigssec6}, as explored in section~\ref{sec:primary}, but also secondary and tertiary observables. Secondary observables are outlined and analytic estimates are provided for these in section~\ref{subsec:secondary_analytic}. We discuss the numerical results for the secondary observables in section~\ref{subsec:resultsSecondary} and tertiary observables are commented in section~\ref{subsec:tertiary} as well as in appendix~\ref{app:tertiary}.

\subsection{Fit of charged fermion masses and quark mixing}
\label{ssec:cferm}

In order to fix the effective parameters $d_{ij}$, $e_{ij}$ and $f_{ij}$, contained in the charged fermion mass matrices $M_d$, $M_e$ and $M_u$, respectively, we perform a chi-squared fit of the charged fermion masses and quark mixing. As discussed in section~\ref{sec:Yukawas_LQcouplings}, accommodating quark mixing correctly requires to consider scenario B, i.e.~the up-type quark mass matrix has to be of the form given 
in eq.~(\ref{eq:MuparaB}). As the mass of the LQ is selected to be maximally a few TeV, see eq.~(\ref{eq:hatmphi_num}), we fit the charged fermion masses at a scale of $\mu=1 \, \mathrm{TeV}$, taken from reference~\cite{Xing:2007fb}. Quark mixing is fitted to the best-fit values given by the PDG (Particle Data Group)~\cite{ParticleDataGroup:2020ssz}, because RG running effects are small. 

This model contains two Higgs doublets, $H_u$ giving masses to up-type quarks,  and $ H_d$ giving masses to down-type quarks as well as charged leptons. The suppression of the down-type quark and charged lepton masses with respect to those of the up-type quarks (in particular, the top quark) is achieved by taking the VEV of $H_d$ to be much smaller than that of $H_u$ -- recall eq.~(\ref{eq:HdHuVEVsapprox}). Thus, in the chi-squared fit we have (mainly) varied the size of the VEV of $H_d$ such that $\langle H_d^0\rangle $ takes a minimum value of $1.22 \;\mathrm{GeV}$ and a maximum value of $4.86 \;\mathrm{GeV}$, generating several viable data sets. Each of these leads to an excellent fit to the charged fermion masses taken at $\mu=1 \, \mathrm{TeV}$~\cite{Xing:2007fb} and to quark mixing~\cite{ParticleDataGroup:2020ssz}. From these data sets, we do not only extract the values of masses and mixing, but more importantly the unitary matrices $L_d$, $R_d$, $L_e$, $R_e$, $L_u$ and $R_u$, necessary in order to compute the form of 
the LQ couplings ${\bf \hat{x}}$ and ${\bf \hat{y}}$ in the charged fermion mass basis, i.e.~the LQ couplings ${\bf x}$, ${\bf y}$ and ${\bf z}$, according to eqs.~(\ref{eq:xpara}), (\ref{eq:zparaA}) and 
(\ref{eq:yparaA}).  For further details about the implementation of the chi-squared fit in the scan, see appendix~\ref{app:supp6_scanmethod}.

\subsection{Biases from primary scan}
\label{ssec:prelimBias}

We remind that the study of primary observables involves samples of  $4(3)[2]\times10^6$ data points for $\hat{m}_\phi = 2(4)[6]$, sampled as described in section~\ref{subsubsec:primary_preliminaries}. These points  have been filtered to select only those that pass the primary constraints in table~\ref{table:primaryconstraints}, including $g_{\tau_A}/g_A^\text{SM}$ at the $3\, \sigma$ level and $\tau_{B_c}^\text{SM}$ at the $1\, \sigma$ level.

In an unbiased scan, it turns out to be difficult to extract points capable of addressing the three flavour anomalies, while passing the constraints arising from the experimental bounds on the radiative cLFV decays $\tau\to \mu \gamma$ and $\mu\to e \gamma$. From table~\ref{table:primaryconstraints_coupling_coeffs} it becomes evident why, since common effective parameters drive these observables, recalling that $a_{ij}$ and $c_{ij}$ are related via the CKM mixing matrix. For this reason, the biases in table~\ref{tab:pc1} are presented for points satisfying all primary constraints, together with two further numerical restrictions, see eqs.~\eqref{eq:b13frommeg} and~\eqref{eq:b23fromtmg}. The latter intent to address the constraints from $\tau\to \mu \gamma$ and $\mu\to e \gamma$ by biasing the values of the magnitudes of the effective parameters $b_{23}$ and $b_{13}$, respectively.

\begin{table}[t!]
 \renewcommand*{\arraystretch}{1.5}
    \centering
\begin{tabular}{|c|c|c|c|c|c|}
\hline
    $\hat{m}_\phi$ & $|a_{33}|$ & $|b_{32}|$ & $\cos(\text{Arg}(a_{33})-\text{Arg}(b_{32}))$ & $|a_{23}|$& $\cos(\text{Arg}(a_{23})-\text{Arg}(b_{23}))$  \\
    \hline 
        2& $[0.2, 0.7]$ & $[1.1, 2.6]$ & $[0.4, 1.0]$& -- &{$[-1.0, 0.0]$}\\
    4& $[0.2, 1.9]$ & $[1.0, 4.5]$ & $[0.1, 1.0]$& $[1.6, 4.4]$&$[-1.0, -0.5]$\\
        6& $[0.2, 3.6]$ & $[0.8, 4.5]$ & $[0.0, 1.0]$ & $[1.4, 4.4]$&$[-1.0, -0.3]$\\
    \hline
\end{tabular}
\caption{\small {\bf Inputs for biasing in comprehensive scan, derived from primary scan}.  These intervals have been identified in the primary scan as satisfying all primary constraints and explaining at least one of the flavour anomalies in $R({D^{(\star)}})$ or in the AMM of the muon at the $3\, \sigma$ level or better. The fact that there is no entry for $|a_{23}|$ in the case of $\hat{m}_\phi=2$ means that no points have been identified in the primary scan that allow the presented conditions to be met. Therefore, we take $[\lambda, 1/\lambda]$ to be the imposed range for biasing the magnitude of the effective parameter $a_{23}$. Then, we manually input the restricted range for $\cos(\text{Arg}(a_{23})-\text{Arg}(b_{23}))$ for $\hat{m}_\phi=2$ to ensure that the contribution to the AMM of the muon is positive, compare eq.~(\ref{eq:gm2_estimate}). Note that these values are taken together with the inequalities found in eqs.~\eqref{eq:b13frommeg} and~\eqref{eq:b23fromtmg} to also bias the values of the effective parameters $|b_{13}|$ and $|b_{23}|$. 
All shown numbers are rounded to one decimal place.
}
\label{tab:pc1}
\end{table}

For a value of the magnitude of the effective parameter $a_{23}$ chosen according to table~\ref{tab:pc1}, we also impose a restriction on the magnitude of $b_{13}$ in order to pass the experimental bound on the 
BR of $\mu\to e \gamma$, see eq.~(\ref{eq:mueg_estimate}),
\begin{align}
    |b_{13}|\lesssim \frac{1}{|a_{23}|} \left\{
\begin{array}{*{2}{c}}
0.41, & \hat{m}_\phi = 2 \\
1.16,& \hat{m}_\phi = 4 \\
2.22,& \hat{m}_\phi = 6 \\
\end{array} \right\}.
\label{eq:b13frommeg}
\end{align}
Furthermore, respecting the experimental constraint on the BR of $\tau \to \mu \gamma$ enforces that, once a value for the magnitude of $a_{33}$ is chosen according to table~\ref{tab:pc1}, the magnitude of $b_{23}$ is restricted such that
\begin{align}
    |b_{23}|\lesssim \frac{1}{|a_{33}|} \left\{
\begin{array}{*{2}{c}}
0.16, & \hat{m}_\phi = 2 \\
0.45,& \hat{m}_\phi = 4 \\
0.86,& \hat{m}_\phi = 6 \\
\end{array} \right\} \, , \label{eq:b23fromtmg}
\end{align}
compare eq.~(\ref{eq:taumug_estimate}).
This means that determining whether a sample point can rather explain the flavour anomalies in $R(D^{(\star)})$, for which $|a_{33}|$ needs to be quite large, or in the AMM of the muon, for which $|b_{23}|$ must be quite large, is tightly controlled by the bound on the BR of $\tau\to\mu\gamma$. The ranges for $|a_{33}|$ indicated in table~\ref{tab:pc1} are the union of the ranges separately extracted using the $3 \, \sigma$ ranges of $R(D^{(\star)})$ and of the AMM of the muon.

Note that these constraints are imposed on the effective parameters in the charged fermion mass basis, while scanning over effective parameters in the interaction basis in the comprehensive scan. Therefore, we have two related, but distinctly defined, regions of parameter space. The transformations between them are given by the unitary matrices, generated by the chi-squared fit to charged fermion masses and quark mixing, as described in section~\ref{ssec:cferm}. As addressed in section~\ref{sec:phenomenology}, varying the effective parameters in the interaction basis in the range found in eq.~\eqref{eq:unbiased_magnitude} ensures the preservation of the expansion in orders of $\lambda$ used to construct the underlying model. In doing so, the corresponding effective parameters in the charged fermion mass basis, calculated from this scan, may fall outside the range $[\lambda, 1/\lambda]$, compare table~\ref{tab:comprehensiveCoupMag} in appendix~\ref{app:supp6_scanmethod}. 

For practicality in implementing the biases, we assume that the LO relations listed in appendix~\ref{app:relatepara} can be used to translate between bases. In particular, we first assume that
\begin{align}
\label{eq:basismappingMicroSum}
   |a_{33}| = |\hat{a}_{33}|, \;\;\; |b_{32}| = |\hat{b}_{32}|,\;\;\;\text{and} \;\;|b_{23}|= |\hat{b}_{23}| \, ,
\end{align} 
which allows us to directly bias the input values for $ |\hat{a}_{33}|,  |\hat{b}_{32}|,$ and  $ |\hat{b}_{23}|$. All other effective parameters in the interaction basis are varied in the ranges, specified by eqs.~\eqref{eq:unbiased_magnitude} and \eqref{eq:unbiased_phase}. We then bias the magnitudes of $a_{23}$ and $b_{13}$ by first extracting their values from the effective parameters in the interaction basis, 
using the aforementioned unitary matrices, and afterwards enforcing the bounds shown in table~\ref{tab:pc1} and eq.~\eqref{eq:b13frommeg}, respectively. For further details regarding the implementation of the scan, 
see appendix~\ref{app:supp6_scanmethod}.
 
We do not claim to have extensively explored the entire multi-dimensional parameter space of this model, but implement the biases from the primary scan to better identify regions capable of explaining the three flavour anomalies and respecting all considered present constraints.
 
 \begin{table}[t!]
\centering
\resizebox{\textwidth}{!}{
\renewcommand{\arraystretch}{1.5}
    \centering
    \begin{tabular}{|l|l|l|l|l|l|l|}
    \hline
    \multicolumn{7}{|c|}{ \textsc{Spread of results for primary observables in comprehensive scan}} \\
    \hline
\multirow{2}{*}{Observable} &\multicolumn{2}{c|}{$\hat{m}_\phi=2$, $P=5955$}  &\multicolumn{2}{c|}{$\hat{m}_\phi=4$, $P=12570$} &\multicolumn{2}{c|}{$\hat{m}_\phi=6$, $P=39807$} \\
\cline{2-7}
  & [min., max.] & Average & [min., max.] & Average & [min., max.] & Average \\
  \hline 
  $R(D)$ & $[0.302, 0.452]$&$0.333$& $[0.297, 0.442]$&$0.321$&$[0.291, 0.436]$& $0.316$\\
  $R(D^\star)$ &$[0.245, 0.293]$&$0.256$&$ [0.243, 0.288]$&$0.252$&$[0.238, 0.285]$& $0.250$\\
  $\Delta a_\mu \times 10^{10}$ &$[-0.350, 18.26]$& $1.658$
  &$[0.320, 22.40]$& $2.363$
  &$[0.013, 20.21]$& $1.252$\\
 BR($\tau\to\mu\gamma$) $\times 10^8$ &$[0.234, 4.200]$& $3.123$&$[0.099, 4.200]$&$ 2.997$& $[0.040, 4.200]$& $2.834$\\
 BR($\mu\to e\gamma$) $\times 10^{14}$& $[0.014, 42.00]$& $19.96$&$ [0.008, 42.00]$&$ 20.65$ & $[0.002, 42.00]$& $20.10$\\
 BR($\tau\to 3 \mu$) $\times 10^{10}$ &$[0.149, 30.92]$& $1.895$& $[0.078, 31.42]$& $3.754$ & $[0.030, 31.86]$& $3.479$\\
 BR($\tau\to \mu e\bar{e}$) $\times 10^{10}$&$[0.103,59.72]$ & $1.948$ &$[0.036, 28.83]$& $2.465$&$[0.016,18.20]$& $1.891$\\
 BR($\mu\to 3 e$) $\times 10^{16}$&$[0.010, 29.54]$ & $14.04$& $[0.005,29.53]$&$14.52$&[0.001, 29.54]& $14.13$\\
 CR($\mu - e$; Al) $\times 10^{16}$&$[0.009, 14.40]$&$6.341$&$[0.002,13.56]$&$6.480$& $[0.002, 13.37]$& $6.298$\\
 $R_{K^\star}^\nu$ &$[0.672, 2.695]$&$1.171$&$[0.672, 2.694]$&$1.096$&$[0.668, 2.698]$& $1.071$\\
 $(1-g_{\tau_A}/g_A^\text{SM})\times 10^4$& $[-1.646,1.963]$&$-0.203$& $[-1.593,4.791]$&$0.742$& $[- 0.777,9.193]$& $1.986$\\
 $\tau_{B_c}^\text{SM}$ (inferred) [ps]& $[0.499, 0.540]$&$0.506$&$[0.500, 0.524]$&$0.507$&$[0.499, 0.527]$& $0.507$\\
 \hline 
\end{tabular}}
\caption{{\small\textbf{Overview of spread of results for primary observables in comprehensive scan}. We present a summary of the statistics reflecting the distribution of primary observables: the minimum, maximum and average values generated for a sample of $P$ points passing the primary constraints.}} 
\label{tab:primaryTableC}
\end{table}

\subsection{Conventions for plots presented in this section}
\label{ssec:comprehensivePlots}

\begin{figure}[t!]
\centering
\begin{subfigure}{0.49\textwidth}
\centering 
  \includegraphics[width=\textwidth]{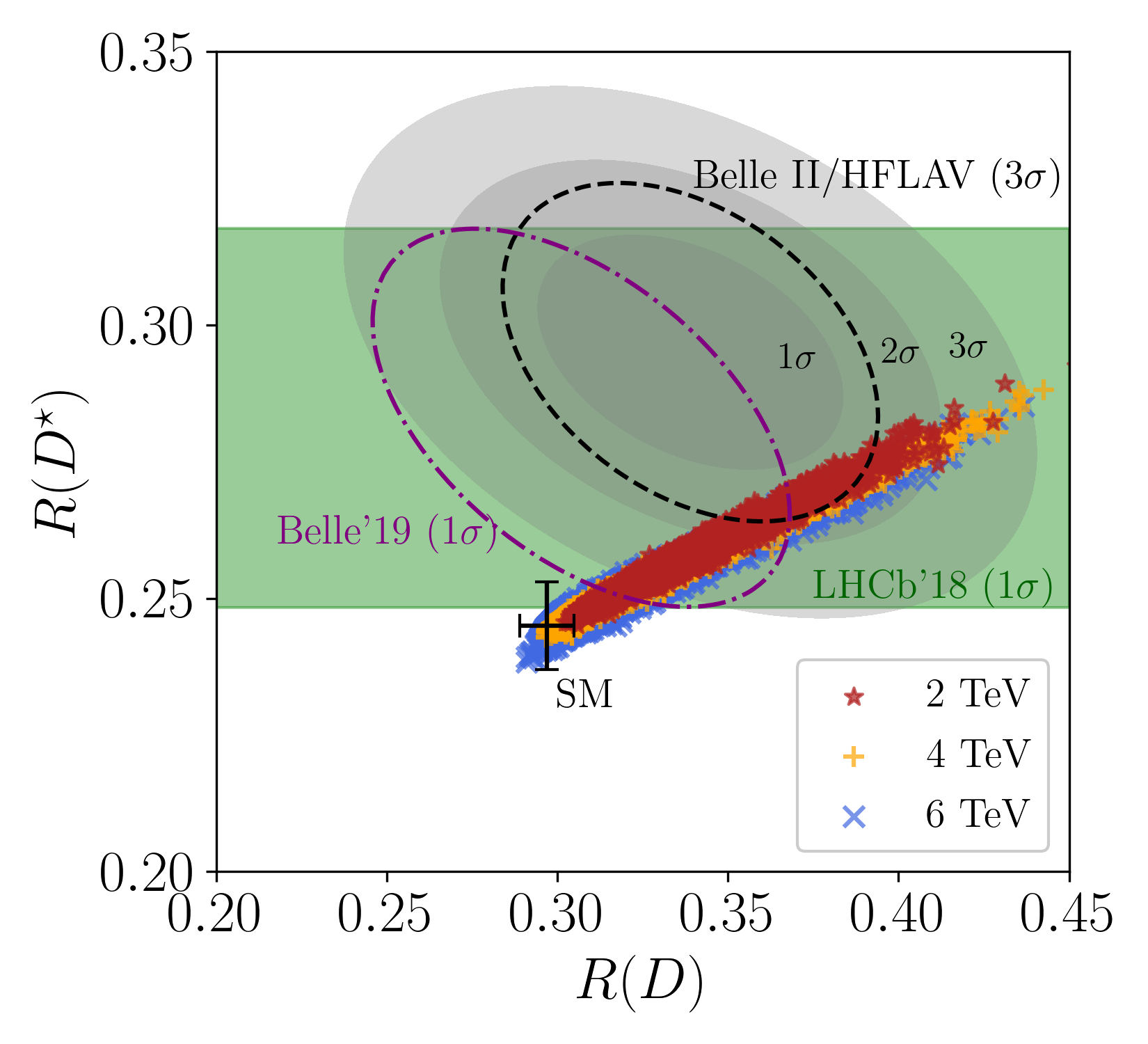}
  \end{subfigure}
  \hspace{1mm}
  \begin{subfigure}{0.49\textwidth}
   \includegraphics[width=\textwidth]{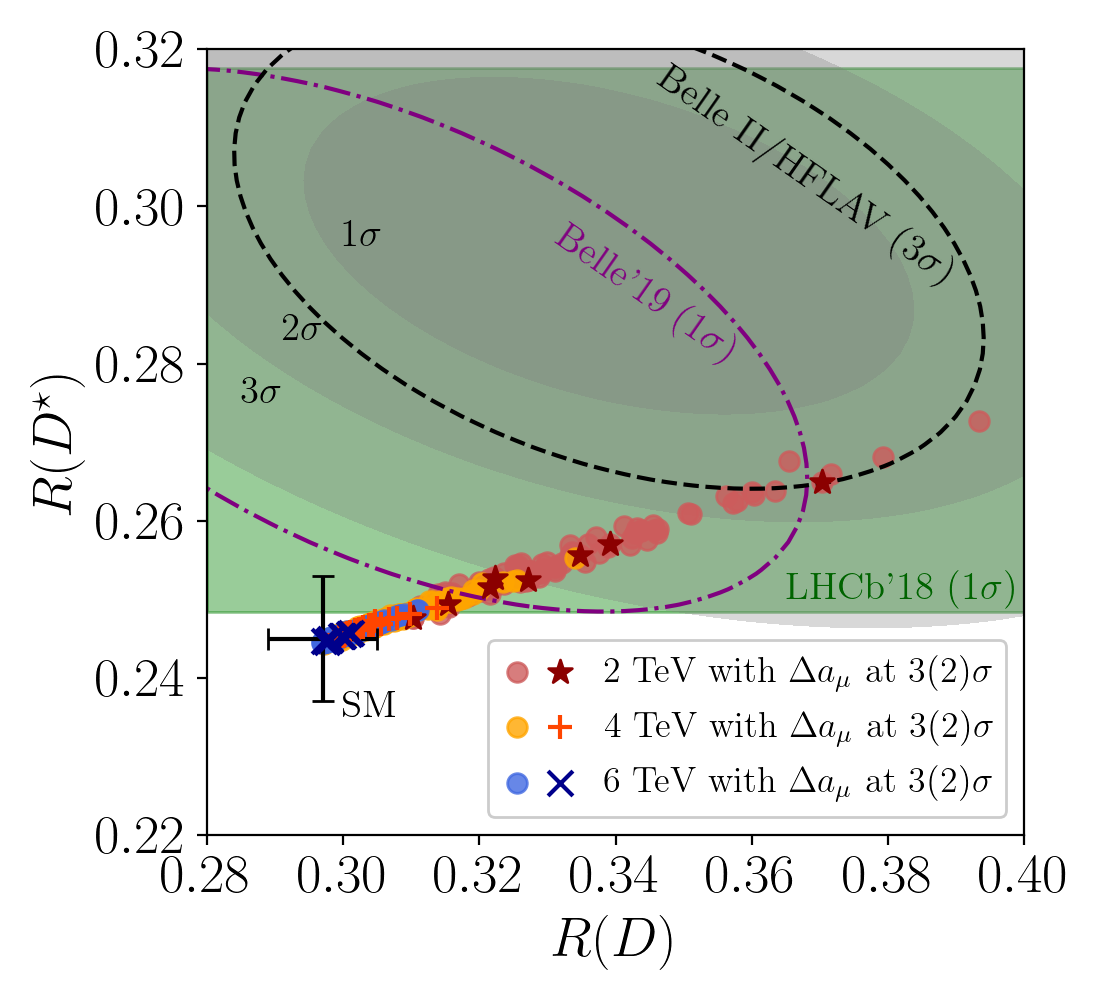}  
\end{subfigure}
\begin{subfigure}{0.49\textwidth}
\centering 
  	\includegraphics[width=\textwidth]{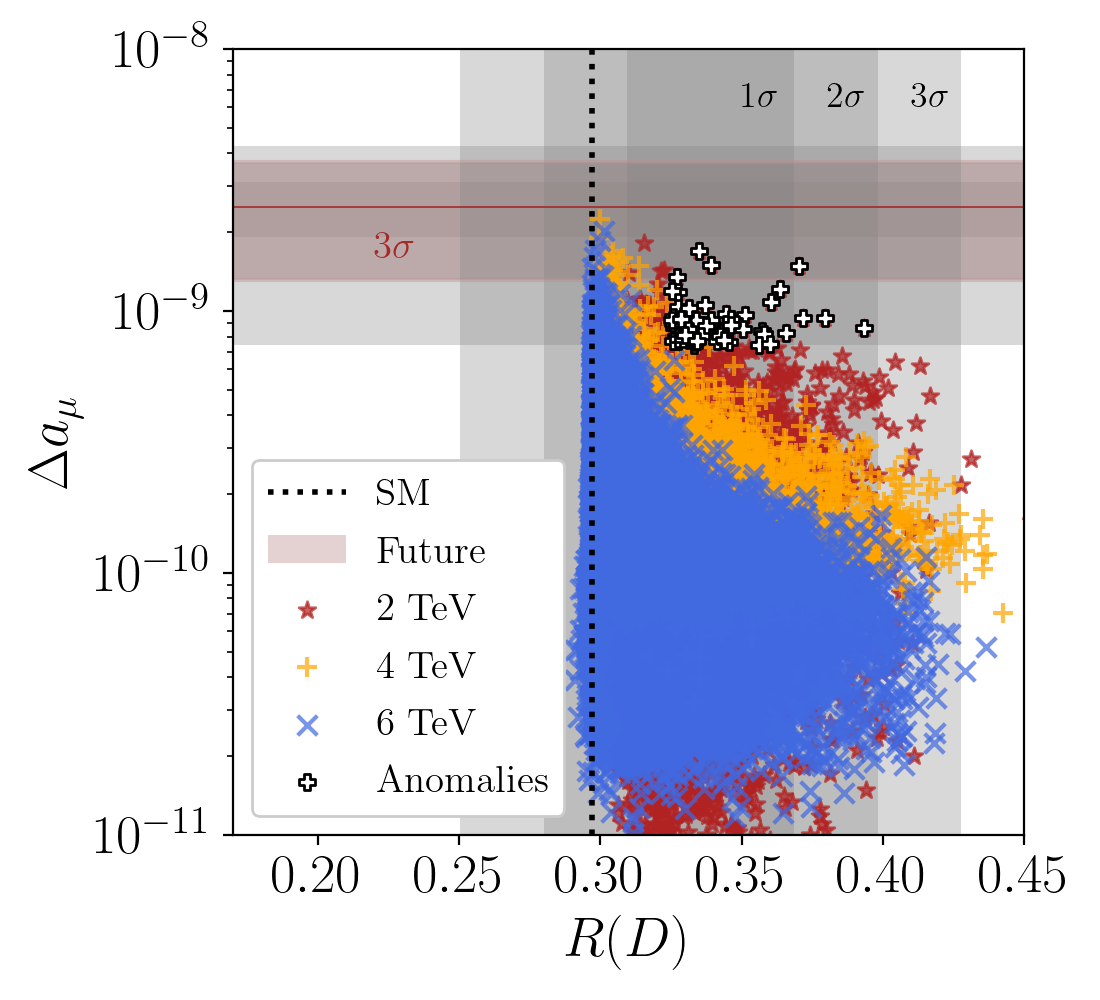} 
\end{subfigure}
\begin{subfigure}{0.49\textwidth}
\centering
	\includegraphics[width=\textwidth]{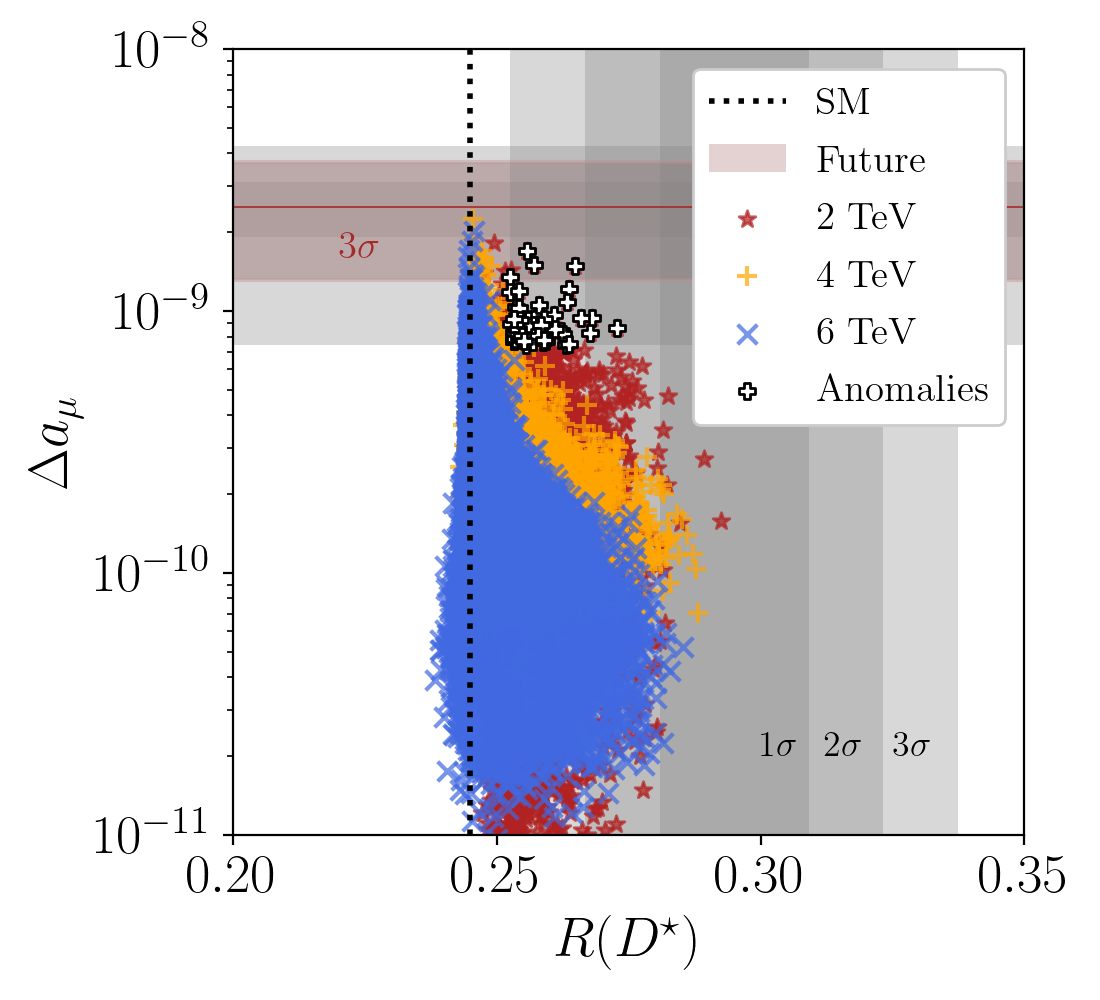}
\end{subfigure}
	\caption{\linespread{1.1}\small{\mathversion{bold}\textbf{Results of the comprehensive scan for the flavour anomalies in $R(D)$, $R(D^\star)$ and in the AMM of the muon.}\mathversion{normal} The top-left plot shows points that pass all considered constraints, while the top-right plot shows points that not only pass these constraints but also satisfy $\Delta a_\mu$ within $3\,\sigma$ (light-coloured circles) or $2\,\sigma$ (dark-coloured other shapes) of the present best-fit value. The scale in the top-right plot is magnified in order to highlight the region populated by the data. In both plots, the purple dot-dashed ellipse shows the $1~\sigma$ contour about the most recent Belle results for $R(D)$ and $R(D^\star)$~\cite{Belle:2019rba}, and a green band shows the $1~\sigma$ region about the most recent LHCb result for $R(D^\star)$~\cite{LHCb:2017smo,LHCb:2017rln}. The black dashed ellipse indicates the prospective $3~\sigma$ reach for $5$ ab$^{-1}$ of data at Belle II~\cite{Forti:2022mti}, assuming the best-fit value from 2021 and the correlation coefficient from the HFLAV collaboration~\cite{Amhis:2019ckw}. The SM values for $R(D)$ and $R(D^\star)$ (with associated uncertainties) are illustrated by a black barred cross in the top plots, and the SM central values are given as black dotted lines in the bottom plots. These SM values are extracted from \texttt{flavio}~\cite{Straub:2018kue,david_straub_2021_5543714}, v2.3. The bottom plots show the AMM of the muon plotted against $R(D)$ (left) and $R(D^\star)$ (right), with the red-brown band showing the $3~\sigma$ projected sensitivity from the Muon g$-$2 experiment~\cite{Muong-2:2015xgu}. This (roughly) overlays the present $2~\sigma$ region, assuming the best-fit value remains the same as the current experimental average (see red-brown solid line). For further information on how to read this figure, see section~\ref{ssec:comprehensivePlots}.}}
	\label{fig:AnomaliesComprehensive}
\end{figure}

Before discussing the results of this comprehensive scan, we first outline the conventions for displaying data in this section. In all plots the displayed coloured points pass all considered constraints -- red stars for $\hat{m}_\phi  = 2 $,  yellow plus signs for $\hat{m}_\phi  = 4$, and blue crosses for $\hat{m}_\phi =6$, as shown in the plot legends. Black dotted lines indicate the central values for SM predictions, black solid lines show present experimental constraints and black dashed lines show prospective bounds.  Where we display a physical observable on an axis, grey shaded regions indicate the 1, 2 and $3~\sigma$ contours about the present experimental best-fit values. If relevant for that observable, a red-brown shaded region indicates a prospective reach, as labelled, with a best-fit value denoted with a solid red-brown line. Where we show an effective parameter (or combination of them) on an axis, the grey shaded band indicates the region of parameter space probed by the primary scan. Overlaid white crosses in each of the displayed plots, labelled `Anomalies' in the legends, are points that can simultaneously address the anomalies in $R({D})$, $R({D^\star})$ and in the AMM of the muon within the $3~\sigma$ range of their present best-fit values. Each of these features can be seen in figure~\ref{fig:AnomaliesComprehensive}. Additional features in plots are defined in the captions.

\subsection{Numerical results for primary observables}
\label{ssec:primaryresultsComprehensive}

For the comprehensive scan, we sample $1.5 \times 10^5$ points for each of the three LQ masses. In the primary scan, approximately 0.005 (0.35) [2.4] percent of sample points have passed the primary constraints for $\hat{m}_\phi = 2(4)[6]$. In contrast, for the comprehensive scan we find that approximately 4(8)[27] percent of sample points pass the primary constraints for $\hat{m}_\phi = 2(4)[6]$.  Therefore, in the comprehensive scan the percentage of viable points has particularly increased for $\hat{m}_\phi = 2$. Below we discuss the efficacy of this biased scan for addressing the flavour anomalies and evading constraints. Once we have imposed all constraints, we identify $58(1)[0]$ points for $\hat{m}_\phi = 2(4)[6]$ that
can generate $R(D)$, $R(D^\star)$ and the AMM of the muon within the $3~\sigma$ range of the present best-fit values. These points are illustrated by white crosses in the plots, as mentioned in section~\ref{ssec:comprehensivePlots}.\footnote{As most of these points correspond to $\hat{m}_\phi = 2$, we do not distinguish between LQ masses for the white crosses.} 

The difference in the parameter space probed by the primary and comprehensive scan, in turn, impacts the resultant ranges of the observables. This may occur due to modifying the sampled region for a particular effective parameter that enters in a dominant contribution according to section~\ref{subsec:primary_analytic} (e.g. smaller accessible values of the magnitude of $b_{13}$, discussed in section~\ref{sec:contrastAnomalies}), or through an enhancement of the effective parameters appearing as subdominant in the primary scan (e.g. enhancement of LFV contributions in processes with neutrinos in the final state, see section~\ref{sec:Comprehensive_hadronic}).  Each of these may be a result of biasing and/or the use of a different basis. We emphasise that for the plots contained in this section, used to contrast the two scans, the coloured points always represent those for which all primary constraints are satisfied. Constraints from secondary and tertiary observables are automatically fulfilled after imposing all primary constraints. Table~\ref{tab:primaryTableC} contains a summary of the spread of the numerical results for the primary observables.

Regarding the computation of the primary observables, we directly employ the analytic expressions for the trilepton decays, i.e.~for BR($\tau\to 3\, \mu$),  BR($\tau\to \mu e \bar{e}$) and BR($\mu\to  3\, e $), from appendix~\ref{sec:trilepton}, and for $Z\to \tau{\tau}$ from appendix~\ref{app:Zdecays}. So, for these observables the calculation method is the same as in the primary scan in section~\ref{sec:primary}. For the other primary observables, the method is different compared to the scan in section~\ref{sec:primary}, since we numerically calculate $R(D)$, $R(D^\star)$ and $R_{K^\star}^\nu$ using the Wilson coefficients in appendix~\ref{app:semi_leptonic}, the \texttt{Wilson} package~\cite{Aebischer:2018bkb} and \texttt{flavio}~\cite{Straub:2018kue,david_straub_2021_5543714}. Furthermore, we compute $\Delta a_\mu$, BR($\tau\to\mu\gamma$),  BR($\mu\to e\gamma$) and CR($\mu - e;$ Al) using \texttt{SARAH} and \texttt{SPheno}~\cite{Porod:2014xia,Porod:2011nf}. 

For  $Z\to \tau{\tau}$, we do not find a discernible difference between the distributions of data from the primary and comprehensive scans. We thus refer to section~\ref{subsubsec:Hadronic_primary} for a discussion of the results. We, however, display the output for $Z\to \tau{\tau}$, when discussing the secondary observable $ g_{\mu_{A}}/g_{A}^{\rm SM}$ in section~\ref{subsec:resultsSecondary}.

\subsubsection{Differences between data sets}

Before comparing the results of the primary and comprehensive scans, we first comment on some important differences between the outputs of the primary scan, discussed in section~\ref{sec:primary}, and the comprehensive one, discussed in this section. In table~\ref{tab:comprehensiveCoupMag} in appendix~\ref{app:supp6_scanmethod} we list the distributions of the unhatted LQ couplings extracted from the comprehensive scan. In this way, we can identify the effective parameters whose magnitude can be (much) smaller than $\lambda$, e.g.~$|b_{13}|$, or (much) larger than $1/\lambda$, e.g. $|a_{22}|$, i.e.~the region sampled in the primary scan. This difference impacts the distribution of the observables influenced by these parameters. 

We look at the effective parameters that dominantly drive the analytic estimates for the primary observables, listed in table~\ref{table:primaryconstraints_coupling_coeffs}. We note that there is complementary influence of the effective parameters $b_{13}$ and $a_{23}$ via the constraints from $\mu\to e$ processes, especially BR($\mu\to e \gamma$). We also see from table~\ref{table:primaryconstraints_coupling_coeffs} that the effective parameter $b_{23}$ is responsible for the dominant contributions to both the AMM of the muon and to cLFV tau decays, especially BR($\tau\to \mu \gamma$). Therefore, as discussed in section~\ref{sec:primary}, the magnitude of $b_{23}$ should not be too large. Table~\ref{tab:comprehensiveCoupMag} shows that $b_{13}$ can take particularly small values in the comprehensive scan, which means that sampled points with larger values of $c_{23}\approx a_{23}$ are capable of avoiding constraints from $\mu\to e$ processes, see eq.~(\ref{eq:b13frommeg}). These larger values of the magnitude of $a_{23}$ then require smaller $b_{23}$ in order to generate large $\Delta a_\mu$ compatible with the experimental indication. At the same time, large contributions to the BR of $\tau\to \mu \gamma$ are avoided. This makes the comprehensive scan more likely to identify a larger sample of viable points consistent with reconciling the anomaly in the AMM of the muon -- which is a challenge for the primary scan. 

Furthermore, we note several instances in which the differences in the two scans result in an amplification of a contribution to a primary observable, identified as subdominant in section~\ref{sec:primary}. Interestingly, this is relevant for the case of LFV contributions to decays with neutrinos in the final state, including $R({D})$, $R({D^\star})$, $\tau_{B_c}$ and $R_{K^\star}^\nu$. As we have seen, cLFV can be considerable in this model, and it is thus reasonable to expect similar violation in decays involving neutrinos.

We discuss the features mentioned above and other differences between the results of these two scans in the following. Throughout this section, we use the term `viable' to denote points that are capable of passing all constraints, but not necessarily addressing the three flavour anomalies. For instance, all data points displayed in plots in this section are \emph{viable} points. 

\subsubsection{Addressing the anomalies}
\label{sec:contrastAnomalies}

In the comprehensive scan, we are able to identify valid points that can explain the anomalies in all three observables, $R({D})$, $R({D^\star})$ and $\Delta a_\mu$, at the $3~\sigma$ level (see light-coloured circles in the top-right plot in figure~\ref{fig:AnomaliesComprehensive}). Indeed, the data reveals that, with the correlation between $R({D})$ and $R({D^\star})$ accommodated, the anomalies in these two observables and the measured value of $\Delta a_\mu$ can be explained at the $2\,\sigma$ level for $\hat{m}_\phi = 2$ (see dark-coloured points in the top-right plot in figure~\ref{fig:AnomaliesComprehensive}). Note that the latter cannot be explicitly seen in the bottom in figure~\ref{fig:AnomaliesComprehensive}, because the $\sigma$-regions about the best-fit values of $R({D})$ and $R({D^\star})$ do not consider the correlation between these two observables.

If we only look at the flavour anomalies in $R(D)$ and $R(D^\star)$, all three considered LQ masses are compatible with viable solutions at the $2~\sigma$ level (see the top-left plot in  figure~\ref{fig:AnomaliesComprehensive}). This is also found in the primary scan in section~\ref{sec:primary}. In case the best-fit values remain the same and the Belle II collaboration increases the precision with which they can probe these observables~\cite{Forti:2022mti}, the prospective $3~\sigma$ contour assuming $5$ ab$^{-1}$ of data is shown in the top in figure~\ref{fig:AnomaliesComprehensive} as black dashed ellipse.\footnote{In generating this projected Belle II chi-squared ellipse, we have assumed the same correlation coefficient and best-fit value as have been reported for the HFLAV averages in 2021~\cite{Amhis:2019ckw}.} We,  therefore, expect that future measurements of $R(D)$ and $R(D^\star)$ at Belle II will provide an important test of this model, and also recognise that the LQ masses we have shown to be viable to explain the anomalies are likely within the reach of upcoming LQ direct-production searches at the LHC.

Furthermore, we indicate the prospects for future measurements of the AMM of the muon. In the bottom in figure~\ref{fig:AnomaliesComprehensive}, the $3~\sigma$ projected sensitivity from the Muon g$-$2 experiment~\cite{Muong-2:2015xgu} is seen to (roughly) overlay the present $2~\sigma$ region for this observable. If, as we have illustrated here, the best-fit value remains the same and the sensitivity is improved, this will challenge the model as an explanation of the three anomalies, but not entirely rule out the viable parameter space. This is consistent with the preceding discussion, and is evident from the distribution of white crosses in the plots in figure~\ref{fig:AnomaliesComprehensive}.

\begin{figure}[t!]
    \centering
\includegraphics[width=0.49\textwidth]{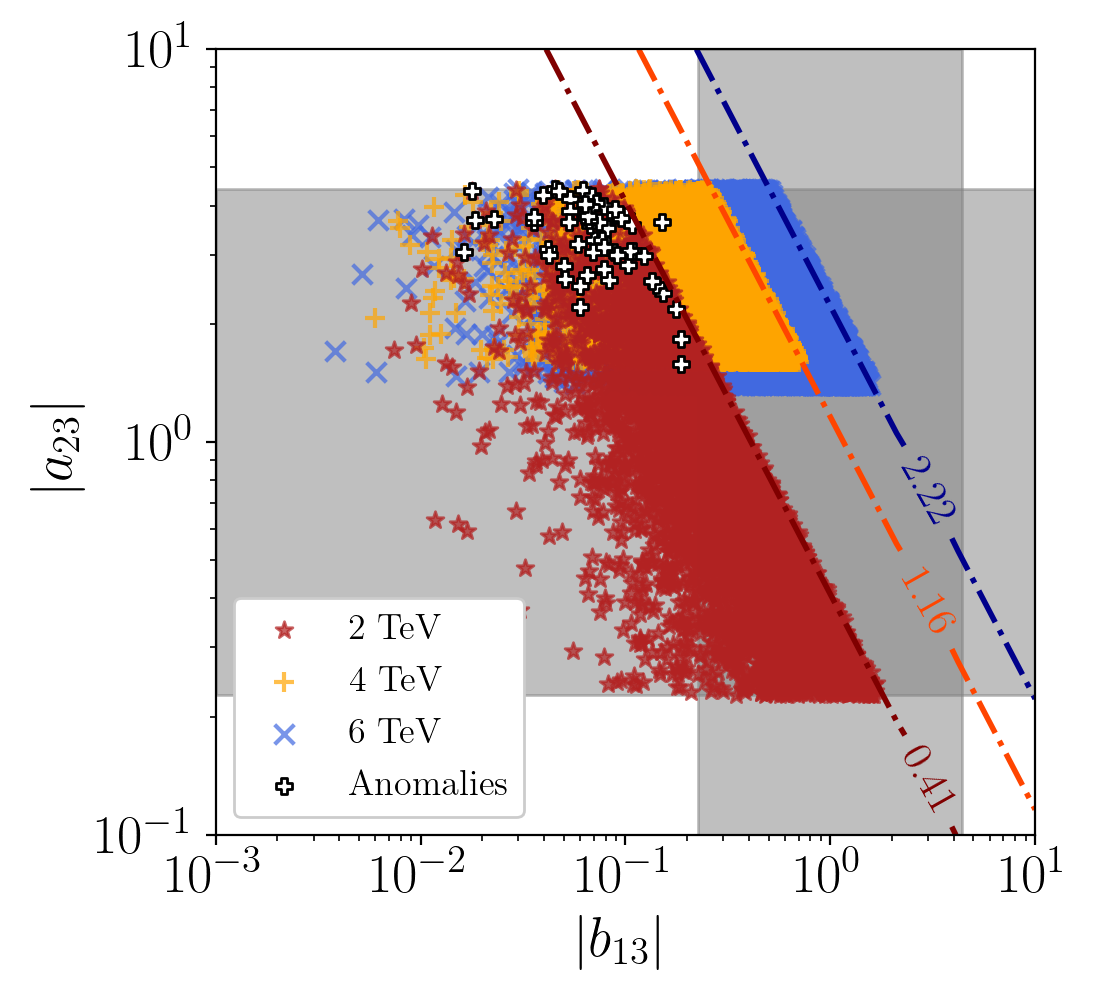}
\includegraphics[width=0.49\textwidth]{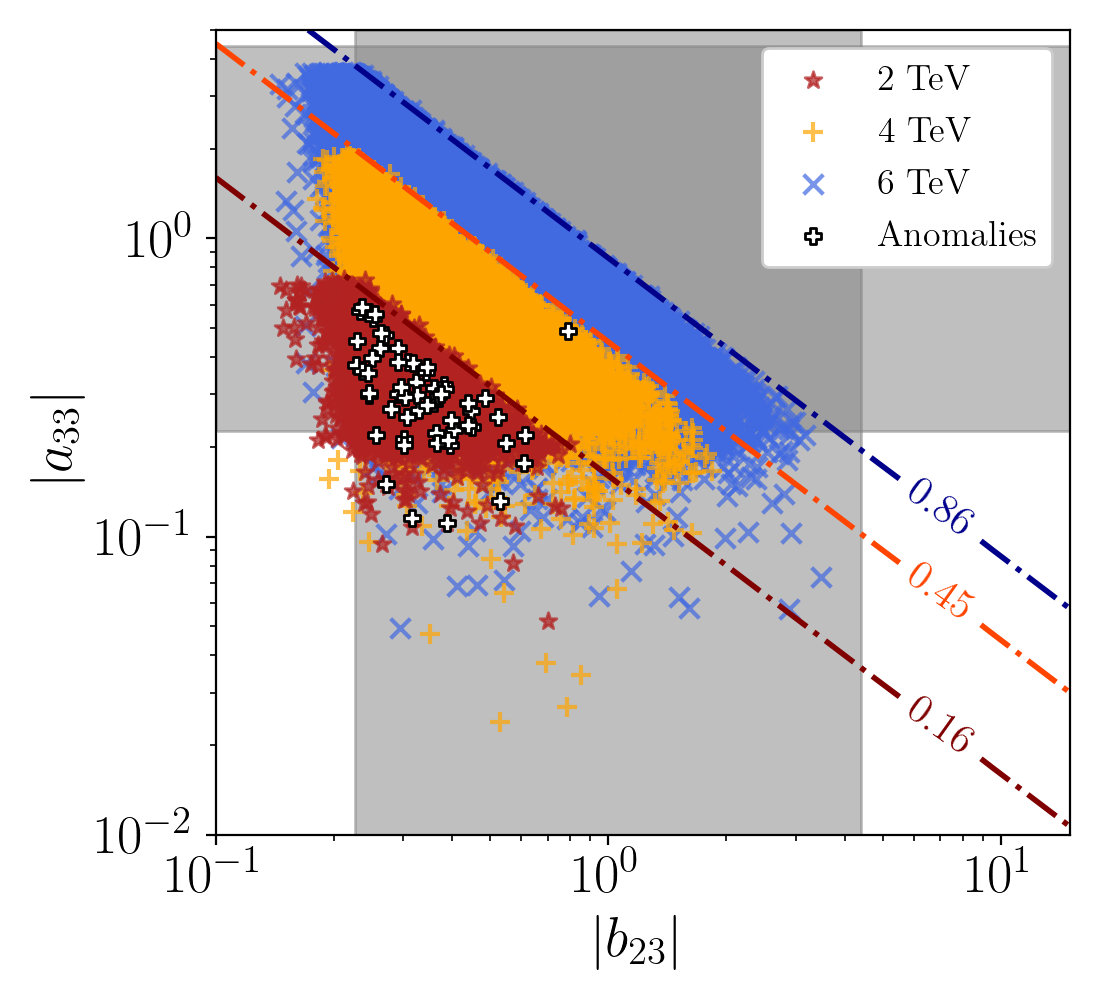}\\
\begin{minipage}{0.49\textwidth}
\includegraphics[width=\textwidth]{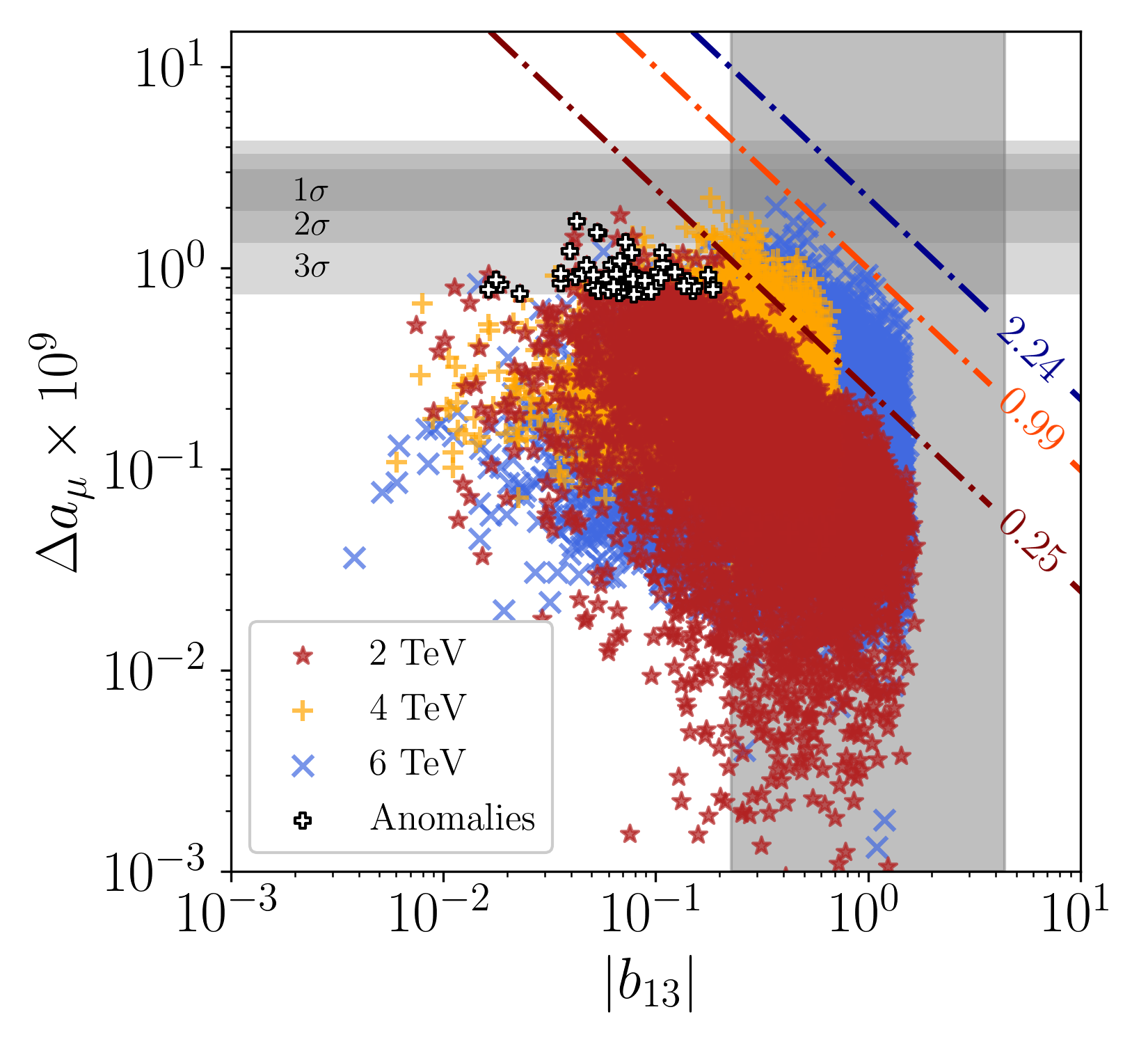}
\end{minipage}\hspace{7mm}
\begin{minipage}{0.45\textwidth}\caption{\linespread{1.1}\small{\textbf{Interplay of imposed biasing and employed basis in comprehensive scan}. In the top, the dot-dashed lines show the imposed bounds from eqs.~\eqref{eq:b13frommeg} and \eqref{eq:b23fromtmg}, in dark red for $\hat{m}_\phi  = 2$, orange for $\hat{m}_\phi  = 4$, and dark blue for $\hat{m}_\phi  = 6$, respectively. The bottom plot shows the influence of this biasing on the viability to explain the anomaly observed in $\Delta a_\mu$ for each LQ mass, and overlaid dot-dashed lines show the approximate bound from eq.~\eqref{eq:delamuEst2}. The grey bands, shown for the effective parameters, illustrate the regions sampled in the primary scan. For further information on how to read this figure, see section~\ref{ssec:comprehensivePlots}.}} \label{fig:compBiasCLFV}\end{minipage}
   \end{figure}
   
\paragraph{Simultaneously addressing all three anomalies.} In the primary scan, it is found to be difficult to identify points that can address all three flavour anomalies simultaneously. A key limiting factor is the interplay  between large contributions to the AMM of the muon and large contributions to the radiative cLFV processes $\tau\to\mu\gamma$ and  $\mu\to e\gamma$. Here we discuss how this limitation is relaxed in the comprehensive scan.

We first analytically show that the maximum value of the AMM of the muon compatible with the primary constraints is (roughly) inversely proportional to the magnitude of the effective parameter $b_{13}$. Recall from the estimate in eq.~\eqref{eq:gm2_estimate} that
\begin{align}
|\Delta a_\mu| \lesssim \frac{|b_{23} c_{23}|}{\hat{m}_\phi ^2} \times 10^{-9} . 
\end{align}
 The present bound on the combination $|b_{13}c_{23}|$  from BR($\mu\to e \gamma$), see eq.~\eqref{eq:approx_bound_mueg}, can be used to parametrise the dependence on $|b_{13}|$ for a viable point. We can also bound the magnitude of $b_{23}$  (roughly) from above by considering the dominant contribution to BR($\tau\to \mu \gamma$), see eq.~\eqref{eq:approx_bound_taumug}, which is proportional to $|b_{23}c_{33}|$, and using the minimum sampled value of the magnitude of $c_{33}\approx a_{33}$ from table~\ref{tab:pc1}. Therefore, an approximate upper bound for the maximally accessible value of $|\Delta a_\mu|$ for each LQ mass can be expressed as
\begin{align}
|\Delta a_\mu| \lesssim \frac{1}{|b_{13}|}  \left\{
\begin{array}{*{2}{c}}
0.249, & \hat{m}_\phi = 2 \\
0.993,& \hat{m}_\phi = 4 \\
2.236,& \hat{m}_\phi = 6 \\
\end{array} \right\}\times 10^{-9} . 
\label{eq:delamuEst2}
\end{align}
This inverse correlation between the maximum value of $|\Delta a_\mu| $ and the magnitude of the effective parameter $b_{13}$ is also influenced by the biasing procedure, which aims to target points that avoid constraints from the two radiative cLFV decays $\mu\to e \gamma$ and $\tau\to \mu \gamma$ using eqs.~\eqref{eq:b13frommeg} and \eqref{eq:b23fromtmg}. 

The influence of biasing the effective parameters and the distribution of those dominating the contributions to radiative cLFV decays are illustrated in the top in figure~\ref{fig:compBiasCLFV}. The top-left plot shows the distribution of values of the magnitude of $b_{13}$ accessible in the primary and comprehensive scans. Whereas the minimum value for $|b_{13}|$ is seen to be as small as $\mathcal{O}(10^{-3})$ in the comprehensive scan, the primary scan only employs values larger than $\lambda \approx 0.22$. From eq.~\eqref{eq:delamuEst2}, we therefore expect larger accessible values of the AMM of the muon, while, at the same time, evading the primary constraints. This is, indeed, the case as shown in the bottom in figure~\ref{fig:compBiasCLFV}, where we show the distribution of $\Delta a_\mu$ plotted against the value of $|b_{13}|$. This effect is especially striking for $\hat{m}_\phi=2$, for which the only points that explain the observed anomaly in $\Delta a_\mu$ within the $3 \, \sigma$ range are found outside the grey-shaded region, i.e~the sampled region of the magnitude of $b_{13}$ in the primary scan. We note that in the bottom in figure~\ref{fig:compBiasCLFV} we also overlay as dot-dashed lines the upper bound for each LQ mass, as shown eq.~\eqref{eq:delamuEst2}. We see reasonable agreement with the data, given that this upper bound has been derived under the assumption that the magnitudes of both $a_{33}$ and $b_{23}$ lie in the interval $[\lambda, 1/\lambda]$, which is,  however, not always
fulfilled in the comprehensive scan, compare top-right plot in figure~\ref{fig:compBiasCLFV}. 

\paragraph{\texorpdfstring{\mathversion{bold}Contributions from lepton flavour violating channels to $R(D)$ and $R(D^\star)$.\mathversion{normal}}{Contributions from lepton flavour violating channels to RD and RDstar}} We comment on the impact of the LFV contribution from $b\to c \tau \nu_\mu$ to $R(D)$ and $R(D^\star)$ in the comprehensive scan. Appealing to the analytic formulae in eqs.~\eqref{eq:RD_estimate} and~\eqref{eq:RDs_estimate}, we see that the contribution from the LFV final state with a muon neutrino is proportional to $|a_{23}b_{32}|$. We recall that the magnitude of $a_{23}$ is biased towards larger values, see table~\ref{tab:pc1}, and we thus expect non-negligible contributions from this decay channel to $R(D)$ and $R(D^\star)$ in the comprehensive scan. Even though this LFV channel does not interfere with the SM contribution, for $\hat{m}_\phi=2$ it can generate an enhancement above the SM value as large as $40$ percent for $R(D)$ and $30$ percent for $R(D^\star)$, respectively, using all viable points. For $\hat{m}_\phi=4, 6$ there is at most a $10$ percent enhancement in either observable above the SM value, which is roughly consistent with the present $1~\sigma$ margin about the best-fit values. Therefore, we emphasise that this effect cannot be neglected, when considering these observables in this model, particularly as future experiments will reach increased sensitivity.

\subsubsection{Leptonic primary constraints}
\label{sec:Comprehensive_leptonic}

\begin{figure}[t!]
    \centering
        \includegraphics[width=0.49\textwidth]{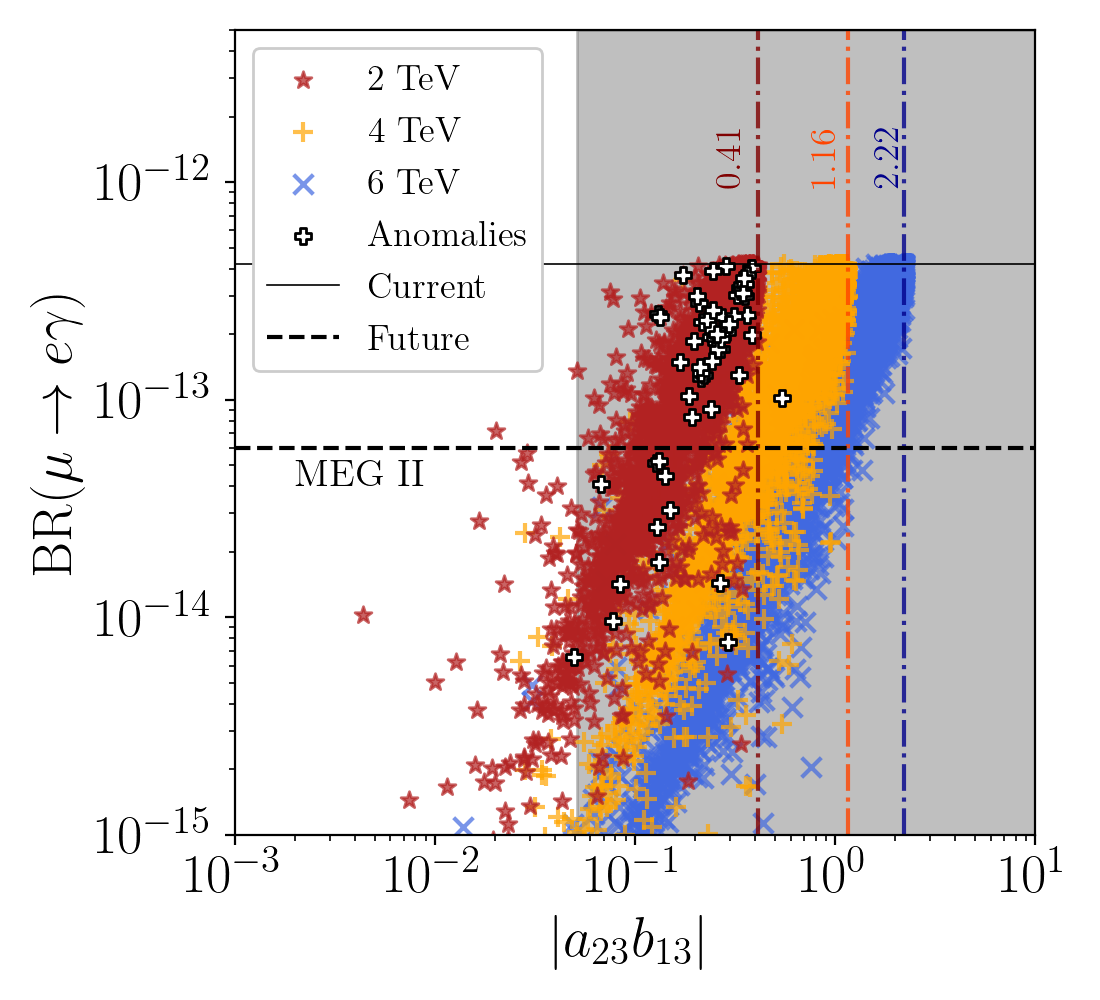}
    \includegraphics[width=0.49\textwidth]{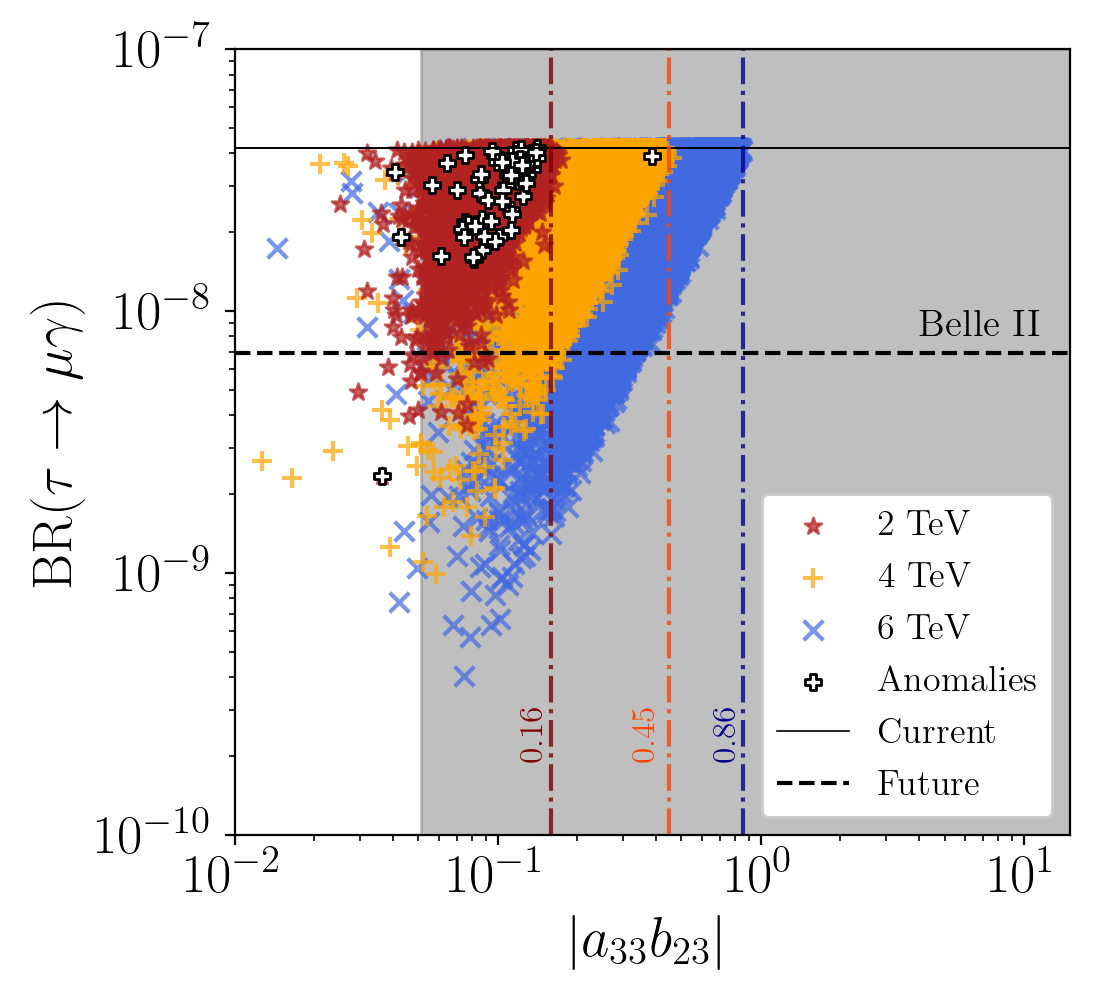}
    \caption{{\small \textbf{Impact of sampling on radiative cLFV decays in comprehensive scan.} The vertical dot-dashed lines show the upper bounds on the respective product of effective parameters taken from eqs.~\eqref{eq:b13frommeg} and~\eqref{eq:b23fromtmg}, in dark red for $\hat{m}_\phi  = 2$, orange for $\hat{m}_\phi  = 4$, and dark blue for $\hat{m}_\phi  = 6$, respectively. For further information on how to read this figure, see section~\ref{ssec:comprehensivePlots}.   
    }}
    \label{fig:Contrasttmgmeg}
\end{figure}

In section~\ref{subsubsec:prelimscan_eiejgamma} it is shown that the strongest present experimental constraints on this model arise from the radiative cLFV decays $\tau\to\mu\gamma$ and $\mu\to e\gamma$. In the following, we discuss the differences between the results of the primary and comprehensive scans for these and other leptonic primary constraints as well as the prospects that these processes offer as signals of this model at future experiments.

\paragraph{Radiative charged lepton flavour violating decays.}
In figure~\ref{fig:Contrasttmgmeg}, we show the impact of biasing on the parameter space in the case of the radiative cLFV decays $\mu\to e\gamma$ and $\tau\to\mu\gamma$. The upper bounds from eqs.~\eqref{eq:b13frommeg} and~\eqref{eq:b23fromtmg} are shown as vertical dot-dashed lines. 

Since in the comprehensive scan the magnitude of the effective parameter $b_{13}$ can be significantly lower than $\lambda$, also the attained values of the product $|a_{23} b_{13}|$ can be smaller than naively
expected, compare coloured points and grey-shaded region in the left plot in figure~\ref{fig:Contrasttmgmeg}. This product being smaller generally corresponds to a suppressed value of BR($\mu\to e\gamma)$, consistent with the analytic estimate in eq.~\eqref{eq:mueg_estimate}. We see that a signal of $\mu \to e\gamma$ is predicted to be observed at MEG II~\cite{MEGII:2021fah} for a large number of viable points, in agreement with the findings of the primary scan, see section~\ref{subsubsec:prelimscan_eiejgamma}. Nevertheless, there are points capable of addressing all three flavour anomalies that remain unconstrained by this observable even with this increased sensitivity.

Similarly, we note that the magnitude of the effective parameter $a_{33}$ can be smaller than $\lambda$, see top-right plot in figure~\ref{fig:compBiasCLFV}, though to a lesser extent than in the case of $b_{13}$. We, thus, expect points in the right plot in figure~\ref{fig:Contrasttmgmeg} to fall below the grey region for the product $|a_{33}b_{23}|$. Smaller accessible values of $a_{33}$ imply viable points with larger magnitude of $b_{23}$, although we see that the present bound from BR($\tau \to \mu\gamma$) is already very constraining on this product. Furthermore, we note that the majority of points able to address the three anomalies, i.e.~the white crosses, lead to a value for BR($\tau \to \mu\gamma$) in the region that can be probed by the future sensitivity of Belle II~\cite{Banerjee:2022xuw}. 

Finally, we mention that the distributions for these radiative cLFV decays with respect to $R(D)$, $R(D^\star)$ and the AMM of the muon can be found in figure~\ref{fig:CompResultsCLFV} in  appendix~\ref{app:addfigssec6}. These can be compared with figure~\ref{fig:constrain_power_taumug} in section~\ref{subsubsec:prelimscan_eiejgamma}.

\begin{figure}[t!]
    \centering
    \includegraphics[width=0.49\textwidth]{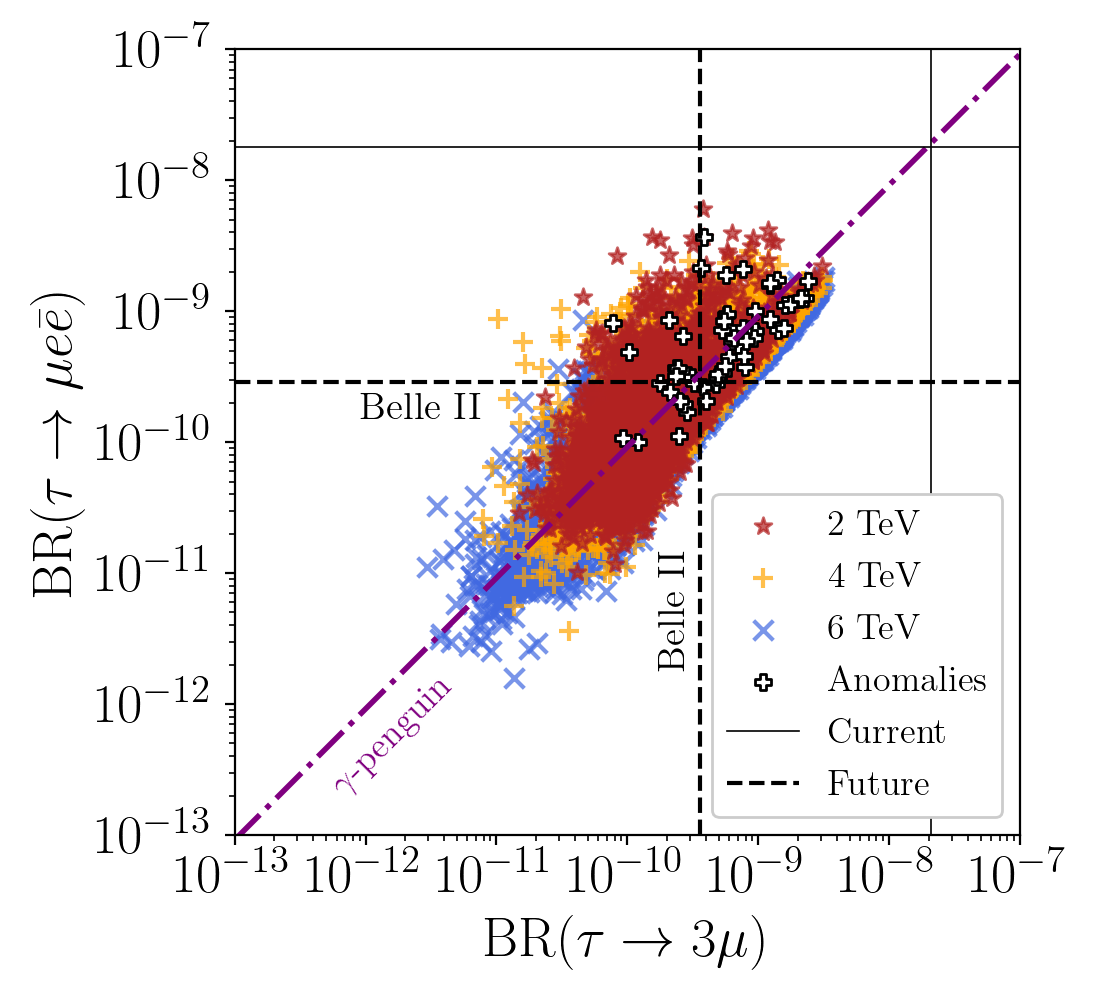} 
 \includegraphics[width=0.49\textwidth]{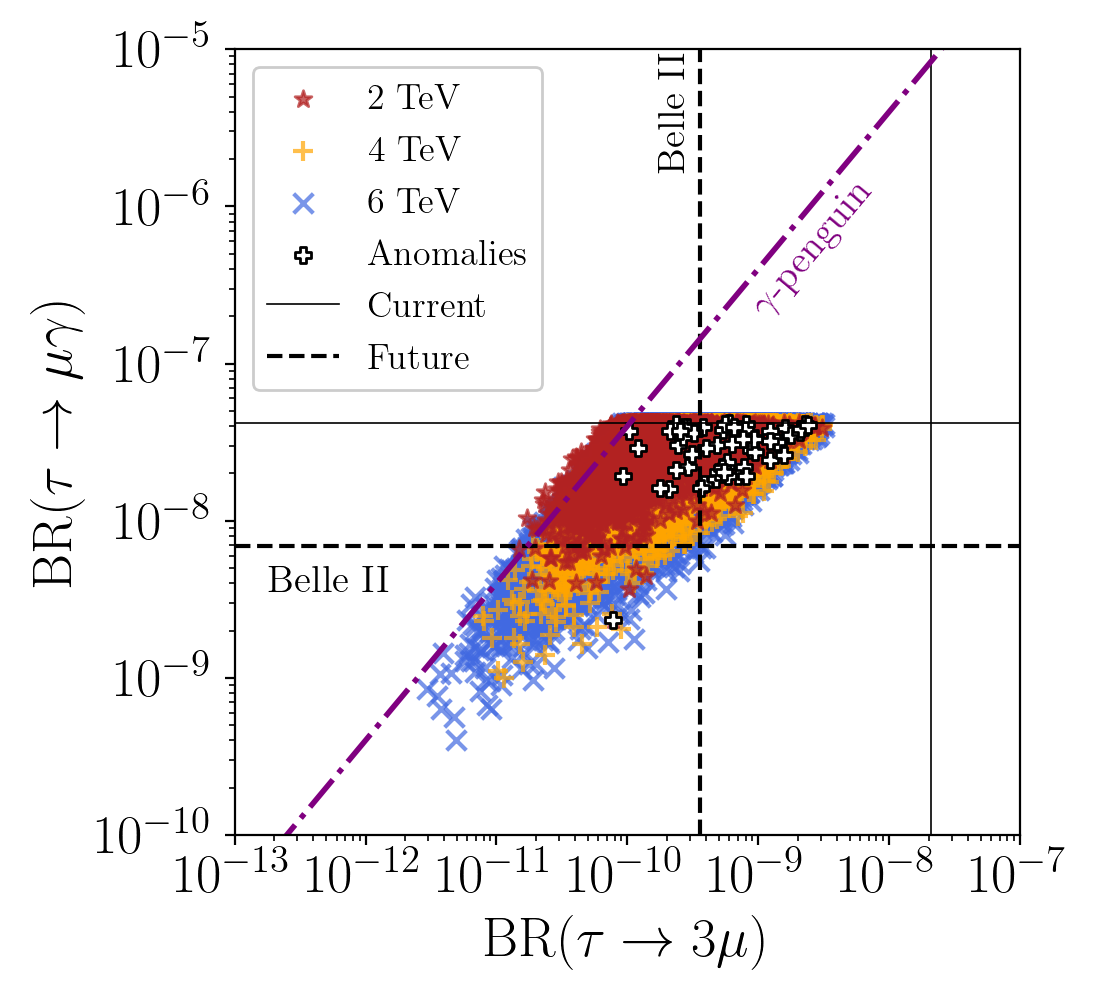}
\includegraphics[width=0.49\textwidth]{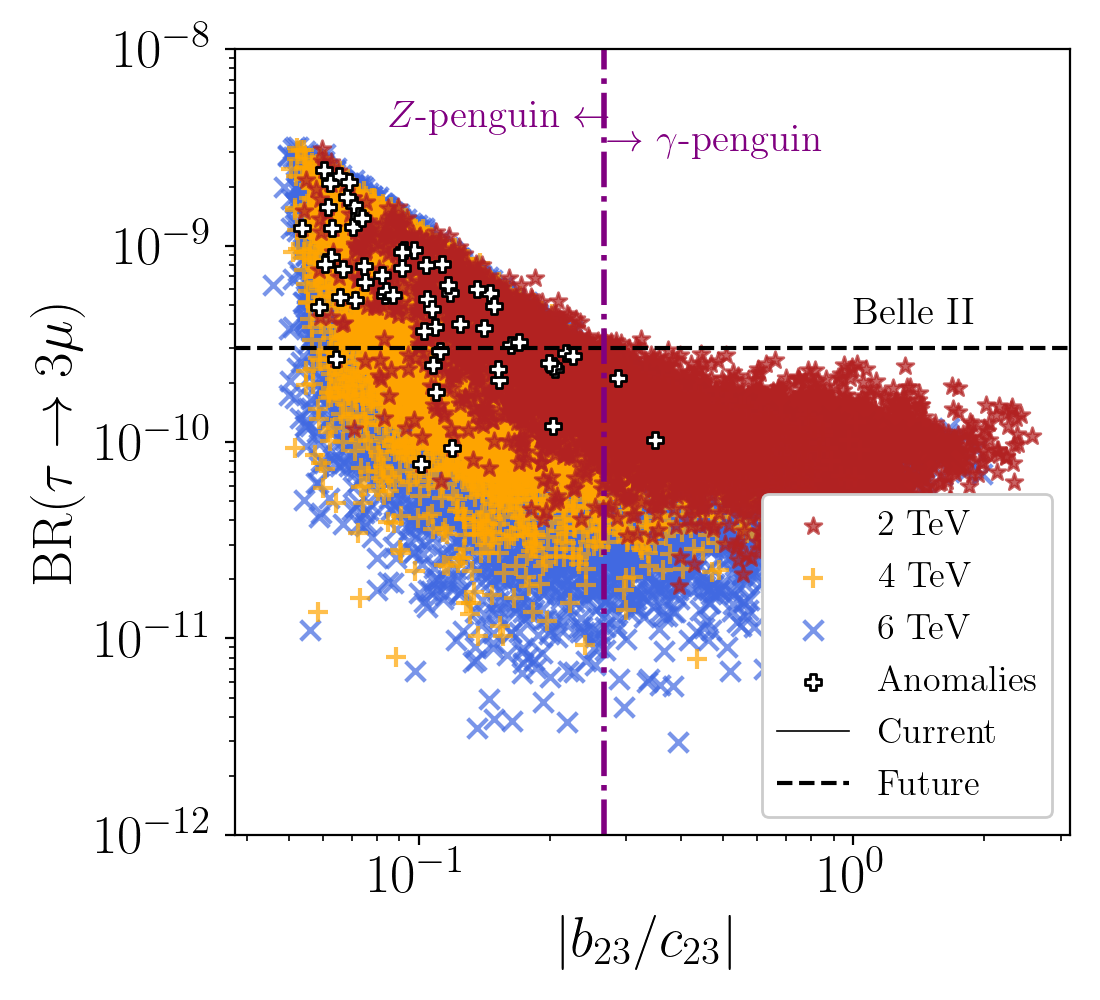}
\includegraphics[width=0.49\textwidth]{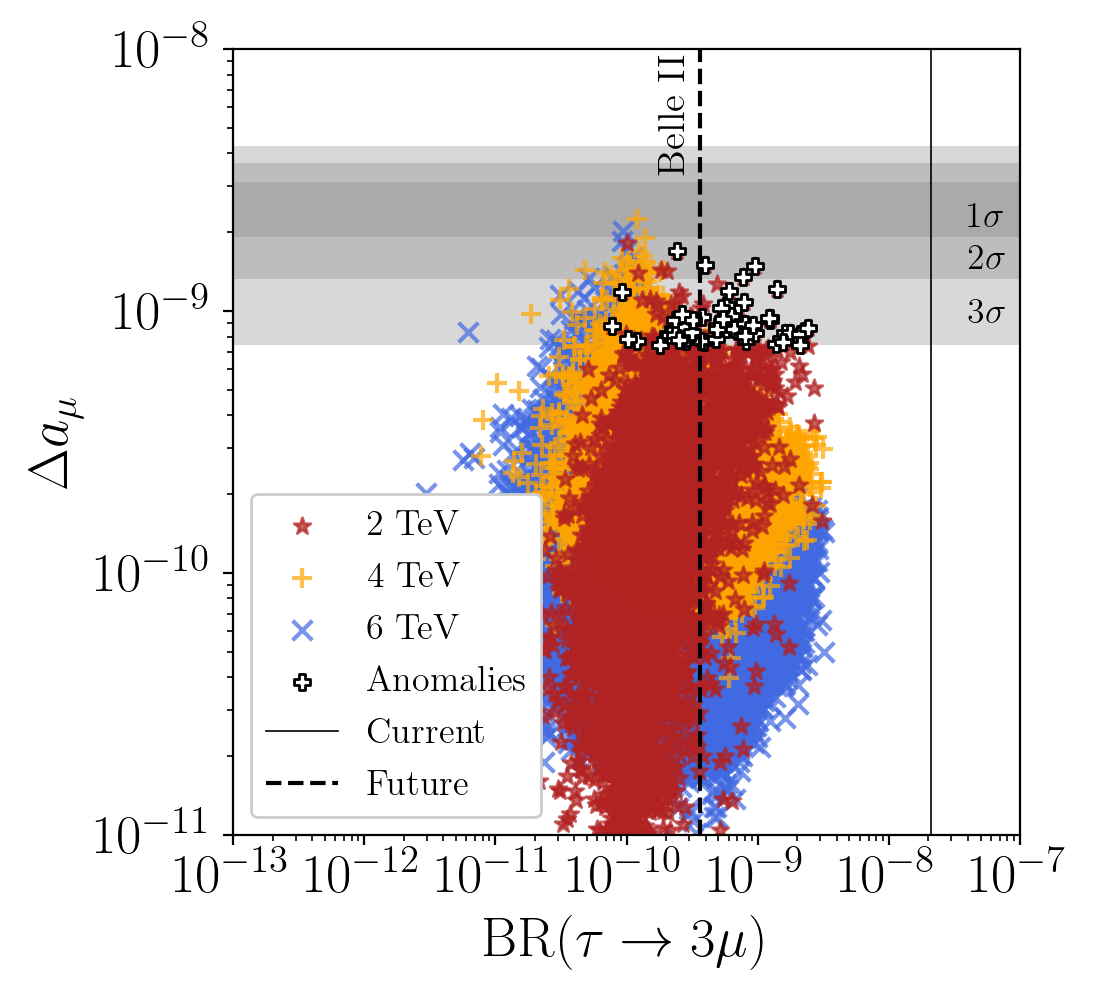}
    \caption{{\small \textbf{Impact of sampling on trilepton tau decays in comprehensive scan.} In the two plots in the top, the purple dot-dashed line, labelled `$\gamma$-penguin', indicates the approximate relation between these observables in the case of $\gamma$-penguin dominance, see eqs.~\eqref{eq:tau_photonpen2} (left) and~\eqref{eq:tau_photonpen} (right). In the bottom-left plot the purple dot-dashed line shows the borderline between $Z$- and $\gamma$-penguin dominance, as labelled. For further information on how to read this figure, see section~\ref{ssec:comprehensivePlots}.}  }
        \label{fig:ContrastTau3lep}
\end{figure}

\paragraph{Trilepton tau decays.} We discuss the results for the primary constraints BR($\tau\to 3\,\mu$) and BR($\tau\to \mu e \bar{e}$) in the comprehensive scan. The relationship between these two observables is displayed in the top-left plot in figure~\ref{fig:ContrastTau3lep}. Unlike in the primary scan in section~\ref{sec:primary}, we in general do not find that BR($\tau\to 3\, \mu$) $\approx$ BR($\tau\to \mu e \bar{e}$), i.e. eq.~\eqref{eq:tau_photonpen2} does not hold for the shown data. This is due to the impact of subdominant contributions to both processes, including the $Z$-penguin contributions explicitly detailed in eqs.~\eqref{eq:tauto3mu_estimate} and~\eqref{eq:tautomuee_estimate}, as well as subleading contributions to both processes, e.g.~contributions due to box diagrams as discussed in appendix~\ref{app:dipole}. We show the relation given in eq.~\eqref{eq:tau_photonpen2}, reflecting $\gamma$-penguin dominance, as an overlaid purple dot-dashed line. We note that there is a strong positive correlation between the observables, but a spread influenced non-trivially by these subdominant contributions. Therefore, both observables should be separately considered with respect to the future prospects for signals at Belle II~\cite{Banerjee:2022xuw}. 

Next, we would like to contrast the results for BR($\tau\to 3\, \mu$) with those of the primary scan. In the top-right plot in figure~\ref{fig:ContrastTau3lep} we show BR($\tau\to 3 \, \mu$) plotted against BR($\tau\to \mu \gamma$).  The overlaid purple dot-dashed line shows the estimate from eq.~\eqref{eq:tau_photonpen}, i.e.~the linear correlation between BR($\tau\to 3\, \mu$) and BR($\tau\to \mu \gamma$) in the case of $\gamma$-penguin dominance. Comparing this plot with the corresponding one in figure~\ref{fig:tau3mu}, we notice a large number of points, particularly for $\hat{m}_\phi=2$, that result in a value for BR($\tau\to3\,\mu$) which
can be probed at Belle II unlike in the case of the primary scan. This includes a sizeable number of the white crosses, points capable of addressing all three anomalies, which motivates further examination of this difference.

The explanation for this difference lies in the relevance of $Z$- and $\gamma$-penguin contributions to BR($\tau\to3 \, \mu$). Eq.~\eqref{eq:tauto3mu_estimate} reveals that both contributions are proportional to $|c_{33}|\approx |a_{33}|$, and also to $|c_{23}|\approx |a_{23}|$ ($Z$-penguin) or $|b_{23}|$  ($\gamma$-penguin), respectively. Smaller $\gamma$-penguin contributions correspond to smaller values of BR($\tau \to \mu\gamma$), see eq.~(\ref{eq:taumug_estimate}), and are thus preferred in the comprehensive scan due to biasing, compare eq.~(\ref{eq:b23fromtmg}). Therefore, in order to generate a significant contribution to the AMM of the muon larger values of the magnitude of the effective parameter $a_{23}\approx c_{23}$ are needed, see eq.~(\ref{eq:gm2_estimate}) and table~\ref{tab:pc1}. This, in turn, typically enhances the $Z$-penguin contributions to BR($\tau\to 3\, \mu$). From eq.~\eqref{eq:tauto3mu_estimate} we can derive the value of the ratio $|b_{23}/c_{23}|$ at which the dominant contribution changes. This is illustrated as a purple dot-dashed line in the bottom-left plot in figure~\ref{fig:ContrastTau3lep}. Consistent with the preceding discussion we see that points preferred by explaining the flavour anomalies overwhelmingly correspond to those which have values of $|b_{23}/c_{23}|$ in the range of $Z$-penguin dominance. In the bottom-right plot in  figure~\ref{fig:ContrastTau3lep}, we also see that, indeed, these same points correspond to significant contributions to $\Delta a_\mu$.

Furthermore, we notice in the bottom-right plot in figure~\ref{fig:ContrastTau3lep} a prominent feature, namely a diagonal cutoff towards the top-right corner of the plot for $\hat{m}_\phi=4,6$, which is not present for $\hat{m}_\phi=2$. In section~\ref{subsubsec:prelimscan_trilepton_mueconv}, this feature has been associated with the inverse proportionality of the maximum contributions to BR($\tau\to\mu\gamma$) and $\Delta a_\mu$, at LO. In the top-right plot in figure~\ref{fig:tau3mu}, we observe this cutoff for all three sampled LQ masses, although less clearly for $\hat{m}_\phi=2$. For larger LQ masses, subdominant contributions, particularly to BR($\tau\to\mu\gamma$) and BR($\tau\to 3 \, \mu$), are more suppressed by the LQ mass, and consequently this cutoff is more pronounced. For $\hat{m}_\phi=2$, such contributions enter and weaken the mentioned inverse proportionality -- leading to the observed spread of points towards the top-right corner of the plot. 

In summary, the flavour anomaly in the AMM of the muon can be explained, while potentially large signals for both $\tau\to 3\, \mu$ and $\tau\to \mu e \bar{e}$ can be observed at Belle II. These signals are driven by largish $Z$-penguin contributions that, unlike contributions due to $\gamma$-penguins, are not constrained by correspondingly large contributions to BR($\tau\to\mu\gamma$). The correlation between larger BRs for trilepton tau decays and sizeable values of the AMM of the muon is enhanced by the biasing, and particularly by the increase in sampled viable points with a large magnitude of the effective parameter $a_{23}$, as discussed in section~\ref{sec:contrastAnomalies}. 

\paragraph{\mathversion{bold}Further $\mu \to e$ processes.\mathversion{normal}}

\begin{figure}[t!]
    \centering
    \begin{minipage}{0.49\textwidth}
\includegraphics[width=\textwidth]{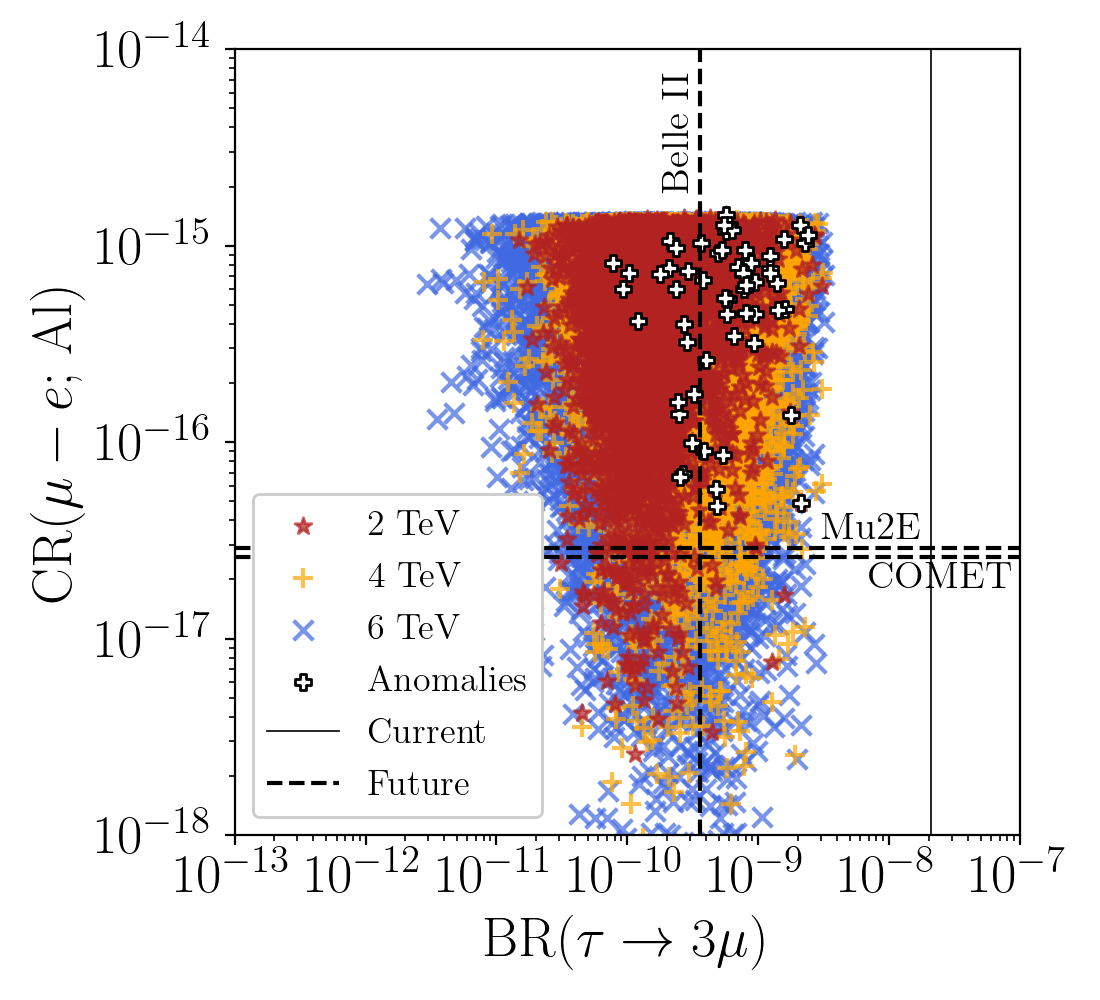}
\end{minipage}\hspace{8mm}
\begin{minipage}{0.45\textwidth}\caption{\linespread{1.1}\small{\mathversion{bold}\textbf{Example of interplay between BR($\tau\to 3\, \mu$) and $\mu \to e$ processes in comprehensive scan.}\mathversion{normal} We plot CR$(\mu - e; \text{Al})$ against BR($\tau\to 3\, \mu$) in order to illustrate the spread of points that correspond to large BR($\tau\to 3\, \mu$), but a small signal in $\mu \to e$ processes. We show for CR$(\mu - e; \text{Al})$ two prospective limits as black dashed lines, from COMET~\cite{COMET:2018auw} and Mu2e~\cite{Mu2e:2014fns}. For further information on how to read this figure, see section~\ref{ssec:comprehensivePlots}. }}    \label{fig:Contrastmue3mu}
\end{minipage}
\end{figure}

 With smaller values of the magnitude of the effective parameter $b_{13}$, see top-left plot in figure~\ref{fig:compBiasCLFV}, in particular the white crosses, we can also access smaller values of the observables dominantly driven by this parameter. These include, as we can see from table~\ref{table:primaryconstraints_coupling_coeffs}, BR($\mu\to e \gamma$), CR$(\mu - e; \text{Al})$ and BR($\mu\to 3 \, e$). Although smaller $b_{13}$ allows to evade present constraints, it does not exclude these channels as means to test this model at future experiments. As the prospects for BR($\mu\to e \gamma$) are already discussed, we focus in the following on the other two processes. From table~\ref{tab:primaryTableC} we see that on average BR($\mu\to 3 \, e) \sim \mathcal{O}(10^{-15})$ which is an order of magnitude larger than the projected sensitivity of Phase 2 of the Mu3e experiment~\cite{Blondel:2013ia}. Still, we find points that can evade this constraint, even with this future sensitivity, including the ones that are capable of addressing the three flavour anomalies. Similarly, we have on average CR$(\mu - e; \text{Al}) \sim \mathcal{O}(10^{-16})$ which is an order of magnitude larger than the future projections for COMET~\cite{COMET:2018auw} and Mu2e~\cite{Mu2e:2014fns}. Compellingly, in figure~\ref{fig:Contrastmue3mu} we see that all white crosses, associated with points that explain the three flavour anomalies at the $3 \, \sigma$ level or better, are within the region of parameter space probed by either future search for $\mu-e$ conversion in aluminium. Therefore, we predict a signal to be observed for this process. Note that the same proportionality between CR$(\mu - e; \text{Al})$ and BR($\mu\to e \gamma$) is found in the comprehensive scan as in the primary scan, illustrated in figure~\ref{fig:mueconv}. So, the upper bound on CR$(\mu - e; \text{Al})$, seen in figure~\ref{fig:Contrastmue3mu}, stems from respecting the present constraint on BR($\mu\to e \gamma$). 

Furthermore, we comment on the relation between $\mu\to e$ processes and BR($\tau\to 3\, \mu$) in this model. This is a continuation of the discussion found in section~\ref{subsubsec:prelimscan_trilepton_mueconv}, in which it is noted that large BR($\tau\to 3\, \mu$), observable at Belle II, is much less likely, if contributions to $\mu\to e$ processes are beyond the reach of future experiments. This relation is found to be less pronounced in the comprehensive scan, because smaller values of the rates of $\mu\to e$ processes can be generally reached, given that their dominant contributions are proportional to the magnitude of $b_{13}$. As this effective parameter can take much smaller values in the comprehensive scan, larger values of the magnitude of $a_{23} \approx c_{23}$ become allowed that can enhance $Z$-penguin
contributions to BR($\tau\to 3\, \mu$). This explains the existence of viable points towards the bottom-right corner in figure~\ref{fig:Contrastmue3mu}, not observed in the corresponding plot of the primary scan, see figure~\ref{fig:tau3mu}. We notice that the plots for BR($\mu\to 3 \, e$) and BR($\mu\to e \gamma$) reveal a very similar behaviour to the one for CR$(\mu - e; \text{Al})$ shown in figure~\ref{fig:Contrastmue3mu}.

\subsubsection{Hadronic primary constraints}
\label{sec:Comprehensive_hadronic}

In the primary scan in section~\ref{sec:primary} the behaviour of the hadronic observables $R_{K^\star}^\nu$ and $\tau_{B_c}$, the $B_c$ lifetime, is explored. Notably, both of these observables involve neutrinos in the final state. Naively, one may be tempted to consider only contributions for which lepton flavour is conserved, and therefore lead to interference with the corresponding SM contributions. However, as pointed out for $R(D)$ and $R(D^\star)$ in section~\ref{sec:contrastAnomalies}, the contributions from LFV channels are found to be non-negligible in the comprehensive scan. We discuss these and other differences between the primary and comprehensive scan for $R_{K^\star}^\nu$ and $\tau_{B_c}$ in the following.

\paragraph{\mathversion{bold}Contributions to observable $R_{K^\star}^\nu$.\mathversion{normal}} 
\begin{figure}[t!]
    \centering
        \includegraphics[width=0.49\textwidth]{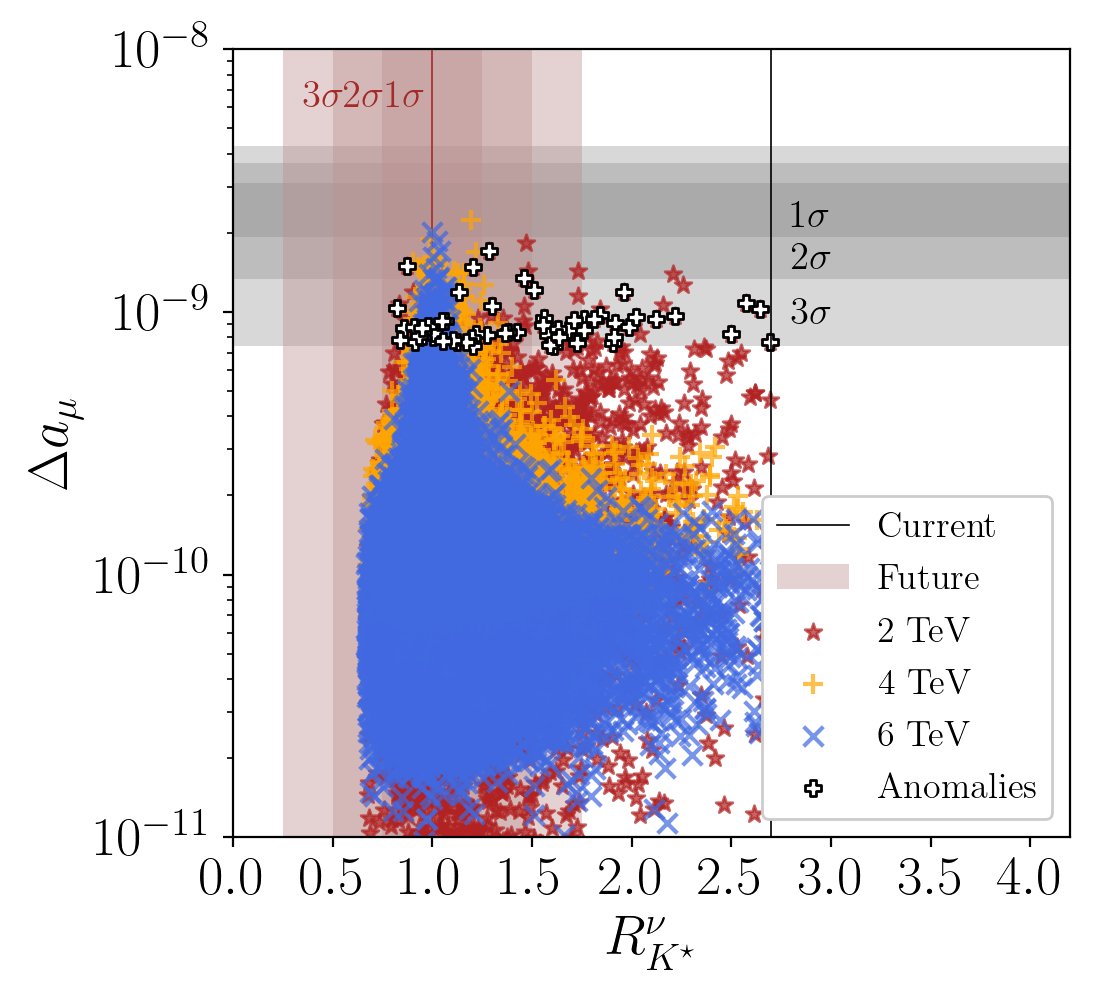}
          \includegraphics[width=0.49\textwidth]{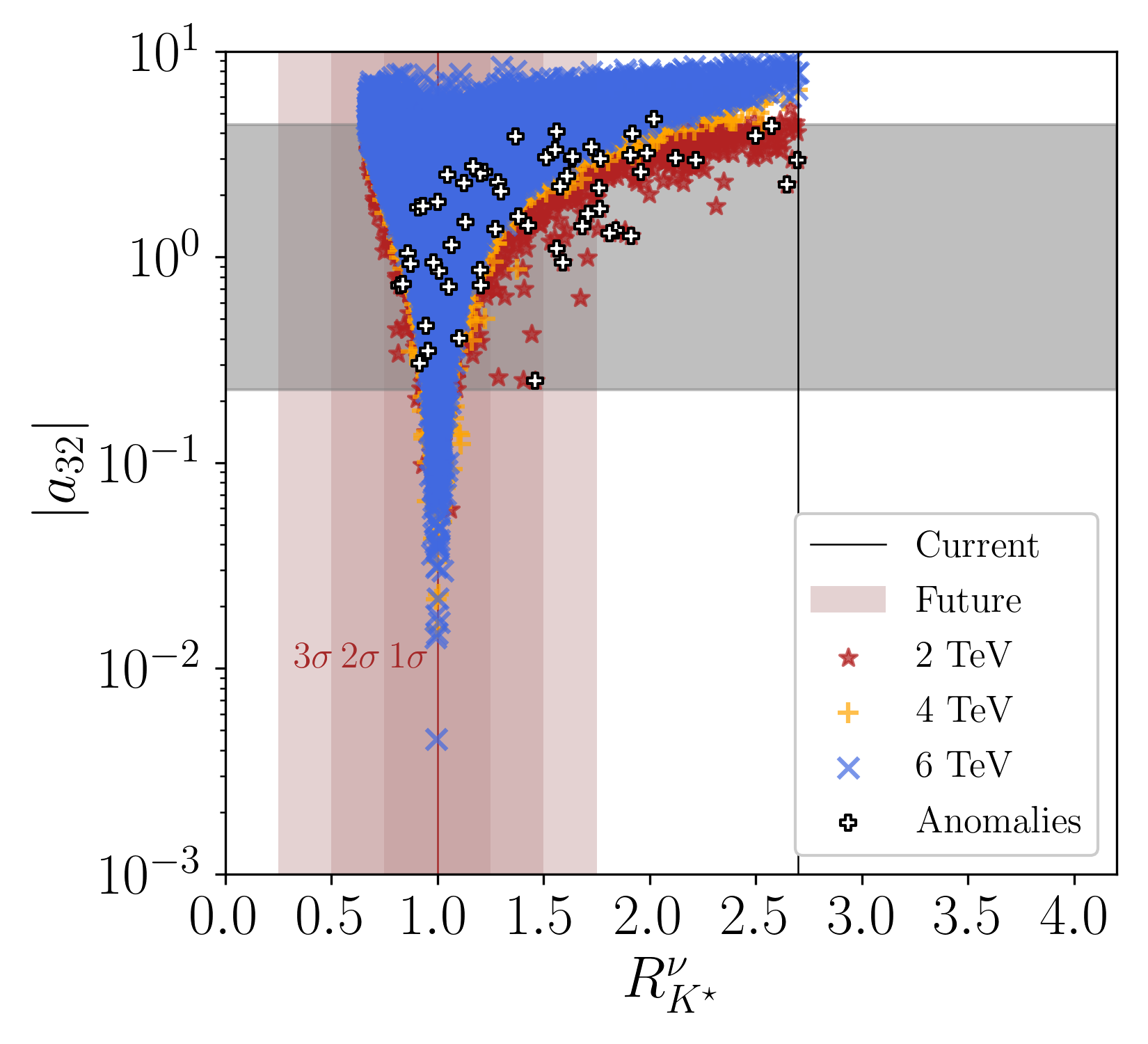}
            \includegraphics[width=0.49\textwidth]{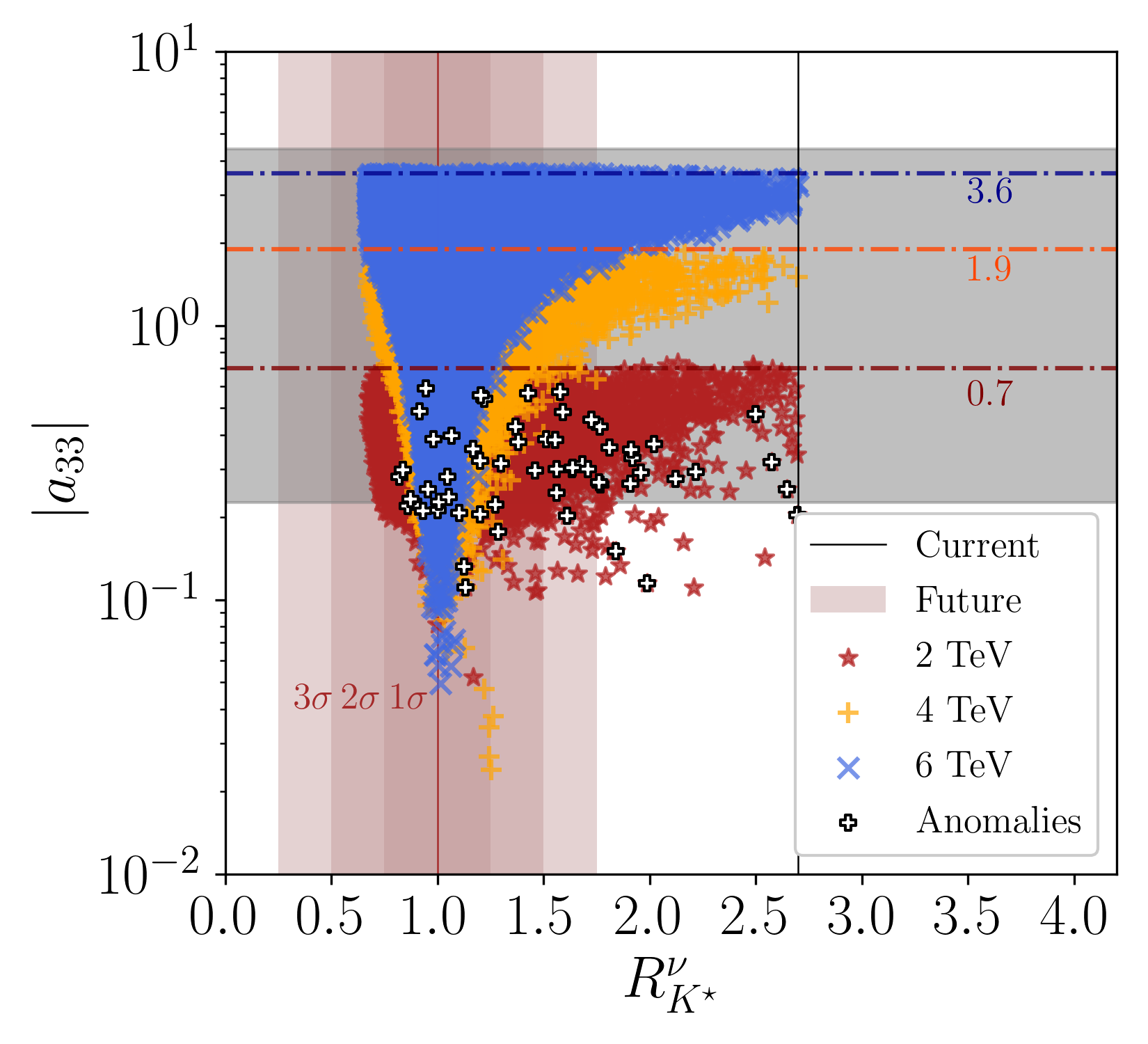}
              \includegraphics[width=0.49\textwidth]{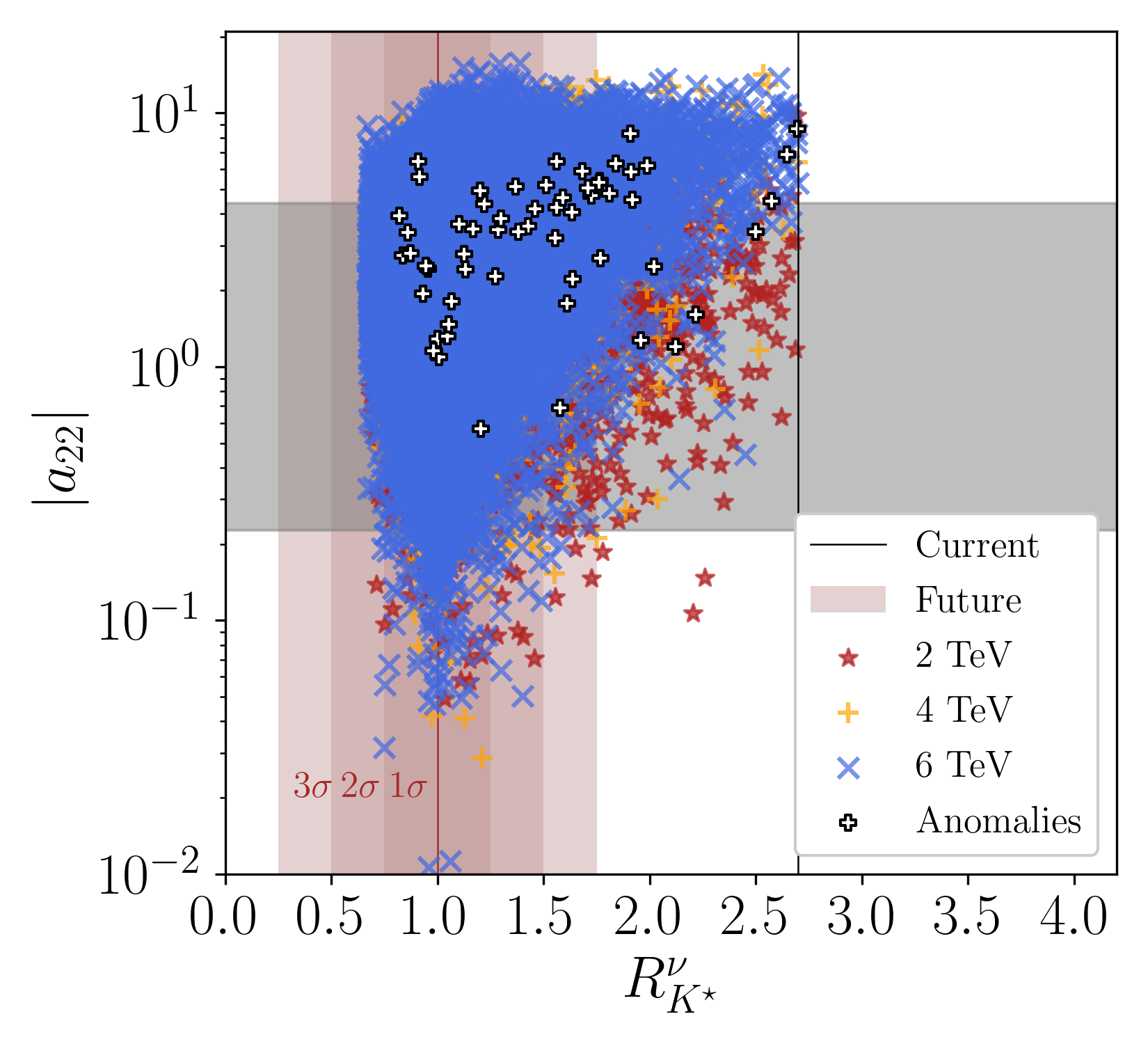}
    \caption{ \small{\textbf{\mathversion{bold}Impact of sampling on $R_{K^\star}^\nu$ in comprehensive scan.\mathversion{normal}} The red-brown shaded regions show the projected reach for $R_{K^\star}^\nu$ at Belle II for 5 ab$^{-1}$ of data, assuming the best-fit value is SM-like. 
The top-left plot shows the parameter space for $R_{K^\star}^\nu$ plotted together with $\Delta a_\mu$. The remaining plots show $R_{K^\star}^\nu$ plotted together with one of the relevant effective parameters, as identified in the analytic estimate in eq.~\eqref{eq:RKstar}. 
For the bottom-left plot, the dot-dashed lines show the upper limit on the magnitude of $a_{33}$ for each LQ mass, taken from table~\ref{tab:pc1}, in dark red for $\hat{m}_\phi  = 2$, orange for $\hat{m}_\phi  = 4$, and dark blue for $\hat{m}_\phi  = 6$, respectively. For further information on how to read this figure, see section~\ref{ssec:comprehensivePlots}.}}
    \label{fig:ComResultRK}
\end{figure}
From section~\ref{subsubsec:Hadronic_primary} we expect the constraint on BR($\tau\to\mu\gamma$), whose dominant contribution is driven by the product $|b_{23}c_{33}|$, to impact the size of the main contributions to $R_{K^\star}^\nu$, driven by the magnitude of $a_{33} \approx c_{33}$, and to the AMM of the muon, depending dominantly on the magnitude of $b_{23}$, see table~\ref{table:primaryconstraints_coupling_coeffs}. 
 This correlation is visible to a certain extent in the primary scan, see bottom-right plot in figure~\ref{fig:tauBc_RK}. In the comprehensive scan, the relation between the effective parameters $a_{33}$ and $b_{23}$ is further enhanced by the biasing, see eq.~\eqref{eq:b23fromtmg}. Thus, the observed correlation is apparent in the top-left plot in figure~\ref{fig:ComResultRK}, in particular for $\hat{m}_\phi =4$ and $\hat{m}_\phi=6$.  As can be seen, the majority of points that imply values for $R_{K^\star}^\nu$ close to its present bound corresponds to smaller values of $\Delta a_\mu$. This trend seems absent for $\hat{m}_\phi =2$, as we 
 discuss in the following.

In order to understand the distribution of values for $R_{K^\star}^\nu$ in the comprehensive scan, we first recall that the dominant contribution in the model has $\nu_\tau\overline{\nu_\tau}$  in the final state, see eq.~\eqref{eq:RKstar}, and is determined by the product $|a_{33} a_{32}|$. In the top-right plot in figure~\ref{fig:ComResultRK}, we see that there is, indeed, a strong correlation between larger values of the magnitude of $a_{32}$ and large values for $R_{K^\star}^\nu$. Numerically we find that $|a_{32}|$ can be as large as $\mathcal{O}(10)$, see table~\ref{tab:comprehensiveCoupMag} in appendix~\ref{app:supp6_scanmethod}. This can be traced back to a potentially large additional contribution to the effective parameter $a_{32}$, originating from the transformation from the interaction to the charged fermion mass basis, compare eq.~\eqref{eq:xpararel}. At the same time, we see in the bottom-left plot in figure~\ref{fig:ComResultRK} that the biasing prefers smaller values of $|a_{33}|\approx |c_{33}|$  for smaller LQ masses, see also table~\ref{tab:pc1}, and that most white crosses also correspond to smaller $|a_{33}|$. For these smaller values, larger values of the magnitude of $b_{23}$ are likely to be compatible with the present bound from BR($\tau\to\mu\gamma$), compare eq.~(\ref{eq:taumug_estimate}). These larger values of $|b_{23}|$ tend to increase $\Delta a_\mu$, see eq.~\eqref{eq:gm2_estimate}, and can push it closer to the present best-fit value. Altogether, we can generally expect larger contributions to $R_{K^\star}^\nu$ to be accessible for $\nu_\tau\overline{\nu_\tau}$ in the final state than in the primary scan, see section~\ref{subsubsec:Hadronic_primary}. However, for $\hat{m}_\phi = 2$
 smaller values of $|a_{33}|$ are usually attained, as we see from the bottom-left plot in figure~\ref{fig:ComResultRK}. Nevertheless, large values for $R_{K^\star}^\nu$ can be obtained.

From eq.~\eqref{eq:RKstar}, we note that there can be sizeable contributions from the channel with $\nu_\mu\overline{\nu_\mu}$ in the final state which are proportional to the product $|a_{23} a_{22}|$ as well as LFV contributions having $\nu_\mu\overline{\nu_\tau}$ and $\nu_\tau\overline{\nu_\mu}$ in the final state that are driven by $|a_{23} a_{32}|$ and $|a_{33} a_{22}|$, respectively. In the comprehensive scan, we find that the magnitude of $a_{22}$ can be as large as $\mathcal{O}(10)$, see bottom-right plot in figure~\ref{fig:ComResultRK} and table~\ref{tab:comprehensiveCoupMag} in appendix~\ref{app:supp6_scanmethod}.
 At the same time, larger values of the magnitude of $a_{23}$ are preferred by the biasing, see table~\ref{tab:pc1}. Therefore, the contribution with $\nu_\mu\overline{\nu_\mu}$ in the final state becomes more significant for $R_{K^\star}^\nu$ in comparison to the primary scan. This argument is supported by the positive correlation in the data between large $|a_{22}|$ and larger $R_{K^\star}^\nu$, shown in the bottom-right plot in  figure~\ref{fig:ComResultRK}. Additionally, the LFV contributions with $\nu_\mu\overline{\nu_\tau}$ and $\nu_\tau\overline{\nu_\mu}$ in the final state can also generate relevant contributions, particularly driven by larger values of the product $|a_{23} a_{32}|$. For the other contribution proportional to the product $|a_{33} a_{22}|$, the biasing prefers smaller values of $|a_{33}|$, see table~\ref{tab:pc1} and bottom-left plot in  figure~\ref{fig:ComResultRK}, in particular for smaller LQ masses, so that we do not expect it to be equally important. 

In summary, the current constraint on $R_{K^\star}^\nu$  is found to genuinely shape the viable parameter space, and the prospective measurement of $B\to K^{(\star)} + \text{invisible}$ at Belle II~\cite{Belle-II:2018jsg} provides a promising avenue to test this model.

\begin{figure}[t!]
\centering
\includegraphics[width=0.49\textwidth]{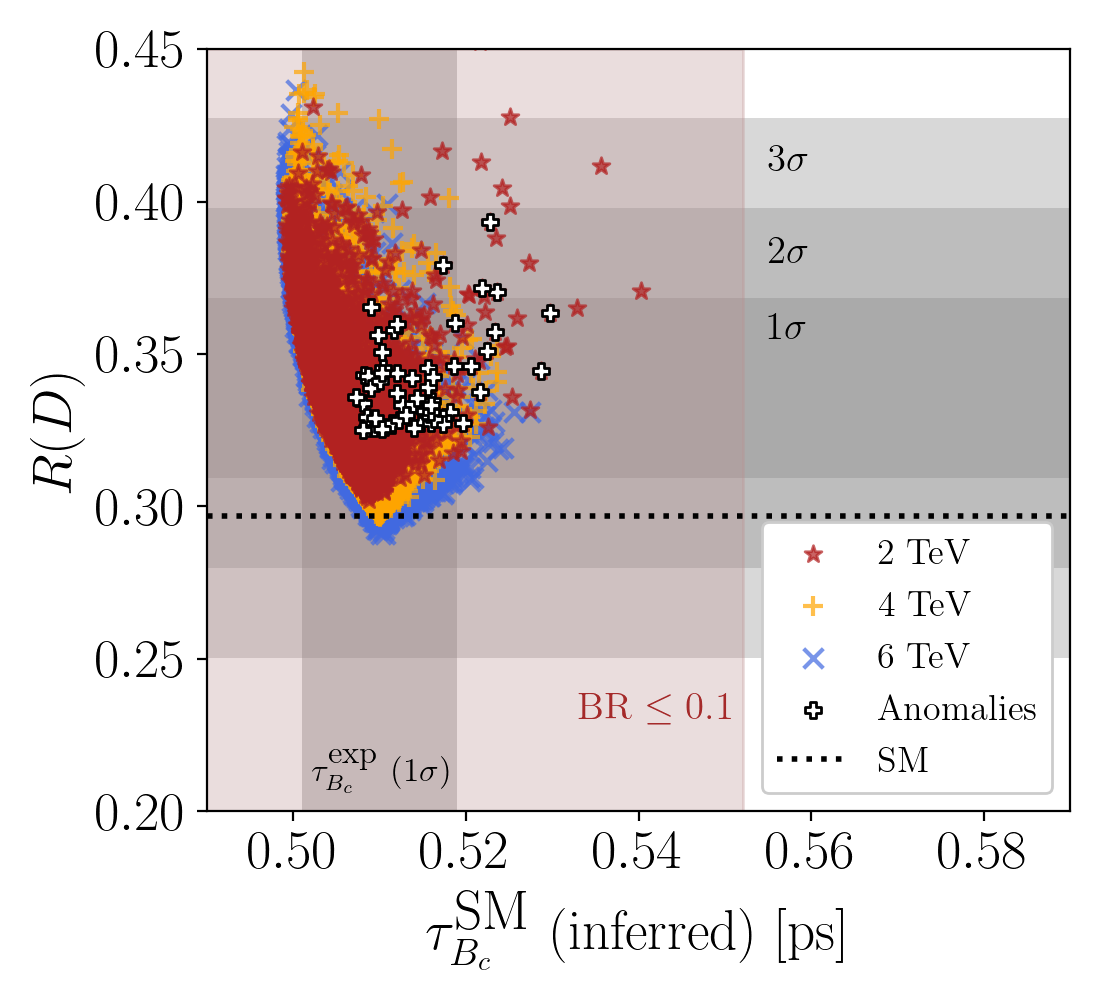}
\includegraphics[width=0.49\textwidth]{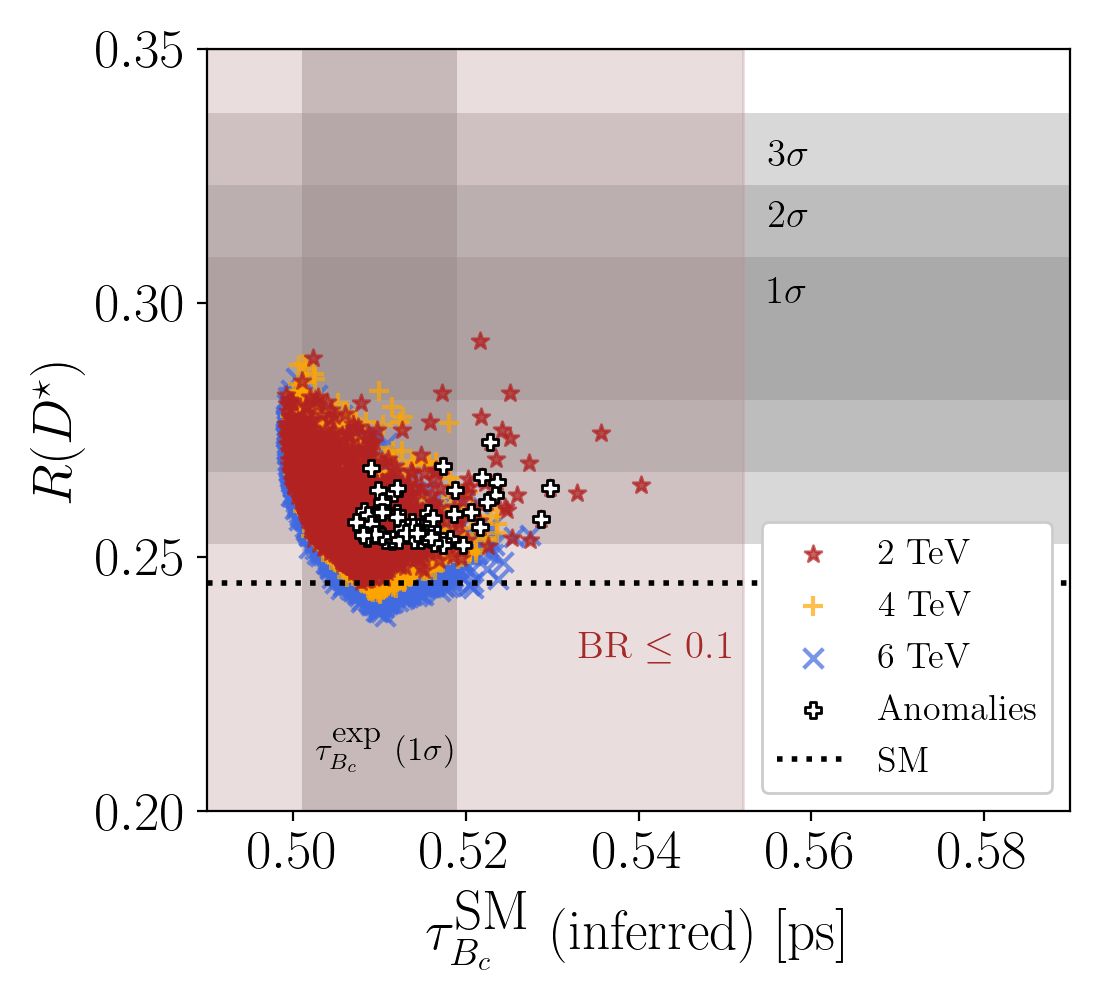}\\
\includegraphics[width=0.49\textwidth]{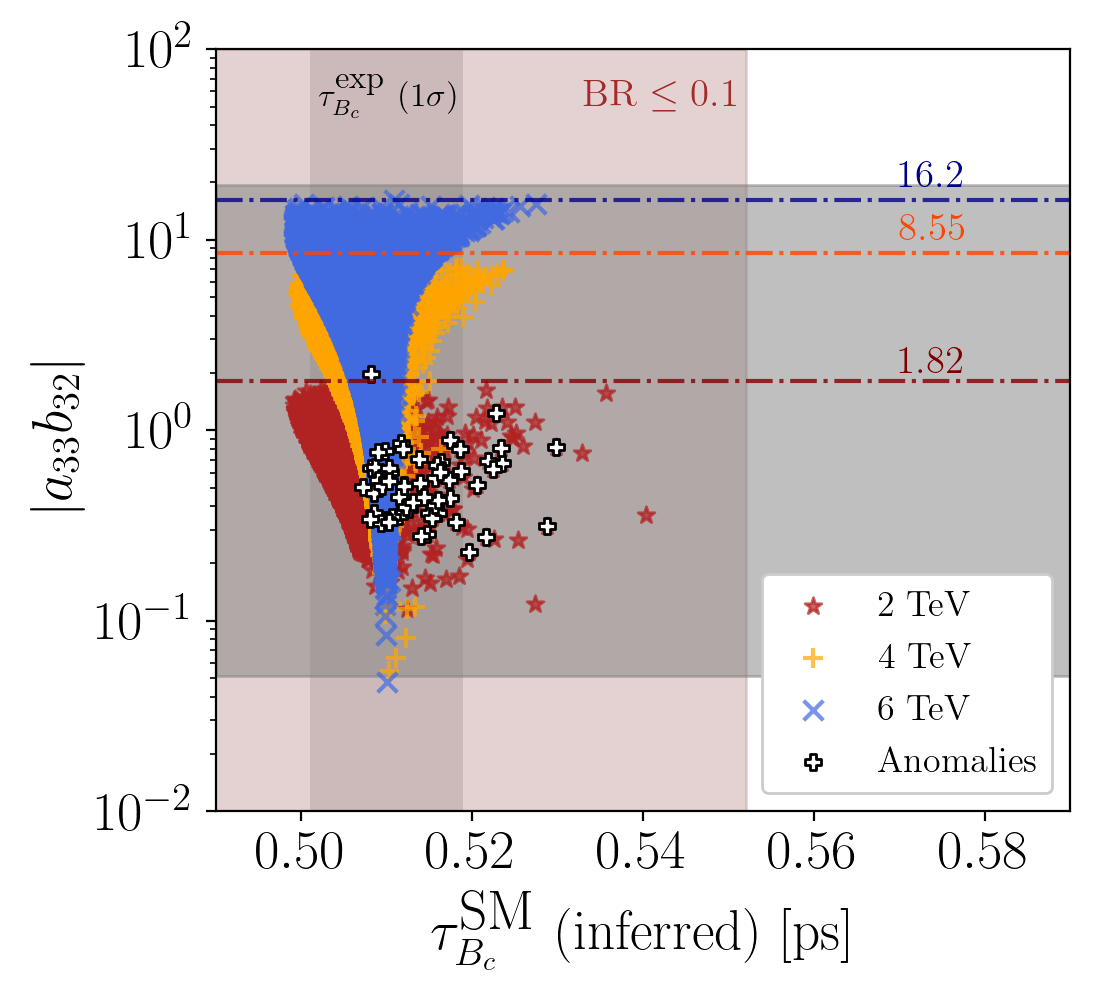}
\includegraphics[width=0.49\textwidth]{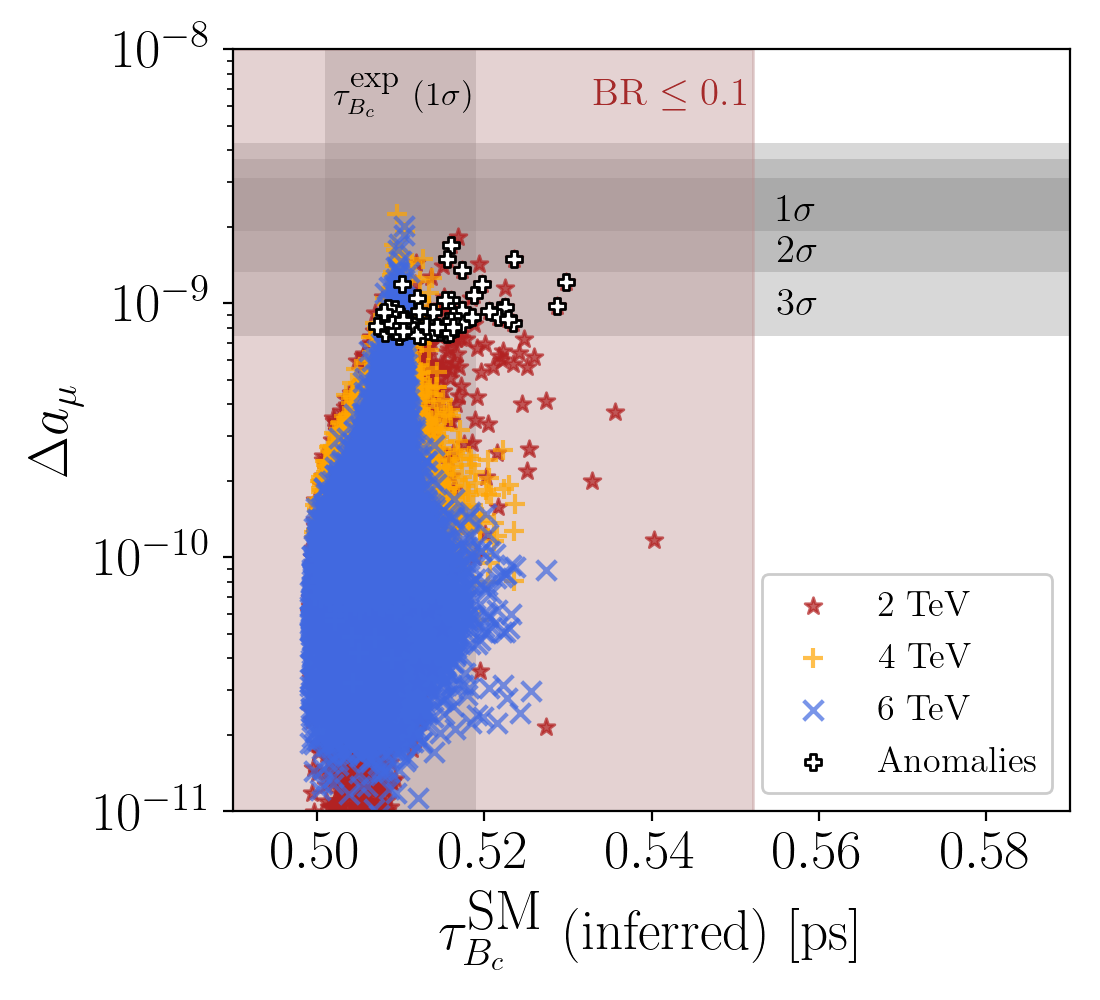}
\caption{{\small \textbf{\mathversion{bold}Impact of sampling on inferred SM contribution to $B_c$ lifetime in comprehensive scan.\mathversion{normal}} For the $B_c$ lifetime, the grey shaded region represents the $1~\sigma$ region about the present experimental best-fit value of $\tau_{B_c}$, and the red-brown shaded region shows the area that corresponds to BR$(B_c\to \tau \nu) \leq 0.1$. For the bottom-left plot, the grey-shaded region represents the sampled parameter range in the primary scan and the dot-dashed lines show the upper limit on the product $|a_{33} b_{32}|$ for each LQ mass, calculated from table~\ref{tab:pc1}, in dark red for $\hat{m}_\phi  = 2$, orange for $\hat{m}_\phi  = 4$, and dark blue for $\hat{m}_\phi  = 6$, respectively.  Further information on how to read this figure can be found in section~\ref{ssec:comprehensivePlots}.}}
\label{fig:tauBcinCom}
\end{figure}

\paragraph{\mathversion{bold}Lifetime of $B_c$ meson.\mathversion{normal}} For this observable, we similarly find differences between the values obtained in the comprehensive and the
 primary scan. While a substantial contribution to the $B_c$ lifetime is associated with values of $R(D)$ and $R(D^\star)$ below their SM predictions in the primary scan, see top in figure~\ref{fig:tauBc_RK} in section~\ref{sec:primary} and discussion in section~\ref{subsubsec:Hadronic_primary}, in the comprehensive scan, especially for $\hat{m}_\phi=2$, points are observed that are in disagreement with this statement and therefore hint at the influence of subdominant contributions. This can be explicitly seen in the top in figure~\ref{fig:tauBcinCom}. 
 
Eq.~\eqref{eq:tauBcSM_estimate} shows that the dominant term for the inferred SM contribution to the $B_c$ lifetime, inferred $\tau_{B_c}^\text{SM}$, is driven by the product $|a_{33} b_{32}|$, which corresponds to the tau neutrino being in the final state and which interferes with the SM contribution. From the bottom-left plot in figure~\ref{fig:tauBcinCom}, we see that this product is sampled over a much smaller range for $\hat{m}_\phi=2$ than for the other LQ masses, due to the biasing imposed, see table~\ref{tab:pc1}. However, for small values of this product, larger values of the inferred $\tau_{B_c}^\text{SM}$ are nevertheless accessible. This indicates the relevance of subdominant contributions. The other contribution, mentioned in eq.~\eqref{eq:tauBcSM_estimate}, is proportional to the product $|a_{23}b_{32}|$, and corresponds to the muon neutrino in the final state. One should recall that the magnitude of $a_{23} \approx c_{23}$ needs to be quite large to explain the observed anomaly in $\Delta a_\mu$ and that we sample more viable points with such larger values due to the biasing, see table~\ref{tab:pc1}. We, thus, can expect an enhancement of the LFV contribution, with the muon neutrino in the final state, to the inferred $\tau_{B_c}^\text{SM}$  -- similarly to the effect found for $R(D)$ and $R(D^\star)$, see section~\ref{sec:contrastAnomalies}. For $\hat{m}_\phi=2$, we see from the bottom in figure~\ref{fig:tauBcinCom} that a sizeable fraction of the points that correspond to small values of the product $|a_{33}b_{32}|$ and larger inferred $\tau_{B_c}^\text{SM}$ also leads to a larger value for the AMM of the muon, compare especially the white crosses.

Despite the differences found in the results of the primary and comprehensive scan, this model still predicts BR($B_c\to\tau\nu$) to be below 0.1. More precise measurements of $B_c\to\tau\nu$ could provide a further test of this model, in particular when considering the complementarity with measurements of $R(D)$ and $R(D^\star)$.

 \begin{table}[t!]\centering
  \def\arraystretch{1.3} 
 \resizebox{\linewidth}{!}{\begin{tabular}{|l|cll|cl|}
 \hline
	 \multicolumn{6}{|c|}{\textsc{ List of Secondary Observables}}\\
 \hline
\multirow{2}{*}{Observable} & \multicolumn{5}{c|}{Experiment}  \\
\cline{2-6}
& \multicolumn{3}{c|}{Current constraint/measurement} & \multicolumn{2}{c|}{Future reach} \\
\hline
$|d_\mu|$ & $<1.5 \times 10^{-19} \, e \, \mathrm{cm}$ & at 90\% C.L. & \cite{Muong-2:2008ebm} & $1000 \, (60) \, [1] \times 10^{-24} \, e \, \mathrm{cm}$ & \cite{Chislett:2016jau,Gorringe:2015cma,Adelmann:2021udj,Semertzidis:1999kv,Farley:2003wt} \\
$ g_{\mu_{A}}/g_{A}^{\rm SM}$ & $ 0.99986\pm 0.00108$   &at $1 \, \sigma$ level & \cite{ALEPH:2005ab,Crivellin:2020mjs} & $\pm 6.3\, (0.63) \times 10^{-5}$ & \cite{Crivellin:2020mjs, Baer:2013cma,FCC:2018evy} \\
$R^{\mu/e}_D$ & $0.995\pm 0.090$ & at $1 \, \sigma$ level & \cite{Belle:2015pkj} &$\pm 0.00995$ & \cite{Krohn:2018edn} \\
$R^{e/\mu}_{D^\star}$ &$1.01\pm0.032$ & at $1 \, \sigma$ level & \cite{Belle:2018ezy} &$\pm 0.0101$ & \cite{Krohn:2018edn} \\
$\mathrm{BR}(B\to\tau\nu)$ & $(1.09 \pm 0.24) \times 10^{-4}$ & at $1 \, \sigma$ level & \cite{ParticleDataGroup:2020ssz} &
$\pm 9\, (4)\%$
at  $5 \, (50)$ ab$^{-1}$  &
\cite{Forti:2022mti}
\\
\hline
 \end{tabular}}
\caption{{\small {\bf List of secondary observables}.
We list the observables that can potentially be used to further constrain and test this model, together with their current experimental constraint/measurement and future sensitivity. In the case of the EDM of the muon, $d_\mu$, the future projection without brackets refers to the reach expected from the Muon g$-$2 experiment at Fermilab~\cite{Chislett:2016jau}, and a similar experimental effort~\cite{Gorringe:2015cma} undertaken at J-PARC. The values in brackets are estimates given for experimental proposals using the frozen-spin technique~\cite{Adelmann:2021udj,Semertzidis:1999kv,Farley:2003wt}.
In the case of the future projections for $g_{\mu_A}$, we assume that the measurements of $g_{\mu_{A}}$ are improved by the same factor as $\sin^2 \theta_\text{eff}$~\cite{Crivellin:2020mjs}; the unbracketed projection is for the ILC~\cite{Baer:2013cma}, and the bracketed value is for the FCC~\cite{FCC:2018evy}. }  } \label{table:secondaryconstraints}
\end{table}

\subsection{Analytic estimates for secondary observables}
\label{subsec:secondary_analytic}
As in section~\ref{sec:primary}, we first discuss analytic estimates for the secondary observables. Present constraints/measurements and future reach for these are summarised in table~\ref{table:secondaryconstraints}.

\subsubsection{Electric dipole moment of muon}
 \label{subsubsec:s_ana_EDMmu}

The contributions to leptonic AMMs and EDMs arise both from the one-loop diagram, shown in figure~\ref{fig:eiejgamma} in section~\ref{subsubsec:p_ana_AMMamu}. In fact, they correspond to the real and imaginary part of the same effective vertex, as can be seen from eq.~\eqref{eq:EDMAMM} in appendix~\ref{app:dipole}.  As we generate large contributions to the AMM of the muon and we allow for complex values for the LQ couplings, we expect that this model can lead to sizeable values for the EDM of the muon. 

Similar to the AMM of the muon, most relevant is the contribution in which a chirality flip occurs via a mass insertion on the internal quark line, and which is thus enhanced by the mass of the top quark. The following expression for $d_\mu$ can be derived, assuming $m_\phi \gg m_t$, 
\begin{align}
  |d_\mu| \approx \frac{ 2|\text{Im}(c_{23} 
   b^*_{23})|}{\hat{m}_{\phi}^2}\times 10^{-22} e\; \text{cm}. \label{EDMmu}
\end{align}
This predicts the value of $d_\mu$ below the current bound, but well within the reach of future experiments, as quoted in table~\ref{table:secondaryconstraints}. This is consistent with the literature for expected correlations between $d_\mu$ and solutions to the present flavour anomaly in the AMM of the muon, particularly for the LQ $\phi$, see e.g.~\cite{Crivellin:2018qmi,Bigaran:2021kmn}.

\mathversion{bold}
\subsubsection{\texorpdfstring{$Z\to \mu{\mu}$}{Zmumu}}
\mathversion{normal}
%%%%%%%%%%%%%%%%%%%%%%%%%%%%%%%%%%%%%%%%%%%%%%%%%%%%%% 
  
In order to achieve sizeable contributions to the AMM of the muon through loops with a top quark, an associated enhanced contribution to the process $Z\to \mu \mu$ is expected, see diagrams in figure~\ref{fig:ZtautauFD}. Similarly to section~\ref{subsubsec:p_ana_leptoZdecays}, we use eq.~\eqref{deltag_ZZ} in appendix~\ref{app:Zdecays} to parametrise the contribution to the effective axial-vector 
coupling of $Z$ to muons in this model.
Following appendix~\ref{app:Zdecays} for the definition of $g_A^{\rm SM}$, $g_A^{\rm SM} <0$, and taking lepton flavour to be conserved for SM couplings, i.e.~$g_A^{\rm SM}$ is the same for all lepton flavours, we find
\begin{align}
\label{eq:Zmumuestimate}
  &g_{\mu_{A}}/g_A^{\rm SM}\approx 1-\left[\left\{
\begin{array}{*{2}{c}}
2.31,\, & \hat m_\phi =2\\ 
0.76,\, & \hat m_\phi =4\\
0.39,\, & \hat m_\phi =6\\
\end{array} \right\}\; |c_{23}|^2 \times  10^{-5} \right].
  \end{align}
If we allow for a $3 \, \sigma$ margin about the best-fit value, given in table~\ref{table:secondaryconstraints}, we obtain upper bounds on the magnitude of $c_{23}$, namely the unbracketed values below
\begin{align}
  &|c_{23}|\lesssim  \left\{\begin{array}{*{2}{c}}
12.1 \, [ 3.8],\, & \hat m_\phi =2\\ 
21.1 \, [ 6.6],\, & \hat m_\phi =4\\
29.5 \, [ 9.2],\, & \hat m_\phi =6\\
\end{array} \right\} \, . \label{eq:ZmumuprojCoup}
  \end{align}
In the comprehensive scan, the values of $c_{23}\approx a_{23}$ typically do not become larger than $1/\lambda$, see table~\ref{tab:comprehensiveCoupMag} in appendix~\ref{app:supp6_scanmethod}, and so the present constraints from this process are not competitive. However, future experiments are projected to be much more sensitive, as can be seen from the values in square brackets in eq.~\eqref{eq:ZmumuprojCoup}. 
These are extracted using the projected sensitivity for the ILC~\cite{Baer:2013cma}, see table~\ref{table:secondaryconstraints}. A further reduction of the error by a factor of ten is expected from the FCC~\cite{FCC:2018evy}, allowing to probe more viable parameter space. Therefore, this observable will be relevant in the future, particularly for $\hat{m}_\phi=2$.

\mathversion{bold}
\subsubsection{\texorpdfstring{Lepton flavour universality ratios $R_D^{\mu/e}$ and $R_{D^\star}^{e/\mu}$}{RDmue and RDstarmue}}
\mathversion{normal}

The observed anomalies in $R(D)$ and $R(D^\star)$ raise the question whether the effects of LFU violation may be evident in other ratios of $b\to c e_i \nu_j$ processes. Two of particular interest are the ratios $R_D^{\mu/e}$ and $R_{D^\star}^{e/\mu}$
\begin{align}
R_D^{\mu/e} = \frac{\Gamma(B\to D \mu \nu) }{\Gamma(B\to D e \nu) }
\;\; \text{and} \;\;
R_{D^\star}^{e/\mu}= \frac{\Gamma(B\to D^\star e \nu) }{\Gamma(B\to D^\star \mu \nu) }
\;.
\end{align}
Using the expressions from appendix~\ref{subsec:RDmueRDstaremu} we arrive at the following estimates at LO
\begin{align}
\label{eq:estimateRDmue}
\frac{R_D^{\mu/e}}{[R_D^{\mu/e}]_\text{SM}}\approx  1+ \left(\frac{2.25\;\text{Re}(b_{22}^* a_{23})+19.7\;\text{Re}(a_{23}^* c_{22})}{\hat{m}_\phi^2}\right)\times {10^{-4}}\; 
\end{align}
and
\begin{align}
\label{eq:estimateRDsemu}
\frac{R_{D^\star}^{e/\mu}}{[R_{D^\star}^{e/\mu}]_\text{SM}}\approx 1-\left(\frac{{0.68\;\text{Re}(b_{22}^*a_{23})}+{19.6\;\text{Re}(a_{23}^* c_{22})}}{\hat{m}_\phi^2} \right)\times {10^{-4}}\;
\;.
\end{align}
The terms proportional to Re$(b_{22}^*a_{23})$ come from the scalar-operator contribution, while the vector-operator contribution is responsible for the dominant terms  proportional to Re$(a_{23}^*c_{22})$. Both contributions arise at the same order in $\lambda$. As shown in sections~\ref{subsubsec:LQcoupinter} and~\ref{subsubsec:LQcoupmass}, in this model the LQ coupling $y_{22}$ turns out to be larger than expected, ${y}_{22}= b_{22} \, \lambda^3$. This coupling enters the estimates for these observables. We also note that both observables depend on the effective parameter $a_{23}\approx c_{23}$, which plays an   important role for addressing the flavour anomaly in the AMM of the muon. Eventually, note that the SM value for both observables is approximately one, with the exact value used in the comprehensive scan being extracted from \texttt{flavio}, v2.3.

\mathversion{bold}
\subsubsection{\texorpdfstring{Leptonic decay $B\to\tau\nu$}{Bpm to tau nu}}
\mathversion{normal}
\label{sec:secBpmtaunu}

In this model, the LQ $\phi$ contributes to the leptonic decay $B\to\tau\nu$, which is CKM-suppressed due to $|V_{ub}|\sim\lambda^3$ in the SM, see eq.~(\ref{eq:LeptonicMesonDecay}) in appendix~\ref{app:pseudoscalarBk} with $u_k = u$ for the full decay width including the contributions from $\phi$. We focus on the case of a tau neutrino in the final state, since its contribution interferes with the SM one. The largest contribution arises for the Wilson coefficient $C^{VLL}_{\nu edu,3331}$, while the Wilson coefficient $C^{SRR}_{\nu edu,3331}$ is suppressed at the scale $\mu = m_\phi$ due to the hierarchy $y_{31}/z_{31} \sim \lambda^2$, see eqs.~(\ref{eq:yparaA}) and (\ref{eq:zparaA}). 
This suppression is only partly compensated by the RG running down to the hadronic scale $\mu=\mu_B = 4.8$ GeV and the chirality enhancement of the scalar-operator contribution. This together results in an enhancement factor of roughly 6.5. We, thus, find
\begin{align}
\frac{\text{BR}(B\to\tau\nu)}{\text{BR}(B\to\tau\nu)_{\text{SM}}} \approx 1 - \frac{0.1}{\hat{m}^2_\phi}\text{Re}(a_{33}c_{31}^*) = 1 - \frac{0.1}{\hat{m}^2_\phi}|a_{33}c_{31}|\cos\big(\text{Arg}(a_{33}) - \text{Arg}(c_{31})\big) \;.
\end{align}
All contributions which are quadratic in Wilson coefficients, induced by the LQ $\phi$, can be neglected. Note that the currently viable parameter space of the model will only be probed by future searches for $B\to\tau\nu$ to an appreciable extent, despite its dependence on the LQ couplings $y_{31}$ and $z_{31}$ which involve quarks of the first generation. 

%%%%%%%%%%%%%%%%%%%%%%%%%%%%%%%%%%%%%%%%%%%%%%%%%%%%%%

\begin{table}[t!]
\centering
\resizebox{\textwidth}{!}{
\renewcommand{\arraystretch}{1.6}
    \centering
    \begin{tabular}{|l|l|l|l|l|l|l|}
    \hline
    \multicolumn{7}{|c|}{ \textsc{Spread of secondary observables in comprehensive scan}} \\
    \hline
\multirow{2}{*}{Observable} &\multicolumn{2}{c|}{$\hat{m}_\phi=2$, $P=5955$}  &\multicolumn{2}{c|}{$\hat{m}_\phi=4$, $P=12570$} &\multicolumn{2}{c|}{$\hat{m}_\phi=6$, $P=39807$} \\
\cline{2-7}
  & [min., max.] & Average & [min., max.] & Average & [min., max.] & Average \\
  \hline 
 $|d_\mu|$  $\times 10^{25}$ [$e$ cm]&$[0.067, 2144]$&$153.4$&$[0.033, 1449]$&$128.1$&$[0.003, 1417]$&$86.51$\\
 $R^{\mu/e}_{D}$ &$[0.983, 1.015]$&$1.000$&$[0.994, 1.004]$&$1.000$&$[0.996, 1.002]$& $1.000$\\
 $R^{e/\mu}_{D^\star}$ &$[0.989, 1.020]$&$1.003$&$[0.999, 1.009]$&$1.003$&$[1.001, 1.007]$&$1.003$\\
  BR($B\to \tau\nu$) $\times 10^5$&$[8.201, 9.505]$&$8.862$&$[8.278, 9.341]$&$8.853$&$[8.216, 9.396]$&$8.845$\\
  $(1-g_{\mu_A}/g_A^\text{SM})\times 10^4$ &$[0.008, 4.223]$&$0.292$&$[0.178, 1.383]$&$0.427$&$[0.069, 0.704]$&$0.214$\\
  \hline 
\end{tabular}}
\caption{{\small\textbf{Overview of spread of secondary observables in comprehensive scan.}  We present a summary of the statistics reflecting the distribution of secondary observables: the minimum, maximum and average values generated for a sample of $P$ points passing the primary constraints.}} \label{tab:secondaryTableC}
\end{table}

\subsection{Numerical results for secondary observables}
\label{subsec:resultsSecondary}

\begin{figure}[t!]
    \centering
 \includegraphics[width=0.49\textwidth]{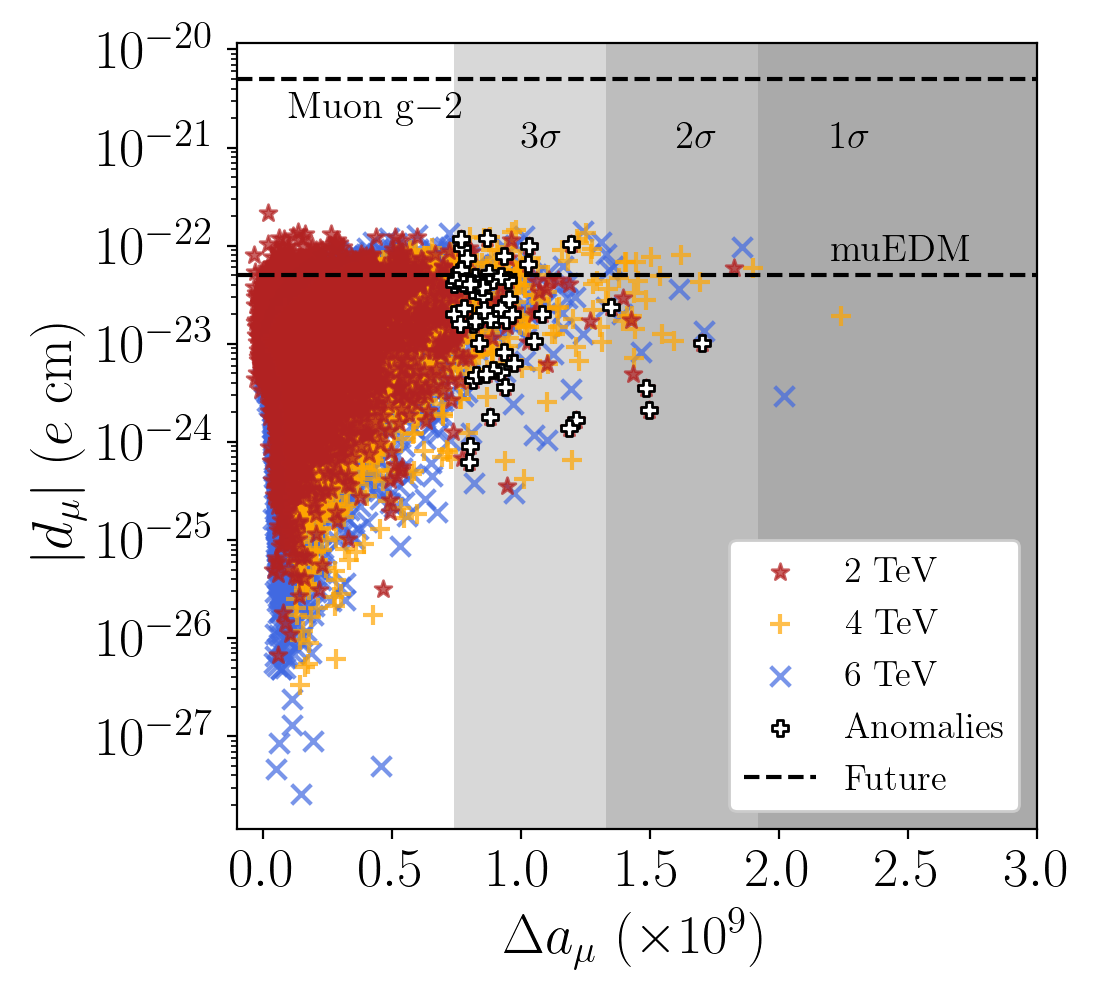}
  \includegraphics[width=0.49\textwidth]{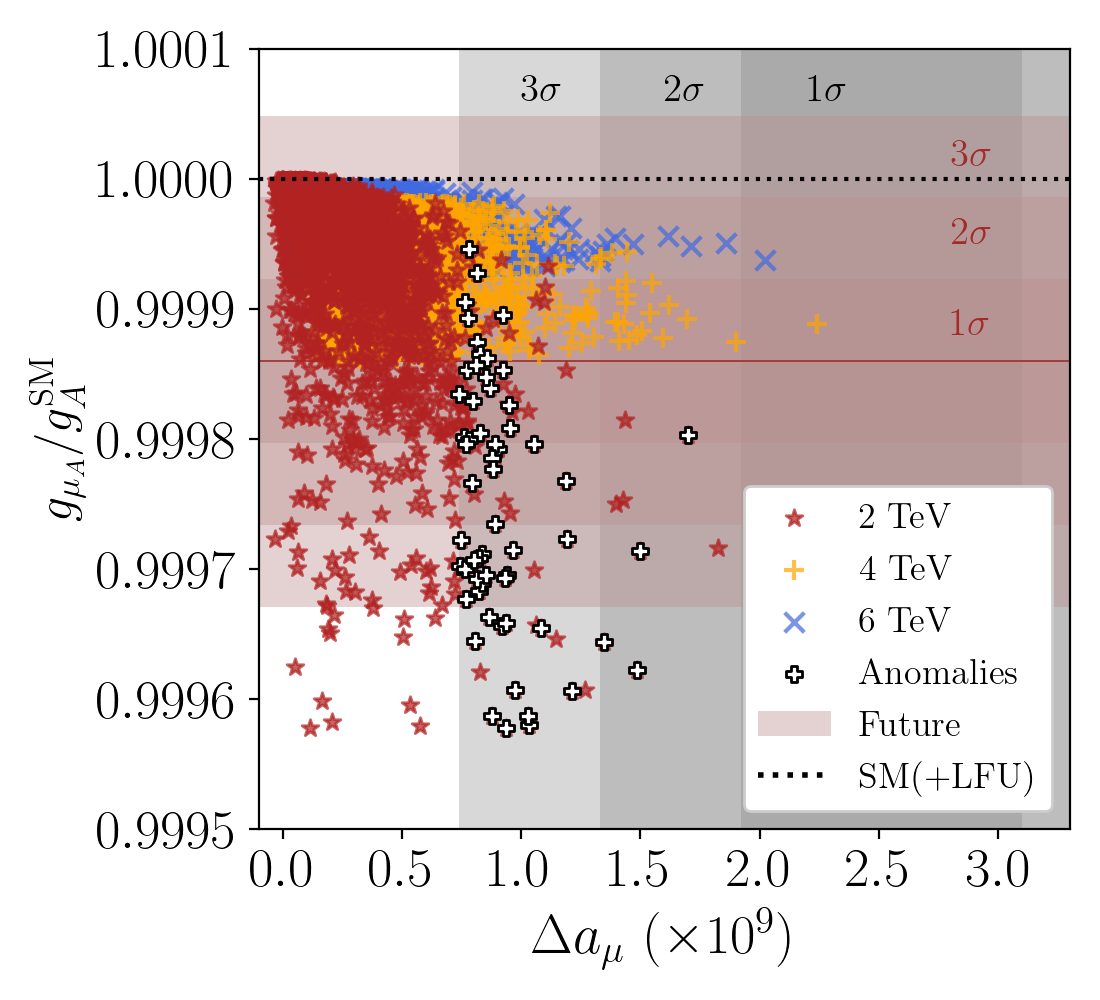}\\
  \begin{minipage}{0.49\textwidth}
  \includegraphics[width=\textwidth]{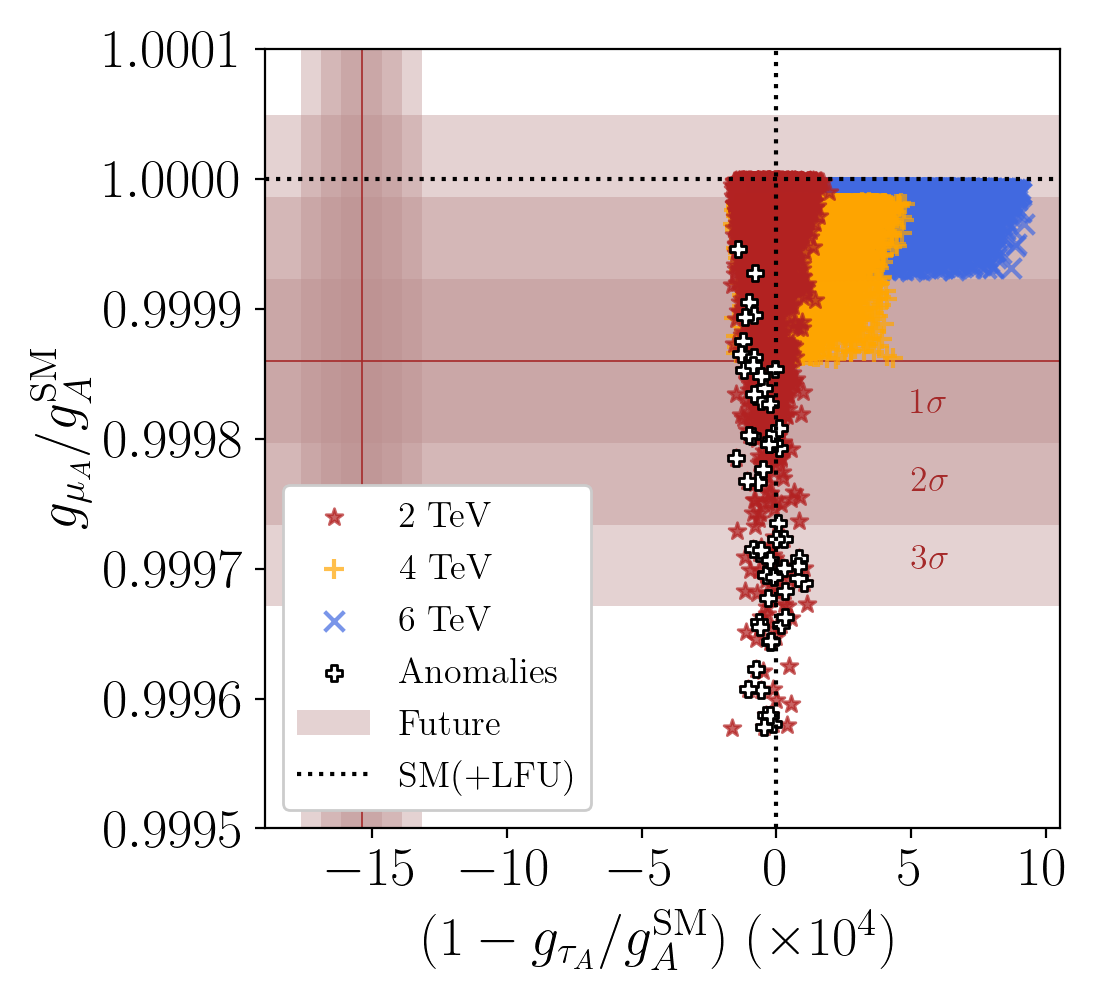}
\end{minipage}\hspace{7mm}
\begin{minipage}{0.45\textwidth} \caption{ \small{\textbf{Future reach of leptonic secondary observables in comprehensive scan.} In the top-left plot we present two constraints on the magnitude of the EDM of the muon, $|d_\mu|$, from the Muon g$-$2 experiment~\cite{Chislett:2016jau} and the muEDM experiment~\cite{Adelmann:2021udj} (as example for the frozen-spin technique), see also table~\ref{table:secondaryconstraints}. For $g_{\mu_A}/g_A^\text{SM}$ and $g_{\tau_A}/g_A^\text{SM}$, the red-brown shaded regions represent the projected sensitivities from the ILC~\cite{Baer:2013cma}, assuming the current best-fit values~\cite{ALEPH:2005ab,Crivellin:2020mjs}, shown as red-brown solid lines. For further information on how to read this figure, see section~\ref{ssec:comprehensivePlots}.}}\label{fig:com_secLep}
\end{minipage}
   \end{figure}

\begin{figure}[t!]
    \centering
 \includegraphics[width=0.49\textwidth]{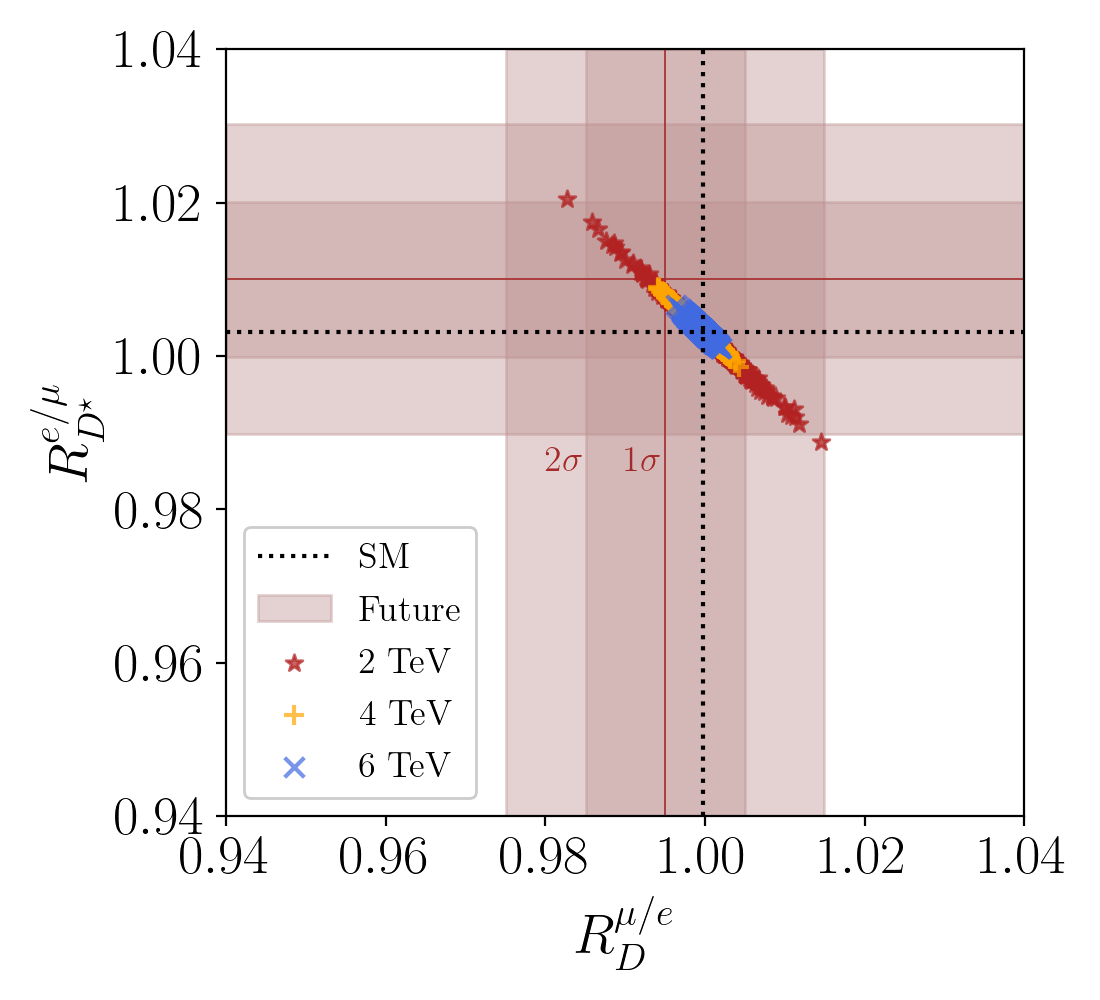}
  \includegraphics[width=0.49\textwidth]{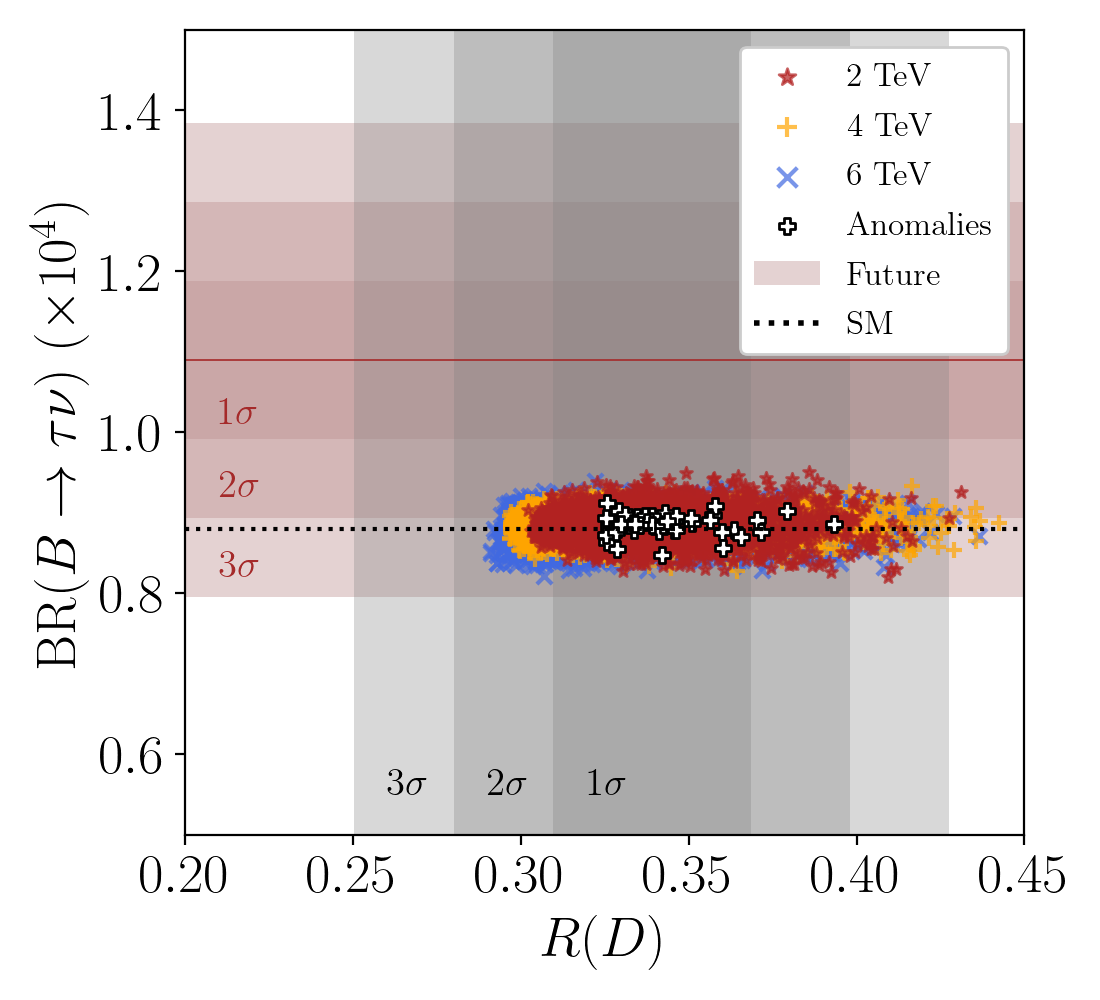}
 \caption{ \small{\textbf{Future reach of hadronic secondary observables in comprehensive scan.} In both plots, we present the projected sensitivities from Belle II for $R^{\mu/e}_{D}$, $R^{e/\mu}_{D^\star}$~\cite{Krohn:2018edn} and BR($B\to \tau\nu$), at 5 ab$^{-1}$,~\cite{Forti:2022mti}. Note that we do not show the white crosses in the left plot, because they would uniformly lie across the entire allowed range. 
 A complementary plot showing BR($B\to \tau\nu$) against $R(D^\star)$  can be found in the right plot in figure~\ref{fig:CompApp_Sec} in appendix~\ref{app:addfigssec6}.
 For further information on how to read this figure, see section~\ref{ssec:comprehensivePlots}. }}\label{fig:com_secHad}
\end{figure}

In this section, we analyse the numerical results for the secondary observables from the comprehensive scan. We first comment on the leptonic observables, illustrated in figure~\ref{fig:com_secLep}, before moving onto the hadronic observables, shown in figure~\ref{fig:com_secHad}. A summary of the spread of the numerical results for the secondary observables is given in table~\ref{tab:secondaryTableC}.

\paragraph{Leptonic secondary observables.} Regarding the EDM of the muon, we see that viable points are capable of generating a maximum of $|d_\mu|\sim \mathcal{O}(10^{-22}) \, e\, \text{cm}$ and that points, associated with explaining the anomaly observed in the AMM of the muon at the $3~\sigma$ level or better, predict the EDM of the muon to lie in the interval $[10^{-25}, 10^{-22}]\, e\, \text{cm}$, see top-left plot in figure~\ref{fig:com_secLep}. While this signal could not be seen at the Muon g$-$2 experiment at Fermilab, some parameter space is expected to be probed at the muEDM experiment at the Paul Scherrer Institute (PSI) and similar experiments using the frozen-spin technique, as indicated by the black dashed lines in the top-left plot in figure~\ref{fig:com_secLep}. We remind that a bias for the difference of the arguments of the effective parameters $a_{23} \approx c_{23}$ and $b_{23}$ is employed, see table~\ref{tab:pc1}, such that more sizeable contributions to the AMM of the muon, and thus a larger real part of the relevant Wilson coefficient, are generated. Therefore, we expect the comprehensive scan to prefer smaller values of the imaginary part of the product of the same effective parameters, see eq.~\eqref{EDMmu}. This, in turn, leads to a distribution of values of $d_{\mu}$ below the analytic estimate of $\mathcal{O}(10^{-22}) \, e\, \text{cm}$. Still, enhanced contributions to both observables are seen to be compatible, and so this effect is limited. Furthermore, note that we do not observe any preference for the sign of the EDM of the muon. Given that the contributions to it turn out to be suppressed relative to present constraints, there is no need for an additional CP symmetry to restrict its size. 

Turning to the effective coupling of $Z$ to muons, we first repeat that according to the analytic estimate the dominant contribution to the ratio $g_{\mu_A}/g_{A}^\text{SM}$ is proportional to $|c_{23}|^2$ 
and negative such that the resulting value of $g_{\mu_A}/g_{A}^\text{SM}$ should always be smaller than one in this model, see eq.~(\ref{eq:Zmumuestimate}). This is consistent with the data, illustrated in the top-right plot in figure~\ref{fig:com_secLep}. Then, we remind that the comprehensive scan prefers larger values of the magnitude of $a_{23} \approx c_{23}$, since this increases the chances to satisfactorily address the flavour anomalies. Such larger values correspond to points with a smaller ratio $g_{\mu_A}/g_{A}^\text{SM}$, as observed in the distribution of white crosses in the top-right plot in figure~\ref{fig:com_secLep}. Present constraints are not competitive enough to be illustrated in this plot, although the present best-fit value may hint at contributions beyond the SM that generate a ratio $g_{\mu_A}/g_{A}^\text{SM}$ smaller than one. From the top-right plot in figure~\ref{fig:com_secLep}, we see that increased precision will allow to probe parts of the viable parameter space of this model. Note that we use the projected ILC bounds~\cite{Baer:2013cma}, but a further reduction of the error by a factor of ten is expected from the FCC~\cite{FCC:2018evy}. For completeness, a plot showing the EDM of the muon and the ratio $g_{\mu_A}/g_{A}^\text{SM}$ can be found in the left plot in figure~\ref{fig:CompApp_Sec} in appendix~\ref{app:addfigssec6}.

Lastly, we observe no correlation between $g_{\mu_A}/g_{A}^\text{SM}$ and $g_{\tau_A}/g_{A}^\text{SM}$. This is expected from the analytic estimates, see eqs.~(\ref{eq:Ztau_coupling_approx}) and (\ref{eq:Zmumuestimate}), since these ratios dominantly depend on distinct effective parameters. The regions of the two ratios in the comprehensive scan are displayed in the bottom plot in figure~\ref{fig:com_secLep} 
 for the three different LQ masses and result from the biasing imposed on the effective parameters $a_{23}\approx c_{23}$ and $a_{33}\approx c_{33}$, respectively, see table~\ref{tab:pc1}.

\paragraph{Hadronic secondary observables.}  The LFU ratios $R^{\mu/e}_{D}$ and $R^{e/\mu}_{D^\star}$ are useful probes for $b\to c e_i \nu_j$ processes that do not involve the tau lepton. The results of the comprehensive scan reveal an anti-correlation in the deviations of these two ratios from the SM values. This is consistent with the analytic estimates in eqs.~\eqref{eq:estimateRDmue} and \eqref{eq:estimateRDsemu}. We note that the effective parameters $b_{22}$ and $c_{22}\approx a_{22}$ which enter the estimates for these observables can be $\mathcal{O}(10)$ in the comprehensive scan, see table~\ref{tab:comprehensiveCoupMag} in appendix~\ref{app:supp6_scanmethod}, while $a_{23}$ is biased towards larger values, see table~\ref{tab:pc1}. This enhancement explains the extent of the distribution of points in the left plot in figure~\ref{fig:com_secHad}. Presently, all predictions are consistent with the measurements, although the experimental sensitivity is expected to considerably improve at Belle II~\cite{Krohn:2018edn}. 

On the other hand, BR$(B \to \tau \nu)$ is interesting as observable, since it probes the process $b u \to \tau \nu$, which is sensitive to the LQ coupling between the bottom quark and tau neutrino common with $b\to c \tau \nu$, but is suppressed by the small coupling between the up quark and tau lepton in this model, $z_{31}= c_{31}  \, \lambda^3$, see eq.~(\ref{eq:zparaA}). For this BR, we predict a value consistent within $2$ to $3~\sigma$ of the projected sensitivity, assuming that the best-fit value of this measurement remains the current one. In the right plot in figure~\ref{fig:com_secHad} we illustrate this observable plotted against $R(D)$. As one can see, we find no correlation between these two observables. A plot for $R(D^\star)$ shows a similar result, and can be found in the right plot in figure~\ref{fig:CompApp_Sec} in appendix~\ref{app:addfigssec6}. We note that the scalar-operator contribution to this observable remains suppressed by the size of the coupling $y_{31} = b_{31} \, \lambda^5$, see eq.~(\ref{eq:yparaA}), consistent with the analytic estimate in section~\ref{sec:secBpmtaunu}. For the effective parameter $b_{31}$, an enhancement only slightly above $1/\lambda$ is found in the comprehensive scan, see table~\ref{tab:comprehensiveCoupMag} in appendix~\ref{app:supp6_scanmethod}, which is not sufficient to make this contribution competitive with the one from the vector operator. Furthermore, we note that, indeed, the effective parameter $c_{31} \approx a_{31}$ can take rather large values.  

%%%%%%%%%%%%%%%%%%%%%%%%%%%%%%%%%%%%%%%%%%%%%%%%%%%%%%
\subsection{Comment on tertiary observables}
\label{subsec:tertiary}
%%%%%%%%%%%%%%%%%%%%%%%%%%%%%%%%%%%%%%%%%%%%%%%%%%%%%%  

In the following, we briefly comment on the results for the tertiary observables extracted from the comprehensive scan. We find that none of these observables, listed in table~\ref{tab:app_tercal} in appendix~\ref{app:tertiary}, provides a signal within the reach of current and planned experiments. Thus, any observation of new physics in these allows to falsify this model. We relegate 
detailed ranges for these observables for each LQ mass to table~\ref{tab:app_ter} in appendix~\ref{app:tertiary} and only make a few comments below. Note that in table~\ref{tab:app_tercal} also the 
 present experimental constraints and calculation method employed in the comprehensive scan are found, while table~\ref{tab:app_ter} also displays the prospective future reach for these observables. 

For processes involving electrons, we first observe that the effective coupling of $Z$ to electrons is suppressed by small LQ couplings of $\mathcal{O}(\lambda^9)$, see eqs.~(\ref{eq:zparaA}) and (\ref{eq:yparaA}). Thus, we do not expect large LQ contributions to $Z\to ee$ in this model, as reflected in the data in table~\ref{tab:app_ter}. We see that these contributions are up to eleven orders of magnitude below future sensitivities. Similarly, the contribution to the AMM of the electron generated in this model for each LQ mass is $\mathcal{O}(10^{-21})$. Present measurements of $\Delta a_e$ hint at a preference for $|\Delta a_e| \sim 10^{-12}$~\cite{Parker2018,Morel:2020dww}, although these two measurements indicate deviations from the SM value with opposite sign and comparable magnitude. We, therefore, note that this model would be incapable of addressing this anomaly, but could be revisited in case the present discrepancy in the experimental results is resolved. Likewise, the results for the EDM of the electron show that a detection in future experiments~\cite{aggarwal2018measuring} should not be expected.

We predict the BRs for $B_s \to \tau \tau$, $D_s \to \tau \nu$ and $D_s \to \mu \nu$ to be only slightly beyond the projected sensitivity to these observables. Furthermore, the future sensitivities for tau decays to a muon and light mesons, i.e. BR$(\tau \to [\rho, \phi, \pi]\mu$), are only one or two orders of magnitude above the maximum value generated for these observables in the comprehensive scan. These decays can, thus, be of interest when considering a next generation of experiments, beyond what is currently found in the literature.

%%%%%%%%%%%%%%%%%%%%%%%%%%%%%%%%%%%%%%%%%%%%%%%%%%%%%%
\section{Summary and outlook}
\label{sec:concl}
%%%%%%%%%%%%%%%%%%%%%%%%%%%%%%%%%%%%%%%%%%%%%%%%%%%%%%

 We have considered an extension of the SM with two Higgs doublets $H_u$ and $H_d$ (in the decoupling limit) and one scalar LQ $\phi$
 that transforms as $(3, 1, -\frac 13)$ under the SM gauge group. The main purpose of the LQ $\phi$ is to explain the flavour 
 anomalies in $R (D)$, $R (D^\star)$ and in the AMM of the muon. The flavour structure of this model is
 constrained by the flavour group $G_f=D_{17} \times Z_{17}$. The three scalars $H_u$, $H_d$ and $\phi$
 are singlets under the dihedral group, whereas the three generations of all SM fermion species transform as doublet and singlet, apart
 from the three RH up-type quarks that are all singlets. In this way, the masses of the charged fermions of the third generation
 arise without breaking the dihedral group. 
 
 The flavour symmetry $G_f$ is (mainly) broken by the VEVs of four different
 spurions, called $S$, $T$, $U$ and $W$, that are assigned to doublets of the dihedral group. While the role of $S$
 is to (mainly) generate the LQ couplings ${\bf \hat{x}}$ and ${\bf \hat{y}}$, $T$ and $U$ are responsible for the 
 mass of the second and first generation of both down-type quarks and charged leptons, respectively. The spurion $W$,
 eventually, is necessary in order 
 to give mass to the charm quark and to generate the correct size of the Cabibbo angle.
 The smaller quark mixing angles as well as the up quark mass arise automatically due to the spurions $S$ as well as $T$ and $U$,
 respectively. According to their roles, the VEVs of these spurions are of different order of magnitude in the expansion
 parameter $\lambda$, $\lambda \approx 0.2$, i.e.~$\langle S \rangle \sim \lambda$, $\langle T \rangle \sim \lambda^2$, $\langle U \rangle \sim \lambda^4$
 and $\langle W \rangle \sim  (\lambda^5, \lambda^4)^t$. In order to achieve suitable textures for the LQ couplings ${\bf \hat{x}}$ and ${\bf \hat{y}}$
 and, at the same time, avoid too large effects related to quarks and/or leptons of the first generation, a residual symmetry $Z_{17}^{\rm diag}$, being
 the diagonal subgroup of a $Z_{17}$ group, contained in $D_{17}$, and the external $Z_{17}$ symmetry, is preserved by both ${\bf \hat{x}}$ and ${\bf \hat{y}}$  at LO.
 
We have performed analytical and numerical studies of the phenomenology of this model. In doing so, all considered observables have been classified as primary, secondary or tertiary. The primary observables include the flavour anomalies in $R(D)$, $R(D^\star)$ and in the AMM of the muon as well as observables for which the present experimental measurements can (significantly) constrain  the viable parameter space of this model. Secondary observables instead do not currently provide competitive constraints, but mid-term future experiments offer an opportunity to probe them and thus this model. For primary as well as secondary observables analytical estimates are given. Lastly, tertiary observables are not expected to allow to probe the model in the mid-term future, but are discussed lest future measurements bring these into disagreement with the SM. 

In the primary scan, we have focussed on the primary observables and varied the effective parameters of the LQ couplings in the charged fermion mass basis as (mostly) independent complex order-one numbers. In this way, we have identified the two radiative cLFV decays $\mu\to e\gamma$ and $\tau\to\mu\gamma$ as the most stringent constraints on the parameter space of the model. Furthermore, we have extracted  biases on the effective parameters of the LQ couplings which have been used to guide the more thorough comprehensive scan. A simultaneous reconciliation of all three flavour anomalies has proven to be very challenging in the primary scan.

The comprehensive scan has involved primary, secondary and tertiary observables. In contrast to the primary scan, it has been performed over effective parameters in the interaction basis. Thus, a subset of these has been fixed by a chi-squared fit to the charged fermion masses and quark mixing, achieving excellent agreement with the measured values (for scenario B). The remaining parameters, taken to be complex order-one numbers and parametrising the LQ couplings, have been biased using the input from the primary scan. In the comprehensive scan, we have found that this model is compatible with all constraints, while being capable of explaining the observed deviations in $R(D)$, $R(D^\star)$ and $\Delta a_\mu$ from the SM predictions within the $3~\sigma$ ranges of their present best-fit values for LQ masses of $2$ and $4$ TeV. Furthermore, an LQ with a mass of $2$ TeV allows for compatibility with all considered constraints, while reconciling the three flavour anomalies at the $2~\sigma$ level. The secondary observables studied in the comprehensive scan are the EDM of the muon, the effective coupling of $Z$ to muons, the LFU ratios $R_D^{\mu/e}$ and $R_{D^\star}^{e/\mu}$ as well as BR$(B\to\tau\nu)$.

The differences between the parameter space probed by the primary and the comprehensive scan have been discussed in detail. The use of the interaction basis is the main reason for the comprehensive scan being able to reconcile all three flavour anomalies. At the same time, this has shown a considerable preference for one of the effective parameters, namely $b_{13}$, being slightly smaller than expected from the construction of the model. This indicates that an improved version of this model should further suppress this particular LQ coupling by $\lambda$ or $\lambda^2$. Contributions beyond the ones from $\gamma$-penguins can play an important role in several decays such that e.g.~not only the tau decay $\tau\to \mu \gamma$ can be accessible at Belle II, but, at the same time, $\tau\to 3 \, \mu$ and $\tau\to \mu e \bar{e}$ can be measured. For the primary observables with neutrinos in the final state, i.e.~$R(D)$, $R(D^\star)$, $R_{K^\star}^\nu$ and the lifetime of the $B_c$ meson, LFV contributions are found to be relevant, generating effects up to 40 percent in some instances. 
  
There are several interesting directions to expand the current study. On the phenomenological side, it is highly interesting to study the observables $R(J/\psi)$ and $R(\Lambda_c)$ that are (tightly) related to the analysed $b \to c$ transitions as well as the angular distributions of $B\to D^\star e_i \nu$~\cite{Bobeth:2021lya} 
and the longitudinal polarisation of the tau lepton in $B\to D^\star \tau \nu$~\cite{Alonso:2016oyd}. Some of these also reveal a (slight)
disagreement between the current measured value and the SM expectation, e.g.~$R(J/\psi)$~\cite{LHCb:2017vlu}. Other flavour anomalies, such as those observed in $b \to s$ transitions, e.g.~in $R (K)$, $R (K^\star)$ and in the process $B_s\to\mu \mu$, may also be relevant to address, see e.g.~reference~\cite{Crivellin:2022qcj} for a recent concise overview. For this purpose, an additional LQ, for example transforming as $(3, 3, -\frac 13)$ under the SM gauge group, has to be added to the model, see e.g.~references~\cite{Bigaran:2019bqv,Saad:2020ihm,Chen:2022hle,Freitas:2022gqs}. This may have the added effect of simultaneously generating neutrino masses.  A neutrino mass mechanism could be incorporated in many different ways. It could be either one type of seesaw mechanism, e.g.~by adding RH neutrinos to the existing model~\cite{Minkowski:1977sc}, or some radiative 
 generation mechanism, see reference~\cite{Cai:2017jrq} for a review. In the current analysis, it has been 
 assumed, for simplicity, that possible diquark couplings of the LQ $\phi$ are forbidden by a baryon number symmetry. However, it may also be interesting to study the
 efficacy of $G_f$ to suppress these couplings beyond the strong existing bounds from searches for proton decay~\cite{ParticleDataGroup:2020ssz}, see e.g.~references~\cite{deMedeirosVarzielas:2019lgb,Davighi:2022qgb} for studies about also controlling them with the help of a flavour symmetry. 

With non-vanishing neutrino masses, lepton mixing becomes physical and its appropriate description, i.e.~two large mixing angles and one small one~\cite{Esteban:2020cvm}, may require a change in the assignment of the LH lepton doublets to representations of $G_f$ or even the extension or change of $G_f$ itself. The observed lepton
mixing angles are often interpreted as sign of unification of the three generations of LH lepton doublets into an irreducible three-dimensional representation
of the flavour symmetry, for reviews see references~\cite{Ishimori:2010au,King:2013eh,Feruglio:2019ybq,Grimus:2011fk}. 
Prime candidates for such a flavour symmetry are the groups belonging to the series $\Delta (6 \, n^2)$ with $n$ integer and
at least two~\cite{Escobar:2008vc}. As has been shown, they lead to an adequate description of lepton as well as quark mixing, see e.g.~references~\cite{deAdelhartToorop:2011nfg,deAdelhartToorop:2011re,King:2013vna,Hagedorn:2014wha,Ding:2014ora}, and also of
the charged fermion mass hierarchies, if accompanied by an appropriate external symmetry, see e.g.~the supersymmetric model in reference~\cite{Hagedorn:2018bzo}. 
Furthermore, it is interesting to consider adding a CP symmetry to $G_f$, given that this can also constrain the two Majorana phases in the lepton sector~\cite{Feruglio:2012cw} (see also references~\cite{Holthausen:2012dk,Chen:2014tpa,Grimus:1995zi,Ecker:1983hz,Ecker:1987qp,Neufeld:1987wa,Harrison:2002kp,Grimus:2003yn}) and might, at the same time, be beneficial for controlling the amount of CP violation in the LQ couplings.

Eventually, an extension of the SM gauge group similar to the Pati-Salam theory has proven to be useful, since in this way the vector LQ transforming as $(3, 1, \frac 23)$ under the SM gauge group arises automatically, when breaking to the SM, see e.g.~references~\cite{Assad:2017iib,Bordone:2017bld,Bordone:2018nbg,King:2021jeo,King:2022sxb}. 
 This vector LQ is capable of addressing all aforementioned flavour anomalies, assuming an appropriate structure of its couplings to the SM fermions can be achieved. 
 While in the case of a vector LQ the flavour structure is determined by the gauge group of the model, for scalar LQs, explaining (some of) the observed flavour anomalies, it is also worth considering a possible embedding of the model into a (partially) unified theory endowed with a flavour (and CP) symmetry.

%%%%%%%%%%%%%%%%%%%%%%%%%%%%%%%%%%%%%%%%%%%%%%%%%%%%%%%%%%%%%%%%
\section*{Acknowledgements}
%%%%%%%%%%%%%%%%%%%%%%%%%%%%%%%%%%%%%%%%%%%%%%%%%%%%%%%%%%%%%%%%

M.S. and C.H. thank John Gargalionis for providing us with data from the scans found in reference~\cite{Cai:2017wry}.
I.B. and C.H. thank Mark Goodsell, Werner Porod and Avelino Vicente for help with \texttt{SARAH} and \texttt{SPheno}. T.F. and M.S. acknowledge helpful correspondence with Thorsten Feldmann. We also thank Peter Stangl for information on \texttt{flavio}. 
In addition to the software packages cited
in the text, this research has made extensive use of \texttt{matplotlib}~\cite{Hunter:2007,thomas_a_caswell_2021_5194481}.
C.H. has been partly supported by the European Union's Horizon 2020 research and innovation programme under the Marie Sk\l{}odowska-Curie grant agreement No.~754496 (FELLINI programme) as well as is supported by Spanish MINECO through the Ram\'o{}n y Cajal programme RYC2018-024529-I, by the national grant PID2020-113644GB-I00 and by the Generalitat Valenciana through PROMETEO/2021/083. 
C.H. has also received support from the European Union's Horizon 2020 research and innovation programme under the Marie Sk\l{}odowska-Curie grant agreement No.~860881 (HIDDe$\nu$ network). I.B. is supported in part by the Australian Research Council and the Australian Government Research Training Program Scholarship initiative. T.F. and M.S. acknowledge support by the Australian Research Council.
 C.H. would like to thank the Instituto de Fisica Teorica (IFT UAM-CSIC) in Madrid for support via the Centro de Excelencia Severo Ochoa Program under Grant CEX2020-001007-S, during the Extended Workshop ``Neutrino Theories”, where this work developed.

%%%%%%%%%%%%%%%%%%%%%%%%%%%%%%%%%%%%%%%%%%%%%%%%%%%%%%%%%%%%%%%%
%%%%%%%%%%%%%%%%%%%%%%%%%%%%%%%%%%%%%%%%%%%%%%%%%%%%%%%%%%%%%%%%

 %%%%%%%%%%%%%%%%%%%%%%%%%%%%%%%%%%%%%%%%%%%%%%%%%%%%%%%%%%%%%%%%
 \mathversion{bold}
 \section*{Note added on $R(D)$ and $R(D^\star)$}
 \mathversion{normal}
 %%%%%%%%%%%%%%%%%%%%%%%%%%%%%%%%%%%%%%%%%%%%%%%%%%%%%%%%%%%%%%%%

After the completion of this work, the LHCb collaboration has published a combined analysis of $R(D)$ and $R(D^\star)$~\cite{LHCb:2023zxo} using muonic tau reconstruction, resulting in $R(D)=0.441\pm0.060\pm0.066$ and 
$R(D^\star)=0.281\pm0.018\pm0.024$, as well as an updated measurement~\cite{LHCb:2023seminar} utilising hadronic $\tau^+$ decays, $R(D^\star)=0.257\pm0.012\pm0.018$. These results are consistent with the SM predictions in 
table~\ref{table:anomalies}, for $R(D^\star)$ within the 
$1 \, \sigma$ and for $R(D)$ within the $2 \, \sigma$ range, but with relatively large experimental uncertainties. Consequently, the HFLAV averages for $R(D)$ and $R(D^\star)$ have been updated to include the new 
measurements~\cite{HFLAV2023}, giving
  $R(D)=0.356\pm0.029$ and $R(D^\star)=0.284\pm0.013$ (with a correlation of $\rho=-0.37$). Compared with the previous averages, quoted in table~\ref{table:anomalies}, this is closer to the SM prediction for $R(D^\star)$, 
  but further away from the SM value of $R(D)$. Nevertheless, both averages remain in agreement at the $1 \, \sigma$ level with the previous ones. The discrepancy with the SM predictions now amounts to $3.2 \, \sigma$ which is
   only slightly reduced from $3.4 \, \sigma$, see table~\ref{table:anomalies}. Thus, the results have not changed significantly and remain qualitatively the same. 

\begin{figure*}[bt!]
    \centering
    \includegraphics[width=0.49\textwidth]{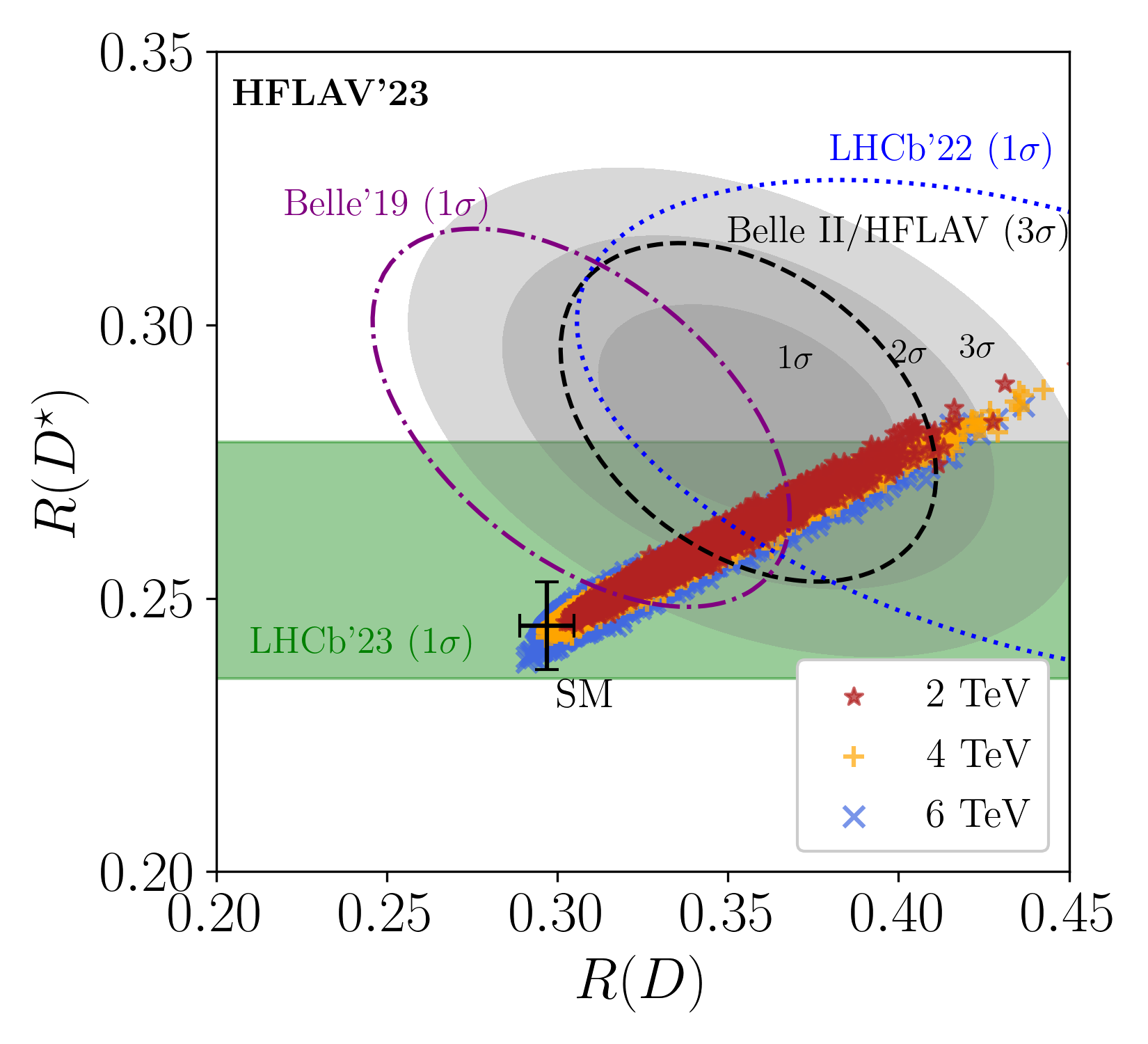}
\includegraphics[width=0.49\textwidth]{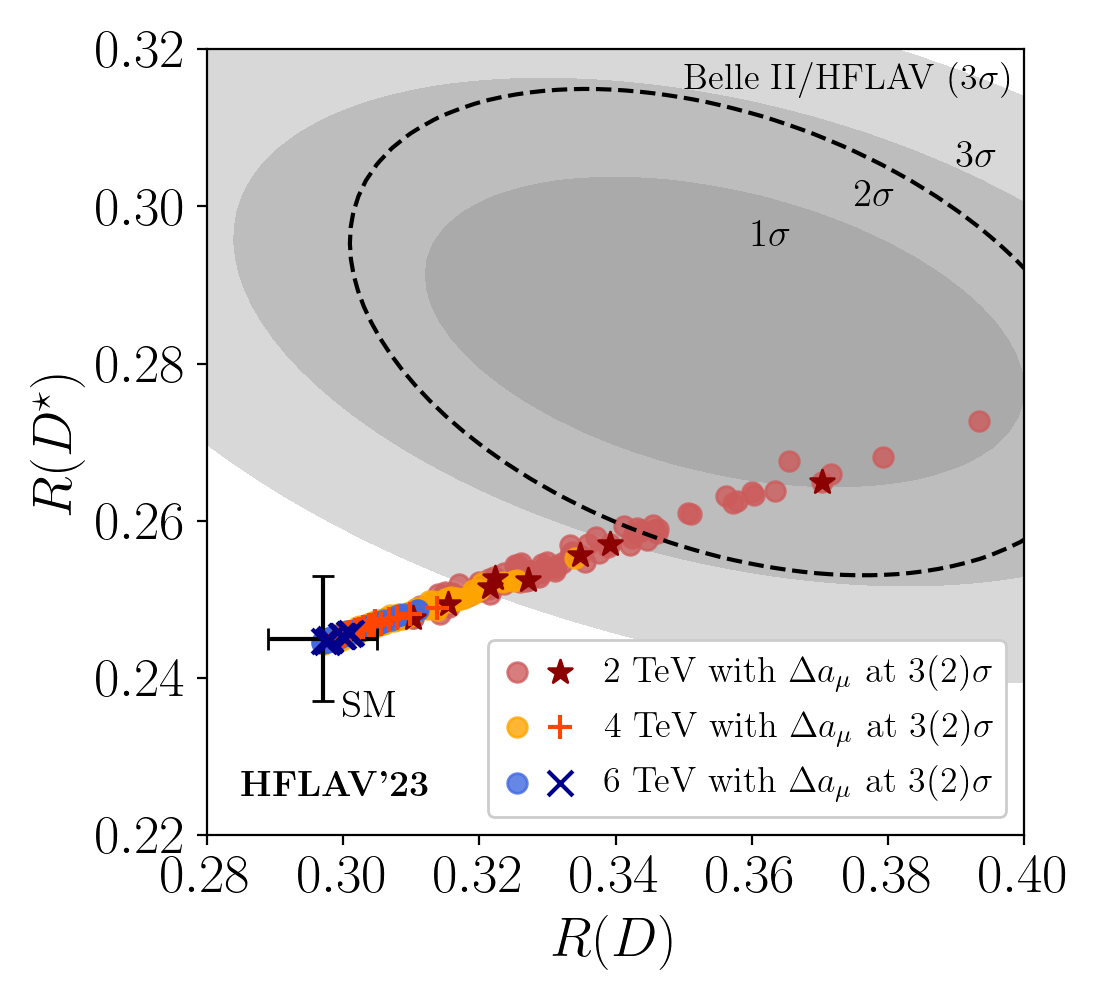}\\
\includegraphics[width=0.49\textwidth]{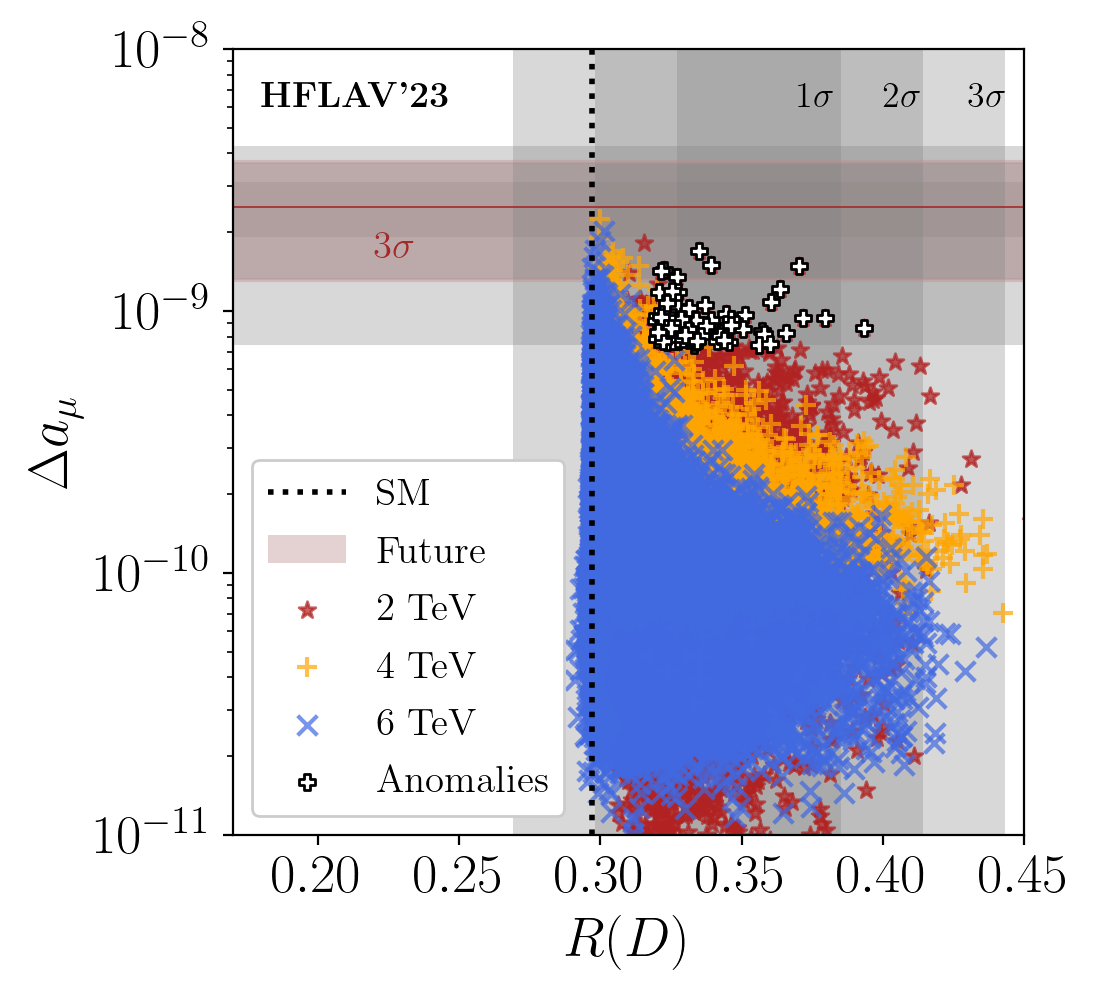}
\includegraphics[width=0.49\textwidth]{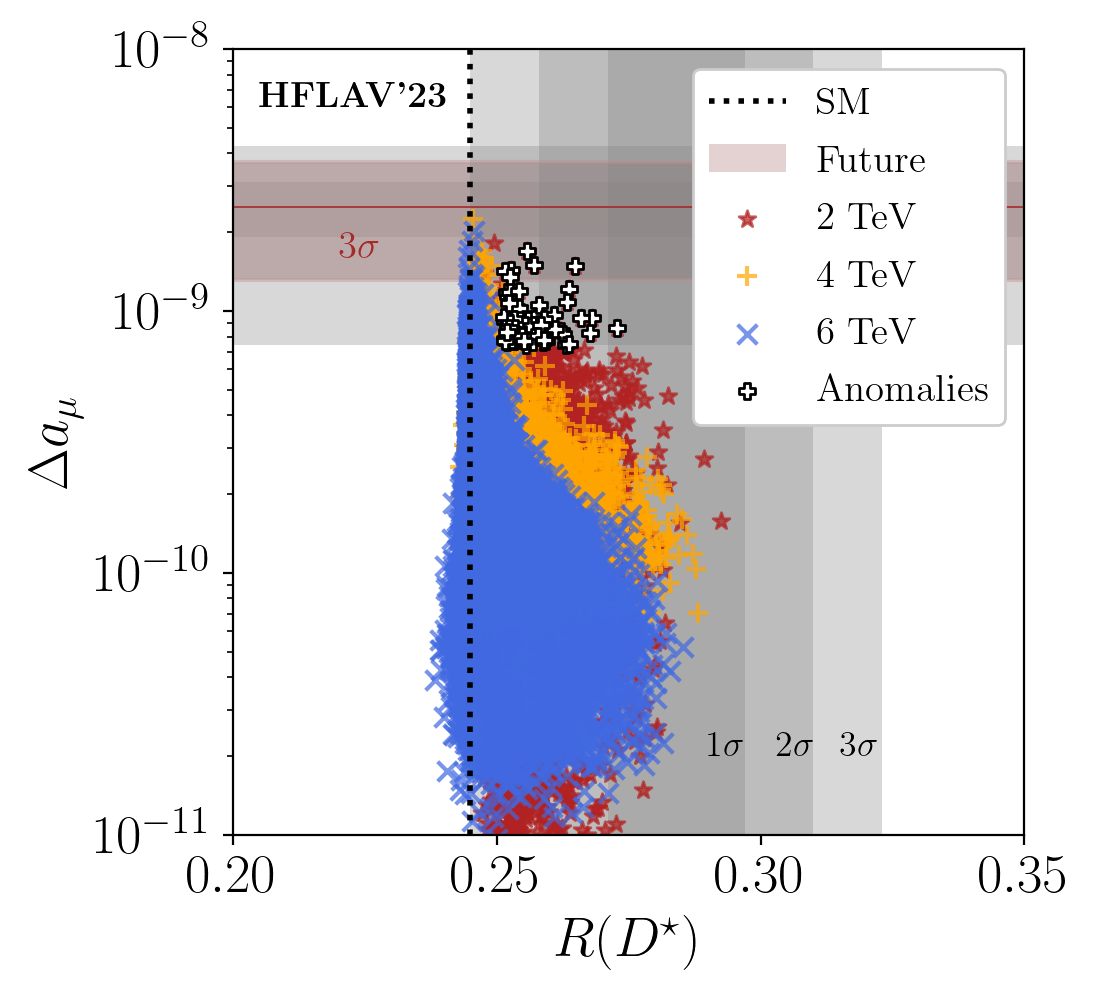}
\caption{\linespread{1.1}\small{\mathversion{bold}\textbf{Results of comprehensive scan for the flavour anomalies in $R(D)$, $R(D^\star)$ and in the AMM of the muon using updated HFLAV averages for $R(D)$ and $R(D^\star)$.}\mathversion{normal} 
This figure is an updated version of figure~\ref{fig:AnomaliesComprehensive}, taking into account the new LHCb combined analysis of $R(D)$ and $R(D^\star)$~\cite{LHCb:2023zxo} and the latest LHCb measurement  of $R(D^\star)$ using hadronic 
$\tau^+$ decays~\cite{LHCb:2023seminar}, which have led to new HFLAV averages for these. In the top-left plot, the blue dotted ellipse represents the $1~\sigma$ contour of the new LHCb combined analysis, while the green band now shows the 
$1~\sigma$ region about the most recent LHCb result for $R(D^\star)$. The black dashed ellipse indicates the prospective $3~\sigma$ reach for $5$ ab$^{-1}$ of data at Belle II~\cite{Forti:2022mti}, assuming the best-fit value from 2023 and the correlation coefficient from the HFLAV collaboration~\cite{HFLAV2023}. In the top-right plot which shows points that not only pass all considered constraints but also satisfy $\Delta a_\mu$ 
within $3\,\sigma$ (light-coloured circles) or $2\,\sigma$ (dark-coloured other shapes) of the present best-fit value we have magnified the scale and, at the same time, removed the results by single experiments for better readability.
For further information on how to read this figure, see section~\ref{ssec:comprehensivePlots}, in particular figure~\ref{fig:AnomaliesComprehensive}.}}
\label{fig:Fig15updated}
\end{figure*}

In any case, it is interesting to confront the outcome of the comprehensive scan with these new averages. In order to do so, we employ the same data set as shown in section~\ref{sec:secondarytertiary} and present 
figure~\ref{fig:Fig15updated} as updated version of figure~\ref{fig:AnomaliesComprehensive}. Once the primary constraints, see table~\ref{table:primaryconstraints}, are enforced in the comprehensive scan, we find $75(5)[0]$ points 
for $\hat{m}_\phi = 2 (4)[6]$ that can now generate $R(D)$, $R(D^\star)$ and the AMM of the muon within the respective $3 \, \sigma$ ranges. In fact, this is a slight quantitative improvement over the $58(1)[0]$ points identified and 
discussed in section~\ref{ssec:primaryresultsComprehensive}. 

%%%%%%%%%%%%%%%%%%%%%%%%%%%%%%%%%%%%%%%%%%%%%%%%%%%%%%%%%%%%%%%%

 \appendix

 %%%%%%%%%%%%%%%%%%%%%%%%%%%%%%%%%%%%%%%%%%%%%%%%%%%%%%
 \mathversion{bold}
\section{\texorpdfstring{Group theory of $D_{17}$}{Group theory of D17}}
\mathversion{normal}
\label{app:D17}
%%%%%%%%%%%%%%%%%%%%%%%%%%%%%%%%%%%%%%%%%%%%%%%%%%%%%%

In this appendix, we briefly summarise the main features of the non-abelian discrete group $D_{17}$~\cite{Blum:2007jz}.
It is a member of the series of dihedral groups $D_n$ that are non-abelian for $n \geq 3$.
It has $34$ distinct elements and contains ten real irreducible representations:
two singlets, the trivial singlet ${\bf 1_1}$ as well as ${\bf 1_2}$,
and eight doublets, called ${\bf 2_\mathrm{{\bf i}}}$ with $\mathrm{{\bf i}}=1,...,8$. All these eight doublets are faithful.
The group $D_{17}$ can be described, like the other dihedral groups, with the help of two generators $a$ and $b$ which
fulfil the following relations
\begin{equation}
\label{eq:D17gens}
a^{17}= e \; , \;\; b^2= e \; , \;\; a \, b \, a = b
\end{equation}
with $e$ denoting the neutral element of the group. The representation matrices $a (\mathrm{{\bf r}})$ and $b (\mathrm{{\bf r}})$ of the two generators $a$ and $b$
read in the different representations $\mathrm{{\bf r}}$
\begin{equation}
\label{eq:absinglets}
a ({\bf 1_1}) = b ({\bf 1_1}) = 1 \;\; \mbox{and} \;\; a ({\bf 1_2}) = 1 \; , \;\; b ({\bf 1_2}) = -1
\end{equation}
as well as
\begin{equation}
\label{eq:abdoublets}
a ({\bf 2_\mathrm{{\bf i}}}) = \left( \begin{array}{cc}
\omega_{17}^\mathrm{i} & 0 \\
0 & \omega_{17}^{17-\mathrm{i}}
\end{array}
\right) \;\; \mbox{and} \;\;
b ({\bf 2_\mathrm{{\bf i}}}) = \left( \begin{array}{cc}
0 & 1 \\
1 & 0
\end{array}
\right)
\; ,
\end{equation}
where $\omega_{17}$ is the 17th root of unity, $\omega_{17}=e^{\frac{2 \, \pi \, i}{17}}$. In this model, we only make use of the doublets ${\bf 2_1}$, ${\bf 2_2}$, ${\bf 2_3}$ and ${\bf 2_4}$.
The most relevant Kronecker products and
Clebsch-Gordan coefficients 
 are presented in the following. 
 The latter have a particularly simple form in the chosen basis. 
Assume $a$ and $b$ are singlets, $\left( \begin{array}{c} c_1 \\ c_2 \end{array} \right)$, $\left( \begin{array}{c} d_1 \\ d_2 \end{array} \right)$ are doublets, then we have~\cite{Blum:2007jz}
\begin{subequations}
\label{eq:CGcoefficientsD17}
\begin{align}
{\bf 1_1} \times {\bf 1_1}&: a \, b \sim {\bf 1_1} \; ,
\\
{\bf 1_1} \times {\bf 1_2}&: a \, b \sim {\bf 1_2} \; ,
\\
{\bf 1_2} \times {\bf 1_2}&: a \, b \sim {\bf 1_1} \; ,
\\
{\bf 1_1} \times {\bf 2_\mathrm{{\bf i}}}&: \left( \begin{array}{c} a \, c_1 \\ a \, c_2 \end{array} \right) \sim {\bf 2_\mathrm{{\bf i}}} \; ,
\\
{\bf 1_2} \times {\bf 2_\mathrm{{\bf i}}}&: \left( \begin{array}{c} a \, c_1 \\ -a \, c_2 \end{array} \right) \sim {\bf 2_\mathrm{{\bf i}}} \; ,
\\
{\bf 2_1} \times {\bf 2_1}&: c_1 \, d_2 + c_2 \, d_1 \sim {\bf 1_1} \; , \;\; c_1 \, d_2 - c_2 \, d_1 \sim {\bf 1_2} \; , \;\; \left( \begin{array}{c} c_1 \, d_1 \\ c_2 \, d_2 \end{array} \right) \sim {\bf 2_2} \; ,
\\
{\bf 2_1} \times {\bf 2_2}&:  \left( \begin{array}{c} c_2 \, d_1 \\ c_1 \, d_2 \end{array} \right) \sim {\bf 2_1} \; , \;\; \left( \begin{array}{c} c_1 \, d_1 \\ c_2 \, d_2 \end{array} \right) \sim {\bf 2_3} \; ,
\\
{\bf 2_2} \times {\bf 2_2}&: c_1 \, d_2 + c_2 \, d_1 \sim {\bf 1_1} \; , \;\; c_1 \, d_2 - c_2 \, d_1 \sim {\bf 1_2} \; , \;\; \left( \begin{array}{c} c_1 \, d_1 \\ c_2 \, d_2 \end{array} \right) \sim {\bf 2_4} \; ,
\\
{\bf 2_1} \times {\bf 2_3}&:  \left( \begin{array}{c} c_2 \, d_1 \\ c_1 \, d_2 \end{array} \right) \sim {\bf 2_2} \; , \;\; \left( \begin{array}{c} c_1 \, d_1 \\ c_2 \, d_2 \end{array} \right) \sim {\bf 2_4} \; ,
\\
{\bf 2_2} \times {\bf 2_3}&:  \left( \begin{array}{c} c_2 \, d_1 \\ c_1 \, d_2 \end{array} \right) \sim {\bf 2_1} \; , \;\; \left( \begin{array}{c} c_1 \, d_1 \\ c_2 \, d_2 \end{array} \right) \sim {\bf 2_5} \; ,
\\
{\bf 2_3} \times {\bf 2_3}&: c_1 \, d_2 + c_2 \, d_1 \sim {\bf 1_1} \; , \;\; c_1 \, d_2 - c_2 \, d_1 \sim {\bf 1_2} \; , \;\; \left( \begin{array}{c} c_1 \, d_1 \\ c_2 \, d_2 \end{array} \right) \sim {\bf 2_6} \; ,
\\
{\bf 2_1} \times {\bf 2_4}&:  \left( \begin{array}{c} c_2 \, d_1 \\ c_1 \, d_2 \end{array} \right) \sim {\bf 2_3} \; , \;\; \left( \begin{array}{c} c_1 \, d_1 \\ c_2 \, d_2 \end{array} \right) \sim {\bf 2_5} \; ,
\\
{\bf 2_2} \times {\bf 2_4}&:  \left( \begin{array}{c} c_2 \, d_1 \\ c_1 \, d_2 \end{array} \right) \sim {\bf 2_2} \; , \;\; \left( \begin{array}{c} c_1 \, d_1 \\ c_2 \, d_2 \end{array} \right) \sim {\bf 2_6} \; ,
\\
{\bf 2_3} \times {\bf 2_4}&:  \left( \begin{array}{c} c_2 \, d_1 \\ c_1 \, d_2 \end{array} \right) \sim {\bf 2_1} \; , \;\; \left( \begin{array}{c} c_1 \, d_1 \\ c_2 \, d_2 \end{array} \right) \sim {\bf 2_7} \; ,
\\
{\bf 2_4} \times {\bf 2_4}&: c_1 \, d_2 + c_2 \, d_1 \sim {\bf 1_1} \; , \;\; c_1 \, d_2 - c_2 \, d_1 \sim {\bf 1_2} \; , \;\; \left( \begin{array}{c} c_1 \, d_1 \\ c_2 \, d_2 \end{array} \right) \sim {\bf 2_8} \; .
\end{align}
\end{subequations}
We note, furthermore, that the Clebsch-Gordan coefficients for combinations, involving conjugated fields, look slightly different, since the generator $a$ is chosen as complex matrix in the two-dimensional
representations ${\bf 2_\mathrm{\bf i}}$, although all these representations are real. For $a$ being a singlet and $\left( \begin{array}{c} c_1 \\ c_2 \end{array} \right)$, $\left( \begin{array}{c} d_1 \\ d_2 \end{array} \right)$ being doublets,
the combinations involving $c^*_{1,2}$ read e.g.~
\begin{subequations}
\label{eq:CGcoefficientsD17complex}
\begin{align}
{\bf 2_\mathrm{{\bf i}}} \times {\bf 1_1}&: \left( \begin{array}{c} c_2^* \, a \\ c_1^* \, a \end{array} \right) \sim {\bf 2_\mathrm{{\bf i}}} \; ,
\\
{\bf 2_\mathrm{{\bf i}}} \times {\bf 1_2}&: \left( \begin{array}{c} c_2^* \, a \\ -c_1^* \, a \end{array} \right) \sim {\bf 2_\mathrm{{\bf i}}} \; ,
\\
{\bf 2_1} \times {\bf 2_1}&: c_1^* \, d_1 + c_2^* \, d_2 \sim {\bf 1_1} \; , \;\; c_1^* \, d_1 - c_2^* \, d_2 \sim {\bf 1_2} \; , \;\; \left( \begin{array}{c} c_2^* \, d_1 \\ c_1^* \, d_2 \end{array} \right) \sim {\bf 2_2} \; ,
\\
{\bf 2_1} \times {\bf 2_2}&:  \left( \begin{array}{c} c_1^* \, d_1 \\ c_2^* \, d_2 \end{array} \right) \sim {\bf 2_1} \; , \;\; \left( \begin{array}{c} c_2^* \, d_1 \\ c_1^* \, d_2 \end{array} \right) \sim {\bf 2_3} \; .
\end{align}
\end{subequations}
The general form of the Kronecker products and Clebsch-Gordan coefficients can be found in~\cite{Blum:2007jz}.

 %%%%%%%%%%%%%%%%%%%%%%%%%%%%%%%%%%%%%%%%%%%%%%%%%%%%%%
 \mathversion{bold}
\section{\texorpdfstring{Relations between Lagrangian and effective parameters}{Relations between Lagrangian and effective parameters in Mu, Md, Me and leptoquark couplings}}
\mathversion{normal}
\label{app:relatepara}
%%%%%%%%%%%%%%%%%%%%%%%%%%%%%%%%%%%%%%%%%%%%%%%%%%%%%%

Here, we collect the relations between the Lagrangian parameters and the effective ones, appearing in the charged fermion mass matrices and LQ couplings assuming real parameters. 

The effective parameters $f_{ij}$, appearing in the up-type quark mass matrix in eq.~(\ref{eq:Mupara}), are related as follows to the Lagrangian parameters $\alpha^u_i$
\begin{eqnarray}
\label{eq:Mupararel}
f_{11}&=& \alpha^u_4 \; ,
\\ \nonumber
f_{12}&=& \alpha^u_2 + \alpha^u_5 \, \lambda^2 + (\alpha^u_5)^\prime \, \lambda^2
+ \alpha^u_{13} \, \lambda^5 + \alpha^u_{18} \, \lambda^3 + \alpha^u_{19} \, \lambda^6 + \alpha^u_{20} \, \lambda^5 + \alpha^u_{27} \, \lambda^7 + \alpha^u_{29} \, \lambda^5 \; ,
\\ \nonumber
f_{13}&=& \alpha^u_{14} \, \lambda^2 + \alpha^u_{15} + \alpha^u_{16} \, \lambda + \alpha^u_{17} \, \lambda^2 + \alpha^u_{21} \, \lambda^3 + \alpha^u_{28} \, \lambda^4 \; ,
\\ \nonumber
f_{21}&=& \alpha^u_{10} \; ,
\\ \nonumber
f_{22}&=& \alpha^u_2 + \alpha^u_5 \, \lambda^2 - (\alpha^u_5)^\prime \, \lambda^2
+ \alpha^u_6 \, \lambda^2 + \alpha^u_9 \, \lambda^8 + \alpha^u_{12} \, \lambda^7 + (\alpha^u_{18})^\prime \, \lambda^5 + (\alpha^u_{20})^\prime \, \lambda^7 \; ,
\\ \nonumber
f_{23}&=& \alpha^u_3 + (\alpha^u_{14})^\prime \, \lambda^9 + (\alpha^u_{16})^\prime \, \lambda^6 + \alpha^u_{30} \, \lambda^{10} \; ,
\\ \nonumber
f_{31}&=& \alpha^u_{22} +  \alpha^u_{23} \; ,
\\ \nonumber
f_{32}&=& \alpha^u_7 +\alpha^u_8 \, \lambda^5 + \alpha^u_{11} \, \lambda^2 + \alpha^u_{26} \, \lambda^7 \; ,
\\ \nonumber
f_{33}&=& \alpha^u_1 + \alpha^u_{24} \, \lambda^{10} + \alpha^u_{25} \, \lambda^{10} \; .
\end{eqnarray}
For the effective parameters $d_{ij}$, used in the down-type quark mass matrix in eq.~(\ref{eq:Mdpara}), we have as relations to the Lagrangian parameters $\alpha^d_i$
\begin{eqnarray}
\label{eq:Mdpararel}
d_{11}&=& \alpha^d_3 + \alpha^d_9 \, \lambda^7 + \alpha^d_{13} \, \lambda^8 \; ,
\\ \nonumber
d_{12}&=& \alpha^d_7 + (\alpha^d_9)^\prime \, \lambda^3 + (\alpha^d_{13})^\prime \, \lambda^2 +  \alpha^d_{16} \, \lambda +  \alpha^d_{24} \, \lambda^4 +  \alpha^d_{25} \, \lambda^3 \; ,
\\ \nonumber
d_{13}&=&  \alpha^d_8 \, \lambda^2 +  \alpha^d_{10} +  \alpha^d_{11} \, \lambda +  \alpha^d_{12} \, \lambda^2 +  \alpha^d_{14} \, \lambda^3 +  \alpha^d_{22} \, \lambda^4 \; ,
\\ \nonumber
d_{21}&=&  \alpha^d_6 \; ,
\\ \nonumber
d_{22}&=&  \alpha^d_2 +  (\alpha^d_{13})^{\prime\prime} \, \lambda^9 +  (\alpha^d_{16})^\prime \, \lambda^6 +  \alpha^d_{21} \, \lambda^{10} \; ,
\\ \nonumber
d_{23}&=&  \alpha^d_4 +  (\alpha^d_8)^\prime \, \lambda^9 +  (\alpha^d_{11})^\prime \, \lambda^6 +  \alpha^d_{23} \, \lambda^{10} \; ,
\\ \nonumber
d_{31}&=&  \alpha^d_{15} \; ,
\\ \nonumber
d_{32}&=&  \alpha^d_5 +  \alpha^d_{17} \, \lambda^6 +  \alpha^d_{20} \, \lambda^6 \; ,
\\ \nonumber
d_{33}&=&  \alpha^d_1 +  \alpha^d_{18} \, \lambda^{10} +  \alpha^d_{19} \, \lambda^{10} \; .
\end{eqnarray}
Likewise, we find for $e_{ij}$, the effective parameters contained in the charged lepton mass matrix $M_e$, see eq.~(\ref{eq:Mepara}), that they are expressed in terms of $\alpha^e_i$, appearing in the
Lagrangians in eqs.~(\ref{eq:chlepopsLO},\ref{eq:chlepopsSLO}), as follows
\begin{eqnarray}
\label{eq:Mepararel}
e_{11}&=& \alpha^e_3 + \alpha^e_6 \, \lambda^7 + \alpha^e_9 \, \lambda^8 \; ,
\\ \nonumber
e_{12}&=& \alpha^e_{12} \; ,
\\ \nonumber
e_{21}&=& \alpha^e_{11} + \alpha^e_{17} \, \lambda^2 + \alpha^e_{22} \, \lambda^3 + \alpha^e_{23} \, \lambda^4 \; ,
\\ \nonumber
e_{22}&=& \alpha^e_2 + (\alpha^e_9)^\prime \, \lambda^9 + \alpha^e_{13} \, \lambda^6 + \alpha^e_{27} \, \lambda^{10} \; ,
\\ \nonumber
e_{23}&=& \alpha^e_4 + \alpha^e_{10} \, \lambda^{11} + \alpha^e_{14} \, \lambda^8 + \alpha^e_{26} \, \lambda^{10} \; ,
\\ \nonumber
e_{31}&=& \alpha^e_7 + \alpha^e_8 \, \lambda^3 + \alpha^e_{15} \, \lambda + \alpha^e_{16} \, \lambda^2 + \alpha^e_{19} \; ,
\\ \nonumber
e_{32}&=& \alpha^e_5 + \alpha^e_{18} \, \lambda^9 + \alpha^e_{20} \, \lambda^6 + \alpha^e_{25} \, \lambda^8 \; ,
\\ \nonumber
e_{33}&=& \alpha^e_1 + \alpha^e_{21} \, \lambda^{10} + \alpha^e_{24} \, \lambda^{10} \; .
\end{eqnarray}
We continue with the relations between the effective parameters $\hat{a}_{ij}$, appearing in the LQ coupling ${\bf \hat{x}}$, see eq.~(\ref{eq:hatxpara}), and the coefficients $\beta^L_i$
\begin{eqnarray}
\label{eq:xhatpararel}
\hat{a}_{11}&=& \beta^L_9 \, \lambda + \beta^L_{10} \, \lambda^2 + \beta^L_{18} \, \lambda^2 + \beta^L_{20} \; ,
\\ \nonumber
\hat{a}_{12}&=& (\beta^L_{10})^\prime \; ,
\\ \nonumber
\hat{a}_{21}&=& \beta^L_6 + \beta^L_7 \, \lambda + \beta^L_{11} \, \lambda^3 + \beta^L_{16} \, \lambda + \beta^L_{19} \, \lambda^4 + \beta^L_{21} \, \lambda^3 \; ,
\\ \nonumber
\hat{a}_{22}&=& \beta^L_4 + (\beta^L_6)^\prime \, \lambda^4 + (\beta^L_{11})^\prime \, \lambda^9 + (\beta^L_{21})^\prime \, \lambda^9 \; ,
\\ \nonumber
\hat{a}_{23}&=& \beta^L_2 + \beta^L_{12} \, \lambda^{11} + \beta^L_{17} \, \lambda^8 + \beta^L_{25} \, \lambda^{10} \; ,
\\ \nonumber
\hat{a}_{31}&=& \beta^L_5 \, \lambda^2 + \beta^L_8 + \beta^L_{13} \, \lambda + \beta^L_{14} \, \lambda^2 + \beta^L_{15} \, \lambda^3 + \beta^L_{24} \, \lambda^4 \; ,
\\ \nonumber
\hat{a}_{32}&=& \beta^L_3 + (\beta^L_5)^\prime \, \lambda^9 + (\beta^L_{13})^\prime \, \lambda^6 + \beta^L_{26} \, \lambda^{10} \; ,
\\ \nonumber
\hat{a}_{33}&=& \beta^L_1 + \beta^L_{22} \, \lambda^{10} + \beta^L_{23} \, \lambda^{10} \; .
\end{eqnarray}
For the LQ coupling ${\bf \hat{y}}$, found in eq.~(\ref{eq:hatypara}), we define the effective parameters $\hat{b}_{ij}$ in terms of the coefficients $\beta^R_i$ as
\begin{eqnarray}
\label{eq:yhatpararel}
\hat{b}_{11}&=& \beta^R_7 + \beta^R_{15} \; ,
\\ \nonumber
\hat{b}_{12}&=& \beta^R_{13} + \beta^R_{14} \, \lambda^3 + \beta^R_{17} \, \lambda + \beta^R_{18} \, \lambda^2 + \beta^R_{21} \; ,
\\ \nonumber
\hat{b}_{13}&=& \beta^R_{10} \, \lambda + \beta^R_{11} \, \lambda^2 + \beta^R_{19} \, \lambda^2 + \beta^R_{23} \; ,
\\ \nonumber
\hat{b}_{21}&=& \beta^R_6 \; ,
\\ \nonumber
\hat{b}_{22}&=& \beta^R_3 + \beta^R_{20} \, \lambda^9 + \beta^R_{24} \, \lambda^6 + \beta^R_{29} \, \lambda^8 \; ,
\\ \nonumber
\hat{b}_{23}&=& \beta^R_2 + \beta^R_9 \, \lambda^4 + \beta^R_{12} \, \lambda^9 + \beta^R_{26} \, \lambda^9 \; ,
\\ \nonumber
\hat{b}_{31}&=& \beta^R_{16} + \beta^R_{22} \; ,
\\ \nonumber
\hat{b}_{32}&=& \beta^R_1 + \beta^R_{27} \, \lambda^{10} + \beta^R_{28} \, \lambda^{10} \; ,
\\ \nonumber
\hat{b}_{33}&=& \beta^R_4 + \beta^R_5 \, \lambda^5 + \beta^R_8 \, \lambda^2 + \beta^R_{25} \, \lambda^7 \; .
\end{eqnarray}
The effective parameters $a_{ij}$ in the LQ coupling ${\bf x}$, given in eq.~(\ref{eq:xpara}), read in terms of the effective parameters $\hat{a}_{ij}$, $d_{ij}$ and $e_{ij}$, as follows
\begin{eqnarray}
\label{eq:xpararel}
a_{11}&=& \hat{a}_{11} +\mathcal{o} (\lambda^3)\; ,
\\ \nonumber
a_{12}&=& -\frac{\hat{a}_{22} e_{11} e_{21}}{e_{22}^2}  + \frac{\hat{a}_{23} d_{23} e_{11} e_{21}}{d_{33} e_{22}^2} + \frac{\hat{a}_{32} e_{11} e_{21} e_{23}}{e_{22}^2 e_{33}} -\frac{\hat{a}_{33} d_{23} e_{11} e_{21} e_{23}}{d_{33} e_{22}^2 e_{33}} + \mathcal{O} (\lambda)\; ,
\\ \nonumber
a_{13}&=& -\frac{\hat{a}_{23} e_{11} e_{21}}{e_{22}^2} + \frac{\hat{a}_{33} e_{11} e_{21} e_{23}}{e_{22}^2 e_{33}} +\mathcal{O}(\lambda^2)\; ,
\\ \nonumber
a_{21}&=& \hat{a}_{21} + \mathcal{O}(\lambda)\; ,
\\ \nonumber
a_{22}&=& \hat{a}_{22} - \frac{d_{23}}{d_{33}} \, \left( \hat{a}_{23} - \frac{\hat{a}_{33} e_{23}}{e_{33}} \right) - \frac{\hat{a}_{32} e_{23}}{e_{33}} +\mathcal{O}(\lambda^2)\; ,
\\ \nonumber
a_{23}&=& \hat{a}_{23} - \frac{\hat{a}_{33} e_{23}}{e_{33}} +\mathcal{O}(\lambda^2)\; ,
\\ \nonumber
a_{31}&=& \hat{a}_{31} - \frac{\hat{a}_{32} d_{12}}{d_{22}} - \frac{\hat{a}_{33} d_{13}}{d_{33}} + \frac{\hat{a}_{33} d_{12} d_{23}}{d_{22} d_{33}} +\mathcal{O}(\lambda)\; ,
\\ \nonumber
a_{32}&=& \hat{a}_{32} - \frac{\hat{a}_{33} d_{23}}{d_{33}} +\mathcal{O}(\lambda^2)\; ,
\\ \nonumber
a_{33}&=& \hat{a}_{33} + \mathcal{O}(\lambda^2) \; .
\end{eqnarray}
Similarly, we can express the effective parameters $c_{ij}$ in the LQ coupling ${\bf z}$ in eq.~(\ref{eq:zparaA}) in terms of $\hat{a}_{ij}$, $e_{ij}$ and $f_{ij}$ and find for scenario A
\begin{eqnarray}
\label{eq:zpararelA}
c_{11}&=& \hat{a}_{11} + \mathcal{O} (\lambda^2)\; ,
\\ \nonumber
c_{12}&=& \frac{\hat{a}_{11} f_{12}}{f_{22}} + \mathcal{O} (\lambda)\; ,
\\ \nonumber
c_{13}&=& -\frac{\hat{a}_{23} e_{11} e_{21}}{e_{22}^2} + \frac{\hat{a}_{33} e_{11} e_{21} e_{23}}{e_{22}^2 e_{33}} + \mathcal{O}(\lambda^2)\; ,
\\ \nonumber
c_{21}&=& -\frac{f_{12}}{e_{33} f_{22} f_{33}} \, \left( \hat{a}_{33} e_{23} f_{23} -\hat{a}_{23} e_{33} f_{23} - \hat{a}_{32} e_{23} f_{33} + \hat{a}_{22} e_{33} f_{33}\right) + \mathcal{O}(\lambda^2)\; ,
\\ \nonumber
c_{22}&=& \hat{a}_{22} - \frac{\hat{a}_{32} e_{23}}{e_{33}} - \left( \hat{a}_{23} - \frac{\hat{a}_{33} e_{23}}{e_{33}}\right) \, \frac{f_{23}}{f_{33}} + \mathcal{O} (\lambda^2)\; ,
\\ \nonumber
c_{23}&=& \hat{a}_{23} - \frac{\hat{a}_{33} e_{23}}{e_{33}} + \mathcal{O}(\lambda^2)\; ,
\\ \nonumber
c_{31}&=& \frac{f_{12} (\hat{a}_{33} f_{23}-\hat{a}_{32} f_{33})}{f_{22} f_{33}} + \mathcal{O} (\lambda^2)\; ,
\\ \nonumber
c_{32}&=& \hat{a}_{32} - \frac{\hat{a}_{33} f_{23}}{f_{33}} + \mathcal{O}(\lambda^2) \; ,
\\ \nonumber
c_{33}&=& \hat{a}_{33} +\mathcal{O}(\lambda^2)\; .
\end{eqnarray}
The effective parameters $b_{ij}$ in the LQ coupling ${\bf y}$, found in eq.~(\ref{eq:yparaA}), read for scenario A in terms of $\hat{b}_{ij}$, $e_{ij}$ and $f_{ij}$
\begin{eqnarray}
\label{eq:ypararelA}
b_{11}&=& \hat{b}_{11} + \mathcal{o}(\lambda^3) \; ,
\\ \nonumber
b_{12}&=& \hat{b}_{12} - \frac{\hat{b}_{22} e_{21}}{e_{22}} - \frac{\hat{b}_{32} e_{31}}{e_{33}} + \frac{\hat{b}_{32} e_{21} e_{32}}{e_{22} e_{33}} +\mathcal{O}(\lambda^2)\; ,
\\ \nonumber
b_{13}&=& \hat{b}_{13} - \frac{\hat{b}_{23} e_{21}}{e_{22}} + \mathcal{O}(\lambda^2)\; ,
\\ \nonumber
b_{21}&=& -\frac{\hat{b}_{22} f_{11} f_{12}}{f_{22}^2} + \frac{\hat{b}_{32} e_{22} e_{23} f_{11} f_{12}}{e_{33}^2 f_{22}^2} + \frac{\hat{b}_{32} e_{32} f_{11} f_{12}}{e_{33} f_{22}^2} + \mathcal{O}(\lambda) \; ,
\\ \nonumber
b_{22}&=& \hat{b}_{22} - \frac{\hat{b}_{32} (e_{22} e_{23}+e_{32} e_{33})}{e_{33}^2} +\mathcal{O}(\lambda^2)\; ,
\\ \nonumber
b_{23}&=& \hat{b}_{23} +\mathcal{O}(\lambda^4)\; ,
\\ \nonumber
b_{31}&=& -\frac{\hat{b}_{32} f_{11} f_{12}}{f_{22}^2} +\mathcal{O}(\lambda) \; ,
\\ \nonumber
b_{32}&=& \hat{b}_{32} + \mathcal{O}(\lambda^6)\; ,
\\ \nonumber
b_{33}&=& \hat{b}_{33} +\frac{\hat{b}_{32} f_{32}}{f_{33}} + \mathcal{O}(\lambda^2)\; .
\end{eqnarray}

\section{Formulae for phenomenology}
\label{app:pheno}

We use the Warsaw basis~\cite{Grzadkowski:2010es} for SM Effective Field Theory (SMEFT) and the JMS basis~\cite{Jenkins:2017jig} below the electroweak scale for low-energy EFT.
 
\subsection{Correction to charged lepton masses}
 \label{app:charged-lepton-mass-correction}
 The LQ contributes via its couplings to the charged lepton self-energies. This results in a correction to the charged lepton masses which is approximately given by~\cite{Crivellin:2020mjs}
\begin{align}
	m_{e_i}  
& =m_{e_i}^{\rm tree}\left( 1 
+ \frac{1}{2}\Sigma _{ii}^{LL}
+ \frac{1}{2}\Sigma _{ii}^{RR} \right)
+ \Sigma _{ii}^{LR}
\end{align}
in terms of the self-energies $\Sigma^{XY}_{ij}$, where $X,Y\in \{L,R\}$ label the chiralities of the charged leptons and $i,j$ the lepton flavour. It can be compactly rewritten as 
\begin{align}
	m_{e_i} & = \left| m_{e_i}^{\rm tree}
	-\frac{3}{16\pi^2} \sum_{j=1}^3 \left(m_{u_j}y_{ij} z_{ij}^* \mathcal{I}_0\left(1,t_{u_j} \right) 
		+\frac{1}{4} m_{e_i}^{\rm tree} \left(|z_{ij}|^2 + |y_{ij}|^2\right) \mathcal{I}_1\left(1, t_{u_j}\right) 
\right)\right|
\;,
	\end{align}
where $t_X$ is defined in eq.~\eqref{eq:tX}.
After removing the UV divergences using minimal subtraction, the loop 
functions take the simple form
\begin{align}
\mathcal{I}_{0}(x,y)&=
1+\ln x+y \ln y\,,
		    &
\mathcal{I}_{1}(x,y)&=
\frac{1}{2}+\ln x -y\,.
\end{align}
The last terms of the loop functions $\mathcal{I}_0$ and $\mathcal{I}_1$ are only numerically relevant for the top quark and can be neglected otherwise. 

\subsection{Leptonic processes}
 \label{app:dipole}
	\subsubsection{Effective interactions at one-loop order}
	The relevant effective Lagrangian using the JMS basis~\cite{Jenkins:2017jig} reads 
	\begin{equation}
		\begin{aligned}
	\mathcal{L} & \supset 
	C_{ee,ijkl}^{VLL} (\overline{e_i} \gamma^\mu P_L e_j)(\overline{e_k} \gamma_\mu P_L e_l)
	+C_{ee,ijkl}^{VRR} (\overline{e_i} \gamma^\mu P_R e_j)(\overline{e_k} \gamma_\mu P_R e_l)
		    +C_{ee,ijkl}^{VLR} (\overline{e_i} \gamma^\mu P_L e_j)(\overline{e_k} \gamma_\mu P_R e_l)
	\\   &
	+\left[C_{ee,ijkl}^{SRR} (\overline{e_i} P_R e_j)(\overline{e_k}  P_R e_l) +\mathrm{h.c.}\right]
	+\left[C_{e\gamma}^{ij}(\overline{e_i} \sigma^{\mu\nu} P_R e_j) F_{\mu\nu} +\mathrm{h.c.}\right]\;.
\end{aligned}
\end{equation}
Note that some of the Wilson coefficients contain redundant indices. 
We define the covariant derivative in QED as $D_\mu=\partial_\mu + i Q e A_\mu$ following the convention in \cite{Jenkins:2017jig}. 
We use \texttt{FeynRules}~\cite{Alloul:2013bka}, \texttt{FeynArts}~\cite{Hahn:2000kx}, \texttt{FormCalc}~\cite{Hahn:1998yk,Hahn:2016ebn}, \texttt{Package-X}~\cite{Patel:2015tea}, and \texttt{ANT}~\cite{Angel:2013hla} to evaluate the amplitudes and match the result to the operator basis.
The Wilson coefficient of the dipole operator is given by 
\begin{align}
\label{eq:Cegamma}
	C_{e\gamma}^{ij} &=-\frac{e}{32 \pi^2 m_\phi^2} \sum_m \Big( \left(m_{e_i} y_{im}^* y_{jm}+ m_{e_j} z_{im}^*z_{jm}\right) \left[f_S(t_{u_m}) -3 f_F(t_{u_m})\right]
		       \\&\qquad\qquad\qquad\qquad\qquad\qquad\qquad\qquad -m_{u_m} z_{im}^* y_{jm} \left[g_S(t_{u_m}) -3 \, g_F(t_{u_m})\right]    \Big) \nonumber
		       \\&
\approx\frac{e}{128 \pi^2 m_\phi^2} \sum_m \left( m_{e_i} y_{im}^* y_{jm}+ m_{e_j} z_{im}^*z_{jm}+ 2 \, m_{u_m} z_{im}^* y_{jm} (7+4\ln t_{u_m})   \right) 
	\;,\nonumber
\end{align}
where $e=|e|$ is the unit electric charge.
The relevant loop functions are given by 
\begin{align}
f_S(x)&= \frac{x+1}{4(x-1)^2}- \frac{x\ln x}{2(x-1)^3},\;\;\;\;
f_F(x)= \frac{x^2-5x-2}{12(x-1)^3}+ \frac{x\ln x}{2(x-1)^4},\\
g_S(x)&=\frac{1}{x-1} -\frac{\ln x}{(x-1)^2},\;\;\;\;\;\;\;\;\;\;\;\;\,
g_F(x)= \frac{x-3}{2(x-1)^2}+\frac{\ln x}{(x-1)^3}.\nonumber
\end{align}
 The contributions to the four-lepton interactions can be split in different parts. The Higgs-penguin contributions are suppressed by the small charged lepton masses and thus negligible.
The $Z$-penguin contributions are given by
\begin{align}
	C_{ee,ijkl}^{VLL,Z} & = \frac{3\sqrt{2}G_F(1-2s_W^2)}{64\pi^2} \sum_m t_{u_m} (1+\ln t_{u_m}) \left( \delta_{il} z_{km}^* z_{jm} +\delta_{ij} z_{km}^* z_{lm} +\delta_{kl} z_{im}^* z_{jm} +\delta_{jk} z_{im}^* z_{lm}\right)\;,
	\\
	C_{ee,ijkl}^{VRR,Z} & =\frac{3\sqrt{2} G_F s_W^2}{32\pi^2 } \sum_m t_{u_m} (1+\ln t_{u_m}) \left( \delta_{il} y_{km}^* y_{jm} +\delta_{ij} y_{km}^* y_{lm} +\delta_{kl} y_{im}^* y_{jm} +\delta_{jk} y_{im}^* y_{lm}\right)\;,
	\\
	C_{ee,ijkl}^{VLR,Z} & = -\frac{3\sqrt{2}G_F}{16\pi^2} \sum_m t_{u_m} (1+\ln t_{u_m}) \left((1-2 \, s_W^2) \, \delta_{ij} y_{km}^*y_{lm} + 2 \, s_W^2 \, \delta_{kl} z_{im}^* z_{jm}
	\right) \;,
\end{align}
where $G_F$ denotes the Fermi constant and $s_W=\sin \theta_W$ the sine of the weak mixing angle, $\theta_W$.

The short-distance $\gamma$-penguin contributions are given by
\begin{align}
	C_{ee,ijkl}^{VLL,\gamma} & = \frac{\alpha_{\rm em} }{96 \pi m_\phi^2} \sum_m(5+4\ln t_{u_m})\left( \delta_{il} z_{km}^* z_{jm} + \delta_{kl} z_{im}^* z_{jm} +\delta_{ij} z_{km}^* z_{lm} + \delta_{jk} z_{im}^* z_{lm} \right)\;, 
	\\
	C_{ee,ijkl}^{VRR,\gamma} & = \frac{\alpha_{\rm em} }{96 \pi m_\phi^2} \sum_m(5+4\ln t_{u_m})\left( \delta_{il} y_{km}^* y_{jm} + \delta_{kl} y_{im}^* y_{jm}+\delta_{ij} y_{km}^* y_{lm} + \delta_{jk} y_{im}^* y_{lm} \right)\;,
	\\
	C_{ee,ijkl}^{VLR,\gamma} & = \frac{\alpha_{\rm em} }{24 \pi m_\phi^2} \sum_m(5+4\ln t_{u_m})\left( \delta_{ij} y_{km}^* y_{lm} + \delta_{kl} z_{im}^* z_{jm}\right)\;,
\end{align}
where $\alpha_{\rm em}$ denotes the fine structure constant.
Finally, the box diagrams also contribute to the four-lepton operators
\begin{align}
	C_{ee,ijkl}^{VLL,\rm box} & =  \frac{3}{256 \pi^2 m_\phi^2} \sum_{m,n}z_{jm} z_{ln}
	\left(z_{in}^* z_{km}^* + z_{im}^* z_{kn}^*\right)\;,
\\
	C_{ee,ijkl}^{VRR,\rm box} & = \frac{3}{256 \pi^2 m_\phi^2} \sum_{m,n}y_{jm} y_{ln}
	\left(y_{in}^* y_{km}^* + y_{im}^* y_{kn}^*\right)\;,
	\\
	C_{ee,ijkl}^{VLR,\rm box} & = 
	\frac{3}{64 \pi^2 m_\phi^2} \sum_{m,n}y_{ln} z_{jm} y_{kn}^* z_{im}^*\;.
\end{align}	
\mathversion{bold}
\subsubsection{\texorpdfstring{Radiative charged lepton flavour violating decays $e_i\to e_j\gamma$}{Radiative charged lepton flavour violating decays ei to ej gamma}}
\mathversion{normal}
The BR for $e_i\to e_j\gamma$ can be expressed in terms of the dipole Wilson coefficients
\begin{equation}
	\mathrm{BR}(e_i\to e_j\gamma) = \frac{m_{e_i}^3}{4\pi \Gamma_{e_i}} \left(| C_{e\gamma}^{ji}|^2 + |C_{e\gamma}^{ij}|^2\right)
	\;,\label{eq:eiejgamma}
\end{equation}
where $\Gamma_{e_i}$ denotes the full decay width of the charged lepton $e_i$.

\subsubsection{Dipole moments}
The electromagnetic current of a particle of mass $m$ coupling to a real on-shell photon can be parametrised in terms of three form factors $F_i$, see e.g.~\cite{Nowakowski:2004cv,Itzykson:1980rh},
\begin{equation}
	\begin{aligned}
	\braket{p_1|j^\mu(0)|p_2}=\bar u(\mathbf{p_1}) \Bigg[
		F_1(q^2) \gamma^\mu
		+ F_2(q^2) \frac{i\sigma^{\mu\nu}}{2m} q_\nu 
		+ F_3(q^2)  \frac{\sigma^{\mu\nu}}{2m}\gamma_5 q_\nu 
	\Bigg] u(\mathbf{p_2})
	\;,
\end{aligned}
\end{equation}
where $q^\mu = p_1^\mu-p_2^\mu$. At zero squared momentum transfer, $q^2=0$, the form factors can be identified with the electric charge $e F_1(0)$, the AMM $a=F_2(0)$, and the EDM $d=-eF_3(0)/2m$. Taking into account the definition of the covariant derivative, we find for the contributions of the dipole operator to the AMM and the EDM of the charged lepton $e_i$ with $F_1(0)=-1$ 
\begin{equation}
	\begin{aligned}
a_{e_i} & = \frac{4m_{e_i}}{e}\, \mathrm{Re}(C_{e\gamma}^{ii})\;,
	       &
d_{e_i} & = 2\, \mathrm{Im}(C_{e\gamma}^{ii})
\;,
\end{aligned}\label{eq:EDMAMM}
\end{equation}
respectively.

\mathversion{bold}
\subsubsection{\texorpdfstring{Trilepton decays $e_i\to e_j e_k \overline{e_m}$}{Trilepton decays ei to ej ek bar em}}
\mathversion{normal}
\label{sec:trilepton}
We have recalculated trilepton decays due to discrepancies in the literature~\cite{Angel:2013hla,Mandal:2019gff} and make use of the recent calculation in terms of EFT~\cite{Calibbi:2021pyh} and earlier references~\cite{Kuno:1999jp,Brignole:2004ah}.
Note there are no redundant indices in reference~\cite{Calibbi:2021pyh}, and thus there are additional symmetry factors. 
The BR for $e_i\to e_je_j\overline{e_j}$ is
\begin{align}\label{eq:trilepton_jjj}
	\mathrm{BR}(e_i\to e_je_j \overline{e_j})= \frac{m_{e_i}^5}{3(16\pi)^3 \Gamma_{e_i}}
	\Bigg[ &16|C^{VLL}|^2+ 16 |C^{VRR}|^2 + 8 |C^{VLR}|^2 + 8 |C^{VRL}|^2
				\\&
			+\frac{256 e^2}{m_{e_i}^2} \left(\ln \frac{m_{e_i}^2}{m_{e_j}^2} -\frac{11}{4}\right) \left(|C_{e\gamma}^{ji}|^2 + |C_{e\gamma}^{ij}|^2\right)
				\nonumber \\&
	-\frac{64 e}{m_{e_i}} \mathrm{Re}\left[(2\,C^{VLL}+C^{VLR}) C_{e\gamma}^{ji*} + ( 2 \,C^{VRR} + C^{VRL})C_{e\gamma}^{ij}\right]\Bigg]\;,\nonumber
\end{align}
where the coefficients in the decay rate are given in terms of the Wilson coefficients
\begin{align}
	C^{VLL} & = 2 \, C_{ee,jijj}^{VLL}\;, &
	C^{VRR} & = 2 \, C_{ee,jijj}^{VRR}\;, &
	C^{VLR} & = C_{ee,jijj}^{VLR}\;, &
	C^{VRL} & = C_{ee,jjji}^{VLR}
	\;.
\end{align}

The BR for $e_i\to e_je_k\overline{e_k}$ is
\begin{align}\label{eq:trilepton_jkk}
	\mathrm{BR}(e_i\to e_je_k \overline{e_k})= \frac{m_{e_i}^5}{3(16\pi)^3 \Gamma_{e_i}}
	\Bigg[ &8|C^{VLL}|^2+ 8 |C^{VRR}|^2 + 8 |C^{VLR}|^2 + 8 |C^{VRL}|^2
				\\&
			+\frac{256 e^2}{m_{e_i}^2} \left(\ln \frac{m_{e_i}^2}{m_{e_j}^2} -3\right) \left(|C_{e\gamma}^{ji}|^2 + |C_{e\gamma}^{ij}|^2\right)\nonumber
				\\\nonumber&
	-\frac{64 e}{m_{e_i}} \mathrm{Re}\left[(C^{VLL}+C^{VLR}) C_{e\gamma}^{ji*} + (  C^{VRR} + C^{VRL})C_{e\gamma}^{ij}\right]\Bigg]\;,
\end{align}
where the coefficients in the decay rate are given in terms of the Wilson coefficients
\begin{align}
	C^{VLL} & = 4 \, C_{ee,jikk}^{VLL}\;, &
	C^{VRR} & = 4 \, C_{ee,jikk}^{VRR}\;, &
	C^{VLR} & = C_{ee,jikk}^{VLR}\;, &
	C^{VRL} & = C_{ee,kkji}^{VLR}\;.
\end{align}
The BR for $e_i\to e_ke_k\overline{e_j}$ is
\begin{align}
	\mathrm{BR}(e_i\to e_ke_k \overline{e_j})= \frac{m_{e_i}^5}{3(16\pi)^3 \Gamma_{e_i}}
	\Bigg[ 16|C^{VLL}|^2+ 16 |C^{VRR}|^2 + 8 |C^{VLR}|^2 + 8 |C^{VRL}|^2\Bigg]\;,
\end{align}
where the coefficients in the decay rate are given in terms of the Wilson coefficients
\begin{align}
	C^{VLL} & = 2 \, C_{ee,kikj}^{VLL}\;, &
	C^{VRR} & = 2 \, C_{ee,kikj}^{VRR}\;, &
	C^{VLR} & = C_{ee,kikj}^{VLR}\;, &
	C^{VRL} & = C_{ee,kjki}^{VLR}\;.
\end{align}
The Higgs-penguin contribution is neglected, because it is suppressed by small charged lepton Yukawa couplings and thus no scalar operators are induced at leading order. 

 \subsection{Semi-leptonic processes}
 \label{app:semi_leptonic}

 \subsubsection{Effective Lagrangian}
The effective Lagrangian relevant for semi-leptonic interactions is
\begin{align}
	\mathcal{L}
&\supset 
C_{eq}^{VLL} (\bar e \gamma^\mu P_L e)  (\bar q \gamma_\mu P_L q)
+ C_{eq}^{VRR} (\bar e \gamma^\mu P_R e) (\bar q \gamma_\mu P_R q)
	 \\&
+ C_{eq}^{VLR} (\bar e \gamma^\mu P_L e)  (\bar q \gamma_\mu P_R q)
+C_{qe}^{VLR}  (\bar q \gamma_\mu P_L q) (\bar e \gamma^\mu P_R e) 
	\nonumber \\&
	 + \left[ C_{eq}^{SRR} (\bar e P_R e)  (\bar q P_R q)
		+C_{eq}^{SRL}(\bar e P_R e)  (\bar q P_L q)
+ C_{eq}^{TRR}  (\bar e \sigma^{\mu\nu} P_R e)  (\bar q \sigma_{\mu\nu} P_R q) 
	+ \mathrm{h.c.}\right]
	\nonumber  \\&
+	C_{\nu q}^{VLL} (\bar \nu \gamma^\mu P_L \nu)  (\bar q \gamma_\mu P_L q)
+ C_{\nu q}^{VLR} (\bar \nu \gamma^\mu P_L \nu)  (\bar q \gamma_\mu P_R q)
	\nonumber   \\&
	+\Big[ 
		C_{\nu e du}^{VLL} (\bar \nu \gamma^\mu P_L e) (\bar d \gamma_\mu P_L u) 
		+C_{\nu e du}^{VLR} (\bar \nu \gamma^\mu P_L e) (\bar d \gamma_\mu P_R u) 
	\nonumber  \\&\quad
		+C_{\nu e du}^{SRR} (\bar \nu P_R e) (\bar d P_R u) 
		+C_{\nu e du}^{SRL} (\bar \nu P_R e) (\bar d P_L u) 
		+C_{\nu e du}^{TRR} (\bar \nu \sigma^{\mu\nu} P_R e) (\bar d \sigma_{\mu\nu} P_R u) 
	+\mathrm{h.c.}\Big]\, , \nonumber 
\end{align}
where the first three lines describe neutral-current interactions between charged leptons and quarks, the fourth line describes neutral-current interactions between neutrinos and quarks, and the last two lines describe charged-current interactions. The flavour indices are suppressed in the above equation. In the following discussion, they are indicated as subscripts, e.g.~$C^{VLL}_{eq,ijkl} (\overline{e_i}\gamma^\mu P_L e_j)(\overline{q_k}\gamma_\mu P_L q_l)$.

The dominant RG corrections are due to QCD. Their correction at one-loop order to the Wilson coefficients of operators with two quarks and two leptons is described by
\begin{align}
	C^{VXY}(\mu=m_b) & = C^{VXY}(\mu=m_\phi) \;,\\ \nonumber 
	C^{SXY}(\mu=m_b)&=\left[ \frac{\alpha_s(m_t)}{\alpha_s(m_b)} \right]^{-\frac{3C_F}{\beta_0^{(5)}}}\left[ \frac{\alpha_s(m_\phi)}{\alpha_s(m_t)} \right]^{-\frac{3C_F}{\beta_0^{(6)}}}
C^{SXY}(\mu=m_\phi)
\;,
\\ \nonumber 
	C^{TXY}(\mu=m_b)&=\left[ \frac{ \alpha_s(m_t)}{\alpha_s(m_b)} \right]^{\frac{C_F}{\beta_0^{(5)}}}\left[ \frac{ \alpha_s(m_\phi)}{\alpha_s(m_t)} \right]^{\frac{C_F}{\beta_0^{(6)}}}
C^{TXY}(\mu=m_\phi)\;,
\end{align}
where $X,Y\in\{L,R\}$ denote the chiralities of the fermion bilinears. The 
Casimir invariant $C_F$ and $\beta_0^{(n_f)}$ which parametrises the one-loop RG equation of the strong coupling are 
\begin{align}\label{eq:CFbeta}
	C_F&=(N_c^2-1)/(2N_c)=4/3\;,
	   &
	\beta_0^{(n_f)}&=11-2n_f/3
\end{align}
with $N_c=3$ colours and $n_f$ flavours.

 \subsubsection{Tree-level matching}
\label{sec:TreeLevelMatching}
Here, we provide the matching to relevant operators in the low-energy EFT at tree level.  At this level, the interactions of the LQ $\phi$ induce Wilson coefficients with two quarks and two leptons. The non-zero Wilson coefficients for neutral-current interactions are given by
\begin{align}\label{eq:treeNC}
	C^{VLL}_{\nu d,ijkl} & = \frac{x_{jl}x_{ik}^*}{2m_\phi^2}\;, 
	&
	C^{VLL}_{e u,ijkl} & = \frac{z_{jl}z_{ik}^*}{2m_\phi^2} \;,
	&
	C^{VRR}_{e u,ijkl} & = \frac{y_{ik}y_{jl}^*}{2m_\phi^2} \;,
	\\\nonumber
	C^{SRR}_{eu,ijkl} &= \frac{z_{ik}^*y_{jl}}{2m_\phi^2}\;,
	&
	C^{TRR}_{eu,ijkl} &= - \frac14\frac{z_{ik}^*y_{jl}}{2m_\phi^2}
	\end{align}
	and the ones for charged-current interactions are 
\begin{align}\label{eq:treeCC}
	C^{VLL}_{\nu e du,ijkl} &= -\frac{x_{ik}^*z_{jl}}{2m_\phi^2}\;,
				 &
	C^{SRR}_{\nu e du,ijkl} &=-\frac{x_{ik}^*y_{jl}}{2m_\phi^2}\;,
	&
	C^{TRR}_{\nu e du,ijkl} &=\frac14\frac{x_{ik}^*y_{jl}}{2m_\phi^2}\;.
\end{align}
For the charged-current observables, involving the quark flavour transition $b\to c$ and defined in the following, the RG running of the contributions to the Wilson coefficients due to the LQ $\phi$ between the scale, set by the LQ mass, and the hadronic scale, $\mu = \mu_B = 4.8$ GeV, is accounted for as 

\begin{align}\label{eq:treeCCrunning}
\frac{C^{VLL}_{\nu edu,\beta\alpha 32}(\mu_B)}{C^{VLL}_{\nu edu,\beta\alpha 32}}
\approx &
\left\{
\begin{array}{*{2}{c}}
1.016, & \hat{m}_\phi = 2 \\ \relax
1.018, & \hat{m}_\phi = 4 \\ \relax
1.019, & \hat{m}_\phi = 6
\end{array}
\right\}, \quad
\frac{C^{SRR}_{\nu edu,\beta\alpha 32}(\mu_B)}{C^{SRR}_{\nu edu,\beta\alpha 32}}
\approx \left\{
\begin{array}{*{2}{c}}
1.675, & \hat{m}_\phi = 2 \\ \relax
1.736, & \hat{m}_\phi = 4 \\ \relax
1.770, & \hat{m}_\phi = 6
\end{array}
\right\},
\\ \nonumber
&
\frac{C^{TRR}_{\nu edu,\beta\alpha 32}(\mu_B)}{C^{TRR}_{\nu edu,\beta\alpha 32}}
\approx \left\{
\begin{array}{*{2}{c}}
0.860, & \hat{m}_\phi = 2 \\ \relax
0.852, & \hat{m}_\phi = 4 \\ \relax
0.848, & \hat{m}_\phi = 6
\end{array}
\right\} \, ,
\end{align}
where the numerical values in brackets have been extracted using the \texttt{Wilson} package~\cite{Aebischer:2018bkb}.

\mathversion{bold}
\subsubsection{\texorpdfstring{$R(D^{(\star)})$}{RD(star)}}
\mathversion{normal}
\label{subsec:RDRDstar}

We define
\begin{align}
\label{RD_full}
G^D_\alpha \approx & \sum_{\beta=1}^3\Bigg(\left\{
\begin{array}{*{2}{c}}
0.500, & \alpha = 1 \\ \relax
0.500, & \alpha = 2 \\ \relax
1.000, & \alpha = 3
\end{array}
\right\}\left|(1+\delta)\cdot 2\sqrt{2}G_FV_{cb}\delta_{\alpha\beta} - \,C^{VLL}_{\nu edu,\beta\alpha 32}(\mu_B)\right|^2 \\
& \quad + \left\{
\begin{array}{*{2}{c}}
0.596 \\ \relax
0.593 \\ \relax
1.120
\end{array}
\right\}\left|C^{SRR}_{\nu edu,\beta\alpha 32}(\mu_B)\right|^2 + \left\{
\begin{array}{*{2}{c}}
0.272 \\ \relax
0.272 \\ \relax
0.662
\end{array}
\right\}\left|C^{TRR}_{\nu edu,\beta\alpha 32}(\mu_B)\right|^2 \nonumber\\
& \quad - \left\{
\begin{array}{*{2}{c}}
0.000 \\ \relax
0.079 \\ \relax
1.563
\end{array}
\right\}\text{Re}\left(\left((1+\delta)\cdot 2\sqrt{2}G_FV_{cb}\delta_{\alpha\beta} - \,C^{VLL}_{\nu edu,\beta\alpha 32}(\mu_B)\right)C^{SRR*}_{\nu edu,\beta\alpha 32}(\mu_B)\right) \nonumber\\
& \quad - \left\{
\begin{array}{*{2}{c}}
0.000 \\ \relax
0.084 \\ \relax
0.959
\end{array}
\right\}\text{Re}\left(\left((1+\delta)\cdot 2\sqrt{2}G_FV_{cb}\delta_{\alpha\beta} - \,C^{VLL}_{\nu edu,\beta\alpha 32}(\mu_B)\right)C^{TRR*}_{\nu edu,\beta\alpha 32}(\mu_B)\right)\Bigg)\nonumber
\end{align}
and
\begin{align}
\label{RDstar_full}
G^{D^\star}_\alpha \approx & \sum_{\beta=1}^3\Bigg(\left\{
\begin{array}{*{2}{c}}
0.501, & \alpha = 1  \\ \relax
0.499, & \alpha = 2  \\ \relax
1.000, & \alpha = 3 
\end{array}
\right\}\left|(1+\delta)\cdot 2\sqrt{2}G_FV_{cb}\delta_{\alpha\beta} - \,C^{VLL}_{\nu edu,\beta\alpha 32}(\mu_B)\right|^2 \\
& \quad + \left\{
\begin{array}{*{2}{c}}
0.039 \\ \relax
0.039 \\ \relax
0.053
\end{array}
\right\}\left|C^{SRR}_{\nu edu,\beta\alpha 32}(\mu_B)\right|^2 + \left\{
\begin{array}{*{2}{c}}
6.372 \\ \relax
6.364 \\ \relax
15.347
\end{array}
\right\}\left|C^{TRR}_{\nu edu,\beta\alpha 32}(\mu_B)\right|^2 \nonumber\\
& \quad - \left\{
\begin{array}{*{2}{c}}
0.000 \\ \relax
-0.012 \\ \relax
-0.139
\end{array}
\right\}\text{Re}\left(\left((1+\delta)\cdot 2\sqrt{2}G_FV_{cb}\delta_{\alpha\beta} - \,C^{VLL}_{\nu edu,\beta\alpha 32}(\mu_B)\right)C^{SRR*}_{\nu edu,\beta\alpha 32}(\mu_B)\right) \nonumber\\
& \quad - \left\{
\begin{array}{*{2}{c}}
-0.001 \\ \relax
-0.261 \\ \relax
-5.620
\end{array}
\right\}\text{Re}\left(\left((1+\delta)\cdot 2\sqrt{2}G_FV_{cb}\delta_{\alpha\beta} - \,C^{VLL}_{\nu edu,\beta\alpha 32}(\mu_B)\right)C^{TRR*}_{\nu edu,\beta\alpha 32}(\mu_B)\right)\Bigg)\;. \nonumber
\end{align}
Here, $\alpha$ ($\beta$) denotes the flavour of the charged lepton (neutrino) in the final state. The numbers in the first (second) [third] entry of the vectors in curly brackets encode the hadronic form factors employed by \texttt{flavio} \cite{Straub:2018kue,david_straub_2021_5543714,Bordone:2019vic} (since v2.0), and the integrated-out phase space for $\alpha = 1$ (2) [3]. These numbers can be compared to the ones that are found in reference~\cite{Fleischer:2021yjo}. The correction $\delta$, $\delta = 0.007$, accounts for QED running of the SM contribution to $C^{VLL}_{\nu edu}$ from the $Z$-boson mass scale down to the hadronic scale, $\mu = \mu_B = 4.8$ GeV. We employ the best-fit value for $V_{cb}$ from the PDG, $V_{cb} \approx 0.0405$~\cite{ParticleDataGroup:2020ssz}. These general formulae then constitute $R(D)$ and $R(D^\star)$
\begin{align}\label{eq:RDRDs}
\frac{R(D)}{R(D)_{\text{SM}}} = \frac{G^D_3}{G^D_2 + G^D_1}, \qquad \frac{R(D^\star)}{R(D^\star)_{\text{SM}}} = \frac{G^{D^\star}_3}{G^{D^\star}_2 + G^{D^\star}_1}\,.
\end{align}
Using the values $R(D)_{\text{SM}} = 0.297\pm0.008$ and $R(D^\star)_{\text{SM}} = 0.245\pm0.008$ given by \texttt{flavio}, v2.3, we find that the results obtained from the expressions above deviate from those obtained from \texttt{flavio} only by up to 0.5 percent in the ranges of $R(D^{(\star)})$ displayed in the plots.

\mathversion{bold}
\subsubsection{\texorpdfstring{$R_{D}^{\mu/e}$ and $R_{D^{\star}}^{e/\mu}$}{RDmue and RDemu(star)}}
\mathversion{normal}
\label{subsec:RDmueRDstaremu}
Similarly to $R(D^{(\star)})$ in appendix~\ref{subsec:RDRDstar}, these observables can be calculated using eqs.~\eqref{RD_full} and~\eqref{RDstar_full}, such that
\begin{align}
R^{\mu/e}_{D} = \frac{\Gamma(B\to D\mu\nu)}{\Gamma(B\to De\nu)}   = \frac{G^D_2}{G^D_1},
\end{align}
 and 
\begin{align}
R^{e/\mu}_{D^\star} = \frac{\Gamma(B\to D^\star e\nu)}{\Gamma(B\to D^\star \mu\nu)} = \frac{G^{D^\star}_1}{G^{D^\star}_2} \, .
\end{align}

\mathversion{bold}
\subsubsection{\texorpdfstring{Leptonic pseudoscalar meson decays $B_k\to \tau\nu$}{Leptonic pseudoscalar meson decays Bk -> tau nu}}
\mathversion{normal}
\label{app:pseudoscalarBk}

A pseudoscalar meson $B_k$, constituted by a bottom quark $b$ and an up-type quark $u_k$, decays into a tau lepton and a neutrino with a rate \cite{ParticleDataGroup:2020ssz,Gonzalez-Alonso:2016etj}
\begin{equation}\label{eq:LeptonicMesonDecay}
	\Gamma_{B_k\to\tau\nu} = \frac{G_F^2}{8\pi}m_{B_k}f_{B_k}^2|V_{u_kb}|^2m_\tau^2\left(1 - \frac{m_\tau^2}{m_{B_k}^2}\right)^2\sum_{\beta = 1}^3\left|(1+\delta)\cdot\delta_{3\beta} - \frac{1}{2\sqrt{2}G_FV_{u_kb}}C^\phi_{\nu edu,\beta 33k}(\mu_B)\right|^2
\end{equation}
where 
\begin{equation}\label{eq:LeptonicMesonDecayNP}
C^\phi_{\nu edu,\beta 33k}(\mu_B) = C^{VLL}_{\nu edu,\beta 33k}(\mu_B) - \frac{m_{B_k}^2}{m_\tau\big(m_{u_k}(\mu_B) + m_b(\mu_B)\big)}C^{SRR}_{\nu edu,\beta 33k}(\mu_B) \, .
\end{equation}
Here, $m_{B_k}$ and $f_{B_k}$ are the mass and decay constant of the meson, respectively. The correction $\delta = 0.007$ accounts for QED running of the SM contribution to $C^{VLL}_{\nu edu}$ from the $Z$-boson mass scale down to the hadronic scale, $\mu = \mu_B= 4.8$ GeV.

From eq.~(\ref{eq:LeptonicMesonDecay}), one may define
\begin{align}\label{eq:tauBcNPcontribution}
	\Gamma_{B_c}^\phi & = \frac{G_F^2}{8\pi}m_{B_c}f_{B_c}^2V_{cb}^2 \, m_\tau^2\left(1 - \frac{m_\tau^2}{m_{B_c}^2}\right)^2 \\
	& \quad \times
	\left(\sum_{\beta = 1}^3\left|(1+\delta)\cdot\delta_{3\beta} - \frac{1}{2\sqrt{2}G_FV_{cb}}C^\phi_{\nu edu,\beta 332}(\mu_B)\right|^2 - (1+\delta)^2\right)\nonumber 
\end{align}
which vanishes in the absence of contributions to $B_c\to\tau\nu$ from the LQ $\phi$. Then, rearranging eq.~(\ref{eq:fullBcLifetime}) yields the following inferred SM contribution to the $B_c$ lifetime
\begin{align}\label{eq:tauBcSM}
	\tau^\text{SM}_{B_c} & = \Bigg[\frac{1}{\tau^\text{exp}_{B_c}} - \frac{G_F^2}{8\pi}m_{B_c}f_{B_c}^2V_{cb}^2 \, m_\tau^2\left(1 - \frac{m_\tau^2}{m_{B_c}^2}\right)^2 \\
	& \quad \times \left(\sum_{\beta = 1}^3\left|(1+\delta)\cdot\delta_{3\beta} - \frac{1}{2\sqrt{2}G_FV_{cb}}C^\phi_{\nu edu,\beta 332}(\mu_B)\right|^2 - (1+\delta)^2\right)\Bigg]^{-1}.\nonumber
\end{align}
We require that the result for $\tau^\text{SM}_{B_c}$ lies in the interval $[0.4,0.7]$ ps, following the estimate in reference~\cite{Beneke:1996xe}, at the $1\,\sigma$ level, and neglect all other uncertainties against the broadness of this range. Furthermore, we use the PDG values $\tau^\text{exp}_{B_c} = 0.510$ ps, $m_{B_c} = 6.2745$ GeV, $m_\tau = 1.7769$ GeV, $V_{cb} \approx 0.0405$~\cite{ParticleDataGroup:2020ssz} as well as $f_{B_c} = 434$ MeV~\cite{Colquhoun:2015oha} and the quark masses $m_c(\mu_B) = 0.9023$ GeV and $m_b(\mu_B) = 4.0945$ GeV, as output by \texttt{flavio}, v2.3.

\mathversion{bold}
\subsubsection{\texorpdfstring{$B\to K^{(\star)}\nu\overline{\nu}$}{B-> K(*) nu nu}}
\mathversion{normal}

The BR of the decay $B\to K^{(\star)}$ plus missing energy is normalised to the SM prediction in the ratio $R^\nu_{K^{(\star)}}$. As $C^{VLR}_{\nu d}$ is not generated at one-loop order in the SM, the decay $B\to K^{(*)} \nu\bar\nu$ is dominated by $C^{VLL}_{\nu d,\alpha\beta 23}$. In contrast to the SM case, the flavours of the neutrinos do not have to coincide for the contribution due to the LQ $\phi$. Following reference~\cite{Buras:2014fpa}, we obtain
\begin{eqnarray}\label{eq:RKnu}
\begin{aligned}
R^\nu_{K^{(\star)}} = \frac{1}{3}\sum_{\alpha,\beta=1}^{3}\left|\delta_{\alpha\beta} + \frac{C^{VLL}_{\nu d,\alpha\beta23}}{C^{VLL}_{\nu d,23,\text{SM}}}\right|^2.
\end{aligned}
\end{eqnarray}
We use $C^{VLL}_{\nu d,23,\text{SM}} \approx (1.01-0.02 \, i)\times(10\,\text{TeV})^{-2}$ which is the value given by \texttt{flavio}, v2.3, see also~\cite{Brod:2010hi}, converted to the JMS basis and evaluated at the hadronic scale, $\mu = \mu_B= 4.8$ GeV.

\mathversion{bold}
\subsubsection{\texorpdfstring{Relevant Wilson coefficients for $b\to s e_i \overline{e_j} $}{Relevant Wilson coefficients for b to s ll'}}
\mathversion{normal}
\label{app:bsee}
The relevant Wilson coefficients for $b\to s e_i \overline{e_j}$ at one-loop order can be obtained from eq.~(A.6) in~\cite{Gherardi:2020qhc}
\begin{align}
	C^{VLL}_{ed,ij 23} & =-\frac{1}{64\pi^2 m_\phi^2} \sum_{m,n} x_{m2}^* x_{m3} z_{jn} z_{in}^*+\frac{\sqrt{2} G_F}{16\pi^2}  
	V_{ts}^* V_{tb} z_{j3} z_{i3}^* \,t_t \;,
	\\
	C_{de,23ij}^{VLR} & =-\frac{1}{64\pi^2 m_\phi^2}\sum_{m,n}x_{m2}^*x_{m3} y_{jn}y_{in}^* -\frac{\sqrt{2} G_F}{32\pi^2}  
	V_{ts}^* V_{tb} y_{j 3} y_{i3}^* \,t_t \left(\ln t_t+\frac32\right)
	\;.
\end{align}
Note that the contributions from the up and the charm quark have been neglected.

For lepton flavour conserving interactions, we have to additionally consider the down-type quark dipole operator
\begin{align}
	\mathcal{L} & \supset C_{d\gamma}^{ij} (\overline{d_i}\sigma_{\alpha\beta} P_R d_j) F^{\alpha\beta} + \mathrm{h.c.}
\end{align}
with the Wilson coefficients
\begin{align}
	C_{d\gamma}^{23} &= -\frac{e\, m_b}{576\pi^2 } \sum_m \frac{x_{m2}^*x_{m3} }{m_\phi^2}\;,
	&
	C_{d\gamma}^{32} &= -\frac{e\, m_s}{576 \pi^2} \sum_m\frac{x_{m2} x_{m3}^*} {m_\phi^2}\;.
\end{align}
These calculated contributions are used in the numerical evaluation of the tertiary constraints in the comprehensive scan.

\mathversion{bold}
\subsubsection{\texorpdfstring{$\mu-e$ conversion rate}{mu e conversion rate}}
\mathversion{normal}
\label{app:mue_conversion}

Not taking into account next-to-leading-order corrections (in the loop expansion), we obtain at one-loop order the following contributions in the limit of vanishing external masses and momenta in addition to the tree-level contributions discussed above.
The short-distance $\gamma$-penguin contributions result in
\begin{align}
	C_{eq,ijkk}^{VLL,\gamma} = C_{eq,ijkk}^{VLR,\gamma} & = -\frac{Q^{q} e^2}{96\pi^2 m_\phi^2} \sum_m z_{im}^* z_{jm}\left(5+4\ln t_{u_m}\right)
	\;,
	\\
	C_{eq,ijkk}^{VRR,\gamma} = C^{VLR,\gamma}_{qe,kkij} & = -\frac{Q^{q}e^2}{96\pi^2 m_\phi^2} \sum_m y_{im}^* y_{jm}\left(5+4\ln t_{u_m}\right)
\;,
\end{align}
where $Q^q$ denotes the electric charge of the quark.
The $Z$-penguin diagrams generate the Wilson coefficients
\begin{align}
	C_{eq,ijkk}^{VLL,Z} & = -\frac{6\sqrt{2} G_F(T_3-Q^{q}s_W^2)}{16\pi^2} \sum_m z_{im}^* z_{jm} t_{u_m} \left(1+\ln t_{u_m}\right)\;,
	\\
	C_{eq,ijkk}^{VLR,Z} & = \frac{6\sqrt{2} G_F Q^{q} s_W^2}{16\pi^2} \sum_m z_{im}^* z_{jm} t_{u_m} \left(1+\ln t_{u_m}\right)\;,
	\\
	C^{VLR,Z}_{qe,kkij} & = \frac{6\sqrt{2} G_F (T_3-Q^{q}s_W^2)}{16\pi^2} \sum_m y_{im}^* y_{jm} t_{u_m} \left(1+\ln t_{u_m}\right)\;,
	\\
	C_{eq,ijkk}^{VRR,Z} & = -\frac{6\sqrt{2} G_F Q^{q}s_W^2}{16\pi^2} \sum_m y_{im}^* y_{jm} t_{u_m} \left(1+\ln t_{u_m}\right)
	\;.
\end{align}
For up-type quarks, there are no contributions from box diagrams. For down-type quarks, there are only box contributions to vector operators. Thus for $\mu-e$ conversion the only relevant contribution is to vector operators with down quarks which, neglecting all Yukawa couplings apart from the one of the top quark, $y_t$, are given by
\begin{align}
	C_{ed,ijkk}^{VLL,\rm box} =&
	\sum_{m,n}\frac{|x_{nk}|^2z_{im}^*z_{jm}}{64\pi^2 m_\phi^2}
	+\frac{\sqrt{2} G_F x_{ik}^* x_{jk} }{16\pi^2}t_W \ln t_W
	-\frac{|V_{td_k}|^2 y_t^2 z_{j3} z_{i3}^* }{32 \pi^2 m_\phi^2}\left[\frac{1}{t_t-t_W}+\frac{t_W\ln \frac{t_W}{t_t}}{(t_t-t_W)^2} \right]
	\\\nonumber&
	-\frac{\sqrt{2} G_F }{16\pi^2}
\sum_m(z_{im}^* x_{jk}V_{u_md_k}^*
+x_{ik}^* z_{jm}V_{u_md_k})
\frac{t_W(t_W\ln t_W -t_{u_m} \ln t_{u_m})}{t_W-t_{u_m}}
	\\\nonumber &
	+\frac{\sqrt{2} G_F }{16\pi^2}\sum_{m,n}z_{in}^* z_{jm}V_{u_m d_k} V_{u_nd_k}^*
	\\\nonumber&
	\qquad\qquad\times
	\left[\frac{t_W^3 \ln t_W}{(t_{u_m}-t_W)(t_{u_n}-t_W)}
		+\frac{t_W t_{u_m}^2 \ln t_{u_m}}{(t_W-t_{u_m})(t_{u_n}-t_{u_m})}
		+\frac{t_W t_{u_n}^2 \ln t_{u_n}}{(t_W-t_{u_n})(t_{u_m}-t_{u_n})}
	\right]
\end{align}
\begin{align}
	C_{de,kkij}^{VLR,\rm box}  =&
	\sum_{m,n}\frac{|x_{nk}|^2 y_{im}^* y_{jm}}{64 \pi^2 m_\phi^2}
	-\frac{|V_{td_k}|^2 y_t^2 y_{j3} y_{i3}^*}{64 \pi^2 m_\phi^2}\left[ \frac{t_t}{t_W-t_t}-\frac{t_W^2\ln t_W}{(t_W-t_t)^2} + \frac{(2t_W-t_t)t_t \ln t_t}{(t_W-t_t)^2}\right]
	\\\nonumber &
	-\sum_{m,n}\frac{ \sqrt{2} G_F V_{u_md_k} V_{u_nd_k}^* y_{in}^*y_{jm} t_W t_{u_m}^{1/2} t_{u_n}^{1/2}}{8\pi^2}
	\\\nonumber &
	\qquad\qquad\times
	\left[
		\frac{t_W\ln t_W}{(t_W-t_{u_m})(t_W-t_{u_n})}
	+
	\frac{t_{u_m} \ln t_{u_m}}{(t_W-t_{u_m})(t_{u_n}-t_{u_m})}
	+
	\frac{t_{u_n}\ln t_{u_n}}{(t_W-t_{u_n})(t_{u_m}-t_{u_n})}
	\right]\;.
\end{align}
The $\mu-e$ CR can be obtained from the effective Lagrangian following reference~\cite{Kitano:2002mt}
\begin{align}\label{eq:mueconvrate}
	\omega_{\rm conv} &=
	\left|-\frac{C_{e\gamma,12}}{2 m_\mu} D + \tilde g_{LS}^{(p)} S^{(p)} + \tilde g_{LV}^{(p)} V^{(p)} + (p\to n)\right|^2
	+ \left|- \frac{C_{e\gamma,21}^*}{2m_\mu} D + \tilde g_{RS}^{(p)} S^{(p)} + \tilde g_{RV}^{(p)} V^{(p)} + (p\to n)\right|^2
\end{align}
with the effective coupling constants
\begin{align}
\tilde g_{LS}^{(N)} & =   \sum_i G_{S}^{q_i,N} \left(C_{eq,12ii}^{SRR} + C_{eq,12ii}^{SRL}\right)
			       \\
\tilde g_{RS}^{(N)} & =  \sum_i G_{S}^{q_i,N} \left(C_{eq,21ii}^{SRR*} + C_{eq,21ii}^{SRL*}\right)
\\
\tilde g_{LV}^{(p)} & =2\left(C_{eu,1211}^{VLL} +C_{eu,1211}^{VLR}\right)+\left(C_{ed,1211}^{VLL} +C_{ed,1211}^{VLR}\right)   
			       \\
\tilde g_{RV}^{(p)} & = 2 \left(C_{eu,1211}^{VRR} + C_{ue,1112}^{VLR}\right)
+ \left(C_{ed,1211}^{VRR} + C_{de,1112}^{VLR}\right)
\\
	\tilde g_{LV}^{(n)} & =\left(C_{eu,1211}^{VLL} +C_{eu,1211}^{VLR}\right)+2\left(C_{ed,1211}^{VLL} +C_{ed,1211}^{VLR}\right)   
			       \\
\tilde g_{RV}^{(n)} & =  \left(C_{eu,1211}^{VRR} + C_{ue,1112}^{VLR}\right)
+2 \left(C_{ed,1211}^{VRR} + C_{de,1112}^{VLR}\right)
\end{align}
with $N=p,n$. In the numerical analysis we use the nuclear form factors $G_S^{q_i,N}$ given in reference~\cite{Kosmas:2001mv} and the overlap integrals $D$, $S^{(N)}$ and $V^{(N)}$ and capture rates $\omega_{\rm capt}$ presented in reference~\cite{Kitano:2002mt}.

%%%%%%%%%%%%%%%%%%%%%%%%%%%%%%%%%%%%%%%%%%%%%%%%%%%%%%
\mathversion{bold}
\subsection{\texorpdfstring{$Z$ decays to fermions}{Z  decays to fermions}}
\mathversion{normal}
\label{app:Zdecays}
%%%%%%%%%%%%%%%%%%%%%%%%%%%%%%%%%%%%%%%%%%%%%%%%%%%%%%   
For calculating the contributions to leptonic $Z$ decays due to the LQ $\phi$, we follow the procedure of reference~\cite{Arnan:2019olv}. To parametrise these effects, we consider the effective Lagrangian for the $Z$ boson interaction with an SM fermion $f_i$
\begin{align}
  \mathcal{L}_{\text{eff}}^Z = \frac{g}{\cos\theta_W} \sum_{ i, j} \overline{f_i} \gamma^\mu \left[ g^{ij}_{f_L} P_L + g^{ij}_{f_R} P_R\right] f_j Z_\mu \, ,
\end{align}
where $g$ is the SU(2) gauge coupling, and 
\begin{align}
    g_{f_{L(R)}}^{ij}=  g^{\text{SM}}_{f_{L(R)}} \delta^{ij}+ \delta g^{ij}_{f_{L(R)}}.\label{definedcoup}
\end{align}
At tree level, the SM effective couplings are given by 
\begin{align}
g^{0}_{f_{L}}= T_3^f- Q^f \sin^2 \theta_W,\hspace{2cm}  g^{0}_{f_{R}}=- Q^f \sin^2 \theta_W,
\end{align}  where $Q^f$ is the electric charge of the fermion $f$, and $T^f_3$ is its third component of weak isospin.

For the remainder of this appendix, we focus on the interactions with charged leptons, i.e. $f_i=e_i$. At higher loop order in the SM, these couplings are modified by factors $\rho_f=1.00937$ and $\sin^2 \theta_\text{eff}=0.231533$~\cite{ParticleDataGroup:2020ssz}, 
\begin{align}
g^\text{SM}_{f_{L}}= \sqrt{\rho_f} \, (T_3^f- Q^f \sin^2 \theta_\text{eff}), \hspace{2cm} g^\text{SM}_{f_{R}}=- \sqrt{\rho_f} \, Q^f \sin^2 \theta_\text{eff}. \label{effectivecouplingsZ}
\end{align}
The contributions to the effective couplings for $Z\to f \bar{f}$ are calculated in general for scalar LQ models in reference~\cite{Arnan:2019olv}. We refrain from detailing these results here, but instead recast the dominant contributions in the context of this model. 

For charged leptons, to contrast with existing constraints, we note the relation of the effective couplings $g_{f_{L(R)}}^{ij}$ to those for vector and axial-vector interactions
\begin{align}
    g_{e_{V(A)}}^{ij}=g_{e_L}^{ij}\pm g_{e_R}^{ij}.
    \end{align}
In this model, charged leptons couple solely to up-type quarks and an enhancement via the top quark mass yields the following dominant contribution
\begin{align}
   \delta g_{e_{A(V)}}^{ii} = \delta g_{{e_i}_{A(V)}} \approx& \frac{N_c}{32 \pi^2}  \frac{t_t (t_t-1-\ln t_t)}{(t_t-1)^2}\left(|z_{i3}|^2 \pm |y_{i3}|^2 \right). 
   \label{deltag_ZZ}
   \end{align}
Note that the dependence of the SM value on $\sin^2 \theta_\text{eff}$ motivates the consideration of the future sensitivity of collider experiments, as listed in table~\ref{table:secondaryconstraints}. Prospective sensitivities are quoted from reference~\cite{Crivellin:2020mjs}, where they have assumed that the measurements of $g_{{e_i}_{A}}$ are improved by the same factor as $\sin^2 \theta_\text{eff}$, and $g_{{e_i}_{A}}$ provides the more sensitive probe to new physics than $g_{{e_i}_{V}}$.

 %%%%%%%%%%%%%%%%%%%%%%%%%%%%%%%%%%%%%%%%%%%%%%%%%%%%%%
 \section{Supplementary information for section~\ref{sec:primary}}
 \label{app:plots_wilsoncoeff}
 
 \begin{figure}[pht!]
 	\includegraphics[scale=0.5]{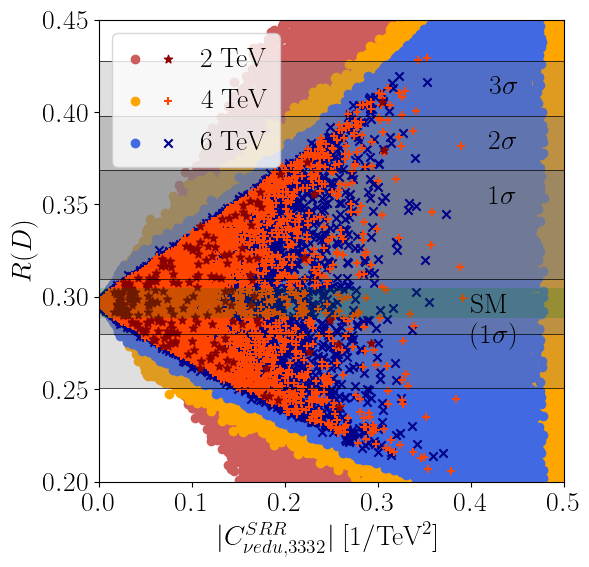} \hspace{1.1cm}
 	\includegraphics[scale=0.5]{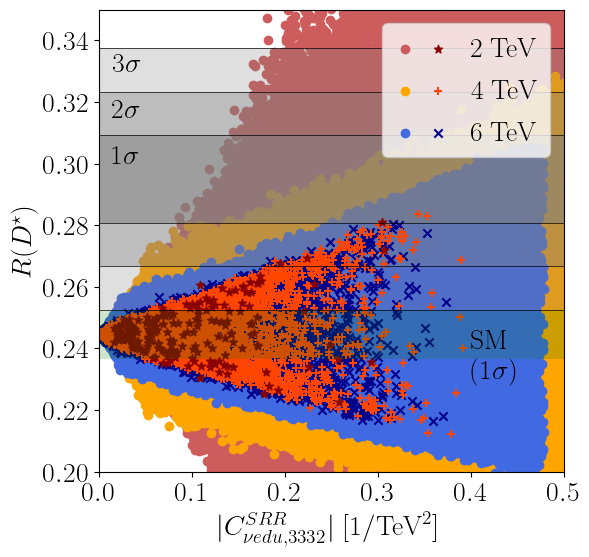}
 	\\
 	\includegraphics[scale=0.5]{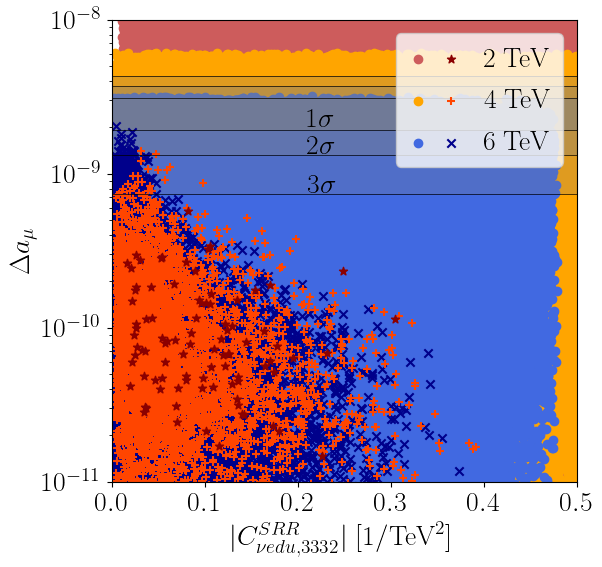}
 	\includegraphics[scale=0.5]{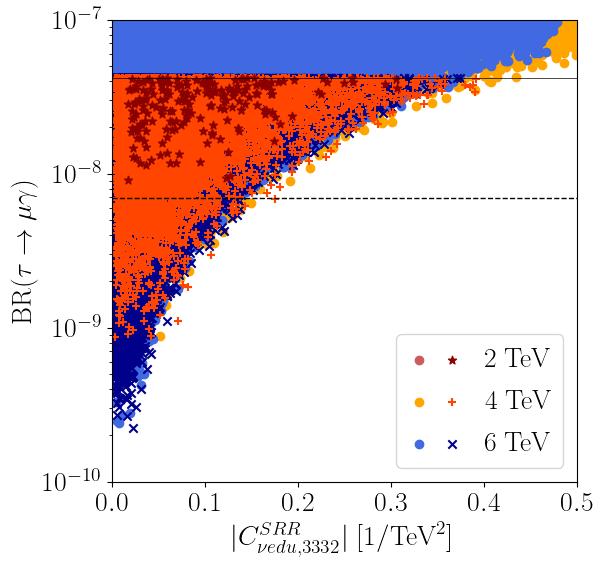}
 	\begin{minipage}[c]{0.5\textwidth}
 		\includegraphics[scale=0.5]{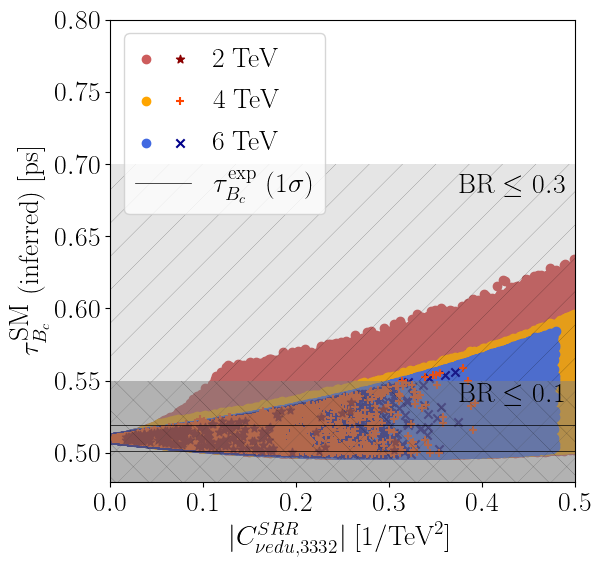}
 	\end{minipage}\hfill
 	\begin{minipage}[c]{0.5\textwidth}
 		\caption{\linespread{1.1}\small{\textbf{\mathversion{bold}Current constraints on and future sensitivity to $|C^{SRR}_{\nu edu,3332}|$\mathversion{normal}.} The Wilson coefficient is computed according to eqs.~(\ref{eq:treeCC}) and (\ref{eq:treeCCrunning}), and the displayed results hold at the hadronic scale, $\mu = \mu_B= 4.8$ GeV. The plots reflect the findings of the primary scan, discussed in section~\ref{sec:primary}. The round points (geometric shapes) indicate that current experimental bounds are violated (respected), see also the main text of section~\ref{subsubsec:primary_preliminaries}.}
 		\label{fig:wilson_coeff}}
 	\end{minipage}
 \end{figure}
In the following, we briefly discuss current constraints on and the projected sensitivity of future experiments to contributions to the magnitude of the Wilson coefficient $C^{SRR}_{\nu edu,3332}$. It constitutes the dominant contribution to the observables $R(D)$, $R(D^\star)$ and $\tau^{\text{SM}}_{B_c}$ in this model. According to eqs.~(\ref{eq:treeCC}) and (\ref{eq:treeCCrunning}), we find $C^{SRR}_{\nu edu,3332} \approx-1.7 \, x_{33} \, y_{32}/(2 \, m_\phi^2) \approx-1.7 \, a_{33} \, b_{32}/(2 \, m_\phi^2)$ at the hadronic scale, $\mu = \mu_B= 4.8$ GeV. 
Note the following statements are directly inferred from the primary scan which is discussed in section~\ref{sec:primary}. A comparison with the comprehensive scan, see section~\ref{sec:secondarytertiary}, only reveals small deviations from the results described below for $\hat{m}_\phi = 2$.
 
As can be seen in the top in figure~\ref{fig:wilson_coeff}, the achievable deviation of $R(D)$ and $R(D^\star)$ from their respective SM values grows linearly with an increase of the magnitude of $C^{SRR}_{\nu edu,3332}$.
Only for $|C^{SRR}_{\nu edu,3332}|\gtrsim 0.2/\text{TeV}^2$, a slight deviation from this trend becomes visible. This confirms that the contributions to $R(D)$ and $R(D^\star)$ which are linear in the Wilson coefficient, since they arise from the interference with the SM contribution, see eqs.~(\ref{eq:RD_estimate}) and (\ref{eq:RDs_estimate}), dominate for smaller values of the LQ couplings. These plots also conveniently illustrate that the anomaly is mainly driven by the experimental data for $R(D^\star)$, that is, explaining $R(D^\star)$ at the $2\, \sigma$ $(1\,\sigma)$ level requires $|C^{SRR}_{\nu edu,3332}|\gtrsim 0.2 \, (0.3)/\text{TeV}^2$.

The centre-left plot in figure~\ref{fig:wilson_coeff} evidences that a correlation between the AMM of the muon, $\Delta a_\mu\propto |b_{23}c_{23}|$, and the Wilson coefficient, $|C^{SRR}_{\nu edu,3332}| \propto |a_{33}b_{32}|$, only arises after imposing the bound on BR$(\tau\to\mu\gamma)$, BR$(\tau\to\mu\gamma) \propto|b_{23}c_{33}|^2\approx|b_{23}a_{33}|^2$. Indeed, the current constraint requires that $|C^{SRR}_{\nu edu,3332}|\lesssim 0.4/\text{TeV}^2$, and the upcoming search for this process at Belle II~\cite{Belle:2021ysv} can strengthen this to $|C^{SRR}_{\nu edu,3332}|\lesssim 0.15/\text{TeV}^2$, see centre-right plot. Note that an efficient test of the capability of the model to explain the AMM of the muon still requires a further refinement of that bound, as is visible in the centre-left plot.

Lastly, one can see that the inferred value of $\tau^{\text{SM}}_{B_c}$ is slightly less sensitive to $|C^{SRR}_{\nu edu,3332}|$ than $R(D)$ or $R(D^\star)$ are.
The distribution of generated sample points for $\hat{m}_\phi = 2$ features a kink which is localised at the upper boundary of the coloured region at $|C^{SRR}_{\nu edu,3332}|\approx 0.13/\text{TeV}^2$, due to the experimental constraint on $b_{32}$, $|b_{32}| < 2.6$, see table~\ref{table:primaryconstraints}. Note that $|C^{SRR}_{\nu edu,3332}|\gtrsim 0.3/\text{TeV}^2$ is necessary to have BR($B_c\to\tau\nu$) exceed approximately $0.1$.

\section{Supplementary information for section~\ref{sec:secondarytertiary}}
\label{app:supp6}

\subsection{Details of method of comprehensive scan}
\label{app:supp6_scanmethod}

In the following, we present details of how the comprehensive scan has been implemented. First note that, although the LQ couplings are sampled in the interaction basis, they are input to \texttt{SPheno} in the charged fermion mass basis, which avoids modifying the hard-coded fermion masses in \texttt{SPheno}. As such, we use the unitary matrices $L_d$, $R_d$, $L_e$, $R_e$, $L_u$ and $R_u$, extracted from the chi-squared fit discussed in section~\ref{ssec:cferm}, to perform this basis transformation. As mentioned in section~\ref{subsubsec:p_ana_AMMamu}, the correction to the muon mass arising from LQ contributions could be corrected for by appropriately redefining the effective parameter $e_{22}$ in the charged lepton mass matrix $M_e$, see eqs.~(\ref{eq:Mepara}) and (\ref{eq:memasses}). Nevertheless, since this redefinition has hardly any effect on the form of the unitary matrices $L_e$ and $R_e$, see analytic expressions in eqs.~(\ref{eq:Leform}) and (\ref{eq:Reform}), it is neglected throughout the scan. 

Furthermore, we notice that we implement the model in the comprehensive scan in a simplified version, considering only one SM-like Higgs doublet that gives masses to all charged fermions. As explained in section~\ref{sec:setup}, the main reason for having two Higgs doublets, $H_u$ and $H_d$, is to facilitate the search for a suitable flavour symmetry. The existence of these two Higgs doublets is, however, not relevant for the explanation of the flavour anomalies, observed in $R(D)$, $R(D^\star)$ and in the AMM of the muon. As a consequence, the suppression of the down-type quark masses and of the charged lepton masses is no longer due to the VEV of $H_d$ being much smaller than that of $H_u$, compare eq.~(\ref{eq:HdHuVEVsapprox}), but becomes encoded in the effective parameters $d_{ij}$ and $e_{ij}$, that must be appropriately rescaled. Such a rescaling only changes the magnitudes of these parameters, but not the results for the unitary matrices $L_d$, $R_d$, $L_e$, $R_e$, $L_u$ and $R_u$, since the latter contain ratios of $d_{ij}$, $e_{ij}$ and $f_{ij}$, respectively. Therefore, this simplification has no impact on the calculated LQ couplings ${\bf x}$, ${\bf y}$ and ${\bf z}$. In addition, considering only one SM-like Higgs doublet allows us to simplify the implementation of this model with the computational tools employed. 

\begin{table}[bt!]
    \centering
\resizebox{11cm}{!}{
\renewcommand{\arraystretch}{1.6}
    \centering
    \begin{tabular}{|c|l|l|l|l|l|l|}
    \hline
    \multicolumn{7}{|c|}{ \textsc{Spread of unhatted LQ couplings in comprehensive scan}} \\
    \hline
\multirow{2}{*}{Parameter} &\multicolumn{2}{c|}{$\hat{m}_\phi=2$}  &\multicolumn{2}{c|}{$\hat{m}_\phi=4$} &\multicolumn{2}{c|}{$\hat{m}_\phi=6$} \\
\cline{2-7}
  & [min., max.] & Average & [min., max.] & Average & [min., max.] & Average \\
  \hline 
 $|a_{11}|$ & $[0.23, 4.41] $&$2.32$& $[0.23, 4.41]$&$2.33$& $[0.23, 4.41]$&$ 2.33$  \\
 $|a_{12}|$ & $[0.01, 5.70]$&$ 0.75$& $[0.003, 7.59]$&$1.09$& $[0.001, 8.62]$&$1.03$  \\
  $|a_{13}|$ & $[0.001, 2.01]$&$ 0.19$& $[0.001, 2.63]$&$ 0.53$& $[0.002, 2.73]$&$ 0.51$  \\
   $|a_{21}|$ & $[0.08, 17.1]$&$ 3.12$& $[0.03, 23.9]$&$ 4.27$& $[0.03, 23.1]$&$ 4.10$  \\
 $|a_{22}|$ & $[0.05, 10.6]$&$2.12$& $[0.03, 14.2]$&$ 3.27$& $[0.01, 15.8]$&$ 3.21$  \\
  $|a_{23}|$ &   $[0.23, 4.40]$&$ 0.91$& $[1.60, 4.40]$&$ 2.34$& $[1.40, 4.40]$&$ 2.30$  \\
   $|a_{31}|$ &  $[0.11, 38.3]$&$ 8.92$& $[0.07, 45.8]$&$ 10.2$& $[0.02, 59.8]$&$ 11.5$  \\
 $|a_{32}|$ &  $[0.06, 5.31]$&$ 2.20$& $[0.02, 6.55]$&$ 2.41$& $[0.005, 8.94]$&$ 2.61$  \\
  $|a_{33}|$ &  $[0.05, 0.73]$&$ 0.37$& $[0.02, 1.90]$&$ 0.85$& $[0.05, 3.62]$&$ 1.58$  \\
  \hline 
   $|b_{11}|$ & $[0.22, 4.43]$&$ 2.33$& $[0.22, 4.43]$&$ 2.31$& $[0.21, 4.44]$&$ 2.32$  \\
 $|b_{12}|$ &   $[0.07, 48.1]$&$ 10.8$& $[0.12, 65.0]$&$ 13.1$& $[0.09, 69.2]$&$ 13.2$  \\
  $|b_{13}|$ &  $[0.007, 1.67]$&$ 0.43$& $[0.006, 0.70]$&$ 0.33$& $[0.004, 1.57]$&$ 0.65$  \\
   $|b_{21}|$ &  $[0.02, 13.3]$&$ 3.58$& $[0.02, 19.9]$&$ 3.70$& $[0.03, 19.6]$&$ 4.06$  \\
 $|b_{22}|$ & $[0.01, 11.2]$&$ 3.63$& $[0.02, 15.5]$&$ 3.97$& $[0.01, 15.5]$&$ 4.20$  \\
  $|b_{23}|$ & $[0.15, 0.80]$&$ 0.31$& $[0.18, 1.84]$&$ 0.38$& $[0.15, 3.48]$&$ 0.43$  \\
   $|b_{31}|$ & $[0.21, 4.07]$&$1.63$& $[0.17, 6.08]$&$ 1.99$& $[0.15, 6.34]$&$ 2.10$  \\
 $|b_{32}|$ & $[1.10, 2.60]$&$ 1.70$& $[1.00, 4.50]$&$2.23$& $[0.80, 4.50]$&$ 2.27$  \\
  $|b_{33}|$ & $[0.03, 12.9]$&$2.68$& $[0.01, 11.1]$&$ 2.25$& $[0.02, 13.5]$&$ 3.02$  \\
 \hline
    \end{tabular}}
    \caption{\small {\bf{Spread of magnitudes of unhatted LQ couplings for viable points in comprehensive scan.}}  Recall that the effective parameters $c_{ij}$ are related to $a_{ij}$ via the CKM mixing matrix. } \label{tab:comprehensiveCoupMag}
\end{table}

We proceed as follows with sampling over the parameter space consistently with the biasing, discussed in section~\ref{ssec:prelimBias}
\begin{enumerate}
    \item Sample $\hat{a}_{33}$, $\hat{b}_{32}$ and $\hat{b}_{23}$ using the biases for the effective parameters in the charged fermion mass basis, see table~\ref{tab:pc1} and eq.~\eqref{eq:b23fromtmg}. Sample all other effective parameters in the interaction basis, including
    $\hat{a}_{23}$ and $\hat{b}_{13}$, with flat priors within the ranges specified in eqs.~\eqref{eq:unbiased_magnitude} and~\eqref{eq:unbiased_phase};
    \item Transform these parameters into the ones in the charged fermion mass basis using the unitary matrices $L_d$, $R_d$, $L_e$, $R_e$, $L_u$ and $R_u$, extracted from the chi-squared fit discussed in section~\ref{ssec:cferm};
    \item Check that the generated values of the effective parameters $a_{23}$ and $b_{13}$ satisfy eq.~\eqref{eq:b13frommeg}, and that $a_{23}$ and the cosine of the arguments of the latter and the effective parameter $b_{23}$ are within the prescribed regions in table~\ref{tab:pc1};
    \item If any of the checks in step 3 fails, return to step 1; otherwise, a valid set of effective parameters is found.
\end{enumerate}
The distribution of the magnitudes of the effective parameters in the charged fermion mass basis, output from the comprehensive scan, is summarised in table~\ref{tab:comprehensiveCoupMag}.
\subsection{Additional plots}
\label{app:addfigssec6}

\begin{figure}[b!]
\centering
	\includegraphics[width=0.49\textwidth]{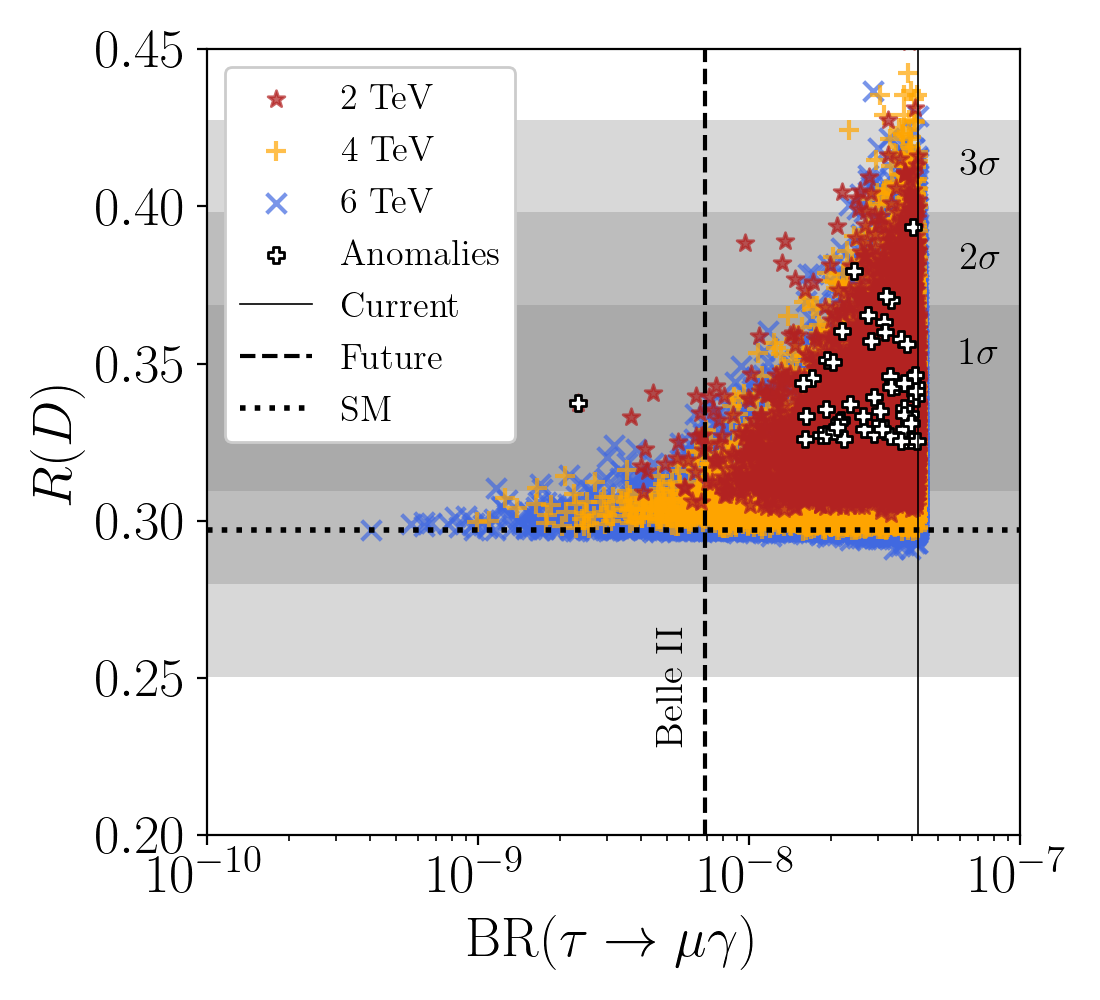}
	\includegraphics[width=0.49\textwidth]{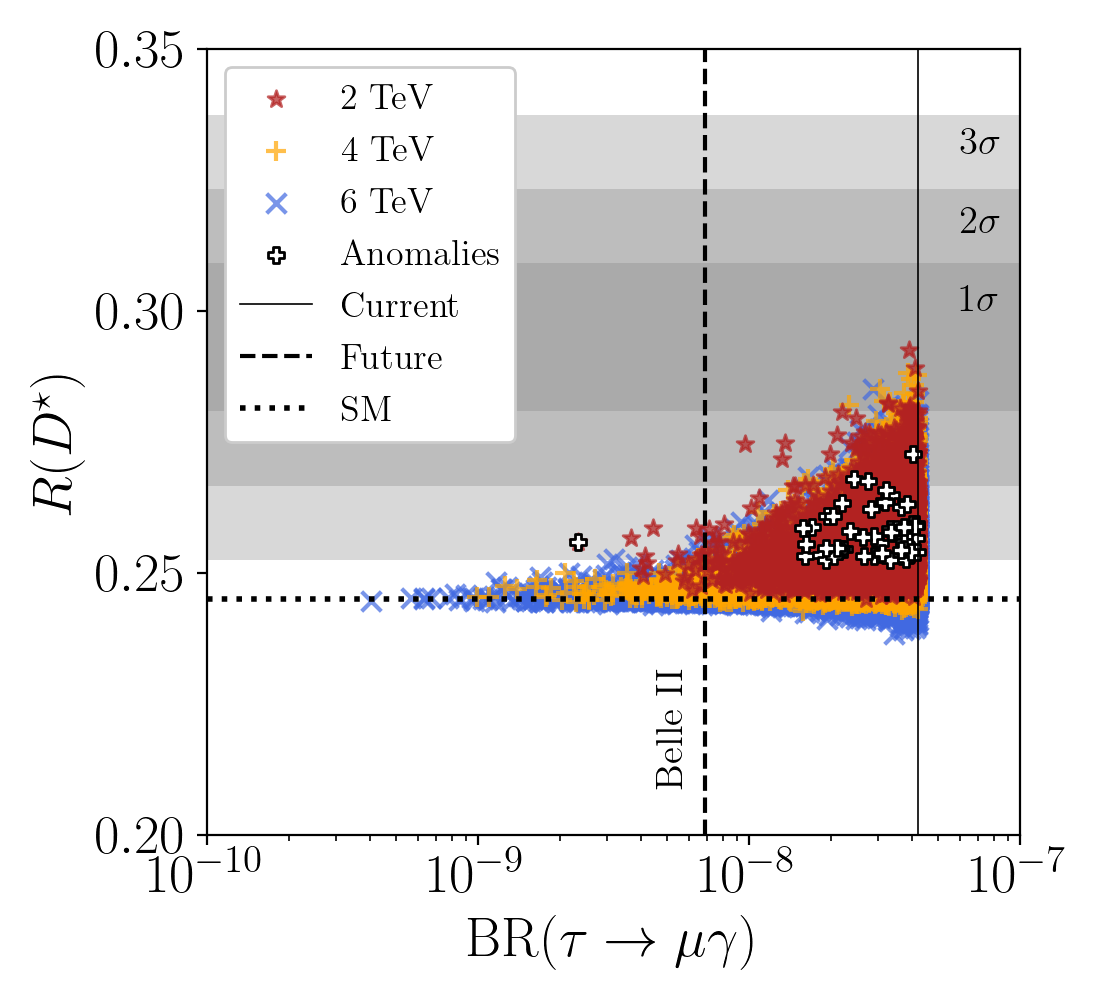}\\
	\includegraphics[width=0.49\textwidth]{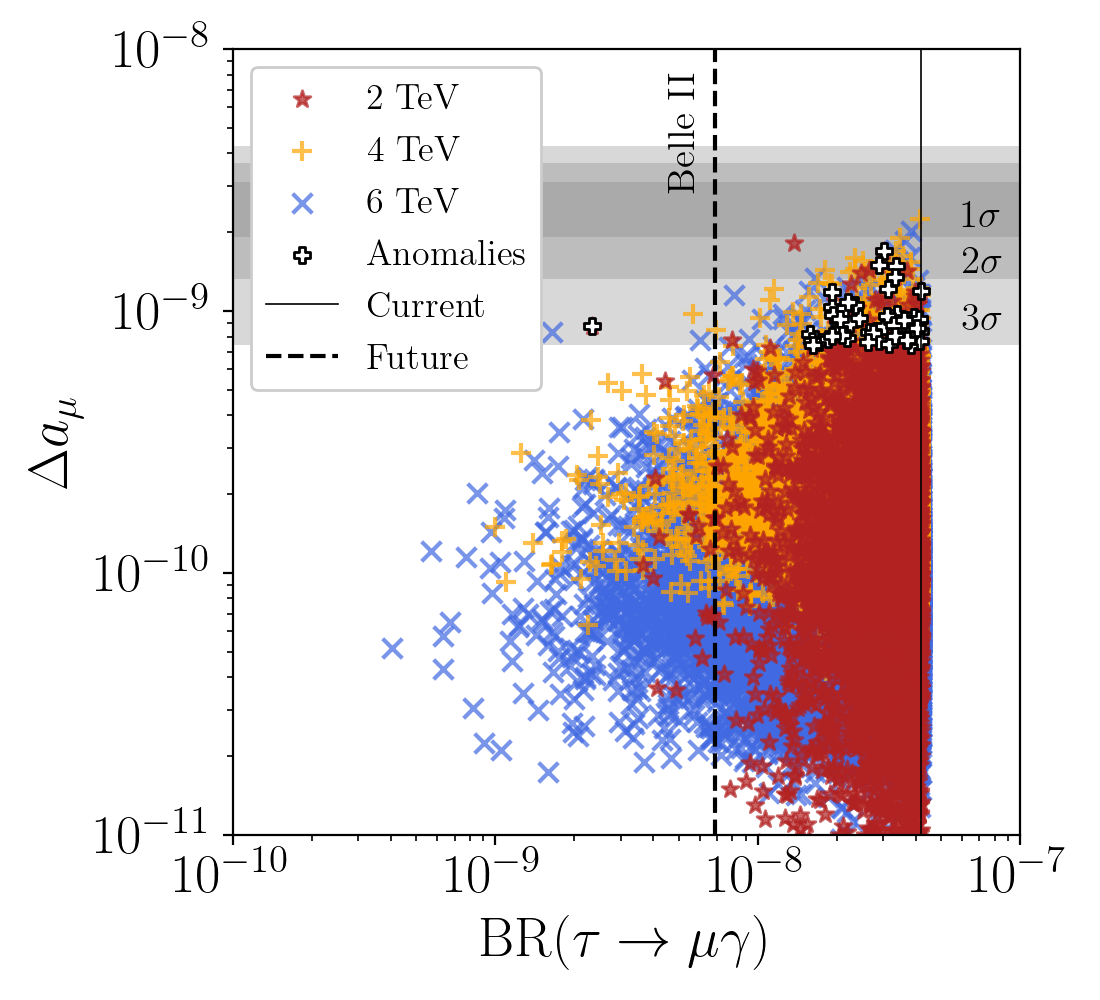}
	\includegraphics[width=0.49\textwidth]{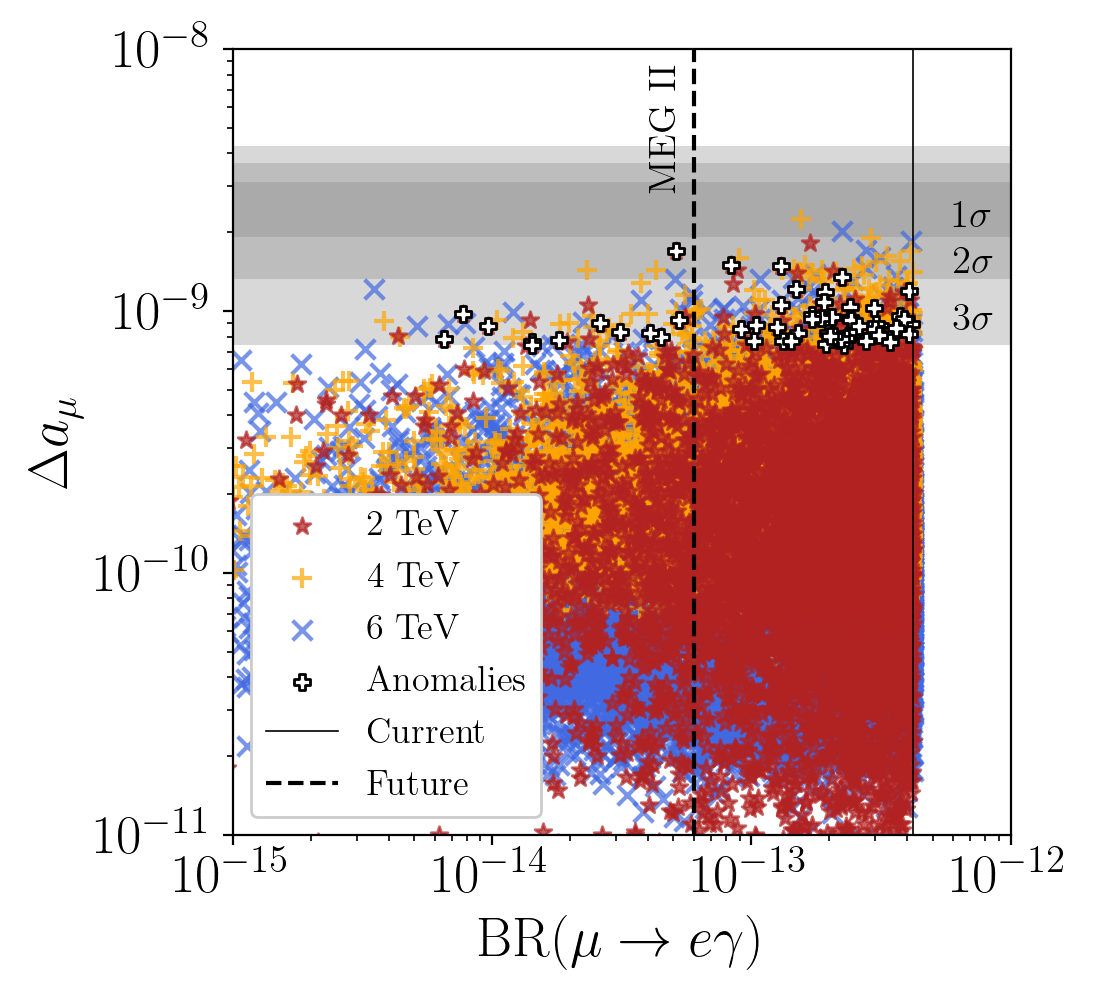}
	\caption{\linespread{1.1}\small{\textbf{Constraining power and future reach of radiative cLFV decays in comprehensive scan.} These plots show the results for the BRs of the radiative cLFV decays $\tau\to \mu \gamma$ and $\mu\to e \gamma$, plotted against the anomalous observables $R(D)$, $R(D^\star)$ and the AMM of the muon. They can be compared with the corresponding plots for the primary scan, displayed in figure~\ref{fig:constrain_power_taumug} in the main text. For further information on how to read this figure, see section~\ref{ssec:comprehensivePlots}.}}
	\label{fig:CompResultsCLFV}
\end{figure}

\begin{figure}[t!]
\centering
  \includegraphics[width=0.49\textwidth]{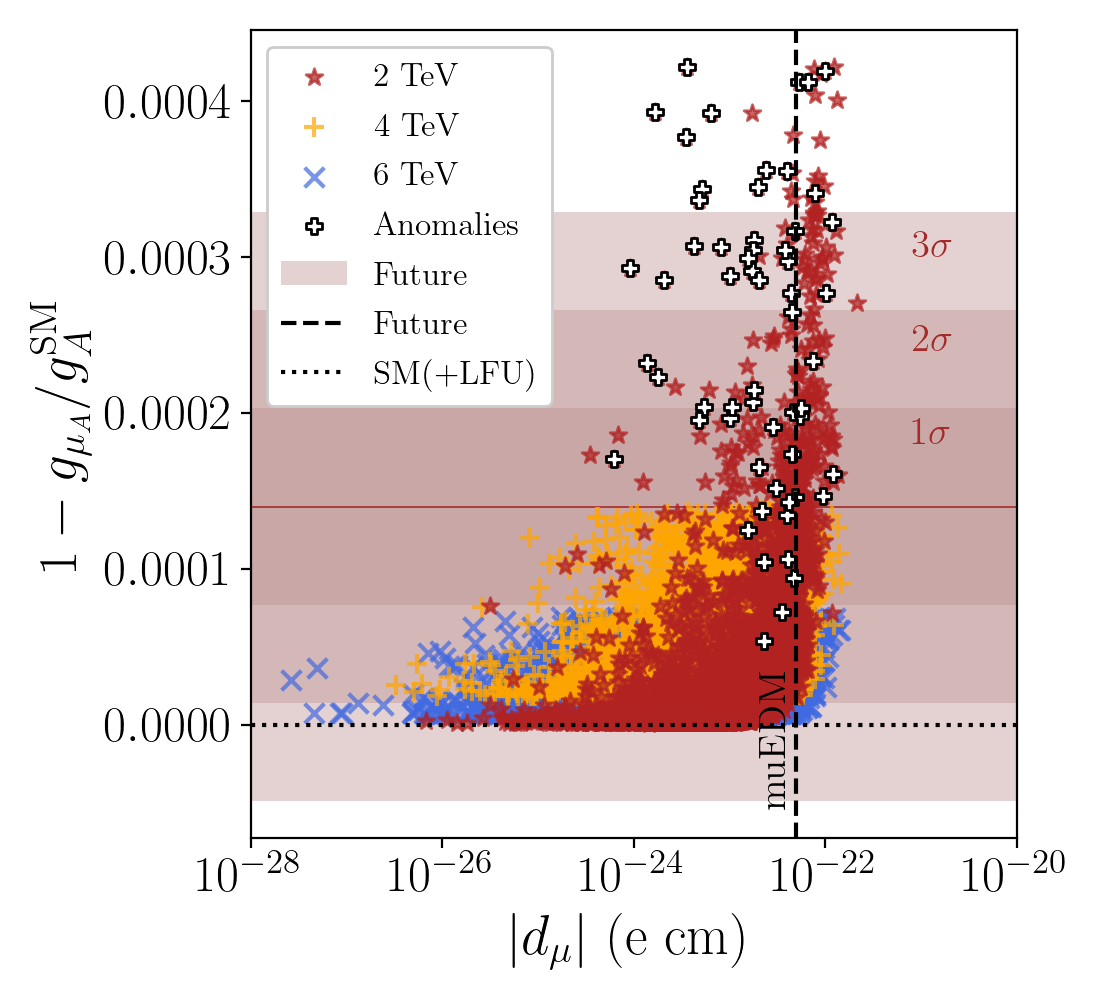}
    \includegraphics[width=0.49\textwidth]{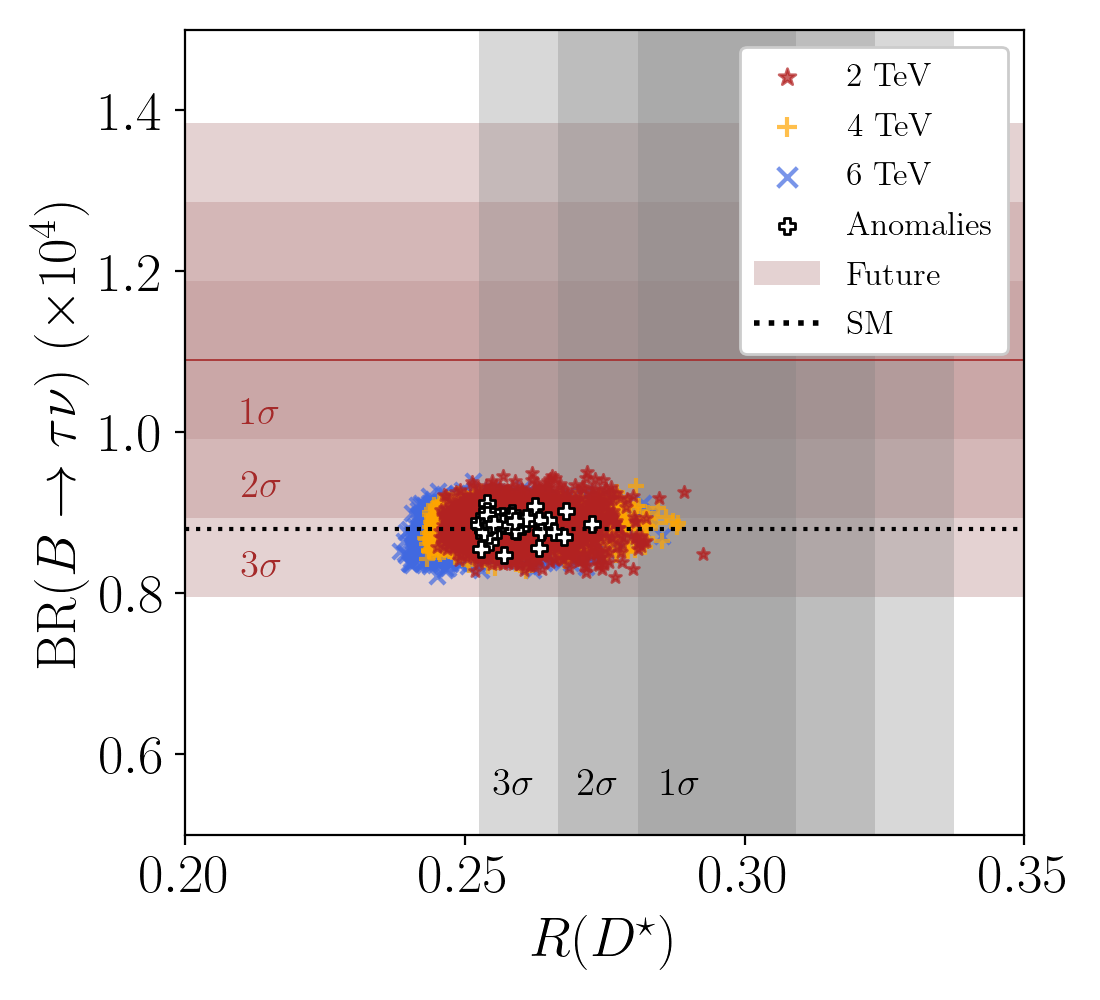}
    \caption{\linespread{1.1}\small{\textbf{Additional plots for secondary observables in comprehensive scan.} In the left plot, for $g_{\mu_A}/g_A^\text{SM}$ the red-brown shaded regions represent the projected sensitivities from the ILC~\cite{Baer:2013cma}, assuming the present best-fit value~\cite{ALEPH:2005ab,Crivellin:2020mjs} shown as red-brown solid line. Furthermore, we show as future constraint on the EDM of the muon the one expected from the muEDM experiment~\cite{Adelmann:2021udj} (as example for the frozen-spin technique). In the right plot, we present the projected sensitivity from Belle II for 5 ab$^{-1}$ for BR($B\to \tau\nu$)~\cite{Forti:2022mti} as red-brown shaded regions about the current best-fit value~\cite{ParticleDataGroup:2020ssz}. For further information on how to read this figure, see section~\ref{ssec:comprehensivePlots}.  }}
    \label{fig:CompApp_Sec}
\end{figure}

In this appendix, we present some supplementary plots showing the distributions of different primary and secondary observables in the comprehensive scan. Figure~\ref{fig:CompResultsCLFV} illustrates the correlation 
between BR($\tau\to \mu \gamma$) and the three different flavour anomalies as well as between BR($\mu\to e \gamma$) and the AMM of the muon. These plots should be compared with the corresponding ones, obtained in the primary scan, see figure~\ref{fig:constrain_power_taumug} in the main text. Observe the effects of biasing in refining the sampled parameter space. Other features and discussion of these observables can be found in section~\ref{sec:Comprehensive_leptonic}. In figure~\ref{fig:CompApp_Sec} we show two additional plots for secondary observables, which complement the discussion in section~\ref{subsec:resultsSecondary}.

\subsection{Tertiary observables}
\label{app:tertiary}

Table~\ref{tab:app_tercal} details the present experimental constraints on the tertiary observables and the calculation method employed for each observable in the comprehensive scan. Table~\ref{tab:app_ter} displays
  a summary of the results for the tertiary observables, mentioning the range for each of them obtained for the sample of $P$ points passing the primary constraints as well as listing the future reach for these observables.

\begin{table}[bht!]
\centering
\resizebox{\textwidth}{!}{
\renewcommand{\arraystretch}{1.3}
    \centering
    \begin{tabular}{|l|l|l||l|l|l|}
    \hline
{Observable} &{Present constraint} & {Calculation method} &{Observable} &{Present constraint} & {Calculation method} \\
        \hline 
            BR($B_s\to \tau\tau$) & $6.8\times 10^{-3}$ & {~\ref{app:bsee} \& \texttt{flavio}}& $\Delta M_{B_s}/\Delta M_{B_s}^{\rm SM}$ & $1.11\pm 0.09$~\cite{Bevan:2014cya} &  \texttt{SPheno}\\
                      BR$(D_s \to \tau \nu)$ & $(5.32\pm 0.11)\times 10^{-2}$ & \texttt{SPheno} &  BR$(D_s \to \mu \nu)$ & $(5.43\pm 0.15)\times 10^{-3}$ & \texttt{SPheno} \\
                      BR$(K^+ \to\pi^+ \nu \nu)$  & $(1.7\pm 1.1)\times 10^{-10}$ & \texttt{flavio} &BR$(K_L \to\pi^0 \nu \nu)$ & $2.6\times 10^{-8}$ &  \texttt{flavio} \\
                $\Delta a_e$ &  - \;\;$\dagger$  & \texttt{SPheno} & $\Delta a_\tau$ & $\lesssim \mathcal{O}(0.01)$ & \texttt{SPheno}\\
         $|d_e|$ [$e$ cm] & $<1.1\times 10^{-29}$ & \texttt{SPheno} & $|d_\tau|$  [$e$ cm] &  $\lesssim \mathcal{O}(10^{-16})$ & \texttt{SPheno}\\
                  BR($B\to X_s \gamma$) &  $(3.32\pm 0.15)\times 10^{-4}$ & \texttt{SPheno} & BR($\tau\to e\gamma$) & $3.3\times 10^{-8}$ &  \texttt{SPheno}\\
          BR($\tau\to 3 \, e$) & $2.7\times 10^{-8}$ & \texttt{SPheno}&BR$(\tau \to \bar{\mu}\mu e)$ & $2.7\times 10^{-8}$  & \texttt{SPheno}\\
                 BR$(\tau \to \bar{e} \mu \mu)$ & $1.7\times 10^{-8}$ & \texttt{SPheno} & BR$(\tau \to \bar{\mu} e e)$ & $1.5\times 10^{-8}$ &  \texttt{SPheno}\\
               BR($\tau \to \pi e$) & $8.8\times 10^{-8}$ & \texttt{SPheno} & BR($\tau \to\pi \mu$)& $1.1\times 10^{-7}$ &~\cite{Mandal:2019gff} \\
                BR($\tau \to\phi e$)& $3.1\times 10^{-8}$& \texttt{SPheno} &BR($\tau \to \rho e$) & $1.8\times 10^{-8}$ & \texttt{SPheno}\\
                BR($\tau \to\phi \mu$) & $8.4\times 10^{-8}$ & \texttt{SPheno} & BR($\tau \to \rho \mu$) & $1.2\times 10^{-8}$& \texttt{SPheno}\\
                $g_{e_{A}}/g_A^{\rm SM} -1$&  $(-3.19\pm 6.98)\times 10^{-4}$\cite{Crivellin:2020mjs} &~\ref{app:Zdecays} &&&\\
                \hline 
    \end{tabular}}
 \caption{{\small\textbf{List of tertiary observables and their present experimental bounds and calculation method in comprehensive scan.}  Constraints quoted without explicit reference are taken from reference~\cite{ParticleDataGroup:2020ssz}. $\dagger$: presently, anomalies in $\Delta a_e$ indicate a preference for $|\Delta a_e| \sim 10^{-12}$~\cite{Parker2018,Morel:2020dww}. However, as the status of these anomalies is unresolved (two separate measurements show deviations from the SM with opposite sign), we take this value to be a future reach rather than a present constraint.}}  \label{tab:app_tercal}
\end{table}

\begin{table}[t!]
\resizebox{16cm}{!}{
\renewcommand{\arraystretch}{1.6}
    \centering
    \begin{tabular}{|r|r|r|r|r|}
    \hline
    \multicolumn{5}{|c|}{ \large \textsc{Spread of tertiary observables in comprehensive scan}} \\
    \hline
\multirow{2}{*}{Observable} & \multirow{2}{*}{Future reach} & $\hat{m}_\phi=2$, $P=5955$ &  $\hat{m}_\phi=4$, $P=12570$ &  $\hat{m}_\phi=6$, $P=39807$ \\
& &   [min., max.] & [min., max.] &   [min., max.]  \\
       \hline 
 BR($B_s\to \tau\tau$) &  $\sim 10^{-6}$ \cite{Grossman:2021xfq}
 	     &$[7.21, 8.49]\times 10^{-7}$ 
             &$[6.22, 9.36]\times 10^{-7}$  
             &$[6.02, 9.61] \times 10^{-7}$\\      
$\Delta M_{B_s}/\Delta M_{B_s}^{\rm SM}$  & -
	&$[0.96, 1.05]$ 
	&$[0.85, 1.11]$
	& $[0.66, 1.36]$\\
 BR$(D_s \to \mu \nu)$ & $(5.49\pm 0.05)\times 10^{-3}$ \cite{Belle-II:2018jsg}
   & $[5.45, 5.46]\times 10^{-3}$
 &$[5.45,5.46]\times 10^{-3}$
 &$[5.45,5.46]\times 10^{-3}$
\\
 BR$(D_s \to \tau \nu)$ & $(5.48\pm 0.12)\times 10^{-2}$ \cite{Belle-II:2018jsg}
 & $[5.28, 5.36]\times 10^{-2}$
 &$[5.30, 5.34]\times 10^{-2}$
 &$[5.30, 5.33]\times 10^{-2}$\\
  $ \frac{\text{BR}(K_L\to\pi^0 \nu \nu)}{\text{BR}(K_L\to\pi^0 \nu \nu)_\text{SM}}$ &$1\pm 0.09$\cite{Goudzovski:2022vbt}&$ [0.99, 1.02] $&$ [0.99, 1.02] $ &$ [0.99, 1.02] $  \\
$\frac{\text{BR}(K^+\to\pi^+ \nu \nu)}{\text{BR}(K^+\to\pi^+ \nu \nu)_\text{SM}}$ &- & $ [0.96, 1.03] $& $ [0.98, 1.01] $& $ [0.99, 1.01] $ \\
BR($B\to X_s \gamma$) & - & $[3.290, 3.292] \times 10^{-4}$ & $[3.290, 3.292] \times 10^{-4}$& $[3.290, 3.292] \times 10^{-4}$\\
 $1-g_{e_{A}}/g_A^{\rm SM} $& $(3.19 \pm 0.041)\times 10^{-4}$ \cite{Crivellin:2020mjs}
 & $[-1.53\times 10^{-13}, 1.55\times 10^{-15}]$
 &$[-8.46\times 10^{-14}, 7.77\times 10^{-16}]$
 &$[-4.62\times 10^{-14}, 7.77\times 10^{-16}]$\\
$\Delta a_e$ & $\mathcal{O}(10^{-12})$ \cite{Parker2018,Morel:2020dww}
 & $[-5.39, 7.33]\times 10^{-21}$
 &$[-3.46, 3.25]\times 10^{-21}$
 &$[-1.68, 2.10]\times 10^{-21}$\\
 $\Delta a_\tau$ & $< \mathcal{O}(10^{-3})$ \cite{Eidelman:2007sb, Kraetzschmar:2021rly}
 & $[-1.28, 1.82]\times 10^{-7}$
 &$[-6.50, 9.87]\times 10^{-8}$
 &$[-7.29, 8.76]\times 10^{-8}$\\
  $|d_e|$ [$e$ cm] &  $\lesssim 5\times 10^{-30}$ \cite{aggarwal2018measuring}
 & $[3.66\times 10^{-36}, 1.18\times 10^{-31}]$
 & $[1.81\times 10^{-37}, 6.28\times 10^{-32}]$
 &$[1.31\times 10^{-38}, 3.30\times 10^{-32}]$\\
   $|d_\tau|$ [$e$ cm] & $\lesssim 10^{-19}$ \cite{Bernreuther:2021elu}
 & $[4.30\times 10^{-26}, 1.12\times 10^{-21}]$
 &$[3.25\times 10^{-27}, 5.10\times 10^{-22}]$
 &$[2.77\times 10^{-27}, 5.70\times 10^{-22}]$ \\
   BR($\tau\to e \gamma$) &
   $< 9\times 10^{-9}$~\cite{Banerjee:2022xuw}
 & $[9.73\times 10^{-19}, 3.31\times 10^{-14}]$
 &$[2.08\times 10^{-19}, 6.24\times 10^{-15}]$
 &$[8.44\times 10^{-20},3.33\times 10^{-14}]$\\
    BR($\tau\to 3 \, e $) &
    $< 4.7\times 10^{-10}$~\cite{Banerjee:2022xuw}
 & $[3.51\times 10^{-20}, 4.47\times 10^{-16}]$
 & $[1.03\times 10^{-20}, 7.66\times 10^{-17}]$
 &$[4.74\times 10^{-21},4.01\times 10^{-16}]$ \\
     BR($\tau\to \bar{\mu} \mu e $) & 
     $< 4.5\times 10^{-10}$~\cite{Banerjee:2022xuw}
 & $[7.63\times 10^{-21}, 1.10\times 10^{-16}]$
 & $[3.34\times 10^{-21}, 2.15\times 10^{-17}]$
 &$[1.38\times 10^{-21}, 7.79\times 10^{-17}]$\\
      BR($\tau\to \bar{e} \mu \mu $) &
      $< 2.6\times 10^{-10}$~\cite{Banerjee:2022xuw}
 & $[1.74\times 10^{-30}, 2.71\times 10^{-21}]$
 & $[1.96\times 10^{-28}, 4.46\times 10^{-21}]$
 & $[8.11\times 10^{-29}, 3.20\times 10^{-21}]$\\
      BR($\tau\to \bar{\mu} e e $) & 
      $< 2.3\times 10^{-10}$~\cite{Banerjee:2022xuw}
 & $[2.45\times 10^{-39}, 1.01\times 10^{-27}]$
 &$[8.78\times 10^{-39}, 9.98\times 10^{-28}]$
 &$[5.32\times 10^{-40}, 2.47\times 10^{-28}]$\\
       BR($\tau\to \pi e $) & 
       $< 7.3\times 10^{-10}$~\cite{Banerjee:2022xuw}
 & $[1.71\times 10^{-23}, 5.11\times 10^{-19}]$
 &$[1.08\times 10^{-24}, 1.00\times 10^{-19}]$
 &$[5.55\times 10^{-25}, 8.78\times 10^{-20}]$\\
 BR($\tau\to \pi \mu $) & $<7.1\times 10^{-10}$~\cite{Banerjee:2022xuw} & $[4.70\times 10^{-19}, 1.09\times 10^{-12}]$& $[2.87\times 10^{-20}, 2.29\times 10^{-13}]$& $[4.83\times 10^{-21}, 7.77\times 10^{-14}]$ \\
         BR($\tau\to \phi e $) & 
         $< 7.4\times 10^{-10}$~\cite{Banerjee:2022xuw}
 & $[1.26\times 10^{-22}, 1.08\times 10^{-17}]$
 & $[5.49\times 10^{-24}, 3.80\times 10^{-18}]$
 & $[4.63\times 10^{-25}, 1.07\times 10^{-18}]$\\
          BR($\tau\to \rho e $) & 
          $< 3.8\times 10^{-10}$~\cite{Banerjee:2022xuw}
 & $[1.41\times 10^{-21}, 5.13\times 10^{-17}]$
 & $[1.57\times 10^{-22}, 1.74\times 10^{-17}]$
 & $[1.33\times 10^{-23}, 4.94\times 10^{-18}]$\\
          BR($\tau\to \phi \mu $) & 
          $< 8.4\times 10^{-10}$~\cite{Banerjee:2022xuw}
 & $[1.28\times 10^{-14}, 3.52\times 10^{-11}]$
 &$[6.77\times 10^{-16}, 1.58\times 10^{-11}]$
 &$[1.38\times 10^{-16}, 4.97\times 10^{-12}]$\\
          BR($\tau\to \rho \mu $) & 
          $< 5.5\times 10^{-10}$~\cite{Banerjee:2022xuw}
 & $[6.55\times 10^{-14}, 2.61\times 10^{-10}]$
 & $[4.67\times 10^{-15}, 1.14\times 10^{-10}]$
  & $[8.44\times 10^{-16}, 3.81\times 10^{-11}]$\\
             \hline 
    \end{tabular}}
    \caption{\small{\textbf{Overview of spread of tertiary observables in comprehensive scan.} We present a summary of the statistics reflecting the distribution of tertiary observables: the minimum and maximum generated  value for a sample of $P$ points passing the primary constraints together with the future reach. For values quoted from Belle II~\cite{Banerjee:2022xuw}, these represent projections for 50 ab$^{-1}$ of data. Prospective reach intervals illustrate the assumption that the best-fit value of the measurement stays the same. The future projection for $\text{BR}(K_L\to\pi^0 \nu \nu)$ is interpreted assuming a future measurement centred at the SM prediction. Note that we quote an extended number of decimal points for  BR($B\to X_s \gamma$) to show that a variation occurs at the third decimal point. A dash~(-) is used to indicate where a future reach could not be identified in the literature. }} \label{tab:app_ter}
\end{table}

%%%%%
\clearpage

\bibliography{refs}
\bibliographystyle{JHEP}

\end{document}